\documentclass[12pt,a4paper]{book}
\usepackage{gensymb} 
\usepackage[utf8]{inputenc}
\usepackage[T1]{fontenc}
\usepackage[default, scale=.95]{opensans} 
\usepackage{amsmath}
\usepackage{amsfonts}
\usepackage{fancyhdr}
\usepackage{bm}
\usepackage{mathtools}
\usepackage{amssymb}
\usepackage{xcolor} 
\definecolor{Prune}{RGB}{99,0,60} 
\definecolor{B1}{RGB}{49,62,72} 
\definecolor{C1}{RGB}{124,135,143}
\definecolor{D1}{RGB}{213,218,223}
\definecolor{A2}{RGB}{198,11,70}
\definecolor{B2}{RGB}{237,20,91}
\definecolor{C2}{RGB}{238,52,35}
\definecolor{D2}{RGB}{243,115,32}
\definecolor{A3}{RGB}{124,42,144}
\definecolor{B3}{RGB}{125,106,175}
\definecolor{C3}{RGB}{198,103,29}
\definecolor{D3}{RGB}{254,188,24}
\definecolor{A4}{RGB}{0,78,125}
\definecolor{B4}{RGB}{14,135,201}
\definecolor{C4}{RGB}{0,148,181}
\definecolor{D4}{RGB}{70,195,210}
\definecolor{A5}{RGB}{0,128,122}
\definecolor{B5}{RGB}{64,183,105}
\definecolor{C5}{RGB}{140,198,62}
\definecolor{D5}{RGB}{213,223,61}
\usepackage{mdframed}
\usepackage{multirow} 
\usepackage{multicol} 
\usepackage{scrextend} 
\usepackage{tikz}
\usepackage{graphicx}
\usepackage[absolute]{textpos} 
\usepackage{colortbl}
\usepackage{array}
\usepackage{geometry}
\usepackage{titlesec}
\usepackage[citecolor=blue]{hyperref}%
\usepackage{cite}
\usepackage{caption}
\hypersetup{ 
	colorlinks=true,
	linkcolor=black,
	urlcolor=purple}

\newcommand{\bea}{\begin{eqnarray}}
\newcommand{\eea}{\end{eqnarray}}
\newcommand{\be}{\begin{equation}}
\newcommand{\ee}{\end{equation}}

\begin{document}
	
	\begin{titlepage}

		\newgeometry{left=6cm,bottom=2cm, top=1cm, right=1cm}
		
		\tikz[remember picture,overlay] \node[opacity=1,inner sep=0pt] at (-13mm,-135mm){\includegraphics{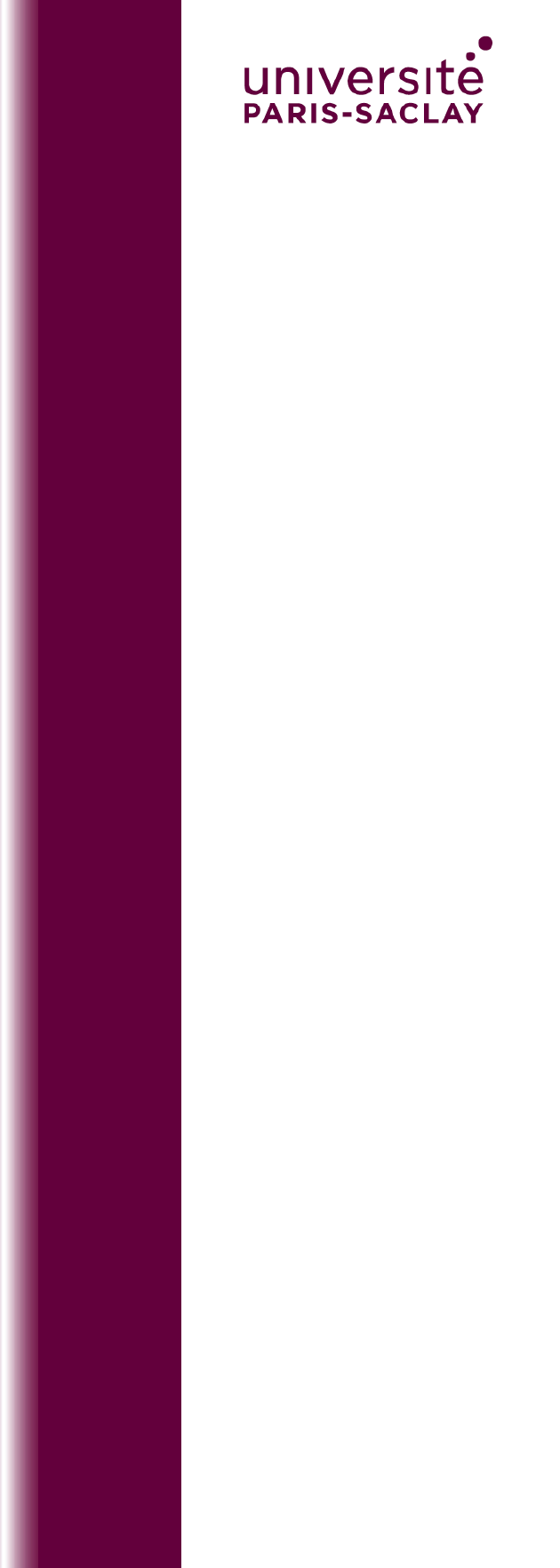}};
		
		
		\color{white}
		
		\begin{picture}(0,0)
			\put(-152,-743){\rotatebox{90}{\Large \textsc{THESE DE DOCTORAT}}} \\
			\put(-120,-743){\rotatebox{90}{NNT :  2023UPASP127}}
		\end{picture}
		
		
		
		\flushright
		\vspace{10mm} 
		\color{Prune}
		\fontfamily{cmss}\fontseries{m}\fontsize{22}{26}\selectfont
		\Huge Extreme value statistics and optimization problems in stochastic processes \\
		
		\normalsize
		\color{black}
		\Large{\textit{Statistiques d'extrêmes et problèmes d'optimisation de processus stochastiques}} \\
		
		\fontfamily{fvs}\fontseries{m}\fontsize{8}{12}\selectfont
		
		\vspace{1.5cm}
		
		\normalsize
		\textbf{Thèse de doctorat de l'université Paris-Saclay} \\
		
		\vspace{6mm}
		
		\small École doctorale n$^{\circ}$ 564, Physique en Ile-de-France (PIF)\\
		\small Spécialité de doctorat: Physique\\
		\small Graduate School : Physique. Référent : Faculté des Sciences d'Orsay \\
		\vspace{6mm}
		
		\footnotesize Thèse préparée dans l'unité de recherche {\bf LPTMS (Université Paris-Saclay, CNRS)}, sous la direction de {\bf Grégory SCHEHR}, Directeur de Recherche,\\
   et la co-direction de {\bf Satya N. MAJUMDAR}, Directeur de Recherche.\\
		\vspace{15mm}
		
		\textbf{Thèse soutenue à Paris-Saclay, le 16 octobre 2023, par}\\
		\bigskip
		\Large {\color{Prune} \textbf{Benjamin DE BRUYNE}} 
		
		\vspace{3em} 

		\flushleft
		\small {\color{Prune} \textbf{Composition du jury}}\\
{\color{Prune} \scriptsize {Membres du jury avec voix délibérative}} \\
		\vspace{2mm}
		\scriptsize
		\begin{tabular}{|p{7cm}l}
			\arrayrulecolor{Prune}
            \textbf{Cécile MONTHUS} & Présidente  \\ 
			Directrice de Recherche, Université
Paris-Saclay  &   \\ 
   			\textbf{Eric BERTIN} &  Rapporteur \& Examinateur \\ 
			Directeur de Recherche, Université Grenoble Alpes  &   \\
			\textbf{Pierpaolo VIVO} &  Rapporteur \& Examinateur  \\ 
			Professeur, King’s College London &   \\ 
   			\textbf{Martin R. EVANS} &  Examinateur \\ 
			Professeur, University of Edinburgh  &   \\ 
   			\textbf{Kilian RASCHEL} &  Examinateur \\ 
			Directeur de Recherche, Université d'Angers    &   \\
			
		\end{tabular} 
		
	\end{titlepage}

\ifthispageodd{\newpage\thispagestyle{empty}\null\newpage}{}
\thispagestyle{empty}
\newgeometry{top=1.5cm, bottom=0cm, left=2cm, right=2cm}
\fontfamily{rm}\selectfont

\lhead{}
\rhead{}
\rfoot{}
\cfoot{}
\lfoot{}

\noindent 
\includegraphics[height=2.45cm]{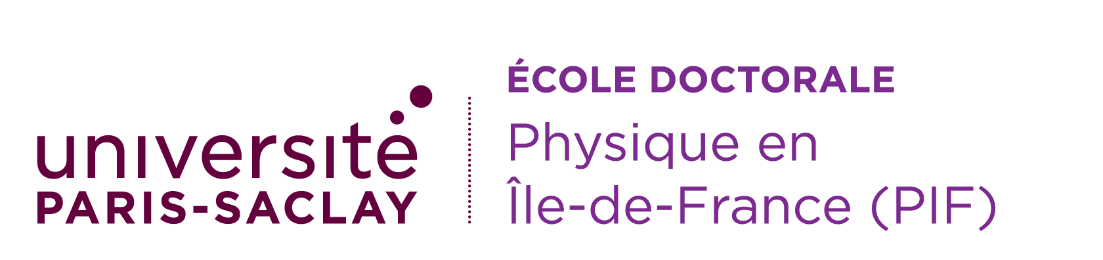}
\vspace{1cm}
\fontfamily{cmss}\fontseries{m}\selectfont

\small

\begin{mdframed}[linecolor=Prune,linewidth=1]

	\textbf{Titre:} Statistiques d'extrêmes et problèmes d'optimisation de processus stochastiques
	
	\noindent \textbf{Mots clés:} Processus stochastiques, statistiques d'extrêmes, mouvement Brownien
	
	\vspace{-.5cm}
	\begin{multicols}{2}
		\noindent \textbf{Résumé: Cette thèse est consacrée à l'étude des statistiques des valeurs extrêmes dans les processus stochastiques et de leurs applications.\\ Dans la première partie, nous obtenons des résultats analytiques exacts sur les statistiques des valeurs extrêmes des marches aléatoires en temps discret et en temps continu. En particulier, nous nous concentrons sur les statistiques de gaps des marches aléatoires et révélons leur universalité asymptotique par rapport à la distribution des sauts dans la limite d'un grand nombre de pas. De plus, nous calculons le comportement asymptotique de la valeur moyenne du maximum de marches aléatoires en présence d'une contrainte de pont et révélons un comportement riche dans leur correction de taille finie. De plus, nous calculons la longueur moyenne de l'enveloppe convexe du mouvement Brownien confiné dans un disque et montrons qu'elle converge lentement vers le périmètre du disque avec une décroissance exponentielle étirée. \\ Dans la deuxième partie, nous nous intéressons à l'échantillonnage numérique des trajectoires rares des processus stochastiques. Nous introduisons une méthode efficace pour échantillonner des marches aléatoires en temps discret. Nous l'illustrons et l'appliquons sur divers exemples. Nous étendons ensuite la méthode à d'autres processus stochastiques, à la fois markoviens et non markoviens. Nous appliquons notre méthode pour échantillonner des particules survivantes en présence d'un environnement de piégeage périodique.\\ Enfin, nous discutons plusieurs problèmes d'optimisation dans les processus stochastiques impliquant des statistiques de valeurs extrêmes. En particulier, nous introduisons une nouvelle technique pour contrôler de manière optimale les systèmes dynamiques soumis à une politique de réinitialisation.}

	\end{multicols}
	
\end{mdframed}

\vspace{2mm}

\begin{mdframed}[linecolor=Prune,linewidth=1]
	
	\textbf{Title:} Extreme value statistics and optimization problems in stochastic processes
	
	
	\begin{multicols}{2}
		\noindent \textbf{Abstract: This thesis is devoted to the study of extreme value statistics in stochastic processes and their applications.\\ In the first part, we obtain exact analytical results on the extreme value statistics of both discrete-time and continuous-time random walks. In particular, we focus on the gap statistics of random walks and exhibit their asymptotic universality with respect to the jump distribution in the limit of a large number of steps. In addition, we  compute the asymptotic behavior of the expected maximum of random walks in the presence of a bridge constraint and reveal a rich behavior in their finite-size correction. Moreover, we compute the expected length of the convex hull of Brownian motion confined in a disk and show that it converges slowly to the perimeter of the disk with a stretched exponential decay. \\ In the second part, we focus on numerically sampling rare trajectories of stochastic processes. We introduce an efficient method to sample bridge discrete-time random walks. We illustrate it and apply it to various examples. We further extend the method to other stochastic processes, both Markovian and non-Markovian. We apply our method to sample surviving particles in the presence of a periodic trapping environment.\\ Finally, we discuss several optimization problems in stochastic processes involving extreme value statistics. In particular, we introduce a new technique to optimally control dynamical systems undergoing a resetting policy. }

	\end{multicols}
\end{mdframed}

\titleformat{\chapter}[hang]{\bfseries\Large\color{Prune}}{\thechapter\ -}{.1ex}
{\vspace{0.1ex}
}
[\vspace{1ex}]
\titlespacing{\chapter}{0pc}{0ex}{0.5pc}

\titleformat{\section}[hang]{\bfseries\normalsize}{\thesection\ .}{0.5pt}
{\vspace{0.1ex}
}
[\vspace{0.1ex}]
\titlespacing{\section}{1.5pc}{4ex plus .1ex minus .2ex}{.8pc}

\titleformat{\subsection}[hang]{\bfseries\small}{\thesubsection\ .}{1pt}
{\vspace{0.1ex}
}
[\vspace{0.1ex}]
\titlespacing{\subsection}{3pc}{2ex plus .1ex minus .2ex}{.1pc}

\newgeometry{top=4cm, bottom=4cm, left=2cm, right=2cm}

\tableofcontents

\frontmatter

\chapter*{Acknowledgments}
\label{chap:ack}
\addcontentsline{toc}{chapter}{\nameref{chap:ack}}

First, I express my profound gratitude to Grégory Schehr who supervised this thesis with great care and invaluable advice. Along with Satya Majumdar, they have spent a generous amount of time guiding me. Time flew by so fast along their side, both in and out of the office. I also thank their collaborators and in particular Pierre Le Doussal for sharing with me his drive for research.

I am indebted to Pierpaolo Vivo and Eric Bertin for accepting to review this thesis. I extend my sincere thanks to Martin Evans, Kilian Raschel, and Cécile Monthus for being part of the jury of this thesis.

I thank all the members of the LPTMS for their hospitality and kindness. I warmly thank Claudine Le Vaou and Karolina Kolodziej for their valuable administrative support, and the lab directors Alberto Rosso and Emmanuel Trizac. I address special thanks to Francesco Mori, with whom I shared the rewarding pleasure of working together. LPTMS has a pleasant atmosphere which is in particular sustained by the permanents and by the students Alessandro, Charbel, Federico, Félix, Flavio, Gabriel, Jules, Lara, Lenart, Li, Lorenzo, Louis, Luca, Lucas, Marco, Mathieu, Matteo, Mert, Pierre. Special thanks to Lenart for hosting countless pool parties with Fabian, Marco, Mauro, Miha, Saptarshi, Saverio, Valerio, Vanja, and Vincenzo. I am also grateful to Ana with whom I shared more than an office. 

I thank the members of my ``comité de suivi'', Christophe Texier, Frédéric Van Wijland, and in particular Olivier Bénichou, who has become a collaborator. I acknowledge the LPTHE for hosting me and Andriani, Léo, Mathis, Max, Réa, and Yann for sharing the office. Special thanks to Andrei and Francesco for the pleasant coffee breaks. I also thank friends from the neighboring labs, Aurélien, Jérémie, Léo, Ludwig, Marc, and Gabriel. I am grateful to my collaborators David Dean, Henri Orland, Julien Randon-Furling, with whom it was a pleasure to work, and in particular Gaia Pozzoli, who I was fortunate to meet and work with her after a summer school. I am also grateful to Pierpaolo Vivo for inviting me to King's College and thank him and Bertrand for the pleasant time there. I have a special thought for Jean-Michel Gillet and Sid Redner who initiated me to research.

I thank my friends who I met in Paris, Andrea, Enrico, Juan, Lorena, Lorenzo, Michelle, Pati, and Tommaso, for encouraging me to pursue my direction. I thank El Mehdi and Théo for the pleasant math workshops. I thank friends from the CDI, Alexandre, Aurélien, Erwan, Maxime, Mazzarine, and Sophie. I extend my thanks to friends from Belgium, Adrien, Charlotte, Donovan, Emelyne, Gilles, Lucas, Marianne, Pieter, Sophie, and Vincent for their unconditional support. Special thanks to Emma and Xavier who hosted me countless times in an always kind atmosphere. I extend my thanks to friends who I met in Canada, Alex, Anne-Catherine, Beata, Elizabeth, Haoxing, Jérémy, Julia, Maxence, Rémi, and Tim who I hope will understand some parts of this thesis. 

I thank my family for their unfaltering love and support, in particular my parents who advised me to follow my path and my sister who regularly challenged me. I thank Simon, Antoine, Lou Anne, Jean-Marc, France, Igne, Jan, Christophe, Lisa, Margo, Michael for the pleasant family times in Belgium. I have a particular thought for my grandparents, Jacques, Juliette, and Walther. I extend my thanks to my host families and friends in New Zealand, who are far but close to my heart, Hamish, Chris, John, Rhonda, Michael, Michelle, John, Sharon, Merv, Jill, and Juliette. Finally, I thank Fanny from the bottom of my heart. She has filled the last year with love and compassion. This thesis is dedicated to her.\\

\chapter*{List of publications}
\label{chap:pub}
\addcontentsline{toc}{chapter}{\nameref{chap:pub}}

\begin{enumerate}
  \item \textbf{Optimization and growth in first-passage resetting},  B. De Bruyne, J. Randon-Furling and S. Redner, 
J.~Stat.~Mech., 013203 (2020).

\item \textbf{Survival probability of a run-and-tumble particle in the presence of a drift}, B. De Bruyne, S. N. Majumdar and G. Schehr, 
J.~Stat.~Mech., 043211 (2021).

\item \textbf{Wigner function for noninteracting fermions in hard wall potentials}, B. De Bruyne, D. S. Dean, P. Le Doussal, S. N. Majumdar and G. Schehr, Phys.~Rev.~A {\bf 104},  013314 (2021).

\item \textbf{Generating discrete-time constrained random walks and Lévy flights}, B. De Bruyne, S. N. Majumdar and G. Schehr,
Phys.~Rev.~E {\bf 104},  024117 (2021).

\item \textbf{Expected maximum of bridge random walks \& Lévy
flights}, B. De Bruyne, S. N.. Majumdar and G. Schehr,
J.~Stat.~Mech., 083215 (2021).

\item \textbf{Generating constrained run-and-tumble trajectories}, B. De Bruyne, S. N. Majumdar and G. Schehr,
J.~Phys.~A:~Math.~Theor. {\bf 54}, 385004 (2021).

\item \textbf{A Tale of Two (and More) Altruists}, B. De Bruyne, J. Randon-Furling and S. Redner,
J.~Stat.~Mech.,  103405 (2021).

\item \textbf{Survival probability of random walks leaping over
traps}, G. Pozzoli and B. De Bruyne,
J.~Stat.~Mech.,  123203 (2021).

\item \textbf{Generating stochastic trajectories with global dynamical constraints}, B. De Bruyne, S. N. Majumdar, H. Orland and G. Schehr,
J.~Stat.~Mech.,  123204 (2021).

\item \textbf{Statistics of the maximum and the convex hull of a Brownian motion in confined geometries}, B. De Bruyne, O. Bénichou, S. N. Majumdar and G. Schehr,
J.~Phys.~A:~Math.~Theor. {\bf 55}, 144002 (2021).

\item \textbf{Resetting in Stochastic Optimal Control}, B. De Bruyne and F. Mori,
Phys.~Rev.~Res. {\bf 5}, 013122 (2023).

\item \textbf{Optimal Resetting Brownian Bridges}, B. De Bruyne, S. N. Majumdar and G. Schehr,
Phys.~Rev.~Lett. {\bf 128}, 200603 (2022).

\item \textbf{First-Passage-Driven Boundary Recession}, B. De Bruyne, J. Randon-Furling and S. Redner, \\J.~Phys.~A:~Math.~Theor. {\bf 55}, 354002 (2022).

\item \textbf{Universal order statistics for random walks \& Lévy flights}, B. De Bruyne, S. N. Majumdar and G. Schehr,
J. Stat. Phys. {\bf 190}, 320 (2023).

\item \textbf{Transport properties of diffusive particles conditioned to survive in trapping environments}, G. Pozzoli and B. De Bruyne,
J. Stat. Mech., 113205 (2022).
\end{enumerate}

\newgeometry{top=4cm, bottom=4cm, left=3cm, right=3cm}
\mainmatter

\makeatletter\@openrightfalse
\chapter{Introduction and presentation of the main results}
\label{chap:int}
\section{Introduction}
One of the remarkable facts in Physics is that one can attempt to find simple laws to describe natural phenomena at a given scale without knowing the microscopic laws governing its constituents at smaller scales. This allows us to advance our understanding of systems on a macroscopic scale, and to achieve technological breakthroughs in society, without having a complete understanding of the fundamental laws that govern the world in which we live. This beautiful concept is one of the main reasons why many laws in Physics work remarkably well in describing what we observe. The universal nature of those laws is rather appealing and is one of the reasons why Physics is worth studying. 

The reasons underlying the concept of universality are rather deep and have fascinated the Physics community for a long time. The seminal work of Kadanoff in 1966, and further pursued by Wilson in 1971, shed some light on this phenomenon and explained it as a consequence of an extensively large number of elements interacting together  \cite{KLP66,KGW71}. Their discovery constituted a major step forward in our understanding of the phases of matter and transitions between them, such as when water boils and turns into vapor. Since then, this emerging universality has been widely observed in a variety of complex systems and constitutes a cornerstone of modern statistical physics.  Anderson sums it up very well in his seminal paper entitled ``More is different'' \cite{PWA72}, where he shows that the behavior of large systems cannot be simply extrapolated by the properties of its constituents, and that entirely new properties appear at  each level of complexity. He particularly highlights the importance of symmetry in the laws of nature.

Large and complex systems appear in a wide variety of fields in natural sciences as well as applied sciences. Although intrinsically different from each other, they share the common pattern of being driven by many interacting parts with a high number of degrees of freedom. Despite being deterministic systems on the microscopic scale, i.e. at the level of the interacting parts, they effectively behave stochastically at the coarse-grained level. Extracting relevant information on the macroscopic behavior of such systems is one of the rather challenging tasks to which the field of Statistical Physics is devoted. While the field was originally focused on systems originating from Physics, it has now become an interdisciplinary field with applications ranging from biology to finance.  Paradigmatic examples of complex systems are the financial markets which have become an active field of research in the community \cite{BJP18}. 

While complex systems usually behave in a typical way, they sometimes display atypical behaviors which can yield to extreme events such as earthquakes, extreme floods, and large wildfires. These events, which are ubiquitous in nature, can have devastating consequences, which makes them worth studying. Some natural questions about them are: (a) what is the magnitude of the largest event?~(b) when does it occur?~(c) is it isolated, or are there other similar events? The field of extreme value statistics (EVS) is devoted to the study of such questions and has found a wide variety of applications ranging from environmental sciences \cite{gumbel,katz} to finance \cite{embrecht,bouchaud_satya}. EVS also plays a key role in physics, especially in the description of disordered systems \cite{bm97,PLDCecile,Dahmen,sg,DerrA,DerrB}, fluctuating interfaces \cite{Shapir,GHPZ,Majumdar04Flu1,Satya_Airy2,SOS_Airy}, and random matrices \cite{TW,SMS14} (for a recent review see \cite{SMS14r,Vivo15,reviewMPS}). On the practical side, extreme events are related to situations of serious hazard for which it is important to gauge the risks, such as in the construction industry, the energy sector, agriculture, territorial planning, logistics, and the financial markets \cite{Lucarini16}. On prominent example is the case of weather and climate extremes, whose number of occurrences has increased in the recent years \cite{Coumou12, IPCC12,IPCC14}. Sometimes, these extremes arise as a result of strongly correlated events, such as when intense precipitations occur at the same time within the same river basin and cause dangerous floods \cite{Davison12,Cooley12}.
On the theoretical side, the EVS of independent and identically distributed (i.i.d.)~random variables have been thoroughly investigated \cite{Arnold,Nagaraja}, but much less is known about the EVS of strongly correlated random variables, which often appear in practical contexts. Several specific models of correlated random variables have been studied and have shown to exhibit very rich behaviors as well as universal features \cite{Feller,dean_majumdar,PLDCecile,pld_carpentier,Satya_Airy2,gyorgyi,satya_yor,comtet_precise,schehr_rsrg,SM12,MMS13,MMS14,BM17,Lacroix,BSG21a,BSG21b,PT20,PT21,Mori1,Mori2,Mori3}.
In particular, it was shown that one-dimensional discrete-time random walks constitute a very useful playground to investigate EVS of strongly correlated random variables \cite{SM12,MMS13,MMS17,MMS14,BM17,Lacroix,PT20,PT21,Feller,BSG23,Erdos46,Darling56,VVI94,Pollackzek,Wendel,Spitzer57}.

When it is too hard to obtain analytical results on EVS, a natural way to proceed is to study them numerically. However, this is not an easy task as their occurrences are rare by definition. A natural question that arises then is: ``How do we efficiently sample rare events?''. In general, rare trajectories are important as they capture specific information about the system that cannot be seen in the typical trajectories where observables concentrate around their mean. For instance, in the context of glasses, rare trajectories are key to understanding the slow structural relaxation dynamics close to the glass transition where fluctuations are important \cite{Gar2018}. Numerical methods to sample them are of primary interest and several methods, such as Monto-Carlo Markov chains and importance sampling, have been developed for both equilibrium and out-of-equilibrium systems \cite{Metro,DLB20,AKH02,HAK11,APN18,BCDG2002,GKP2006,GKLT2011,KGGW2018,Causer21,Doob,Pitman,CT2013,Rose21,MajumdarEff15,BSG21ca,BSG21cb,BSG21cc,Das21,Oakes20}. 

Closely related to extreme value questions, the concept of \textit{first-passage} has been extensively investigated both in mathematics \cite{aurzada2015persistence,SA} and in physics \cite{bray2013persistence,redner2001guide,benichou2011intermittent,majumdar2007brownian,majumdar2010universal,Hanggi90}. This concept refers to the rather general problem of finding the time it takes for a particular event to happen for the first time. It plays a crucial role in various phenomena such as chemical reactions, animals searching for food, financial stocks reaching a stop price, or rivers overflowing their banks. As for the EVS, this observable is usually quite difficult to compute analytically. However, when it is possible, it usually exhibits rich features. For instance, interesting behaviors already arise at the level of one of the simplest stochastic processes, namely the one-dimensional Brownian motion, for which the first-passage event is certain but will take on average an infinite amount of time to happen. This apparent paradox arises from the fact the probability distribution of the first-passage time is normalized but has a power law tail such that the first moment is infinite \cite{bray2013persistence,redner2001guide}. One way to alter this behavior is to introduce \textit{resetting} in the dynamics, in which the process is reset to its starting position at a constant rate \cite{ES11,ES11b,ESG20}. This renders the mean first-passage time finite and even minimized at a critical resetting rate. This feature has found natural applications in the optimization of search
processes, where the search begins anew if the target is
not found within a certain amount of time \cite{KG14,KG15,CM15,BBR16,EM16}. More generally, resetting alters the motion in fundamental ways and has sparked much research on its rich consequences \cite{BS14,CS15,Reuveni15,MSS15,R16,PR17,B18,BCS19}.

In this thesis, we obtain new analytical results on the EVS of a class of stochastic processes that are representatives
of strongly correlated random variables. In some cases, we obtain universal results, which have the merit of remaining valid for a wide range of models. Furthermore, they allow for a better understanding of the relevant features that govern the EVS and sometimes reveal unexpected transitions. Additionally, we provide new methods to numerically sample rare trajectories for a wide class of stochastic processes. These methods are illustrated in numerous examples and are shown to be very efficient in practice. Finally, we present a few applications of EVS in some stochastic optimization problems. All the new results that were published in this thesis are surrounded by a box. Since most of these results are analytical, they sometimes require rather lengthy computations. We decided not to provide the full details of the derivations in this thesis but rather to give an overview and some perspectives on the results. Many parts of this thesis have been taken from the published articles and we will regularly refer to them where detailed calculations can be found.

\section{Overview of the thesis}
This thesis is organized as follows. In the remaining of Chapter \ref{chap:int}, we present a selection of the main results of this thesis. We present some analytical results on the order statistics of discrete-time random walks. We reveal their asymptotic universal behavior in the limit of a large number of steps. Then, we present further analytical results on the expected maximum of bridge random walks. We discuss their asymptotic limit, as well as their finite-size correction, which exhibits rich features depending on the tail of the jump distribution. Finally, we introduce an efficient method to generate bridge random walks, and discuss generalizations to other types of rare trajectories. 

In Chapter \ref{chap:ext}, we focus on the EVS of some class of stochastic processes. In Section \ref{sec:int}, we provide an introduction to classical results in EVS. In Section \ref{sec:ord}, we investigate the order statistics of discrete-time random walks and sketch the derivation of their universal behavior in the limit of a large number of steps. In Section \ref{sec:exp}, we study the expected maximum of bridge discrete-time random walks and discuss its rich asymptotic behavior. In Section \ref{sec:con}, we turn to continuous-time processes and derive the distribution of the length of the convex hull of Brownian motion in a confined domain. The results in this chapter have led to several publications whose abstracts can be found on p.~\pageref{chap:A14}, p.~\pageref{chap:A5}, and p.~\pageref{chap:A10}.

In Chapter \ref{chap:con}, we are interested in numerically sampling rare trajectories. In Section \ref{sec:conB}, we recall some known results to generate rare trajectories for Brownian motion. In Section \ref{sec:conD}, we introduce an efficient method to sample bridge discrete-time random walks. We illustrate our method and apply it to various examples. In Section \ref{sec:gen}, we generalize our method to other types of rare trajectories and extend it to other types of stochastic processes, both Markovian and non-Markovian. In Section \ref{sec:app}, we apply our method to sample diffusive particles in the presence of a periodic trapping environment. We briefly discuss the effective transport properties of the surviving particles. The results in this chapter have led to several publications whose abstracts can be found on p.~\pageref{chap:A4}, p.~\pageref{chap:A9}, p.~\pageref{chap:A6} and p.~\pageref{chap:A15}.

In Chapter \ref{chap:sto}, we discuss several optimization problems in stochastic processes. In Section \ref{sec:opt}, we introduce a resetting Brownian bridge as a simple model to study search processes in the presence of a bridge constraint. We highlight a surprising mechanism induced by resetting that enhances the fluctuations of the process. In Section \ref{sec:res}, we combine the notion of resetting and optimal control into an analytical framework, analogous to the Hamilton-Jacobi-Bellman paradigm, to optimally control dynamical systems undergoing a resetting policy. We illustrate our method with various examples. In Section \ref{sec:fpr}, we investigate classic diffusion with the added feature that a diffusing particle is reset to its
starting point each time the particle reaches a specified threshold. We define and solve a non-trivial optimization in which a cost is incurred whenever the particle is reset and a
reward is obtained when the particle stays near the reset point. The results in this chapter have led to several publications whose abstracts can be found on p.~\pageref{chap:A12}, p.~\pageref{chap:A11}, and p.~\pageref{chap:A1}.

Due to the number of articles written during this thesis, it would not have been reasonable to present all of them. A choice had to be made and some papers have been left aside in the Appendix. We briefly mention them here and will not discuss them further in this thesis. 

\begin{itemize}
  \item The paper entitled ``Survival probability of a run-and-tumble particle in the presence of a drift'' discusses the survival probability of a persistent random walk with an arbitrary speed distribution, not necessarily symmetric, in the presence of an absorbing boundary located at the origin. We obtain a general formula, which we apply to the case of a two-states particle with velocity $\pm v_0$ in the presence of a constant drift, known as the drifted run-and-tumble particle, and obtain rich behaviors with three distinct phases depending on the intensity of the drift. The abstract can be found on p.~\pageref{chap:A2}.
  
  \item The paper entitled ``Survival probability of random walks leaping over traps'' studies the survival probability of a random walk in the presence of finite-size traps over which it can jump. We show that the decay rate of the survival probability depends non-trivially on the size of the trap. We generalize the model to random walks with power-law distributed waiting times and derive some diffusive limits of the model. The abstract can be found on p.~\pageref{chap:A8}.
  
  \item The paper entitled ``A Tale of Two (and More) Altruists'' presents a minimalist dynamical model of wealth evolution and wealth sharing among $N$ agents. We compare the effects of an altruistic policy versus an individualistic one. We show that the best policy depends on the criterion chosen. While altruism leads to more global median wealth, the longest-lived individualists accumulate most of the wealth and live longer that the altruists. The abstract can be found on p.~\pageref{chap:A7}.
  
   \item The paper entitled ``First-Passage-Driven Boundary Recession'' investigates a moving boundary problem for a Brownian particle on the
semi-infinite line in which the boundary moves by a distance proportional to the time between successive collisions of the particle and the boundary. We find that the tail of the distribution of the $n^\text{th}$ hitting time becomes progressively heavier as $n$ increases. In addition, we find a slow double logarithmic growth of the number of hits. The abstract can be found on p.~\pageref{chap:A13}.
  
  \item The paper entitled ``Wigner function for noninteracting fermions in hard wall potentials'' discusses the quantum fluctuations in phase space of $N$ non-interacting fermions in a $d$ dimensional box. We obtain scaling functions of these fluctuations close to the Fermi surf in the limit of $N\to \infty$ and show that they are universal with respect to the dimension $d$ of the box. The abstract can be found on p.~\pageref{chap:A3}.
\end{itemize}
\newpage
\section{Presentation of a selection of the main results}
In this section, we present a selection of the main results obtained in this thesis. These results concern discrete-time random walks (RWs). We present the model and the observables in the remainder of this section and discuss the main results in the following sections of this chapter.

In its simplest form, a one-dimensional discrete-time random walk $x_m$ evolves according to the Markov rule
\begin{align}
  x_m = x_{m-1} + \eta_m\,,\label{eq:xm}
\end{align}
starting from $x_0=0$ where the jumps $\eta_m$'s are independent and identically (i.i.d.)~distributed random variables drawn from a jump distribution $f(\eta)$. The random walk defined in (\ref{eq:xm}) is a rather general model and the sequence of positions $x_m$'s could correspond to a wide variety of physical observables, such as the trajectory of a tagged particle in a random environment, the evolution of the height of a river, the logarithm of a stock price, etc. Even if the random jumps $\eta_m$'s are rarely i.i.d.~in practical applications, this model serves as a basis to obtain exact analytical results that constitute a benchmark for more complex models. 

The jump distribution $f(\eta)$ in the RW defined in (\ref{eq:xm}) is kept generic but is assumed to be symmetric $f(\eta)=f(-\eta)$. It could be a Gaussian distribution $f(\eta)=e^{-\eta^2/2}/\sqrt{2\pi}$, a discrete distribution such as $f(\eta)=\frac{1}{2}\delta(\eta-1)+\frac{1}{2}\delta(\eta+1)$, a Cauchy distribution $f(\eta)=1/[\pi(1+\eta^2)]$, etc. In particular, it may not have a finite second moment, such as in L\'evy flights where $f(\eta) \sim |\eta|^{-1-\mu}$ for large $|\eta|$, with $0<\mu<2$ denoting the L\'evy index. Generically, we refer to these discrete-time Markov jump processes as ``random walks''. Discrete-time RWs with finite variance jump distributions converge to Brownian motion in the limit $n\rightarrow \infty$ due to the Donsker theorem, which is a functional extension of the central limit theorem. When the variance is not finite, they converge to L\'evy processes and are generally much harder to study. While Brownian motion and L\'evy processes are interesting by themselves, RWs are also important to study, especially in the limit of large but finite $n$ where they display finite-size effects that are lost in the $n\rightarrow\infty$ limit. 

Discrete-time RWs constitute a very useful playground to investigate EVS of strongly correlated variables. Indeed, the set of positions $\{x_1,\ldots,x_n\}$ is strongly correlated and constitutes a simple, yet non-trivial, example of correlated random variables. For instance, for a random walk with a jump distribution with a finite variance $\sigma^2$,  the two-point correlation function is given by $\langle x_m x_n\rangle=\sigma^2 \min(m,n)$.
While the set of positions $\{x_1,\ldots,x_n\}$ are usually ordered in the direction of time, we need to order them in the direction of space to study their EVS. We arrange them in decreasing order of magnitude and define the $k^\text{th}$ maximum $M_{k,n}$ of the set of positions $\{x_1,\ldots,x_n\}$ with $k=1, 2, \ldots, n$ such that (see figure \ref{fig:model}) 
\begin{align} 
M_{1,n} > M_{2,n} > \ldots> M_{n,n} \;.
\label{def_order}
\end{align}
\begin{figure}[t]
  \begin{center}
    \includegraphics[width=0.33\textwidth]{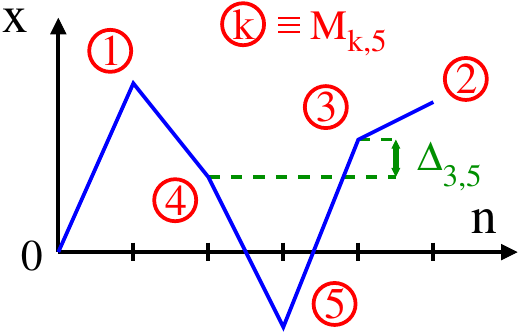}\hspace{5em}\includegraphics[width=0.33\textwidth]{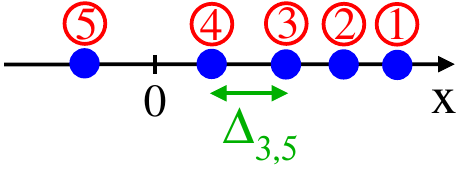}
    \caption{\textbf{Left panel:} A trajectory of a random walk of $n=5$ steps. The positions are ordered by ascending order $M_{1,n} > \ldots> M_{n,n}$, where $M_{k,n}$ is the $k^\text{th}$ maximum of the set of positions $\{x_1,\ldots,x_n\}$. Note that the initial position $x_0=0$ is not included.  \textbf{Right panel:} One-dimensional representation along the $x$-axis of the ordered positions $M_{k,n}$ of the trajectory of the random walk in the left panel. The gap $\Delta_{k,n}$ is the difference between two consecutive maxima $\Delta_{k,n} = M_{k,n} - M_{k+1,n}$. }
    \label{fig:model}
  \end{center}
\end{figure}
Therefore $M_{1,n}$ and $M_{n,n}$ are respectively the global maximum and minimum of the walk (excluding the initial position $x_0=0$). While the marginal distribution of the global maximum and minimum are well-known and have been thoroughly studied \cite{Erdos46,Darling56}, for instance by using the Pollaczek-Spitzer formula \cite{Pollackzek,Spitzer57}, the statistics of the $k$-th maximum are much less known.  They can however be characterized through the Pollaczek-Wendel identity \cite{Pollackzek,Wendel}, which interestingly relates, in distribution, the $k^\text{th}$ maximum to the global maximum and minimum of two independent copies of the random walk (see equation \ref{eq:wid} below). This identity provides an explicit expression, valid for any continuous and symmetric jump distribution, for the Fourier transform of the double generating function, with respect to $k$ and $n$, of the marginal probability distributions of $M_{k,n}$ \cite{Pollackzek,Wendel}. This identity has been revisited in several other works and has been employed to study $\alpha$-quantiles, i.e., the statistics of $M_{k,n}$ with $k = \alpha n$ in the limit $n \to \infty$ with $0<\alpha<1$ \cite{Chaumont,Port,Dassios,Embrechts,Dassiosa}.  

While the statistics of the global maximum $M_{1,n}$ is already of great interest on its own, it is sometimes necessary to know whether this maximum is an isolated event or if there exist other events with a similar magnitude. A natural observable to describe how close the global maximum is to the second maximum is the gap between them $\Delta_{1,n}=M_{1,n}-M_{2,n}$. More generally, we define the $k^\text{th}$ gap $\Delta_{k,n}$ as the difference between the $k^{\text{th}}$ and $(k+1)^{\text{th}}$ maxima:
\begin{align}
  \Delta_{k,n} = M_{k,n} - M_{k+1,n}\,,\quad k=1,\ldots,n-1\,.\label{eq:defD}
\end{align}
By definition, $\Delta_{k,n}$ are positive random variables. Along the $x$-axis, the ordered positions $M_{k,n}$ can be seen as a one-dimensional gas of strongly interacting particles with the gaps $\Delta_{k,n}$ being the inter-particle distances (see the right panel in figure \ref{fig:model}). 

In addition to being discrete in time, some RWs are also constrained. A prominent example is the \textit{bridge} random walk $X_m$ which is a RW that evolves locally as in (\ref{eq:xm}) with the constraint that it has to return to the origin after a fixed amount of time $n$ (see figure \ref{fig:bridge}):
\begin{align}
  X_0 = X_n = 0\,.\label{eq:bridgec}
\end{align}
\begin{figure}[t]
  \begin{center}
    \includegraphics[width=0.5\textwidth]{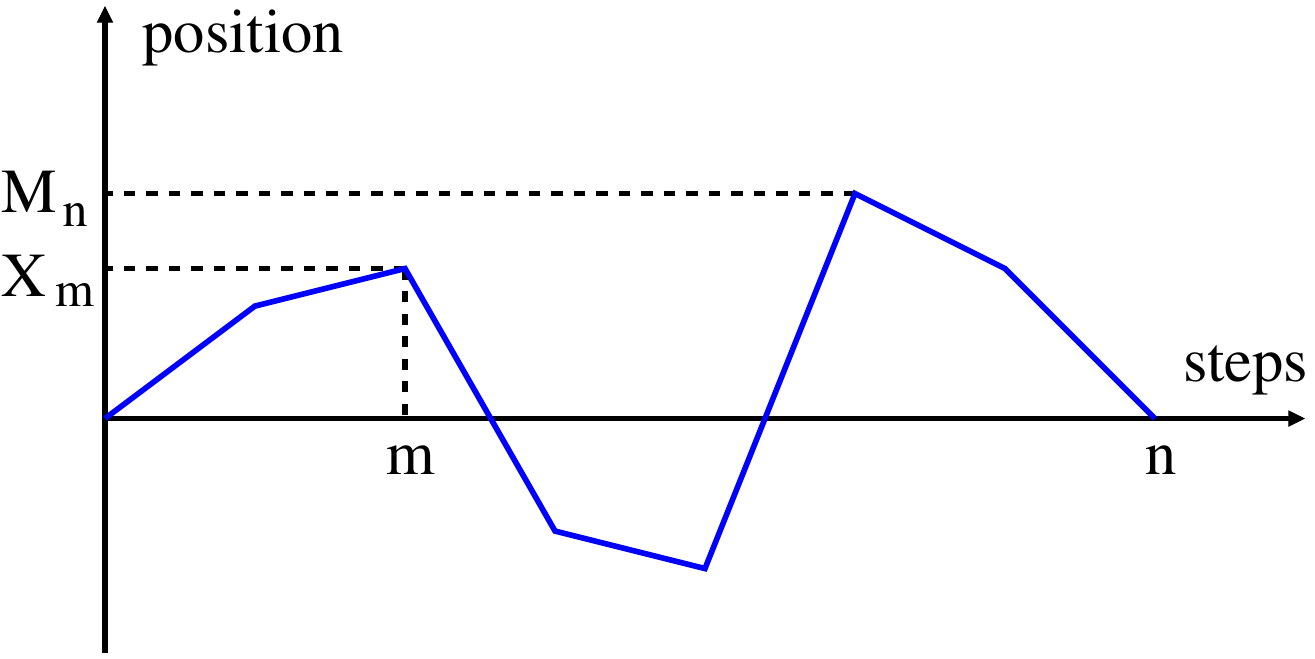}
    \caption{A bridge RW of $n$ steps is a RW that is constrained to start at the origin and return to the origin after $n$ steps. The maximum of the bridge RW of $n$ steps is denoted $M_n$.  }
    \label{fig:bridge}
  \end{center}
\end{figure} 
 For example, bridge random walks appear in many applications ranging from computer science to graph theory \cite{majumdar2007brownian,KnuthThe98,FlajoletOn98,MajumdarExact02,HararyDyna97,Takacs91,Schehr10Area,Perret12,Janson07}. Bridge random walks also appear frequently in physics problems such as in fluctuating interfaces \cite{Shapir,Antal,Majumdar04Flu1,Satya_Airy2,SOS_Airy,MD2006,gyorgyi,Burk}, record statistics in time series \cite{GodrecheMecords15,GodrecheMecords17} or in anomalous diffusion of cold atoms \cite{Barkai1,Barkai2}.

In the remainder of this section, we present three main results regarding the EVS of discrete-time RWs. In Section \ref{sec:uni}, we present some results on the statistics of the gaps $\Delta_{k,n}$ in (\ref{eq:defD}) and unveil their universal behavior in the limit of a large number of steps. In Section \ref{sec:expi}, we discuss the effect of the bridge constraint (\ref{eq:bridgec}) on the expected value of the global maximum $M_{1,n}$ and show that it displays a rich behavior in the limit of $n\to \infty$. Finally, in Section \ref{sec:geni}, we introduce an efficient method to numerically sample RW bridge trajectories as well as other types of rare trajectories.

\subsection{Universal order statistics for random walks}
\label{sec:uni}
In this section, we present our results on the distribution of the gaps of discrete-time RWs. We assume a general expansion of the Fourier transform of the jump distribution $\hat f(q)$ of the form 
\begin{align}
  \hat f(q) = \int_{-\infty}^\infty d\eta\, e^{i\eta q} f(\eta)  \sim 1 - |q|^\mu + O(|q|^{2\mu})\,,\quad q\rightarrow 0\,,\label{Fourier}
\end{align}
where $1\leq \mu\leq 2$ is the L\'evy index and where we have set the typical jump to one by rescaling the distribution. While $\mu=2$ corresponds to finite variance distributions, $\mu<2$ corresponds to infinite variance distributions, with heavy tails that decay like $f(\eta)\sim\eta^{-1-\mu}$ for $\eta\rightarrow\infty$. We leave aside jump distributions with $\mu< 1$ as they yield to transient random walks where the gap statistics behave quite differently \cite{BSG23}.

Obtaining the distributions of the gaps $\Delta_{k,n}$ in (\ref{eq:defD}) is quite challenging as it does not suffice to know the marginal distributions of $M_{k,n}$ and $M_{k+1,n}$ to devise their distribution of $\Delta_{k,n}$ as $M_{k,n}$ and $M_{k+1,n}$ are correlated. Recently, it has nevertheless been possible to obtain some analytical results for the statistics of $\Delta_{k,n}$.  By using the linearity of the expectation value, one can connect the expected gap to the expected $k^\text{th}$ maximum through $ \langle \Delta_{k,n}\rangle=\langle M_{k,n}\rangle-\langle M_{k+1,n}\rangle$, where $\langle \cdot\rangle$ denotes the average over all random walk trajectories. Using this identity, it was shown that, for symmetric and continuous jump distributions with finite variance ($\mu=2$), the expected gap has a well-defined stationary limit $\langle \Delta_{k}\rangle$ given by \cite{SM12}
\begin{align}
   \langle \Delta_{k}\rangle=\lim_{n\to \infty}\langle \Delta_{k,n}\rangle =\frac{1}{k} \int_0^\infty \frac{dq}{\pi q^2}\left[1-\hat f(q)^k\right]\,,\label{eq:statg}
\end{align}
where $\hat f (q)$ is the Fourier transform of the jump distribution defined in (\ref{Fourier}). This expression was later proved rigorously and shown to remain valid for heavy-tailed distributions as long as $\mu>1$ \cite{PT20}.
The expression (\ref{eq:statg}) is strikingly simple and can be computed explicitly for a few notable distributions, e.g., 
\begin{align}
\begin{array}{lll}
  \langle \Delta_{k}\rangle &= \dfrac{\Gamma \left(k+\frac{1}{2}\right)}{\sqrt{\pi } k
   \Gamma (k)}\,, & \text{for} \quad \hat f(q)=\dfrac{1}{1+q^2}\quad \text{(double-sided exponential) \;,}\\[1em]
    \langle \Delta_{k}\rangle &= \dfrac{1}{\sqrt{\pi k}}\,, & \text{for} \quad \hat f(q)=e^{-q^2}\quad \text{(Gaussian)\;,}\\[1em]
    \langle \Delta_{k}\rangle &=\Gamma\left(1-\dfrac{1}{\mu}\right)\dfrac{k^{\frac{1}{\mu}-1}}{\pi}\,, & \text{for} \quad \hat f(q)=e^{-|q|^\mu}\quad \text{(stable with $\mu>1$)\;,}
    \end{array}
\end{align}
where $\Gamma(z)$ is the standard Gamma function.
Although the expression in (\ref{eq:statg}) depends on the full details of the jump distribution $f(\eta)$, it becomes universal in the limit $k\to\infty$ and behaves as
\begin{align}
   \langle \Delta_{k}\rangle   & \sim \Gamma\left(1-\frac{1}{\mu}\right)\frac{k^{\frac{1}{\mu}-1}}{\pi}\,,\quad k\to \infty\,,\label{eq:avgGass}
\end{align}
which depends only on the L\'evy index $\mu>1$ of the jump distribution.
 This universal decay is quite remarkable and raises the question of whether this universal behavior of the first moment extends to the full PDF $P_{k,n}(\Delta)$. This question turns out to be quite challenging given the absence of analytical tools to go beyond the first moment. In Ref.~\cite{SM12}, the full gap distribution $P_{k,n}(\Delta)$ was computed in the special case of the double-sided exponential distribution $f(\eta) \propto e^{-|\eta|}$, using a backward Fokker-Planck approach. It was indeed shown in this case that $P_{k,n}(\Delta)$ converges towards a limiting distribution as $n \to \infty$, i.e.,
\begin{align} \label{limiting_pkn}
\lim_{n \to \infty} P_{k,n}(\Delta) = P_k(\Delta) \;,
\end{align}
where the generating function of $P_k(\Delta)$ (with respect to $k$) was computed explicitly. Furthermore, in the scaling limit $k \to \infty$, $\Delta \to 0$ keeping $\sqrt{k}\, \Delta$ fixed, it was shown that the stationary PDF $P_k(\Delta)$ takes the scaling form \cite{SM12}
\begin{align}
  P_k(\Delta) \sim \sqrt{k}\mathcal{P}_2\left(\sqrt{k}\Delta\right)\,,\quad k\to \infty\,,\label{eq:Pscal2}
\end{align}
where $\mathcal{P}_2(x)$ is given by  
\begin{equation}\label{exact_F} 
\mathcal{P}_2(x) = 2\left[\frac{2}{\sqrt{\pi}}(1+x^2) -
e^{x^2}x(2x^2+3) {\rm erfc}(x)\right] 
\,, 
\end{equation} 
with ${\rm erfc}(z) = (2/\sqrt{\pi})\int_z^\infty e^{-t^2} \, dt$ being the complementary error function. For $x\to \infty$, the scaling function behaves as $\mathcal{P}_2(x)\sim (3/\sqrt{\pi })\, x^{-4}$.

Remarkably, it was conjectured in \cite{SM12}, based on numerical simulations, that the limiting behaviors in (\ref{limiting_pkn}) and (\ref{eq:Pscal2}) actually hold for any continuous and symmetric jump distribution with finite variance ($\mu=2$), with the \textit{same} universal scaling function $\mathcal{P}_2(x)$ as
given in (\ref{exact_F}). This conjecture was later on corroborated by exact analytical results for a rather wide class of jump distributions, namely Erlang distributions of the form $f(\eta)\propto |\eta|^p e^{-|\eta|}$, with $p$ being an integer \cite{BM17}.   

This conjecture attracted some attention in the probability literature. In \cite{PT20}, it was proved that the stationary distribution $\lim_{n\to\infty}P_{k,n(\Delta)}=P_k(\Delta)$ as in (\ref{limiting_pkn}) exists for any continuous jump distribution [see Proposition 1.2 in \cite{PT20}]. In addition, for the case of the double-sided exponential distribution, the result in (\ref{eq:Pscal2}) and (\ref{exact_F}) was proved rigorously \cite{PT21} in the framework of fluctuation theory for random walks \cite{Feller}. However, the question of the universality of these results 
in (\ref{eq:Pscal2}) and (\ref{exact_F}) for symmetric and continuous jump distributions with finite variance $\sigma^2$, beyond the cases of Erlang distributions, remained open. Furthermore, much less was known for the case of a jump distribution which has heavy tails with $\mu<2$. 

Our contribution to this line of work is twofold: (i) we showed that the conjecture on the universal behavior of the limiting distribution of the gaps is indeed true for $\mu=2$, (ii) we extended these results and obtained the limiting distribution for heavy-tailed distributions and showed that it is also universal, i.e., it depends only on $1\leq \mu<2$. Furthermore, we unveiled the existence of a ``condensation'' phenomenon which is absent for $\mu=2$.  Our method extends an original idea developed by Spitzer in a paper \cite{Spi56} which seems to have attracted little notice.

As it was conjectured for the case of finite variance distributions, we find that the limiting probability distribution $ P_k(\Delta)$ becomes universal, not only for $\mu=2$ but also for $1\leq \mu\leq 2$. In the scaling limit $k\to\infty$ with $\Delta=O(k^{1/\mu-1})$, we find that, for $\mu>1$, the distribution $ P_k(\Delta)$ behaves as
\begin{align}
\Aboxed{
  P_k(\Delta) &\sim  \frac{1}{k^{\frac{1}{\mu}-1}} \mathcal{P}_\mu\left(\frac{\Delta}{k^{\frac{1}{\mu}-1}}\right)\,,\quad \Delta=O\left(k^{\frac{1}{\mu}-1}\right)\,,\quad k\to \infty\,,
  }\label{eq:Pkas}
\end{align}
where $\mathcal{P}_\mu(x)$ is a universal scaling function given by (see the left panel in figure \ref{fig:Pmu})
  \begin{figure}[t]
  \begin{center}
    \includegraphics[width=0.4\textwidth]{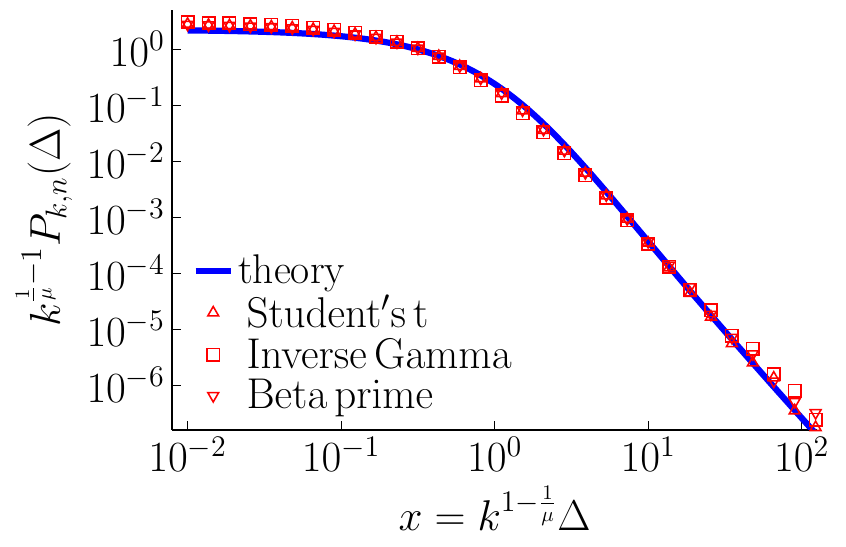}\includegraphics[width=0.4\textwidth]{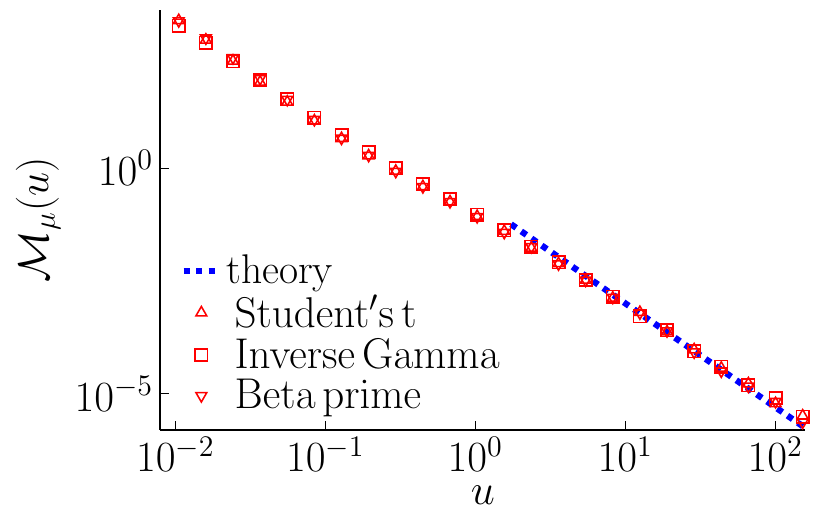}
    \caption{\textbf{Left panel:} Gap distribution $P_{k,n}(\Delta)$ in the stationary limit $n\to \infty$ and in the scaling regime $k\to \infty$ with $\Delta\to 0$ as a function of the scaling variable $x= k^{1-1/\mu}\Delta$. The theoretical prediction (blue line) in (\ref{eq:Px}) is compared to numerical distributions for $k=10^4$, $n=10^6$ obtained for various jump distributions: the Student's t distribution, the inverse gamma distribution and the beta prime distribution. The L\'evy index has been set to $\mu=1.8$ and the histograms have been obtained by sampling over $10^7$ realizations. \textbf{Right panel:} Scaling function $\mathcal{M}_\mu(u)$ as a function of $u$ for $\mu=1.3$. The theoretical prediction for the tail (dashed blue line) in (\ref{eq:expMmu}) is compared to numerical simulations for $k=10^2$ and $n=10^5$ obtained for various jump distributions: the Student's t distribution, the inverse gamma distribution and the beta prime distribution. The histograms have been obtained by sampling over $10^7$ realizations.}
    \label{fig:Pmu}
  \end{center}
\end{figure}
\begin{align}
\Aboxed{
\mathcal{P}_\mu(x) = \frac{\mu B_\mu}{(\mu-1)^2}\left[\mu\, E_{\frac{\mu-1}{\mu},-\frac{1}{\mu}}(-B_\mu x)+(2\mu-1)E_{\frac{\mu-1}{\mu},\frac{\mu-1}{\mu}}(- B_\mu x)\right]\,,
}\label{eq:Px}
\end{align}
where 
\begin{align}
B_\mu=\left[\sin\left(\frac{\pi}{\mu}\right)\right]^{-1} \quad \quad {\rm and} \quad \quad E_{\alpha,\beta}(z)=\sum_{k=0}^\infty \frac{z^k}{\Gamma(\alpha k+\beta)}\,,
\label{eq:MittagLeff}
\end{align} 
is the two-parameter Mittag-Leffler function (sometimes called Wiman's function). The expression (\ref{eq:Px}) simplifies for $\mu=2$ and we recover the expression in (\ref{exact_F}). The asymptotic behaviors of the universal scaling function ${\cal P}_\mu(x)$, for $1 < \mu < 2$, are given by
\begin{align}
  \mathcal{P}_\mu(x) \sim \left\{\begin{array}{ll}
    \dfrac{2 }{\sin \left(\frac{\pi }{\mu }\right)\Gamma
   \left(2-\frac{1}{\mu }\right)}\,, & \; x\to 0\,,
   \\
   \\
    \dfrac{2 \sin ^2\left(\frac{\pi }{\mu }\right)}{
   \Gamma \left(\frac{2}{\mu }-1\right)} \dfrac{1}{x^3} \,,& \; x\to \infty\,.\\[1em]
  \end{array}\right.\label{eq:Pxa}
\end{align}
Note that the $1/x^3$ tail is universal for all $1 < \mu < 2$. However, exactly at $\mu = 2$, the tail behaves as $1/x^4$, as in (\ref{exact_F}). Thus there is a discontinuous jump in the exponent from $3$ to $4$ as $\mu$ approaches $2$. This is consistent with the fact that in the second line of  (\ref{eq:Pxa}) the amplitude vanishes as $\mu \to 2$ and then the leading order decay comes from the subleading term scaling as $1/x^4$. 

When $\mu = 1$, the scaling form in  (\ref{eq:Pkas}) is no longer valid and the typical scale of $\Delta$ changes from $k^{1/\mu-1}$ to $1/\ln k$ as $\mu \to 1$. In this case, the scaling form of the distribution $P_k(\Delta)$ reads
\begin{align}
\Aboxed{
 P_{k}(\Delta) \sim  \ln(k)\, \mathcal{P}_1\left(\ln(k)\Delta\right)\,,\quad \Delta=O[\ln(k)^{-1}]\,,\quad k\to\infty\,,}\label{eq:pkmu1}
\end{align}
where we find that the scaling function ${\cal P}_1(x)$ is given explicitly by
\begin{align}
\Aboxed{
  \mathcal{P}_1(x) = \frac{2\pi^2}{(\pi+x)^3}\,.}\label{eq:mathcalp1}
\end{align}
One can check that the scaling function ${\cal P}_\mu(x)$ is normalized to unity, i.e., $\int_0^\infty {\cal P}_\mu(x) dx = 1$, which indicates that ${\cal P}_\mu(x)$ indeed describes the \textit{typical} behavior of $P_k(\Delta)$, corresponding to $\Delta = O(k^{1/\mu-1})$ for large $k$. However, the first moment of ${\cal P}_\mu(x)$ reads
\begin{align}
  \int_0^\infty dx \,x \mathcal{P}_\mu(x) = \frac{\sin\left(\frac{\pi}{\mu}\right)}{\Gamma\left(\frac{1}{\mu}\right)}\,,\label{eq:avgPmu}
\end{align}
which, together with the scaling form in (\ref{eq:Pkas}) does not yield the correct value for the expected stationary gap given in (\ref{eq:avgGass}) for $1<\mu<2$. This apparent paradox can be resolved by noticing that, for $\mu <2$, there are atypically large gaps of scale $k^{1/\mu}$ that are not captured by the scaling form in (\ref{eq:Pkas}), which only describes the typical gaps, of order $k^{1/\mu-1}$. More precisely, we find that for $1 < \mu <2$, the distribution of the gap for large $k$ has two parts:  a typical one where $\Delta=O(k^{\frac{1}{\mu}-1})$ described as in (\ref{eq:Pkas}) and an additional atypical one where $\Delta=O(k^{\frac{1}{\mu}})$, e.g.,
\begin{align}
  P_k(\Delta) \sim \left\{\begin{array}{lll}
  \frac{1}{k^{\frac{1}{\mu}-1}} \mathcal{P}_\mu\left(\frac{\Delta}{k^{\frac{1}{\mu}-1}}\right)\,,  & \Delta = O(k^{\frac{1}{\mu}-1})\,,&\text{(typical gap/``fluid'')}\\
  \frac{1}{k^{1+\frac{1}{\mu}}} \mathcal{M}_\mu\left(\frac{\Delta}{k^{\frac{1}{\mu}}}\right)\,,   & \Delta = O(k^{\frac{1}{\mu}})\,,&\text{(atypical large gap/``condensate'')}
  \end{array}\right.\quad k\to \infty\,,  \label{eq:Pksum}
\end{align}
where $\mathcal{P}_\mu(x)$ is given in (\ref{eq:Px}) and $\mathcal{M}_\mu(u)$ is a universal scaling function (see the right panel in figure \ref{fig:Pmu}). Note that the contribution of $\mathcal{M}_\mu(u)$ to the normalization vanishes in the limit $k\to\infty$, while its contribution to the average value is of the same order as the scaling function (\ref{eq:Pkas}). This is actually reminiscent of the ``condensation'' phenomena in the distribution of the mass in a class of mass transport models, such as in zero-range processes \cite{EH05,MEZ05,EMZ06,Satya_Houches} or even in the active run-and-tumble particles and related models \cite{Gradenigo,MGM21,MoriPRE,Smith}. Here, the gaps play the role of ``mass'' at a given site in the transport models or the ``runs'' in run-and-tumble models. In this condensed phase, there is a coexistence of a ``fluid'' regime where the masses (or the runs) are of the typical size and a ``condensation'' part which contains atypically large masses or runs. In our model, for $\mu <2$, we also see the coexistence of the ``fluid'' regime, consisting of typical gaps of order $\Delta = O(k^{1/\mu-1})$ and a ``condensate'' consisting of large gaps of order $\Delta = O(k^{1/\mu})$. The scaling function ${\cal M}_\mu(u)$ in  (\ref{eq:Pksum}) that describes the condensate part is hard to compute analytically. Fortunately, we still managed to extract the asymptotic tail of $\mathcal{M}_\mu(u)$ which behaves as
\begin{align}
\Aboxed{
  \mathcal{M}_\mu(u)\sim \frac{  \mu }{\Gamma \left(1-\frac{\mu }{2}\right)^2}\frac{1}{u^{\mu+1}}\,,\quad u\to \infty\,.}\label{eq:expMmu}
 \end{align}
  Interestingly, the amplitude in (\ref{eq:expMmu}) is the same as the one for the distribution of the first gap obtained in \cite{MMS13, MMS14}. As in (\ref{eq:Pxa}), the amplitude in (\ref{eq:expMmu}) vanishes for $\mu\to 2$ signaling a different tail behavior (for the double-sided exponential jump distribution, corresponding to $\mu=2$, this tail is actually exponential and not algebraic \cite{SM12,BM17}).

\subsection{Expected maximum of discrete-time bridge random walks}
\label{sec:expi}
In this section, we present our results on the expected maximum of \textit{bridge} RWs (see figure \ref{fig:bridge}). Let us first recall some results on the maximum of \textit{free} RWs. We define the maximum $M_n $ after $n$ steps 
 \begin{align}
  M_n = \text{max}\{x_0,\ldots,x_n\}\,.\label{eq:maxdef}
\end{align}
At variance with our previous definition in (\ref{def_order}), we include the initial position $x_0$ in the set of positions. The maximum $M_n$ is an observable that is commonly studied in the mathematics literature \cite{Kac54,Darling56,Spitzer56,Doney87,Doney08,CD10,Kuznetsov11} as well as in natural and practical contexts, such as animal foraging where the spatial extent of their territory can be characterized by the extreme points of their trajectories \cite{Dumonteil13,Majumdar21convex,Randon09Convex,Majumdar10Random,GrebenkovConvex17,Schawe18Large,Cauchy32}. 
For a free random walk starting from the origin $x_0=0$, it is well-known that the expected maximum $\langle M_n\rangle$ is simply given by \cite{Kac54,Spitzer56}
\begin{align}
  \langle M_n\rangle
  &=\sum_{m=1}^n \frac{1}{m}
   \int_{0}^\infty dy\, y\, p_m(y) \,,\label{eq:maxfree}
\end{align}
where $p_m(y)$ is the forward propagator of the free random walk, i.e., the probability density that the position $y$ is reached in $m$ steps given that it started at the origin. It is simply given~by
 \begin{align}
   p_m(y) = \int_{-\infty}^\infty \frac{dk}{2\pi} \,\left[\hat f(k)\right]^m\,e^{-i\,k\,y}\,,\label{eq:PI}
\end{align}
 where $\hat f(k)$ is the Fourier transform of the symmetric jump distribution $f(\eta)$. For jump distributions with finite variance, such that $\hat f(k)$ behaves as in (\ref{Fourier}) with $\mu=2$, the expected maximum $\langle M_n\rangle$ grows, to leading order for large $n$, as $2\sqrt{n/\pi}$, a result that can be easily obtained from the diffusive (i.e., Brownian) limit. In contrast, the second leading order term in the asymptotic limit of the expected maximum is non-trivial to obtain and contains the leading finite-size correction \cite{comtet_precise}. This finite-size correction is actually important as it appears as a leading order term in the thermodynamic limit of various geometrical properties such as the difference between the expected maximum $\langle M_n\rangle$ and the average absolute position of the RW $\langle|x_n|\rangle$ \cite{comtet_precise}. In addition, it turns out that it also has applications in various algorithmic problems \cite{Coffman93,Coffman98}. This leading finite-size correction was obtained in \cite{comtet_precise} where it was shown that, for finite variance jump distributions (with additional regularity properties \cite{comtet_precise}), this correction is a constant $\gamma$ such that
\begin{align}
  \langle M_n\rangle\sim 2\sqrt{\frac{n}{\pi}}+\gamma \,,\quad n\rightarrow \infty\,,\label{eq:mnfree}
\end{align}
where $\gamma$ is given by \cite{comtet_precise}
\begin{align}
 \gamma =  \frac{1}{\pi}\,\int_0^\infty \frac{dk}{k^2}\,\ln\left[\frac{1-\hat f\left(k\right)}{k^2}\right]\,,\label{eq:gamma}
\end{align}
where $\hat f(k)$ is the Fourier transform of the jump distribution. Interestingly, the constant $\gamma$ depends on the full details of the jump distribution and takes non-trivial values (e.g., for Gaussian jump distributions, $\gamma= \zeta(1/2)/\sqrt{\pi}$ where $\zeta$ is the Riemann zeta function). The analog of the asymptotic expansion (\ref{eq:mnfree}) for free RWs with heavy tails is given in \cite{comtet_precise,GrebenkovConvex17}.

Our contribution to this line of work is threefold: (i) we obtain an expression for the expected maximum after $n$ steps of bridge RWs which is valid for all $n$ and extends the expression (\ref{eq:maxfree}), (ii) we extract its asymptotic limit for $n$ large and obtain the analog for bridge RWs of the expression (\ref{eq:mnfree}), (iii) we extend our results to heavy-tailed distributions with Lévy index $\mu<2$.

In the remaining of this section, $M_n$ refers to the maximum of a \textit{bridge} random walk (\ref{eq:bridgec}) as in figure \ref{fig:bridge}, and no more to the one of a free random walk. We find that the expected maximum of a bridge RW after $n$ steps is given by
\begin{align}
\Aboxed{
 \langle M_n\rangle = \sum_{m=1}^n \frac{1}{m}
   \int_{0}^\infty dy\, y\, \frac{p_m(y)  p_{n-m}(y)}{p_n(0)}\,,}\label{eq:maxbridgeI}
   \end{align}
where now the average is over all the \textit{bridge} trajectories and where $p_m(y)$ is the propagator of the free random walk (\ref{eq:PI}). This expression nicely extends the one for the expected maximum of a free random walk (\ref{eq:maxfree}). Our derivation of the expected maximum (\ref{eq:maxbridgeI}) is based on the celebrated Spitzer's formula \cite{Spitzer57} (see equation \ref{eq:spitzer} below). It recovers, using a different method, a result obtained in the mathematics literature \cite{Kac54} derived using combinatorial arguments.
 
For jump distributions whose Fourier transform behaves as in (\ref{Fourier}), we find that the expected maximum of bridge RWs grows for large $n$ as
\begin{align}
\Aboxed{
   \langle M_n\rangle   &\sim h_1(\mu)\,n^{\frac{1}{\mu}} \,,}\label{eq:Emnintro1}
\end{align}
where the amplitude $h_1(\mu)$ is given by
\begin{align}
\Aboxed{
  h_1(\mu) = \frac{\mu\pi}{8\,\Gamma\left(1+\frac{1}{\mu}\right)}\,,}\label{eq:h1intro}
\end{align}
where $\Gamma(z)$ is the standard Gamma function. Note that the asymptotic expression (\ref{eq:Emnintro1}) is universal with respect to the jump distribution, i.e.~it only depends on the Lévy index $\mu$. A plot of this function $h_1(\mu)$ is shown in figure \ref{fig:h1}. It is interesting to compare the asymptotic large $n$ expansion of the expected maximum of a bridge RW (\ref{eq:Masall}) with the one of a free RW obtained previously in \cite{comtet_precise,GrebenkovConvex17}. While the leading power in $n$ are identical in both cases, the results for the bridge RW differ from the free RW in two ways: (i) $\langle M_n\rangle$ for the bridge RW is finite even for L\'evy flights with L\'evy exponent $0<\mu \leq 1$ while it is infinite for a free RW: this 
is due to the bridge constraint that pins the initial and final positions to the origin, (ii) the amplitude $h_1(\mu)$ is different from the one obtained  for the free random walks $h^*_1(\mu)$ given by \cite{comtet_precise,GrebenkovConvex17}
\begin{align}
  h^*_1(\mu) &= \frac{\mu}{\pi}\,\Gamma\left(1-\frac{1}{\mu}\right)\,.\label{eq:hfree1}
\end{align} 
Comparing the amplitude $h_1(\mu)$ and $h_1^*(\mu)$ (see figure \ref{fig:h1}), we see that the amplitude of the free RW (for $\mu > 1$) is larger than the one for the bridge RW, which is expected as the bridge RW cannot go as far as the free RW due to its constraint to return to the origin. 
\begin{figure}[t]
  \begin{center}
    \includegraphics[width=0.5\textwidth]{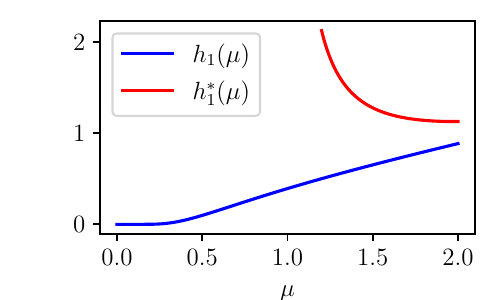}
    \caption{Amplitudes $h_1(\mu)$ in  (\ref{eq:h1intro}) and $h_1^*(\mu)$ in  (\ref{eq:hfree1}) of the leading order term of the large $n$ limit of the expected maximum $\langle M_n\rangle$ of a bridge random walk and a free random walk, respectively as a function of the L\'evy index $0 < \mu \leq 2$. Contrary to free RWs, bridge RWs with $\mu<1$, have a well-defined expected maximum due to the bridge constraint that pins the initial and final positions to the origin. This explains why the blue curve is defined for $\mu<1$ while the red one diverges upon approaching $\mu=1$.}
    \label{fig:h1}
  \end{center}
\end{figure}
\begin{figure}[t]
  \begin{center}
    \includegraphics[width=0.3\textwidth]{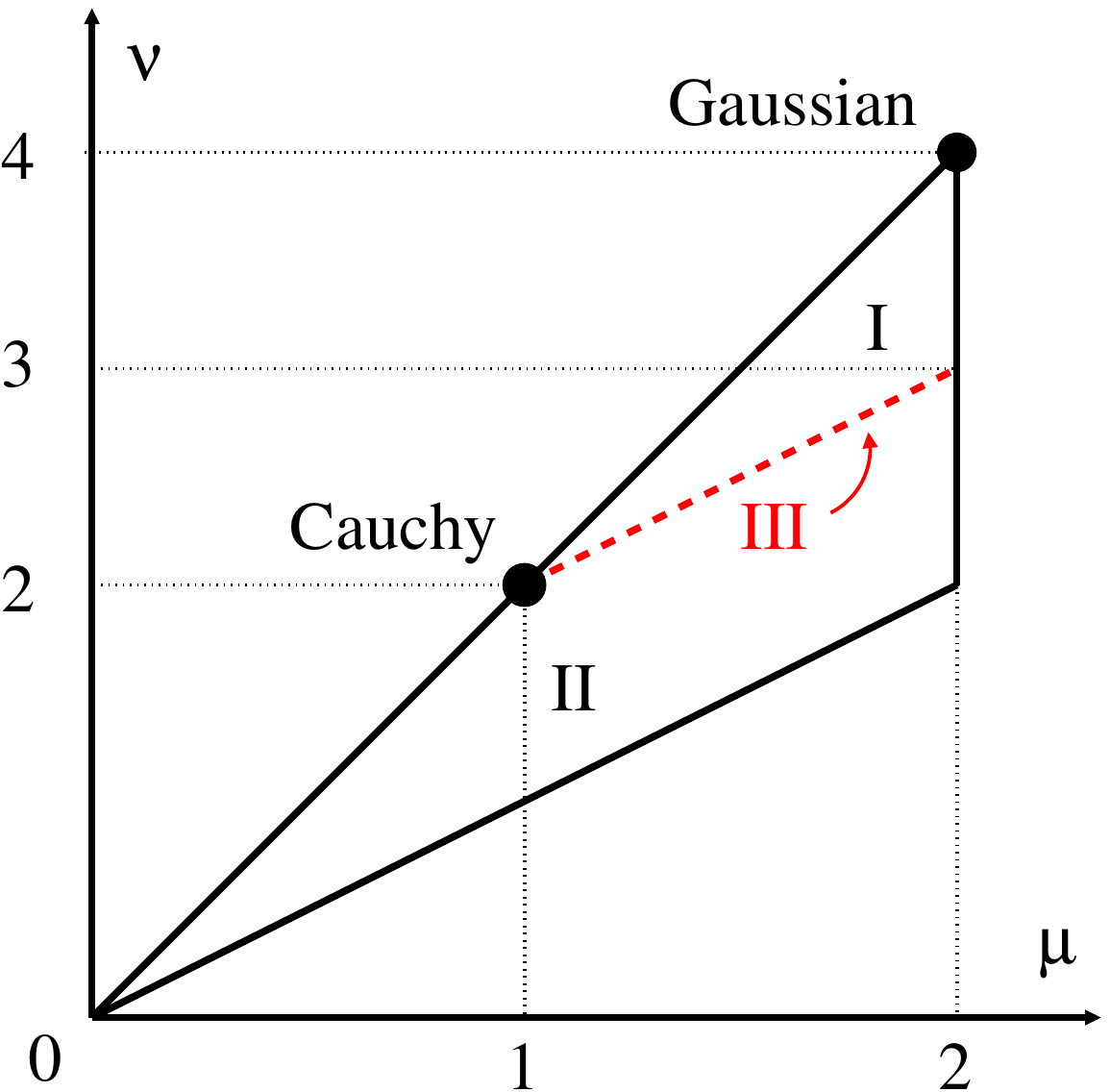}
    \caption{The parameters $\mu$ and $\nu$ are the exponents of the expansion of the Fourier transform of the jump distribution $\hat f(k)$ such that $\hat f(k) \sim 1 - |k|^\mu + b\,|k|^{\nu}$ for $k\rightarrow 0$. The second leading order term in the large $n$ limit of the expected maximum $\langle M_n\rangle$ depends on the phase in which the couple $(\mu,\nu)$ is located (I, II or III). The second leading order term of each phase is given in (\ref{eq:Masall}). As a reference, the couple $(\mu,\nu)$ for Gaussian and Cauchy  jump distributions are indicated at $(\mu=2,\nu=4)$ and $(\mu=1,\nu=2)$, respectively.}
    \label{fig:phaseDiagram}
  \end{center}
\end{figure}

The second leading order term in the large $n$ limit of the expected maximum $\langle M_n\rangle$ is slightly more subtle as it depends on the next order terms in the expansion of the Fourier transform of the jump distribution in (\ref{Fourier}). We assume that $\hat f(k)$ behaves as 
\begin{align}
  \hat f(k) \sim 1 - |k|^\mu + b\,|k|^{\nu} + O(|k|^{2\,\mu})\,,\quad k\rightarrow 0\,,\label{eq:f}
\end{align}
where $0<\mu\leq 2$, $\mu<\nu\leq 2\,\mu$, and $b$ is a constant. 
We find that the second leading order term of the expected maximum $\langle M_n\rangle$ displays a rich behavior depending on the two exponents $\mu$ and $\nu$ (see figure \ref{fig:phaseDiagram}):\begin{subequations}
\begin{itemize}
  \item  Phase I ($1<\mu\leq 2$ and $\mu+1<\nu\leq 2 \mu$)
\begin{align}
\Aboxed{
 \langle M_n\rangle   &\sim h_1(\mu)\,n^{\frac{1}{\mu}} 
+ \frac{1}{2\pi}\int_{-\infty}^\infty  \frac{dk_1}{k_1^2} \ln\left(\frac{1-\hat f(k_1)}{|k_1|^\mu}\right) \,,\quad n\rightarrow \infty\,, }\label{eq:MasI}
\end{align}
\item Phase II ($0<\mu\leq 2$ and $\mu<\nu \leq 2\,\mu$ and  $\nu<\mu+1$)
\begin{align}
\Aboxed{
  \langle M_n\rangle  &\sim h_1(\mu)\,n^{\frac{1}{\mu}} 
  + b \,h_2(\mu,\nu)\,n^{\frac{1+\mu-\nu}{\mu}}\,,\quad n\rightarrow \infty\,,}\label{eq:MasII}
\end{align}
\item Phase III ($1\leq\mu\leq 2$ and $\nu=\mu+1$)
\begin{align}
\Aboxed{
 \langle M_n\rangle &\sim h_1(\mu)\,n^{\frac{1}{\mu}}  -\frac{b}{\pi\,\mu}\,\ln(n)\,,\quad n\rightarrow \infty\,,}\label{eq:MasIII}
    \end{align}
\end{itemize}
\label{eq:Masall}
\end{subequations}
where the amplitude $h_1(\mu)$ is given in (\ref{eq:h1intro}) and $h_2(\mu,\nu)$ is given by 
\begin{align}
\Aboxed{
  h_2(\mu,\nu) = \frac{-1}{2\,\Gamma\left(1+\frac{1}{\mu}\right)}\left(\frac{\pi \, \Gamma \left(\frac{\nu +1}{\mu }\right)}{4\, \Gamma
   \left(1+\frac{1}{\mu }\right)} + \frac{  (\nu-\mu  )  \csc
   \left(\frac{\pi  \nu }{\mu }\right) }{\mu  
   \Gamma \left(-\frac{\nu }{\mu }\right)}\int_0^\infty dv \frac{\left(v^{\nu }-1\right)\left(v^{\mu -\nu
   }-1\right)}{\left(v^2-1\right) \left(v^{\mu }-1\right)}\right)\,.}\label{eq:h2s}
\end{align}
Note that the amplitude $h_2(\mu,\nu)$ diverges as $\nu$ approaches $\mu+1$ as expected as it corresponds to approaching the red line in figure \ref{fig:phaseDiagram}. The amplitude $h_2(\mu,\nu)$ can be exactly evaluated for $\mu=1$ and $\mu=2$, leading to
\begin{subequations}
\begin{align}
  h_2(\mu=1,\nu) &=  \frac{ \pi ^2-2 (\nu -1) [\pi  (2 \nu -1)
   \cot (\pi  \nu )+\pi  \csc (\pi  \nu )-2]}{8\sin (\pi  \nu ) \Gamma
   (-\nu )}\, ,\label{eq:h21}\\
h_2(\mu=2,\nu) &=  \frac{\sqrt{\pi } (\nu -2) (\nu -1) \sec
   \left(\frac{\pi  \nu }{2}\right)}{4\,\Gamma \left(-\frac{\nu
   }{2}\right)}-\frac{1}{2} \Gamma \left(\frac{\nu +1}{2}\right)\,.\label{eq:h22}
\end{align}    
\end{subequations}

In the specific case of a jump distribution with finite variance $\sigma$ in phase I, the asymptotic limit of the expected maximum of the bridge random walk is given by
\begin{align}
  \langle M_n\rangle   &\sim \frac{1}{2}\sqrt{\pi\,n} + \gamma\, \quad, \quad n \to \infty\,,\label{eq:maxga}
\end{align}
where, remarkably, $\gamma$ is the same constant correction (\ref{eq:gamma}) as in the expected maximum of the free RW obtained in \cite{comtet_precise} discussed in the introduction. 

\subsection{Generating discrete-time constrained trajectories}
\label{sec:geni}
In this section, we discuss a method that we have introduced to generate bridge RWs, i.e.~RWs that evolve locally as in (\ref{eq:xm}) but with the global constraint (\ref{eq:bridgec}) such that they have to return to the origin at a time $n$ (see figure \ref{fig:bridge}). This method is the discrete-time version of the continuous-time effective Langevin dynamics developed in \cite{MajumdarEff15,Doob,CT2013}, which is discussed in Section \ref{sec:conB}. Generating bridge RWs is a challenging problem since a general prescription is not known for arbitrary jump distribution $f(\eta)$ in (\ref{eq:xm}). In the special case when the jump distribution is a pure Gaussian, i.e., $f(\eta) = e^{-\eta^2/2}/\sqrt{2 \pi}$, one can still generate bridge trajectories by using the discrete-time path transformation
 \begin{align} \label{bcbm_discr}
 X_m = x_m - \frac{m}{n} x_n \;\,,
 \end{align}
 which maps a free trajectory $x_m$ to a bridge trajectory $X_m$ of length $n$. One can check that the right-hand side in (\ref{bcbm_discr}) indeed vanishes at $m=n$ and that the two-point correlation function corresponds to the bridge one. As this is a linear transformation between two Gaussian processes, this guarantees that $X_m$ is indeed a Gaussian bridge RW.
 However, this prescription does not work when $f(\eta)$ is not Gaussian. Another example where one can easily generate a bridge configuration corresponds to the $\pm 1$ random walk, where the jump distribution is $f(\eta) = (1/2) \delta(\eta+1) + (1/2) \delta(\eta - 1)$ \cite{Rose21,ChafaiRW12}. It is therefore important to develop an algorithm that does not depend on the specific form of the jump distribution. One possibility is to perform Markov chain Monte-Carlo simulations which consist in sampling the full joint jumps distribution $\{\eta_0\,,\ldots,\eta_{n-1}\}$ with the global bridge constraint that the total sum of the jumps is $\sum_{m=0}^{n-1} \eta_m=0$ (see e.g.~\cite{Schehr10Area}). This Monte-Carlo method can also be computationally costly and requires advanced techniques to probe the tails of distributions as the Monte-Carlo algorithm sometimes struggles to equilibrate the system. Given this absence of generic and efficient methods to generate constrained random walks, it is then highly desirable to derive an effective discrete-time jump process valid for arbitrary jump distributions $f(\eta)$. Our main contribution to this line of work is that we found that bridge trajectories can be generated by effective dynamics
\begin{align}
  X_m = X_{m-1} + \eta_{m,n}(X_{m-1})\,,\label{eq:bridgeeff}
\end{align}
 where the effective jump $\eta_{m,n}(X_{m-1})$ at time $m$ of the bridge of length $n$ is drawn from the effective jump distribution $\tilde f(\eta\,|\,X,m,n)$ which is given by 
\begin{align}
\Aboxed{
  \tilde f(\eta\,|\,X,m,n) = f(\eta) \,\frac{Q(X+\eta,n-m-1)}{Q(X,n-m)}\,,} \label{eq:eff}
\end{align}
where $Q(x,m)$ is the backward propagator of the free process defined by
\begin{align}
   Q(x,m) = \int_{-\infty}^\infty \frac{dk}{2 \pi}\,\left[\hat f(k)\right]^m\,e^{i\,k\,x}\;,\label{eq:QI}
\end{align}
where $\hat f(k)$ is the Fourier transform of the jump distribution $f(\eta)$ (which is not necessarily symmetric).
The effective distribution is therefore the free distribution that is modified in such a way that steps that take the walker closer to its final destination (here the origin) are more likely to happen. Note that this effective distribution is parametrized by the current position $X$ of the bridge and moreover is non-stationary, i.e., it depends on the current time $m$ and also the total duration $n$. At the last step, when $m=n-1$, the numerator in (\ref{eq:eff}) simplifies to $Q(X+\eta,0)=\delta(X+\eta)$ and constraints the particle to return to the origin. Note that  (\ref{eq:eff}) can be viewed as an explicit representation of the generalized Doob transform which has been used previously \cite{CT2013,Rose21}. 

For certain specific jump distributions $f(\eta)$, it is possible to compute $\tilde f(\eta\,|\,X,m,n)$ explicitly and sample directly from it. For such cases where one has a direct sampling method to sample from $\tilde f(\eta\,|\,X,m,n)$, one can easily draw a random number $\eta$ from this distribution to generate the bridge configuration numerically. However, in some cases, no direct sampling methods exist. In these cases, one can rely 
upon ``acceptance-rejection sampling'' (see e.g. \cite{Gilks92}). As an example, we sampled Cauchy random walk bridges. The normalized free jump distribution is symmetric with divergent moments and is given by
\begin{align}
  f(\eta) = \frac{1}{\gamma\,\pi}\,\frac{1}{\left[1+\left(\frac{\eta}{\gamma}\right)^2\right]}\,,\label{eq:cauchyDist}
\end{align} 
where $\gamma$ is a parameter that provides the typical scale of the jumps. 
The effective step distribution at the $m^\text{th}$ step (\ref{eq:eff}) is therefore given by 
\begin{align}
 \Aboxed{ \tilde f(\eta\,|\,X,m,n) = \frac{1}{\gamma\,\pi}\,\frac{n-m}{\left(n-m-1\right)}\,\frac{1}{\left[1+\left(\frac{\eta}{\gamma}\right)^2\right]}\,\frac{1+\left(\frac{X}{\gamma(n-m)}\right)^2}{\left[1+\left(\frac{X+\eta}{\gamma(n-m-1)}\right)^2\right]} \;.}\label{eq:effC}
  \end{align}
Note that, unlike the free distribution $f(\eta)$ in (\ref{eq:cauchyDist}), the effective distribution in (\ref{eq:effC}) is asymmetric, has a power law tail 
$\tilde f(\eta\,|\,X,m,n)\propto 1/\eta^4$ as $|\eta| \to \infty$ and consequently has a finite second moment. In figure \ref{fig:bridgeC}, we show a sampled bridge Cauchy random walk (left panel) and a comparison between the numerical and theoretical position distribution at an intermediate time to check the sampling method. 

In Chapter \ref{chap:con}, we extend the method to other types of constrained RWs, and in particular to RWs with a fixed area below their trajectory. We also generalize the method to a non-Markovian process and discuss applications to a survival problem.

\begin{figure}[t]
 \includegraphics[width=0.5\textwidth]{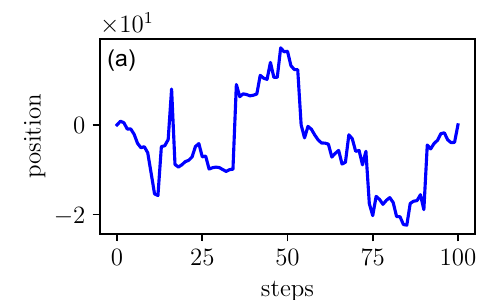}%
\hfill
  \includegraphics[width=0.5\textwidth]{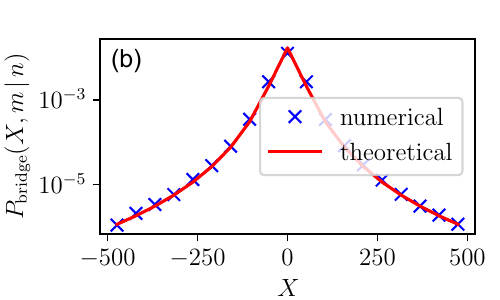}%
\hfill
\caption{\textbf{Left panel (a):} Typical trajectories of a Cauchy bridge random walk of $n=100$ steps generated using the acceptance-rejection sampling method on the effective distribution in (\ref{eq:effC}). \textbf{Right panel (b):} Position distribution at $m=75$ for a Cauchy bridge of $n=100$ steps generated from the effective distribution. The position distribution $P_{\rm bridge}(X,m|n)$ obtained numerically by sampling $10^5$ trajectories using the acceptance-rejection sampling on the effective  distribution in (\ref{eq:effC}) is compared with the theoretical prediction. }\label{fig:bridgeC}
\end{figure}

\chapter{Extreme value statistics in stochastic processes}
\label{chap:ext}
In this chapter, we focus on the EVS of some class of stochastic processes. We first provide an introduction and present some classical results. Then, we discuss the order statistics and the expected maximum of random walks and sketch the derivation of the results presented in the introduction. Finally, we turn to continuous time stochastic processes and study the convex hull of Brownian motion in confined geometries.

\section{Introduction to extreme value statistics}
\label{sec:int}
\subsection{Independent and identically distributed random variables}
Let us start by recalling some well-known results on the EVS of a set $\{\eta_1,\ldots,\eta_n\}$ of i.i.d.~random variables drawn from a continuous probability distribution $f(\eta)$. The cumulative distribution of the maximum of the set $\max\{\eta_1,\ldots,\eta_n\}$ reads
\begin{align}
  \text{Prob.}\left(\max\{\eta_1,\ldots,\eta_n\}<M\right) = \text{Prob.}\left(\eta_1<M,\ldots,\eta_n<M\right) = \left(\int_{-\infty}^M d\eta \,f(\eta)\right)^n\,,\label{eq:maxii}
\end{align}
where we used that the random variables are i.i.d in the last equality. One can analyze the expression (\ref{eq:maxii}) to find that the maximum of the set of i.i.d.~random variables asymptotically behaves, in the limit of $n\to \infty$, as 
\begin{align}
 \max\{\eta_1,\ldots,\eta_n\} \sim a_n + b_n\,\chi\,,\quad n\to \infty\,,\label{eq:asiid}
\end{align}
where $a_n$ and $b_n$ are scaling coefficients, and $\chi$ is a random variable of order $O(1)$. Note that the asymptotic relation in (\ref{eq:asiid}) between the random variables $\max\{\eta_1,\ldots,\eta_n\}$ and $\chi$ is valid ``in distribution''. The coefficients $a_n$ and $b_n$ respectively describe the typical value of the maximum and the typical size of its fluctuations. While these scaling coefficients naturally depend on the parent distribution $f(\eta)$, the random variable $\chi$ turns out to be universal, in the sense that it does not depend on the full details of $f(\eta)$ but only on its tail behavior. It turns out that there exist exactly three distributions for $\chi$ which are 
\begin{itemize}
  \item the Gumbel distribution, $\text{Prob.}\left(\chi < z\right) = e^{-e^{-z}}$, if the parent distribution has unbounded support with a tail that decays faster than any power law,
  \item the Fréchet distribution, $\text{Prob.}\left(\chi < z\right) =\Theta(z)e^{-z^{-\mu}}$, if the parent distribution has unbounded support with a tail that decays as a power law $f(\eta)\sim \eta^{-1-\mu}$ as $\eta \to \infty$,
  \item Weibull distribution, $\text{Prob.}\left(\chi < z\right) =\Theta(z)e^{-z^{\alpha}}$, if the parent distribution has a finite edge in $\eta_*$ such that $f(\eta) \propto (\eta_*-\eta)^{\alpha-1}$ as $\eta\to \eta_*$,
\end{itemize} 
where we denoted by $\Theta(z)$ the Heavyside step function such that $\Theta(z)=1$ if $z>0$ and $\Theta(z)=0$ otherwise. Note that in the case where $f(\eta)$ has a finite support and decays faster than a power law at the upper edge, the distribution of the maximum is given by the Gumbel law. Further discussions on the maximum of i.i.d.~random variables, and in particular on how to find the coefficients $a_n$ and $b_n$ can be found in the original work of Gumbel \cite{gumbel} and in the pedagogical reviews \cite{SMS14r,reviewMPS}. In addition, note that (\ref{eq:asiid}) only describes the ``typical fluctuations'' of the maximum but does not describe ``atypical'' fluctuations which are rather described by large deviation tails (see \cite{Vivo15} for a pedagogical review).

A natural extension of the results above is to not only consider the global maximum of the set $\{\eta_1,\ldots,\eta_n\}$ but also the second maximum, the third, etc \cite{Arnold,Nagaraja}. Let us arrange the set of variables in decreasing order of magnitude
and define the $k^{\text{th}}$ maximum $M_{k,n}$ of the set of variables $\{\eta_1,\ldots,\eta_n\}$ with $k=1,\ldots,n$ such that
\begin{align}
  M_{1,n}> M_{2,n}>\ldots>M_{n,n}\,.
\end{align}
Obviously, $M_{1,n}$ and $M_{n,n}$ are respectively the global maximum and minimum of the set. The cumulative distribution of the $k^\text{th}$ maximum $M_{k,n}$ reads
\begin{align}
  \text{Prob}.\left(M_{k,n}<M\right) = \sum_{m=0}^{k-1}\binom{n}{m}\left(\int_{M}^\infty d\eta f(\eta)\right)^{m}\left(\int_{-\infty}^M d\eta f(\eta)\right)^{n-m}\,,\label{eq:Mkniid}
\end{align}
which simply states that for the $k^\text{th}$ maximum to be less than $M$, there must be at most $k-1$ variables above $M$ and the remaining ones below $M$. One can analyze the expression (\ref{eq:Mkniid}) and find that the $k^\text{th}$ maximum of the set of i.i.d.~random variables asymptotically behaves, in the limit of $n\to \infty$, as 
\begin{align}
  M_{k,n} \sim a_n + b_n \chi_k\,,\quad n\to \infty\,,\label{eq:Mkniid2}
\end{align}
where $a_n$ and $b_n$ are scaling coefficients and $\chi_k$ is a random variable of order $O(1)$. Remarkably, the random variable $\chi_k$ turns out to be universal and its cumulative distribution is given by
\begin{align}
  \text{Prob.}\left(\chi_k<z\right) = \text{Prob.}\left(\chi<z\right)\sum_{j=0}^{k-1} \frac{\left[-\ln\left( \text{Prob.}\left(\chi<z\right)\right)\right]^{j}}{j!}\,,\label{eq:mkndid}
\end{align}
where $\chi$ is the random variable describing the fluctuations of the global maximum in (\ref{eq:asiid}). In particular, for $k=1$, the expression (\ref{eq:mkndid}) recovers the distribution of the global maximum.

Another natural aspect to study is how close the maxima are to each other. One observable that describes this is the gap $\Delta_{k,n}$ between two consecutive maxima defined as
\begin{align}
  \Delta_{k,n} = M_{k,n} - M_{k+1,n}\,,\quad k=1,\ldots,n-1\,.\label{eq:dkndef}
\end{align}
In the limit $n\to \infty$, it asymptotic behaves as
\begin{align}
  \Delta_{k,n} \sim b_n \delta_k\,,\quad n\to\infty\,,\label{eq:dkniid}
\end{align}
where $b_n$ is the same scaling coefficient as in (\ref{eq:Mkniid2}) and $\delta_k$ is a random variable of order $O(1)$ whose cumulative probability distribution is given by \cite{SMS14r}
\begin{align}
  \text{Prob.}\left(\delta_k < z\right) = \frac{\Theta(z)}{(k-1)!}\int_{-\infty}^\infty dx  g'(x)\left[-\ln\left(g(x)\right)\right]^{k-1}\left[1-\frac{g(x-z)}{g\left(x\right)}\right] \,,
\end{align}
where $g(x)=\text{Prob.}\left(\chi<x\right)$ and $\chi$ is the random variable describing the fluctuations of the global maximum in (\ref{eq:asiid}). In particular, if $\chi$ is Gumbel distributed, the distribution of the gap is simply exponential $\text{Prob.}\left(\delta_k < z\right) = \Theta(z)\left(1 -e^{-kz}\right)$.

Let us end this section with three comments. First, the asymptotic limit in (\ref{eq:Mkniid}) and in (\ref{eq:dkniid}) are such that $n\to \infty$ with $k$ fixed. It turns out that there exists also another interesting limit where $n\to \infty$ with $k=\alpha n$ with $0<\alpha<1$ (see for instance \cite{BLth}). Second, the gaps defined in (\ref{eq:dkndef}) have a nice telescopic structure such that their sum gives
\begin{align}
\sum_{k=1}^{n-1} \Delta_{k,n} = M_{1,n} - M_{n,n}\,.
\end{align}
This property has been exploited in several works to investigate the statistics of sums of strongly correlated random variables (see for instance the pedagogical review \cite{CB08}). Finally, physical systems are rarely described by i.i.d.~random variables. Correlations are ubiquitous in nature. Fortunately, for the case of \textit{weakly} correlated random variables, one can still make use of the above results to study their EVS. To see this, let us consider the problem of finding the distribution of the maximum of a set of dependent random variables $\{x_1,\ldots,x_n\}$ with short-range correlations $\langle x_i x_j \rangle -\langle x_i \rangle \langle x_j\rangle \propto e^{-|i-j|/\xi}$, where $\xi$ is a finite correlation length. The key idea is to group the random variables in blocks of length $\xi$, such that random variables belonging to different blocks are considered to be approximately uncorrelated. Then, we see that the problem reduces to finding the distribution of the maximum of $n'=n/\xi$ independent random variables which are the local maxima of each block. Of course, one might ask what happens to the EVS of \textit{strongly} correlated random variables? This will be the topic of the next section.

\subsection{Strongly correlated random variables: the case of random walks}
In the previous section, we discuss the EVS of i.i.d.~random variables and their universality classes. The case of strongly correlated random variables is much less understood and no general theory currently exists. In the absence of a general framework, there has been a common interest to study exactly solvable cases in the hope of gaining some general insights. In this section, we will discuss the EVS of discrete-time random walks which turns out to be a very useful model of strongly correlated random variables. 

Let us consider the one-dimensional discrete-time RW defined in (\ref{eq:xm}) with a continuous and symmetric jump distribution $f(\eta)$ and starting from $x_0=0$. Before discussing the EVS of such model, let us briefly recall some basic results on RWs. One central quantity in the study of RWs is the propagator $p_n(x)$ which is the probability that the random walk is located $x$ at step $n$ given that it started from $x_0=0$ (see figure \ref{fig:xnMn}). 
\begin{figure}[t]
  \begin{center}
    \includegraphics[width=0.5\textwidth]{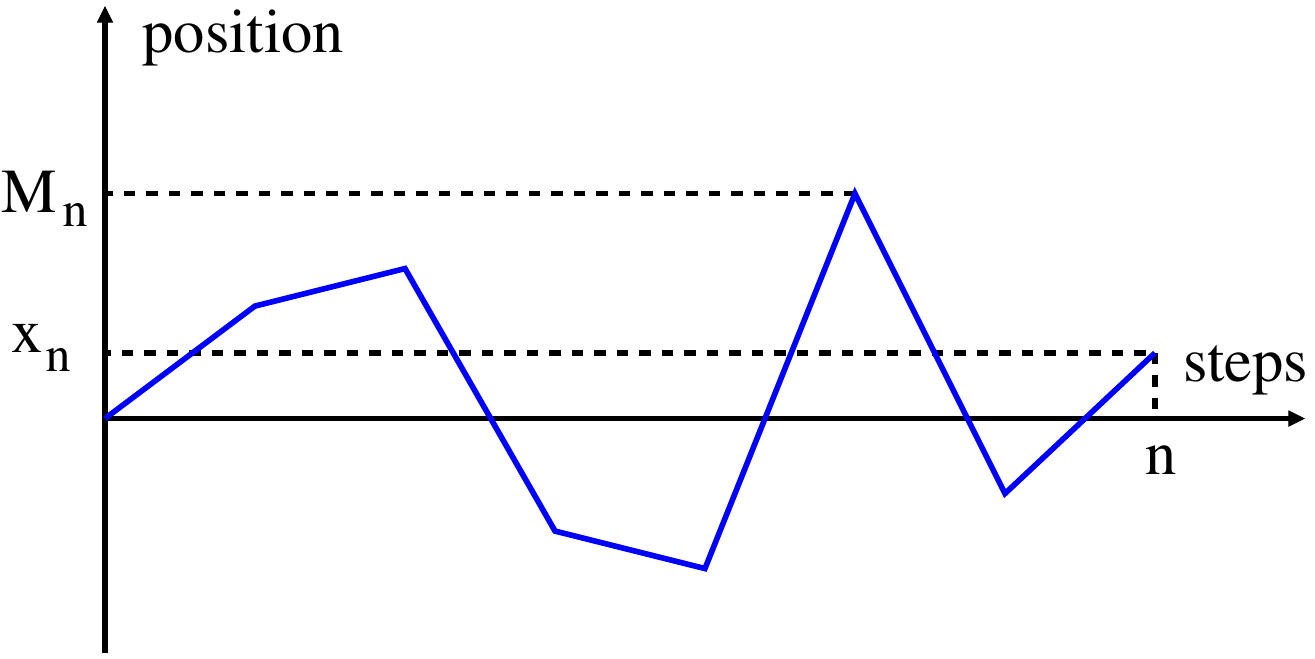}
    \caption{The propagator $p_n(x)$ is the probability that the random walk is located at $x$ at step $n$ given that it started from $x_0=0$. The Spitzer formula in (\ref{eq:spitzer}) gives the generating function of the double Laplace-Fourier transform of the joint distribution of the maximum $M_n$ and the final position $x_n$ of a discrete-time RW of $n$ steps. }
    \label{fig:xnMn}
  \end{center}
\end{figure}
From the Markov rule (\ref{eq:xm}), one can derive the following integral equation for the evolution of the propagator:
\begin{align}
  p_n(x)  = \int_{-\infty}^\infty dy\, p_{n-1}(y) f(x-y)\,, \label{eq:prop}
\end{align}
which simply states that for the particle to be located at $x$ at step $m$, it must have been at some $y$ at step $n-1$ and jumped from $y$ to $x$. The probability of this event is then summed over all $y$ and weighted by the jump distribution $f(\eta)$. The initial condition of (\ref{eq:prop}) is $p_{0}(x) =\delta(x)$ which fixes the particle at the origin initially. The integral equation (\ref{eq:prop}) is easily solved in Fourier space and gives (\ref{eq:PI}). For regular jump distributions, whose Fourier transform behaves as $\hat f(q) \sim  1 - |q|^\mu$ when $q\to 0$ (\ref{Fourier}), one can analyze the expression (\ref{eq:PI}) in the large $n$ limit and find that it takes the scaling form 
\begin{align}
  p_n(x) \sim \frac{1}{n^{\frac{1}{\mu}}}\mathcal{L}_\mu\left(\frac{x}{n^{\frac{1}{\mu}}}\right)\,,\quad n\to \infty\,,\quad \text{with}\quad \mathcal{L}_\mu(z) = \int_{-\infty}^\infty \frac{dk}{2\pi} e^{-ikz - |k|^\mu}\,,
\end{align} 
where $\mu$ is the Lévy index of the jump distribution (\ref{Fourier}). For $\mu=2$, which corresponds to finite variance jump distributions, the scaling function $\mathcal{L}_2(z)$ is a Gaussian distribution $\mathcal{L}_2(z) = 1/\sqrt{2\pi}e^{-z^2/2}$, which a consequence of the Central Limit Theorem. When $\mu=1$, which corresponds to jump distributions which decay as $f(\eta)\sim |\eta|^{-2}$ when $\eta\to\pm \infty$, the scaling function $\mathcal{L}_1(z)$ is a Cauchy distribution $\mathcal{L}_1(z) = 1/[\pi(1+z^2)]$. In general, the scaling function $\mathcal{L}_\mu(z)$ does not have an explicit form and is referred to as a stable law of index $\mu$. Random walks with different Lévy index $\mu$ behave very differently as $\mu$ becomes lower than $\mu=2$, as the second moment of the jump distribution is infinite, and as it becomes lower than $\mu=1$ as the first moment becomes infinite as well (see figure \ref{fig:trajmu}). One salient feature is that random walks with Lévy index $\mu\geq 1$ are recurrent, which means that they return to the origin infinitely many times with probability one, whereas random walks with $\mu<1$ are transient, which means that the probability that they return to the origin infinitely many times is zero.

\begin{figure}[t]
  \begin{center}
    \includegraphics[width=0.3\textwidth]{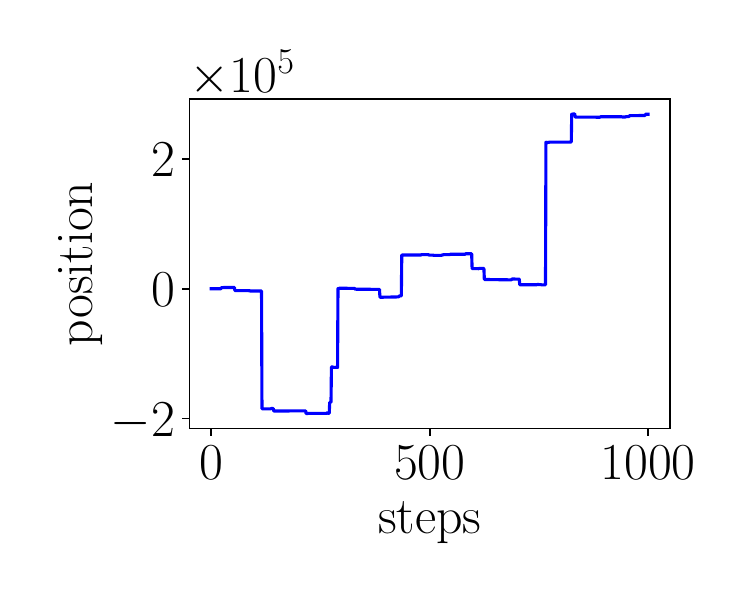}\includegraphics[width=0.3\textwidth]{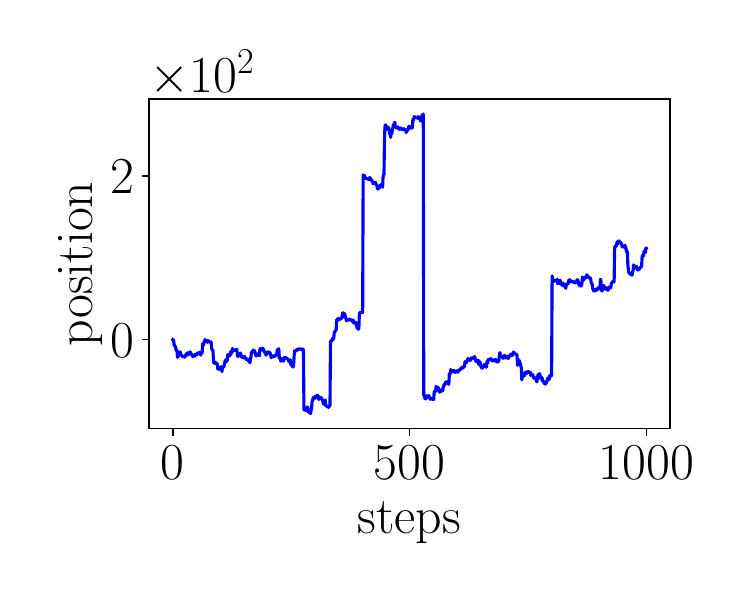}\includegraphics[width=0.3\textwidth]{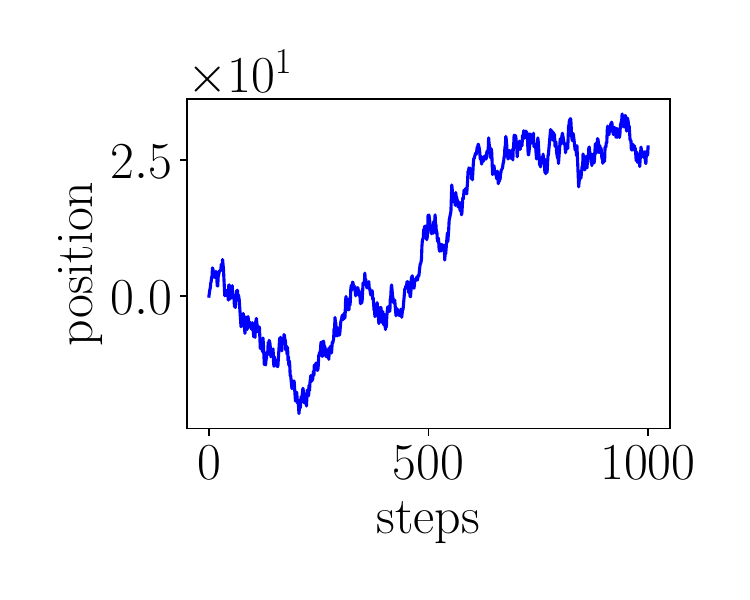}
    \caption{Sample paths of random walks during $n=10^3$ steps with different Lévy index $\mu$ [see equation (\ref{Fourier})]. When $0<\mu<1$, the jump distribution has an infinite first moment and trajectories are dominated by big jumps (left panel). When $1<\mu<2$, the jump distribution has a finite first moment but an infinite second moment and the trajectories look like a combination of jumps and continuous paths (center panel). When $\mu=2$, the jump distribution has a finite second moment, and the trajectories resemble a continuous path (right panel). The trajectories in the left, center, and right panels were generated with a Student distribution of index $\mu=0.5$, a Student distribution with index $\mu=1.5$, and a Gaussian distribution respectively.}
    \label{fig:trajmu}
  \end{center}
\end{figure}

One of the simplest EVS questions that one can ask is ``What is the probability that the minimum of the set of positions $\{x_1,\ldots,x_n\}$ is positive ?''. This is a rather natural question to ask as it corresponds to the fraction of all the RW trajectories that do not cross the origin up to step $n$ (see figure \ref{fig:surv}).
\begin{figure}[t]
  \begin{center}
    \includegraphics[width=0.5\textwidth]{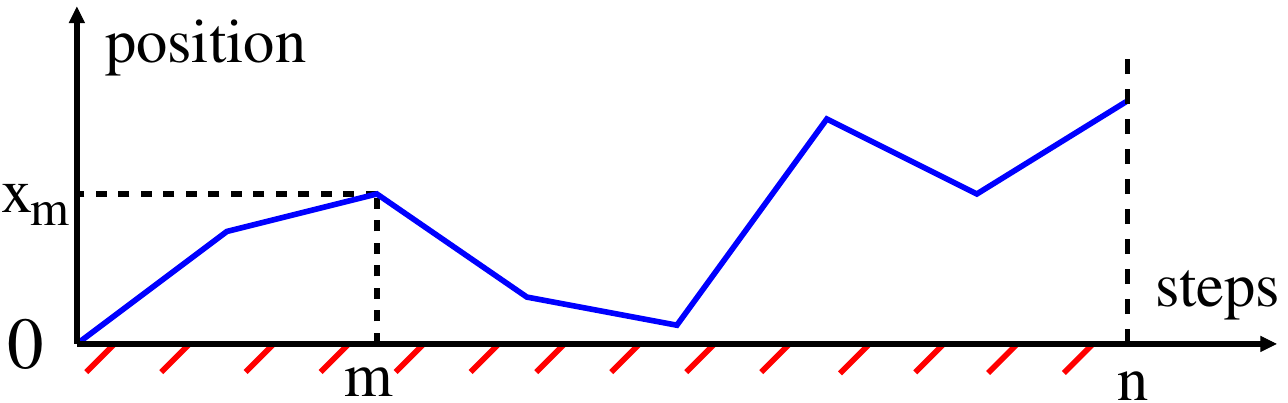}
    \caption{The survival probability $S_n$ of a random walk up to step $n$ is the probability that the random walk did not cross the origin up to step $n$ given that it started from $x_0=0$.}
    \label{fig:surv}
  \end{center}
\end{figure}
 This probability is usually called ``survival probability'' as it is the probability that the RW survives on the positive axis. The survival probability $S_n$ is therefore defined as the probability that all the positions are positive up to step $n$ (see figure \ref{fig:surv}):
\begin{align}
  S_n = \text{Prob.}\left(x_1\geq 0,\ldots,x_n\geq 0\,|\,x_0=0\right)\,.\label{eq:survd}
\end{align}

In terms of the jump distribution $\eta_m$'s defining the jumps in the RW in (\ref{eq:xm}), the survival probability reads explicitly as 
\begin{align}
  S_n = \int_{-\infty}^\infty \left[\prod_{i=1}^n d\eta_i f(\eta_i)\right]\Theta(\eta_1)\Theta(\eta_1+\eta_2) \ldots \Theta(\eta_1+\ldots+\eta_n)\,,\label{eq:Sn}
\end{align}
where the product of Heavyside step functions encodes the constraints that the positions are positives and the term in brackets weights each jump. Naively, by looking at (\ref{eq:Sn}), the survival probability of a RW seems difficult to compute and will \textit{a priori} depend on the jump distribution $f(\eta)$ as it appears explicitly in the multi-dimensional integral. However, a remarkable result in the theory of RWs, which is due to Sparre Andersen \cite{SA}, states that the survival probability in (\ref{eq:Sn}) is universal and given by 
\begin{align}
  S_n = \binom{2n}{n} 2^{-2n}\,.\label{eq:SAres}
\end{align}
For large $n$, the survival probability decays as $S_n\sim 1/\sqrt{\pi n}$.
This result is valid for any symmetric and continuous jump distribution $f(\eta)$. This result is a very nice example of the concept of universality and how symmetry plays a crucial role. To illustrate this, let us check this result for the first values of $n$. For $n=1$, the expression (\ref{eq:Sn}) reads
\begin{align}
  S_1 = \int_{0}^{\infty} d\eta_1 f(\eta_1)\,.
\end{align}
As the jump distribution is symmetric and that it is normalized to unity, the integral directly evaluates to $S_1=1/2$, independently of $f(\eta)$. When $n=2$, the expression (\ref{eq:Sn}) reads
\begin{align}
  S_2 = \int_0^\infty  d\eta_1\int_{-\eta_1}^\infty  d\eta_2 f(\eta_1) f(\eta_2)\,.\label{eq:S2}
\end{align}
The weight function $f(\eta_1)f(\eta_2)$ has two types of symmetries: (i) reflection symmetry $\eta_i\leftrightarrow -\eta_i$ for $i=1,2$ since the jump distribution is symmetric (ii) permutation symmetry $\eta_1 \leftrightarrow \eta_2$ which is due to the fact the weight function takes a product form as the jumps are independent. In the $\eta_1$--$\eta_2$ plane, these symmetries split the plane into $8$ wedges (see figure \ref{fig:SA}). Due to the symmetries, integrating the weight function over any of these wedges yield the same value. Because the weight function is normalized to unity over the whole plane, this gives that the integral over any of these wedges is exactly $1/8$. Because the integration domain in (\ref{eq:S2}) covers exactly three of these wedges, this gives $S_2=3/8$, which matches with the general expression (\ref{eq:SAres}) evaluated at $n=3$. Extrapolating this reasoning to arbitrary $n$, we qualitatively understand that it is the role of the symmetries of the weight function in brackets in (\ref{eq:Sn}) which are strong enough to constraint the value of the integration to become (\ref{eq:SAres}). A full proof for (\ref{eq:SAres}) involving combinatorial arguments can be found in the original paper \cite{SA}.
\begin{figure}[t]
  \begin{center}
    \includegraphics[width=0.4\textwidth]{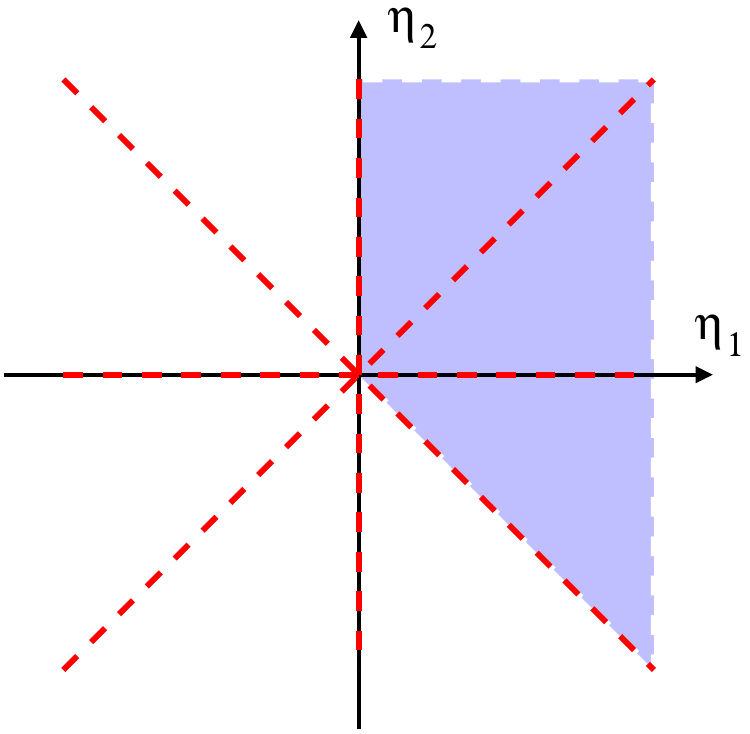}
    \caption{In the $\eta_1$--$\eta_2$ plane, the weight function $f(\eta_1)f(\eta_2)$ has four symmetry lines $\eta_1=0$, $\eta_2=0$, $\eta_1+\eta_2=0$ and $\eta_1-\eta_2=0$ (depicted in red). These lines split the plane into $8$ wedges. The integration domain of the integral in (\ref{eq:S2}) exactly overlaps three of the wedges (shaded blue area).}
    \label{fig:SA}
  \end{center}
\end{figure}

Let us mention a recent extension of the Sparre Andersen result to higher dimensions. It considers a random walk $\bm{x}_m$ in $d$ dimensions which evolves according to 
\begin{align}
  \bm{x}_m =  \bm{x}_{m-1} + \bm{\eta}_m\,,
\end{align}
where $\bm{x}_0=\bm{0}$ and $\bm{\eta}_m$ are i.i.d.~random vectors drawn from a radially symmetric distribution $f(\bm{\eta}_m)$. A natural extension to the survival probability problem in $d$ dimensions is to consider the following question: ``What is the probability that the RW trajectory remains in some open linear half-space?". In other words, this is the probability that the origin is not included in the convex hull of the RW trajectory (the convex hull being the smallest polygon that contains its trajectory). Remarkably, it was shown that this probability is universal and given by \cite{Kabluchko17a,Kabluchko17b}
\begin{align}
  \text{Prob.}\left(\bm{0} \notin \text{Conv}\left(\bm{x}_1,\ldots,\bm{x}_n\right)\right) = \frac{2}{2^n n!}\sum_{k=1}^{\lceil d/2 \rceil}B(n,d-2k+1)\,,\label{eq:kab}
\end{align}
where $B(n,k)$ are the coefficient of the polynomial 
\begin{align}
  (t+1)(t+3)\ldots (t+2n-1) = \sum_{k=0}^{n} B(n,k)t^k\,.\label{eq:Bnk}
\end{align}
Setting $d=1$, we recover the Sparre Andersen result in (\ref{eq:SAres}). In the limit of large $n$, the absorption probability (\ref{eq:kab}) decays as \cite{Kabluchko17b}
\begin{align}
\text{Prob.}\left(\bm{0} \notin \text{Conv}\left(\bm{x}_1,\ldots,\bm{x}_n\right)\right) \sim \frac{1}{2^{d-2}(d-1)!\sqrt{\pi }}\,\frac{\ln(n)^{d-1}}{\sqrt{n}}\,,\quad n\to \infty\,.\label{eq:askab}
\end{align} 
Note that the generalization of Sparre Andersen result in (\ref{eq:kab}) is not much known in the physics literature. In particular, one can use it to study the survival probability of continuous-time persistent random walks and extend some recent results on the universal survival probability of the run-and-tumble particle \cite{MoriL20,MoriE20} (see Appendix \ref{app:convRTP}).

The survival probability is not the only quantity that exhibits universality. For instance, the number of steps spent above the origin $T_n$ is also universal for random walks with symmetric and continuous jump distributions starting from $x_0=0$. One way to see this is to rely on the Sparre Andersen result as we briefly show now. Let us express $T_n$ in terms of the positions $x_n$'s as
\begin{align}
  T_n = \sum_{i=1}^n \Theta(x_i)\,,\label{eq:defTn}
\end{align}
which simply counts the number of the $n$ steps which are above the origin. Let us compute the generating function $\langle z^{T_n}\rangle $ of the random variable $T_n$. By using its definition in (\ref{eq:defTn}), we find
\begin{align}
  \langle z^{T_n}\rangle = \langle \prod_{i=1}^n z^{\Theta(x_i)}\rangle = \langle \prod_{i=1}^n \left[1 + (z-1)\Theta(x_i)\right]\rangle\,,\label{eq:zTn}
  \end{align}
  where we used that $\Theta(x)$ is a binary variable. Expanding the product in (\ref{eq:zTn}), we find
  \begin{align} 
  \langle z^{T_n}\rangle  &= \sum_{k=0}^n (z-1)^k \sum_{0< i_1< \ldots< i_k\leq n} \langle \Theta(x_{i_1})\ldots \Theta(x_{i_k})\rangle\,.\label{eq:zTnc}
\end{align}
Explicitly writing the average over all trajectories gives (see figure \ref{fig:avg})
\begin{figure}[t]
  \begin{center}
    \includegraphics[width=0.5\textwidth]{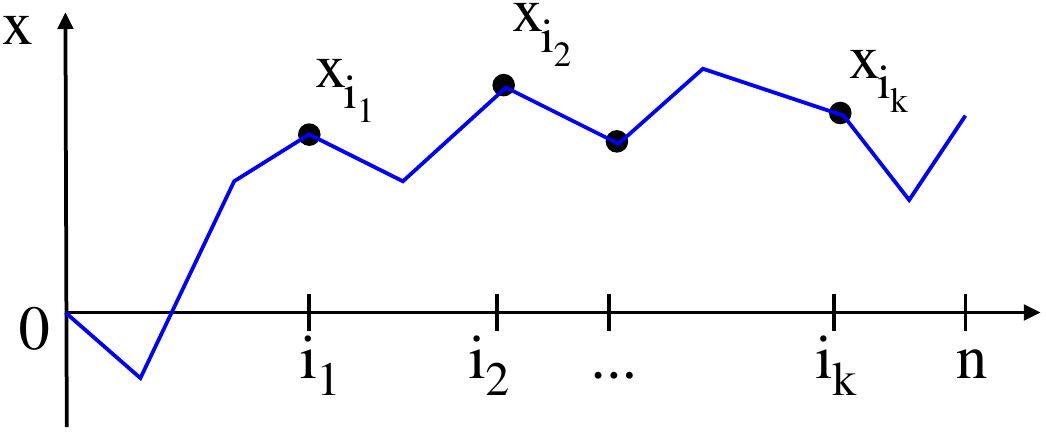}
    \caption{The average $\langle \Theta(x_{i_1})\ldots \Theta(x_{i_k})\rangle$ is the probability that the random walk is above the origin at the intermediate times $i_1,\ldots,i_k$ given that it started at $x_0=0$. Note that the random walk is allowed to propagate freely in between those times. }
    \label{fig:avg}
  \end{center}
\end{figure}
\begin{align}
   \langle \Theta(x_{i_1})\ldots \Theta(x_{i_k})\rangle &=  \int_{0}^\infty dx_{i_1}\ldots dx_{i_k}p_{i_1}(x_{i_1}) p_{i_2-i_1}(x_{i_2}-x_{i_1})\ldots p_{i_k-i_{k-1}}(x_{i_k}-x_{i_{k-1}})\,,\label{eq:Tnk}
\end{align}
where $p_k(x)$ is the propagator of the random walk in (\ref{eq:PI}). Recognizing the convolution structure over the summation indices in (\ref{eq:zTnc}), we take a generating function with respect to $n$ which gives
\begin{align}
  \sum_{n=0}^\infty s^n \langle z^{T_n}\rangle = \sum_{k=0}^\infty \frac{(z-1)^k}{1-s} \int_{0}^\infty dx_{1}\ldots dx_{k} \bar p_s(x_1)\bar p_s(x_2-x_1)\ldots \bar p_s(x_k-x_{k-1})\,,\label{eq:Tnke}
\end{align}
where $\bar p_s(x) = \sum_{n=1}^\infty s^n p_n(x)$ is the generating function of the propagator. We now show that the expression in (\ref{eq:Tnke}) can be interpreted as an effective random walk. To see this, we define an effective jump distribution, parameterized by $s$, which reads
\begin{align}
  \mathrm{f}_s(\eta)= \frac{(1-s)}{s} \bar p_s (\eta)\,,\label{eq:fze}
\end{align}
where the prefactor $(1-s)/s$ in the definition of (\ref{eq:fze}) has been added for normalization. One can check by using that $\int_{-\infty}^\infty d\eta \,\bar p_s(\eta)=\sum_{n=1}^\infty s^n \int_{-\infty}^\infty d\eta\, p_n(\eta)=s/(1-s)$, that the distribution $\mathrm{f}_s(\eta)$ is properly normalized. The expression (\ref{eq:Tnke}) now reads
\begin{align}
  \sum_{n=0}^\infty s^n \langle z^{T_n}\rangle  = \sum_{k=0}^\infty\frac{[s(z-1)]^k}{(1-s)^{k+1}}\int_{0}^\infty dx_{1}\ldots dx_{k}\, \mathrm{f}_s(x_1)\mathrm{f}_s(x_2-x_1)\ldots \mathrm{f}_s(x_k-x_{k-1})\,.\label{eq:Tnkr}
\end{align}
Upon changing variables $\eta_i=x_i-x_{i-1}$, (\ref{eq:Tnkr}) becomes
\begin{align}
  \sum_{n=0}^\infty s^n \langle z^{T_n}\rangle  = \sum_{k=0}^\infty\frac{[s(z-1)]^k}{(1-s)^{k+1}} \int_{-\infty}^\infty \left[\prod_{i=1}^k d\eta_i\, \mathrm{f}_s(\eta_i)\right]\Theta(\eta_1)\Theta(\eta_1+\eta_2) \ldots \Theta(\eta_1+\ldots+\eta_k)\,.\label{eq:Tnkra}
\end{align}
We now recognize that the multiple integrals in (\ref{eq:Tnkra}) is the survival probability after $k$ steps of a random walk with a symmetric and continuous jump distribution $\mathrm{f}_s(\eta)$. Due to the Sparre Andersen theorem, this survival probability is independent of the jump distribution and is given by (\ref{eq:SAres}). Therefore (\ref{eq:Tnkra}) reads
\begin{align}
     \sum_{n=0}^\infty s^n \langle z^{T_n}\rangle   = \sum_{k=0}^\infty \frac{[s(z-1)]^k}{(1-s)^{k+1}} S_k\,,
\end{align}
where $S_k$ is given in (\ref{eq:SAres}). Using that $\sum_{k=0}^\infty z^k S_k=1/\sqrt{1-z}$, we find
\begin{align}
 \sum_{n=0}^\infty s^n \langle z^{T_n}\rangle  =  \frac{1}{\sqrt{(1-s)(1-sz)}}\,,\label{eq:asl}
\end{align}
Note that the double generating function (\ref{eq:asl}) can be inverted and gives
\begin{align}
  \text{Prob.}\left(T_n=m\right) = \binom{2m}{m}\binom{2(n-m)}{n-m} 2^{-2n}\,,\label{eq:Tnfin}
\end{align}
which is the discrete-time version of the arc-sine law \cite{Feller,Levy} and is remarkably independent of the jump distribution. 
A plot of the distribution is shown in figure \ref{fig:arcsine}. 
\begin{figure}[t]
  \begin{center}
    \includegraphics[width=0.5\textwidth]{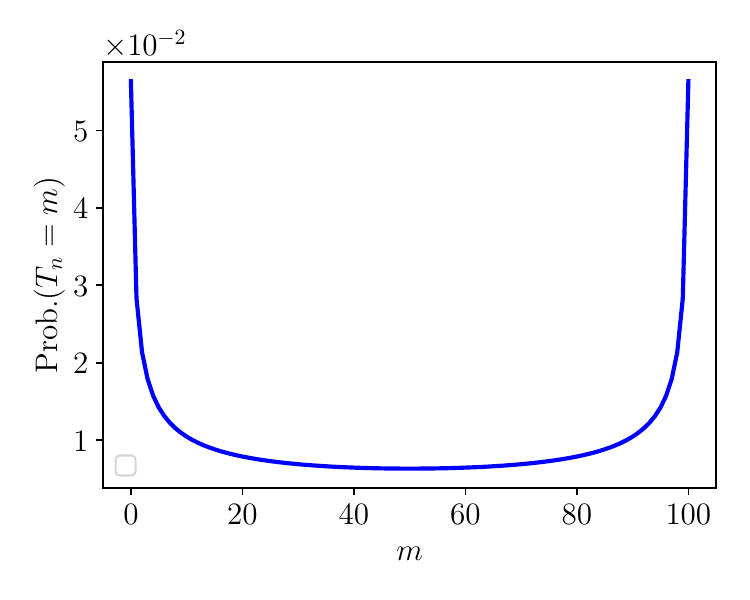}
    \caption{Distribution of the time spent above the origin $T_n$ after $n=100$ steps for a random walk with a continuous and symmetric jump distribution (see equation (\ref{eq:Tnfin})).}
    \label{fig:arcsine}
  \end{center}
\end{figure}
Note that (\ref{eq:Tnfin}) has a well-defined continuous time limit in which $n\to \infty$ and $m=O(n)$, which is the well-known ``arc-sine law'':
\begin{align}
   \text{Prob.}\left(T_n=m\right) \sim \frac{1}{\pi\sqrt{m(n-m)}}\,,\quad n\to \infty\,,\quad m=O(n)\,.\label{eq:Tnfina}
\end{align}
Note that the derivation above, which involves an effective random walk, is original and has not been seen elsewhere up to our knowledge. The idea of using an effective random walk will be used again in the next section.

Another universal quantity is the time $\tau_n$ at which a random walk of $n$ steps reaches its maximum. To see this, let us fix the location of the maximum at $M_n$ and decompose the trajectories into two parts: (I) the first one which goes from the origin to $M_n$ in $\tau_n$ steps while staying below $M_n$, (II) the second one which goes from $M_n$ and stays below it during the remaining $n-\tau_n$ steps (see figure \ref{fig:tmax}). 
\begin{figure}[ht]
  \begin{center}
    \includegraphics[width=0.45\textwidth]{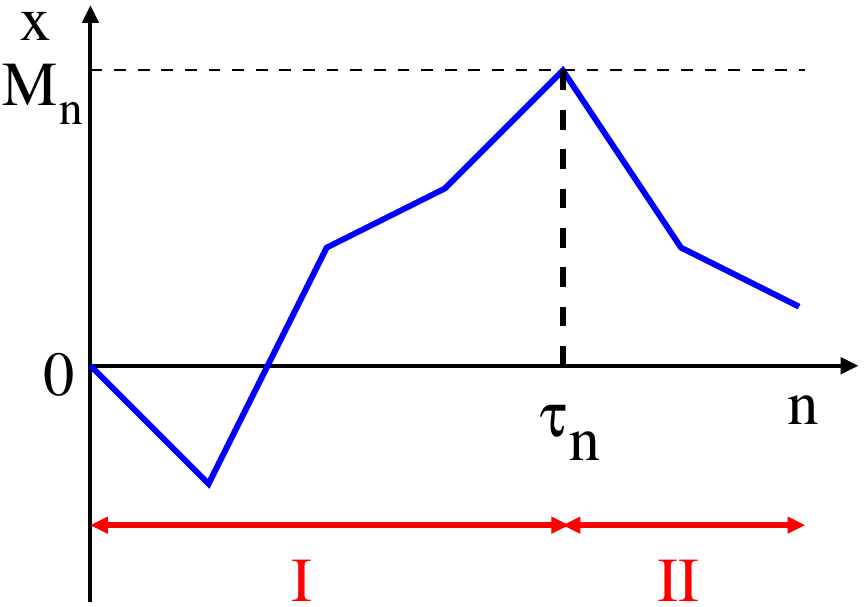}
    \caption{The maximum $M_n$ and the time $\tau_n$ at which it is achieved are two examples of EVS for discrete-time random walks of $n$ steps. Notice that the trajectories can be decomposed into two parts: (I) a path that propagates from the origin to the maximum $M_n$ in $\tau_n$ steps while staying below $M_n$, (II) a second path that stays below $M_n$ during $n-\tau_n$ steps.  }
    \label{fig:tmax}
  \end{center}
\end{figure}
Upon using space translation invariance and time-reversal symmetry, the first part can be seen backward in time as the survival probability of a random walk during $\tau_n$ steps with a initial position located at $M_n$. The second part (seen forward in time) can be seen as the survival probability during $n-\tau_n$ steps starting from $M_n$. Due to the Markov nature of the random walk, the two parts are independent. Upon integrating over all possible values of the maximum $M_n$, we find that the distribution of the time of the maximum $\tau_n$ is given by the product of two survival probabilities:
\begin{align}
  \text{Prob.}\left(\tau_n=m\right) = S_{m}S_{n-m}\,.\label{eq:taum}
\end{align}
Upon inserting the expression of $S_n$ given in (\ref{eq:SAres}), we find  
\begin{align}
  \text{Prob.}\left(\tau_n=m\right) =\binom{2m}{m}\binom{2(n-m)}{n-m} 2^{-2n}\,,
\end{align}
which, interestingly, is the same distribution as for $T_n$ in (\ref{eq:Tnfin}). Note that by requiring that the probability distribution (\ref{eq:taum}) is normalized to unity, one can recover Sparre Andersen result in a simple way \cite{MMS20a}.

We saw that the symmetry condition of the jump distribution is crucial for the Sparre Andersen result in (\ref{eq:SAres}) to hold. A natural question to ask is what happens if the jump distribution is not symmetric?  As expected, the survival probability will not be universal anymore. However, there exists a generalization of (\ref{eq:SAres}) to asymmetric (but still continuous) jump distribution which reads
\begin{align}
  \bar S(s) = \sum_{n=0}^\infty s^n \,S_n = \exp\left[\sum_{n=1}^\infty \frac{s^n}{n}\text{Prob.}\left(x_n\geq 0\right)\right]\,,\label{eq:genSA}
\end{align}
where $\text{Prob.}\left(x_n\geq 0\right)$ is the probability that the $n^{\text{th}}$ step is positive. Note that if the jump distribution is symmetric, this probability is exactly $1/2$ and (\ref{eq:genSA}) becomes $\bar S(s) = 1/\sqrt{1-s}$ which is exactly the generating function of (\ref{eq:SAres}). 

Another way to ``break'' the universality of the Sparre Andersen result in (\ref{eq:SAres}) is to consider a random walk not starting from $x_0=0$ but from a strictly positive initial position $x_0>0$, but still with a symmetric and continuous jump distribution. In this case, we naturally extend the survival probability in (\ref{eq:survd}) to an arbitrary initial position $x_0\geq 0$ as
\begin{align}
  S_n(x_0) = \text{Prob.}\left(x_1\geq 0,\ldots,x_n\geq 0|x_0\right)\,.\label{eq:survdx}
\end{align} 
As for the propagator in (\ref{eq:prop}), the survival probability $S_n(x_0)$ satisfies an integral equation which reads
\begin{align}
  S_n(x_0) = \int_0^\infty dy\, S_{n-1}(y)\,f(y-x_0)\,,\label{eq:survi}
\end{align}
which states that for the particle to survive during $n$ steps starting from $x_0$, it must jump to some positive $y$ and survive from $y$ in the remaining $n-1$ steps. The probability of this event is then summed over all $y\geq 0$ and weighted by the jump distribution $f(\eta)$. The initial condition of (\ref{eq:survi}) is $S_0(x_0)=\Theta(x_0)$ which says that the survival probability is initially one if $x_0\geq 0$ and zero otherwise. The integral equation (\ref{eq:survi}) belongs to the Wiener-Hopf class of integral equations and turns out to be much harder to solve than the one in (\ref{eq:prop}) as the integration only takes place on the positive axis. However, because the kernel $f(\eta)$ is a probability density, it has a solution known given by the so-called Pollaczek-Spitzer formula. The formula for the generating function of the Laplace transform of the survival probability
after $n$ steps for a random walk (with a symmetric jump distribution) starting from $x_0$ is given by 
\begin{align}
  \sum_{n=0}^\infty s^n \int_0^\infty dx_0\,S_n(x_0) e^{-p x_0} = \frac{1}{p\sqrt{1-s}}\exp\left[-\frac{p}{\pi} \int_0^\infty dk\, \frac{\ln(1-s \hat f(k))}{p^2+k^2}\right]\,,\label{eq:SAres2}
\end{align}
where $\hat f(k)$ is the Fourier transform of the jump distribution (\ref{Fourier}). An original derivation of this formula is given in Appendix \ref{app:PSf}. As expected, the expression (\ref{eq:SAres2}) explicitly depends on the jump distribution. While the survival probability starting from $x_0>0$ is not universal anymore for all $n$, one might ask if there are any universal features in the limit of a large number of steps $n\to \infty$? It turns out that the expression (\ref{eq:SAres2}) possesses rather rich limiting behaviors, depending if the initial position $x_0$ is scaled with $n$ or is kept fixed \cite{MMS17}. For smooth jump distributions which behaves as in (\ref{Fourier}), the two scaling behaviors are given by \cite{MMS17}:
\begin{align}
  S_n(x_0) \sim \left\{\begin{array}{ll}
    \frac{1}{\sqrt{n}}U(x_0)\,,& x_0 =O(1)\,,\\
    V_\mu\left(\frac{x_0}{n^{1/\mu}}\right)\,,&x_0=O(n^{\frac{1}{\mu}})\,,
  \end{array}\right.\quad n\to \infty\,,\label{eq:Snxa}
\end{align}
where the function $U(x_0)$ is non-universal and given by 
\begin{align}
  \int_0^\infty dx_0 e^{-\lambda x_0}U(x_0) = \frac{1}{\lambda\sqrt{\pi}}\exp\left[-\frac{\lambda}{\pi}\int_0^\infty \frac{dk}{\lambda^2+k^2}\ln(1-\hat f(k))\right]\,,\label{eq:UMou}
\end{align}
and, in contrast, $V_\mu(z)$ is universal, in the sense that it only depends on the Lévy index $\mu$ of the distribution (\ref{Fourier}), and is given by the following double integral transform \cite{MMS17,Darling56}
\begin{align}
  \int_0^\infty dy e^{-y} y^{\frac{1}{\mu}} \int_0^\infty dz V_\mu(z)e^{-w y^{\frac{1}{\mu}}z} = \frac{1}{w}J_\mu(w)\,,\quad \text{with}\qquad J_\mu(w) =e^{-\frac{1}{\pi}\int_0^\infty \frac{du}{1+u^2}\ln(1+w^\mu u^\mu)}\,.
\end{align}
For finite variance distributions $\mu=2$, the scaling function $V_\mu(z)$ can be computed explicitly and evaluates to $V_2(z)=\text{erf}(z/2)$, which is the well-known survival probability of Brownian motion \cite{bray2013persistence,redner2001guide}. The asymptotic behavior in (\ref{eq:Snxa}) is in agreement at $x_0=0$ with the Sparre Andersen result in (\ref{eq:SAres}) as $U(x_0=0)=1/\sqrt{\pi}$. In addition, the two regimes in (\ref{eq:Snxa}) match smoothly as $U(x_0)$ behaves as $U(x_0)\sim A_\mu x_0^{\mu/2}$ for $x_0\to \infty$ and $V_\mu(z)$ behaves as $V_\mu(z)\sim A_\mu z^{\mu/2}$ for $z\to 0$, with the constant $A_\mu$ given by $A_\mu=1/[\sqrt{\pi}\Gamma(1+\mu/2)]$ \cite{MMS17}.

To make a connection with the results of the previous section on the maximum of i.i.d.~random variables, we would like to know what is the distribution of the maximum $M_n=\max\{x_0,\ldots,x_n\}$ of a random walk of $n$ steps (see figure \ref{fig:xnMn})? The answer to this question turns out to be closely related to the survival probability discussed above. Indeed, notice that for a random walk with a symmetric jump distribution, we have, for $M\geq 0$,
\begin{align}
\text{Prob.}\left(M_n \leq M |x_0=0\right)&=\text{Prob.}\left(x_1\leq M,\ldots,x_n \leq M |x_0=0\right)\nonumber \\
 &= \text{Prob.}\left(x_1\leq 0,\ldots,x_n \leq 0|x_0=-M\right)\nonumber\\
  &=\text{Prob.}\left(x_1\geq 0,\ldots,x_n \geq 0|x_0=M\right)\nonumber\\
  &= S_n(M)\,,\label{eq:idmaxS}
\end{align}
where we used space translation invariance in the second line, space reflection invariance in the third line and recognized the survival probability in (\ref{eq:survd}) starting from $x_0=M$. One can then analyze the Sparre-Andersen formula (\ref{eq:SAres2}) to obtain insights on the statistics of the maximum $M_n$, such as the ones discussed in Section \ref{sec:expi}.

Going beyond the global maximum, one might ask about the statistics of the $k^{\text{th}}$ maximum of the set of positions $\{x_0,\ldots,x_n\}$ which are strongly correlated (see figure \ref{fig:model}). Interestingly, there exists an identity that relates the $k^{\text{th}}$ maximum $M_{k,n}$ with the maximum and the minimum of two independent copies of the same random walk. This identity was explicitly derived by Wendel in \cite{Wendel} (but was probably previously known to others) and reads
\begin{align}
  M_{k,n} = M^{(1)}_{n-k} + m^{(2)}_k\,,\label{eq:wid}
\end{align}
where $M^{(1)}_{n-k}$ is the maximum of a random walk after $n-k$ steps and $m^{(2)}_k$ is the minimum of another independent copy of the random walk after $k$ steps. The equality between the random variables in (\ref{eq:wid}) is valid in distribution. Since the identity (\ref{eq:wid}) involves two independent copies of the global maximum and global minimum, it is sufficient to study the statistics of the global maximum $M_n$ (resp.~global minimum $m_n$) of a single random walk to obtain some results on the $k^{\text{th}}$ maximum $M_{k,n}$ \cite{Chaumont,Port,Dassios,Embrechts,Dassiosa}. 

The statistics of the gaps $\Delta_{k,n}$ between two consecutive maxima $M_{k,n}$ and $M_{k+1,n}$ were not known until recently (see Section \ref{sec:uni}). The main difficulty to study them is that it does not suffice to know the marginal distribution of $M_{k,n}$ and $M_{k+1,n}$ to devise the distribution of the gaps as $M_{k,n}$ and $M_{k+1,n}$ are strongly correlated. Their study required a new method which is discussed in the next section.

Finally, let us mention that there are other extreme observables that are interesting to study in the case of discrete-time random walks. For instance, the study of records, which arise when an event is of a larger magnitude than all the previous one, has attracted a lot of attention recently. The study of records attempts to answer some natural questions: ``How many records occur in a given time?'', ``How long do they survive?'', ``What is the longest and shortest age of a record?'', etc.  While records are well understood for i.i.d.~random variables \cite{VBN04}, much less is known for strongly correlated variables. It turns out that discrete-time random walks serve again as a convenient toy model to study them. Recently, important progress has been made and has shown that they display rather rich and sometimes universal behaviors with respect to the jump distribution (see for instance the pedagogical reviews \cite{SMS14r,GodrecheMecords17}).

\section{Order statistics for random walks}
\label{sec:ord}
In this section, we present the main idea that was used to obtain the results on the order statistics of random walks in Section \ref{sec:uni}. This idea originates from a paper by Spitzer entitled ``On Interval Recurrent Sums of Independent Random Variables'' \cite{Spi56}. We will first explain the problem he was interested in and how he solved it. Then, we will briefly show how we extended his idea to obtain the results presented in Section \ref{sec:uni}. The detailed derivations can be found in the paper whose abstract is given at the end of this section on p.~\pageref{chap:A14}.

In \cite{Spi56}, Spitzer focused on discrete-time random walks as in (\ref{eq:xm}) with continuous and symmetric jump distributions $f(\eta)$ with a Lévy index $\mu \geq 1$ [see equation \ref{Fourier}]. He was interested in the number of times $\nu_n(a,\Delta)$ a random walk of $n$ steps visits the interval centered on $a$ of width $\Delta$ (see figure \ref{fig:Spi}):
\begin{align}
  \nu_n(a,\Delta) = \sum_{i=1}^n \Theta\left(\frac{\Delta}{2} - |x_i-a|\right)\,.\label{eq:nun}
\end{align}
\begin{figure}[t]
  \begin{center}
    \includegraphics[width=0.5\textwidth]{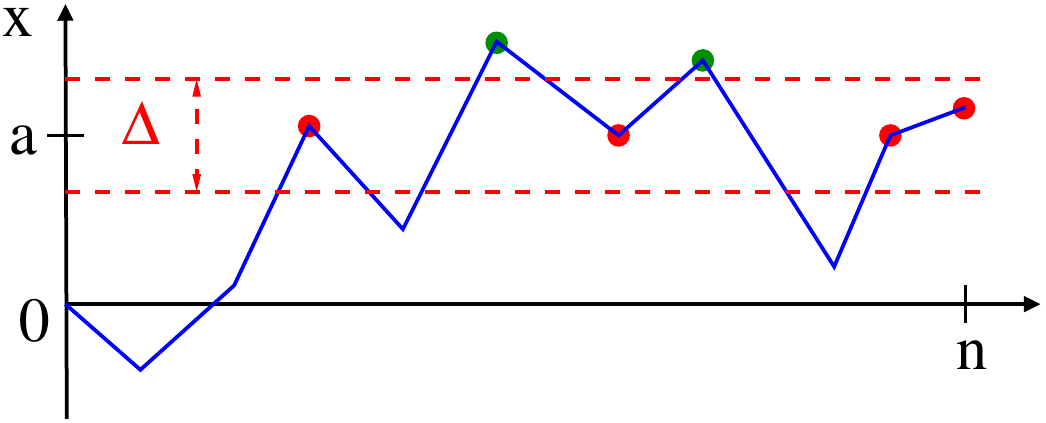}
    \caption{For a random walk of $n$ steps, the quantity $\nu_n(a,\Delta)$ counts the number of times it visits the interval centered on $a$ of width $\Delta$, and the quantity $\kappa_n(a,\Delta)$ counts the number of times it goes above that interval. In the trajectory represented, the random walk of $n=10$ steps visits the interval $\nu_{10}(a,\Delta)=4$ times (red dots) and goes above it $\kappa_{10}(a,\Delta)=2$ times (green dots). }
    \label{fig:Spi}
  \end{center}
\end{figure}
Because random walks with Lévy index $\mu\geq 1$ are recurrent, they will eventually visit any interval at least once with probability $1$, which means that
\begin{align}
 \lim_{n\to\infty} \text{Prob.}\left[\nu_n(a,\Delta) = 0\right] = 0\,.\label{eq:lim1}
\end{align}
The result (\ref{eq:lim1}) holds for any interval of fixed width $\Delta$. However, if one scales $\Delta$ with $n$, such that $\Delta\to 0$ as $n\to \infty$, one should find an interesting limiting behavior.
Spitzer was interested in this limiting behavior and found that
\begin{align}
  \lim_{n\to \infty}\text{Prob.}\left[\nu_n(a,C_{n,\mu}\,\Delta)=0\right] =  {\rm E}_{1-1/\mu}(-\Delta)\,,\label{eq:lims}
\end{align}
where 
\begin{align}
  C_{n,\mu} =
\left\{\begin{array}{ll}
\mu \sin\left( \frac{\pi}{\mu}\right) n^{1/\mu -1} \;, & 1 < \mu \leq 2\,, \\
\pi (\ln n)^{-1} \,, &\mu =1\,,
\end{array}\right.\label{eq:Cnmu}
\end{align}
and $E_p(x)=E_{p,1}(x)$ is the Mittag-Leffler function (\ref{eq:MittagLeff}).
The result (\ref{eq:lims}) is quite remarkable as it does not depend on $a$ (as long as it is fixed) and is universal as it does not depend on the full details of the jump distribution $f(\eta)$, but only on its Lévy index $\mu$. In particular, in the case $\mu = 2$ (corresponding to jump distributions with finite variance), the result (\ref{eq:lims}) reads
\begin{align}
\lim_{n\to \infty}\text{Prob.}\left[\nu_n\left(a,\frac{2\Delta}{\sqrt{n}}\right)=0\right] =  e^{\Delta^2} \,{\rm erfc}(\Delta)  \;,
 \label{res_alpha2}
\end{align}
where ${\rm erfc}(z) = \frac{2}{\sqrt{\pi}}\int_x^\infty e^{-t^2}\, dt$ is the complementary error function. In the case of $\mu=1$, the limiting distribution (\ref{eq:lims}) becomes
\begin{align}
  \lim_{n\to \infty}\text{Prob.}\left[\nu_n\left(a,\frac{\Delta}{\pi\ln(n)}\right)=0\right] = \frac{1}{1+\Delta}\,.\label{eq:lims1}
\end{align}

Let us now sketch how Spitzer showed this result. His derivation is similar to the one that was used to derive the distribution of the occupation time in (\ref{eq:defTn}). We consider the generating function $\langle z^{\nu_n(a,\Delta)}\rangle$ of the random variable $\nu_n(a,\Delta)$. Following similar steps as in (\ref{eq:zTn})-(\ref{eq:zTnc}), we find
\begin{align}
  \langle z^{\nu_n(a,\Delta)}\rangle = \sum_{k=0}^n (z-1)^k \sum_{0<i_1<\ldots<i_k\leq n} \left\langle \Theta\left(\frac{\Delta}{2} - |x_{i_1}-a|\right)\ldots \Theta\left(\frac{\Delta}{2} - |x_{i_k}-a|\right)\right\rangle\,.\label{eq:nunad}
\end{align}
Explicitly writing the average over all trajectories gives (see figure \ref{fig:avgspi})
\begin{align}
  \Bigg\langle \Theta\Bigg(\frac{\Delta}{2} &- |x_{i_1}-a|\Bigg)\ldots \Theta\left(\frac{\Delta}{2} - |x_{i_k}-a|\right)\Bigg\rangle\nonumber\\
  & =\int_{a-\frac{\Delta}{2}}^{a+\frac{\Delta}{2}} dx_{i_1}\ldots dx_{i_k}p_{i_1}(x_{i_1}) p_{i_2-i_1}(x_{i_2}-x_{i_1})\ldots p_{i_k-i_{k-1}}(x_{i_k}-x_{i_{k-1}})\,,\label{eq:nunk}
\end{align}
\begin{figure}[t]
  \begin{center}
    \includegraphics[width=0.5\textwidth]{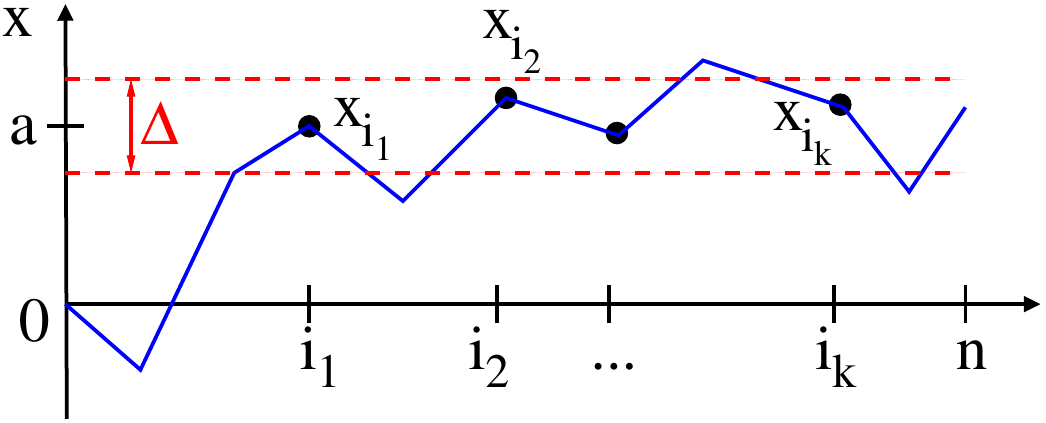}
    \caption{The average $\left\langle \Theta\left(\Delta/2 - |x_{i_1}-a|\right)\ldots \Theta\left(\Delta/2 - |x_{i_k}-a|\right)\right\rangle$ is the probability that the random walk visits the interval located at $a$ of width $\Delta$ at the intermediate times $i_1,\ldots,i_k$ given that it started at $x_0=0$. Note that the random walk is allowed to propagate freely in between those times.}
    \label{fig:avgspi}
  \end{center}
\end{figure}
where $p_k(x)$ is the propagator of the random walk in (\ref{eq:PI}). Recognizing the convolution structure over the summation indices in (\ref{eq:nunad}), we take a generating function with respect to $n$ which gives
\begin{align}
  \sum_{n=0}^\infty s^n \langle z^{\nu_n(a,\Delta)}\rangle  = \sum_{k=0}^\infty \frac{(z-1)^k}{1-s} \int_{a-\frac{\Delta}{2}}^{a+\frac{\Delta}{2}} dx_{1}\ldots dx_{k} \bar p_s(x_1)\bar p_s(x_2-x_1)\ldots \bar p_s(x_k-x_{k-1})\,,\label{eq:Tnkes}
\end{align}
where $\bar p_s(x) = \sum_{n=1}^\infty s^n p_n(x)$ is the generating function of the propagator. We now need to take the limit $n\to \infty$, which corresponds to taking the limit $s\to 1$ in (\ref{eq:Tnkes}). In this limit, the generating function of the propagator in (\ref{eq:PI}) behaves, for $\mu>1$, as 
\begin{align}
  \bar p_s(x) = \sum_{n=1}^\infty s^n \int_{-\infty}^\infty \frac{dk}{2\pi} e^{-ikx}\hat f(k)^n &= \int_{-\infty}^\infty \frac{dk}{2\pi} e^{-ikx}\frac{s\hat f(k)}{1-s\hat f(k)}\nonumber\\
  &\sim \frac{1}{\mu\sin\left(\frac{\pi}{\mu}\right)}\frac{1}{(1-s)^{1-\frac{1}{\mu}}}\,,\quad s\to 1\,,\label{eq:barpas}
  \end{align}
  where we derived the asymptotic behavior by rescaling $k$ by $(1-s)^\frac{1}{\mu}$ and by using the small $k$ expansion of $\hat f(k)$ in (\ref{Fourier}) along with the following identity $
  \int_0^\infty  \frac{dk}{1+k^\mu} = \pi /[\mu \sin\left(\pi/\mu\right)]$.
Note that the asymptotic behavior in (\ref{eq:barpas}) does not depend on $x$ anymore (as long as it is kept fixed). Inserting the asymptotic expansion (\ref{eq:barpas}) into (\ref{eq:Tnkes}), we find, for $\mu>1$,
\begin{align}
  \sum_{n=0}^\infty s^n \langle z^{\nu_n(a,\Delta)}\rangle  &\sim \sum_{k=0}^\infty \frac{(z-1)^k}{1-s} \left(\frac{\Delta}{\mu\sin\left(\frac{\pi}{\mu}\right)}\frac{1}{(1-s)^{1-\frac{1}{\mu}}}\right)^k \,, \quad s\to 1\,.
\end{align}
Inverting the generating function with respect to $s$  by using the Tauberian theorem
\begin{align}
  \sum_{n=0}^\infty z^n a_n &\sim\frac{1}{(1-z)^\alpha}\,,\quad z\rightarrow 1\,,\quad \iff \quad a_n \sim \frac{n^{\alpha-1}}{\Gamma(\alpha)}\,,\quad n\rightarrow\infty\,,\label{eq:taub}
\end{align}
we find, for $\mu>1$,
\begin{align}
  \langle z^{\nu_n(a,\Delta)}\rangle&\sim \sum_{k=0}^\infty \frac{(z-1)^k}{\Gamma\left(1+k-\frac{k}{\mu}\right)} \left(\frac{\Delta}{C_{n,\mu}}\right)^k\sim E_{1-\frac{1}{\mu}}\left(\frac{(z-1)\Delta }{C_{n,\mu}}\right) \,,\quad \Delta \to 0, \quad n\to \infty\,,
\end{align}
where we recognized $C_{n,\mu}$ in (\ref{eq:Cnmu}) and the Mittag-Leffler function $E_p(x)=E_{p,1}(x)$ defined in (\ref{eq:MittagLeff}). Further setting $z=0$, we recover Spitzer's result in (\ref{eq:lims}) for $\mu>1$. Note that the marginal case of $\mu=1$ can be obtained similarly by analyzing the asymptotic behavior of the propagator in (\ref{eq:barpas}) in this case.

Let us now sketch how we used similar ideas to derive our results on the order statistics of random walks presented in Section \ref{sec:uni}. We start by introducing the random variable
\begin{align}
  \kappa_n(a,\Delta) = \sum_{i=1}^n \Theta\left(x_i - a - \frac{\Delta}{2}\right)\,,\label{eq:kapn}
\end{align}
which counts the number of times the random walk has been above the level $a+\Delta/2$ (see figure \ref{fig:Spi}). It turns out that the probability distribution $P_{k,n}(\Delta)$ of the $k^{\text{th}}$ gap of a random walk of $n$ steps, discussed in Section \ref{sec:uni}, is related to the joint distribution of the two counting random variables $\nu_n(a,\Delta)$ and $\kappa_n(a,\Delta)$ defined in (\ref{eq:nun}) and (\ref{eq:kapn}). Indeed, one can show that \cite{BSG23} 
\begin{align} 
P_{k,n}(\Delta) = \partial^2_{\Delta} \int_{-\infty}^\infty  \text{Prob.}\left[\kappa_n(a,\Delta)=k,\nu_n(a,\Delta)=0\right] \;da \;,\quad k=1,\ldots,n-1\,.
\label{rel_SPDF2}
\end{align}
From the relation (\ref{rel_SPDF2}), we see that to obtain the distribution of the gap $P_{k,n}(\Delta)$, we need to study the joint distribution of the two random variables $\kappa_n(a,\Delta)$ and $\nu_n(a,\Delta)$. To do so, we consider the double generating function $\langle z^{\nu_n(a,\Delta)} w^{\kappa_n(a,\Delta)} \rangle$. Following similar steps as in (\ref{eq:zTn})-(\ref{eq:zTnc}), we find
\begin{align}
  \langle z^{\nu_n(a,\Delta)} w^{\kappa_n(a,\Delta)} \rangle &= \sum_{k=0}^n \sum_{l=0}^{n-k}  (w-1)^{k}(z-1)^{l} \sum_{0<i_1<\ldots<i_l\leq n} \sum_{0<j_1<\ldots<j_k\leq n}\nonumber\\
  &\hspace{3em} \Bigg\langle \Theta\left(\frac{\Delta}{2} - |x_{i_1}-a|\right)\ldots \Theta\left(\frac{\Delta}{2} - |x_{i_l}-a|\right)\times\nonumber\\
 &\hspace{4em} \Theta\left(x_{j_1}-a-\frac{\Delta}{2}\right)\ldots\Theta\left(x_{j_k}-a-\frac{\Delta}{2}\right) \Bigg\rangle\,.\label{eq:zwnuka}
\end{align}
\begin{figure}[t]
  \begin{center}
    \includegraphics[width=0.6\textwidth]{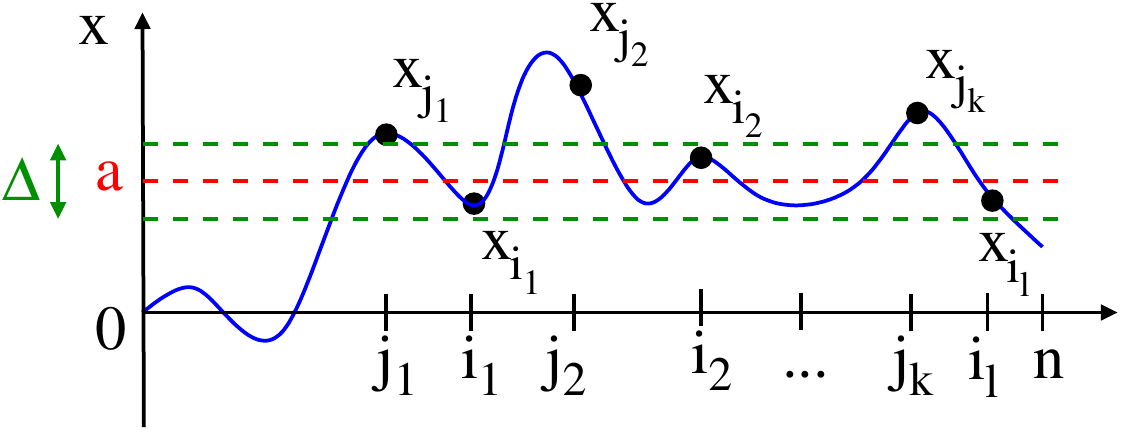}
    \caption{The average in brackets in (\ref{eq:zwnuka}) corresponds to all the trajectories of $n$ steps that start at the origin and are above the level $a+\Delta/2$ at the intermediate times $j_1,\ldots,j_k$ and in the interval $]a-\Delta/2,a+\Delta/2[$ at the intermediate times $i_1,\ldots,i_l$. }
    \label{fig:figDist}
  \end{center}
\end{figure}
The probability on the right-hand side in (\ref{eq:zwnuka}) is the probability that the random walk is above the level $a+\Delta/2$ at the intermediate times $j_1,\ldots,j_k$ and in the interval $]a-\Delta/2,a+\Delta/2[$ at the intermediate times $i_1,\ldots,i_l$ (see figure \ref{fig:figDist}). The expression (\ref{eq:zwnuka}) has some similarities with the generating function of the occupation time above the origin in (\ref{eq:zTnc}).
The additional difficulty here is the presence of the double sum over the $i$'s and the $j$'s, which are ordered among themselves, i.e. $i_1  < \ldots < i_l$ and $j_1<\ldots<j_k$, but not among each other. We choose to order the sum over the $j$'s with respect to the sum over the $i$'s. To do so, we denote by $r_q$ the number of $j's$ between $i_{q-1}$ and $i_{q}$ (with $i_0=1$ and $i_{l+1}=n$). By the Markov property of the random walk, one can write the probability in (\ref{eq:zwnuka}) as
\begin{align}
& \sum_{0<j_1<\ldots<j_k\leq n} \Bigg\langle \Theta\left(\frac{\Delta}{2} - |x_{i_1}-a|\right)\ldots \Theta\left(\frac{\Delta}{2} - |x_{i_l}-a|\right) \times \nonumber\\
&\hspace{15em} \Theta\left(x_{j_1}-a-\frac{\Delta}{2}\right)\ldots\Theta\left(x_{j_k}-a-\frac{\Delta}{2}\right) \Bigg\rangle \nonumber\\
  & = \sum_{r_1+\ldots+r_{l+1}=k}  \int_{a-\frac{\Delta}{2}}^{a+\frac{\Delta}{2}}dx_{i_1}\ldots dx_{i_l}K_{i_1}(x_{i_1},r_1|0) K_{i_2-i_1}(x_{i_2},r_2|x_{i_1})\ldots \nonumber\\
   &\hspace{15em}\times K_{i_l-i_{l-1}}(x_{i_l},r_l|x_{i_{l-1}})H_{n-i_l}(r_{l+1}|x_{i_{l}}) \,,\label{eq:SN2b}
\end{align}
where $K_i(x_2,r|x_1)$ is the propagator from $x_1$ to $x_2$ during $i$ steps with $r$ steps above $a+\frac{\Delta}{2}$ summed over all possible locations of the $r$ steps (see figure \ref{fig:G}), which reads
\begin{figure}[t]
  \begin{center}
    \includegraphics[width=0.6\textwidth]{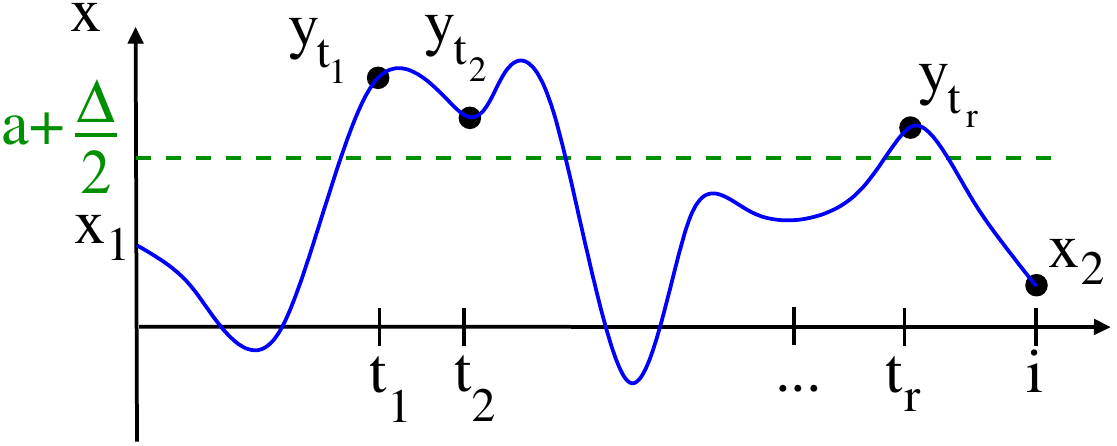}
    \caption{The propagator $K_i(x_2,r|x_1)$ is the probability that a random walk starting from $x_1$ reaches $x_2$ after $i$ steps with $r$ steps at times $t_1,\ldots,t_r$ above $a+\frac{\Delta}{2}$ summed over all possible locations of the $r$ steps. }
    \label{fig:G}
  \end{center}
\end{figure}
\begin{align}
 K_i(x_2,r|x_1)= \sum_{1\leq t_1<\ldots<t_r< i} \int_{a+\frac{\Delta}{2}}^\infty dy_{t_1}\ldots dy_{t_r} &p_{t_1}(y_{t_1}-x_1) p_{t_2-t_1}(y_{t_2}-y_{t_1})\ldots\nonumber\\
 & p_{t_r-t_{r-1}}(y_{t_r}-y_{t_{r-1}}) p_{i-t_r}(x_2-y_{t_{r}})\,,\label{eq:Gprop} 
\end{align}
where we recall that $p_j(x)=\int_{-\infty}^\infty \frac{dq}{2\pi} e^{-iqx} [\hat f(q)]^j$ is the free propagator of the original random walk starting from the origin. In equation (\ref{eq:SN2b}), $H_i(r|x_1)=\int_{-\infty}^{\infty} dx_2 K_i(x_2,r|x_1)$ is the propagator with a free end. Taking advantage of the convolution structure over the $i$'s and the $j$'s, and shifting all integration variables $x_{i_1},\ldots,x_{i_l}$ by $a+\frac{\Delta}{2}$, we take a generating function of (\ref{eq:zwnuka}) with respect to $n$, which gives
\begin{align}
    \sum_{n=0}^\infty s^n\langle z^{\nu_n(a,\Delta)} w^{\kappa_n(a,\Delta)} \rangle&=  \sum_{l=0}^\infty (z-1)^l
  \int_{-\Delta}^{0}dx_{1}\ldots dx_{l}\bar K_s\left(x_{1},w\,\bigg|\,-a-\frac{\Delta}{2}\right) \bar K_s(x_{2},w|x_{1})\ldots \nonumber\\
  &\hspace{12em}\times\bar K_s(x_{l},w|x_{l-1})\bar H_{s}(w|x_{l}) \,,\label{eq:SN2b2}
\end{align}
where $\bar K_s(x_1,w|x_0)$ and $\bar H_s(w|x)$ are the generating functions of $K_i(x_1,r|x_0)$ and $H_i(r|x)$ which, after shifting the integration variables $y_{t_1},\ldots,y_{t_r}$ in (\ref{eq:Gprop}) by $a+\frac{\Delta}{2}$, are given by
\begin{align}
   \bar K_s(x_2,w|x_1) &= \frac{s}{1-s}\sum_{r=0}^\infty u^r\int_{0}^\infty dy_{1}\ldots dy_{r} \mathrm{f}_{s}(y_{1}-x_1)\mathrm{f}_{s}(y_{2}-y_{1})\ldots\mathrm{f}_{s}(x_2-y_{r})\,,\label{eq:GEse}\\
   \bar H_{s}(w|x_1) &= \frac{1}{1-s}\sum_{r=0}^\infty u^r \int_{0}^\infty dy_{1}\ldots dy_{r} \mathrm{f}_{s}(y_{1}-x_1)\mathrm{f}_{s}(y_{2}-y_{1})\ldots \mathrm{f}_{s}(y_{r}-y_{r-1})\,,\label{eq:HEse}
\end{align}
where we again introduced the effective jump distribution $\mathrm{f}_s(\eta)$ from (\ref{eq:fze}) and defined $u=s(w-1)/(1-s)$ to ease notation.  We now interpret the multiple integrals (\ref{eq:GEse}) and (\ref{eq:HEse}). The former one can be expressed in terms of the excursion probability $E_s(r+1,x_2|x_1)$ of $r+1$ steps for a random walk with a jump distribution $\mathrm{f}_{s}(\eta)$ from $x_1$ to $x_2$, i.e., 
\begin{align}
  E_s(r+1,x_2|x_1) &= \int_{0}^\infty dy_{1}\ldots dy_{r} \mathrm{f}_{s}(y_{1}-x_1)\mathrm{f}_{s}(y_{2}-y_{1})\ldots\mathrm{f}_{s}(x_2-y_{r})\,.\label{eq:Esse}
\end{align}
The excursion probability $E_s(r,x_2|x_1)$ is the probability that a random walk, starting from $x_1$, reaches $x_2$ after $r$ steps while remaining above the origin during the intermediate steps
 (see the right panel in figure \ref{fig:survexc}).
 \begin{figure}[t]
 \centering
     \includegraphics[width=0.38\textwidth]{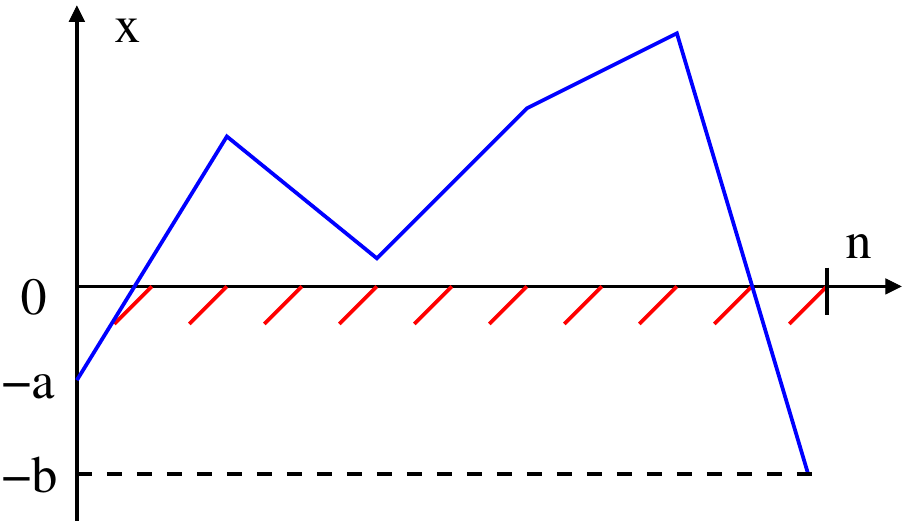}\qquad \includegraphics[width=0.38\textwidth]{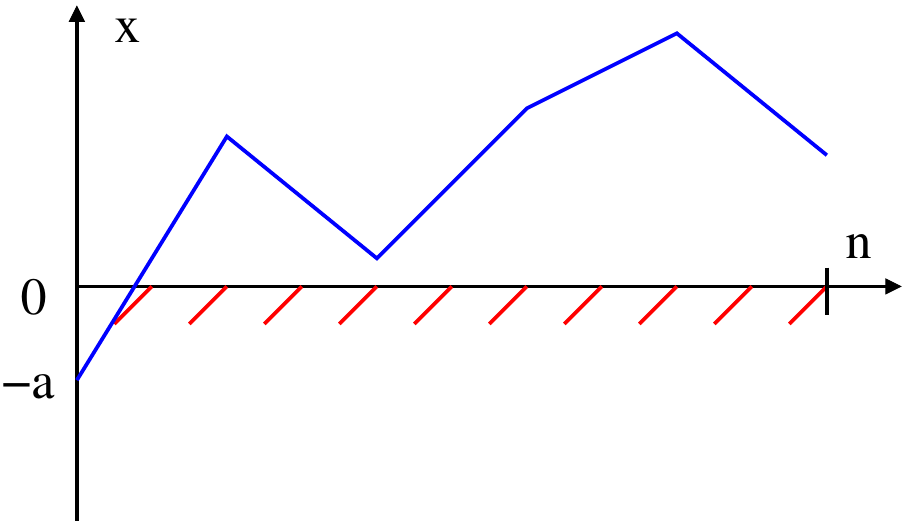}
     \caption{\textbf{Left panel}: The excursion probability $E_s(n,-b\,|-a)$ is the probability that a random walk starting from $-a<0$ with a jump distribution $ \mathrm{f}_s(\eta)$ reaches $-b<0$ after $n$ steps while remaining above the origin during the intermediate steps. \textbf{Right panel}: The survival probability $S_s(n|-a)$, is the probability that a random walk starting from $-a<0$ with a jump distribution $ \mathrm{f}_s(\eta)$ did not cross the origin except at the first step. }
     \label{fig:survexc}
 \end{figure}
  Note that, contrary to the usual definition of the excursion where the initial and final positions are positive $x_1,x_2>0$, the current computation requires extending it to a negative initial and final positions. 
 The latter one in (\ref{eq:HEse}) is the survival probability $S_s(r|x_1)$ during $r$ steps for a random walk with a jump distribution $ \mathrm{f}_s(\eta)$ starting from $x_1$:
\begin{align}
 S_s(r|x_1) &= \int_{0}^\infty dy_{1}\ldots dy_{r} \mathrm{f}_{s}(y_{1}-x_1)\mathrm{f}_{s}(y_{2}-y_{1})\ldots \mathrm{f}_{s}(y_{r}-y_{r-1})\,.\label{eq:Ssse}
 \end{align}
The survival probability $S_s(r|x_1)$ is the probability that the random walk, starting from $x_1$, did not cross the origin during $r$ steps except at the first step (see the right panel in figure \ref{fig:survexc}). Note again that the current computation requires extending it to a negative initial position.
Using their generating functions
\begin{align}
\tilde E_s(w,x_2|x_1)&=\sum_{r=1}^\infty w^r E_s(r,x_2,x_1) \;,\label{GF_E}\\
\bar S_s(w|x_1)&=\sum_{r=0}^\infty w^r S_s(r|x_1)\,,\label{eq:genSszma}
\end{align}
the expressions in (\ref{eq:GEse}) and (\ref{eq:HEse}) can be re-written as \cite{BSG23}
\begin{subequations}
\begin{align}
\bar K_s(x_2,w|x_1)&=\frac{1}{z-1}\tilde E_s\left(u,x_2|x_1\right)\,,\\
\bar H_s(w|x_1)&=\frac{1}{1-s}\bar S_s\left(u|x_1\right)\,,
\end{align}
\label{eq:relGEHS}
\end{subequations}
where we used again the notation $u=s(w-1)/(1-s)$. 
Upon using (\ref{rel_SPDF2}) and performing a few manipulations, one can show that the generating function of the gap distribution is given by
\begin{align}
 \sum_{n=0}^\infty \sum_{k=1}^{n-1} s^n w^k  P_{k,n}(\Delta) &= \partial^2_{\Delta} \Bigg[\frac{u}{(1-s)}\sum_{l=1}^\infty \frac{1}{(1-w)^l}\int_{-\Delta}^0 dx_{1}\dots dx_{l}\bar S_s\left(u,x_{1}\right)\tilde E_s\left(u,x_{2}|x_{1}\right)\ldots \nonumber\\
 &\hspace{15em}\tilde E_s\left(u,x_{l}|x_{l-1}\right)\bar S_s\left(u,x_{l}\right)\Bigg]\,.\label{eq:genPkn}
\end{align}
It turns out that the generating functions $\bar S_s\left(u,x\right)$ and $\tilde  E_s\left(u,x_{2}|x_{1}\right)$ can be explicitly evaluated by using Sparre Andersen result in (\ref{eq:SAres2}) and another formula due to Pollaczek and Spitzer (see Appendix E in \cite{BSG23}). By performing the appropriate asymptotic analysis in (\ref{eq:genPkn}), we obtain the results discussed in Section \ref{sec:uni}. The full derivation can be found in the paper whose abstract is given on p.~\pageref{chap:A14}.
\newcounter{mycounter}
\setcounter{mycounter}{1}

\begin{figure}
\begin{center}
 \fboxsep=10pt\relax\fboxrule=1pt\relax
 \fbox{
   \begin{minipage}{\textwidth}
   
\hspace{2em}
\begin{center}
\LARGE \bf Universal order statistics for random walks \& Lévy flights 
\end{center}

\hspace{2em}
\begin{center}
B. De Bruyne, S. N. Majumdar and G. Schehr,
J. Stat. Phys. {\bf 190}, 320 (2023).
\end{center}

\hspace{2em}
\begin{center}
  {\bf Abstract:} 
\end{center}

We consider one-dimensional discrete-time random walks (RWs) of $n$ steps, starting from $x_0=0$, with
arbitrary symmetric and continuous jump distributions $f(\eta)$, including the important case of L\'evy flights. We study the statistics of the gaps $\Delta_{k,n}$ between the $k^\text{th}$ and $(k+1)^\text{th}$ maximum of the set of positions $\{x_1,\ldots,x_n\}$. We obtain an exact analytical expression for the probability distribution $P_{k,n}(\Delta)$ valid for any $k$ and $n$, and jump distribution $f(\eta)$, which we then analyze in the large $n$ limit. For jump distributions whose Fourier transform behaves, for small $q$, as $\hat f (q) \sim 1 - |q|^\mu$ with a L\'evy index $0< \mu \leq 2$, we find that the distribution becomes stationary in the limit of $n\to \infty$, i.e. $\lim_{n\to \infty} P_{k,n}(\Delta)=P_k(\Delta)$. We obtain an explicit expression for its first moment $\langle \Delta_{k}\rangle$, valid for any $k$ and jump distribution $f(\eta)$ with $\mu>1$, and show that it exhibits a universal algebraic decay $ \langle \Delta_{k}\rangle\sim k^{1/\mu-1} \Gamma\left(1-1/\mu\right)/\pi$ for large $k$. Furthermore, for $\mu>1$, we show that in the limit of $k\to\infty$ the stationary distribution exhibits a universal scaling form $P_k(\Delta) \sim  k^{1-1/\mu} \mathcal{P}_\mu(k^{1-1/\mu}\Delta)$ which depends only on the L\'evy index $\mu$, but not on the details of the jump distribution. We compute explicitly the limiting scaling function $\mathcal{P}_\mu(x)$ in terms of Mittag-Leffler functions. For $1< \mu <2$, we show that, while this scaling function captures the distribution of the typical gaps on the scale $k^{1/\mu-1}$, the atypical large gaps are not described by this scaling function since they occur at a larger scale of order $k^{1/\mu}$. This atypical part of the distribution is reminiscent of a ``condensation bump'' that one often encounters in several mass transport models. 
\end{minipage}
   }
\end{center}
\captionsetup{labelformat=empty}
\caption{\textbf{Abstract of article \themycounter} : Universal order statistics for random walks \& Lévy flights.}
\label{chap:A14}
\addtocounter{mycounter}{1}
\end{figure}

\newpage

\section{Expected maximum of bridge random walks}
\label{sec:exp}
In the previous section, we focused on the EVS of random walks that could evolve freely. In this section, we study the EVS of \textit{constrained} random walks, for which there exist very few results. We will focus on the global maximum $M_{n}$ of bridge random walks, as defined in (\ref{eq:bridgec}), that are constrained to return to the origin (see figure \ref{fig:bridge}). The main results are presented in Section \ref{sec:expi} and we make use of this section to briefly sketch how we obtained them. Our derivation is based on an extension of the Pollaczek-Spitzer's formula presented in (\ref{eq:SAres2}), which involves the joint distribution of the global maximum $M_n=\max\{x_0,\ldots,x_n\}$ and the final position $x_n$ of a random walk of $n$ steps (see figure \ref{fig:xnMn}). The formula provides the generating function of the double Laplace-Fourier transform of the joint distribution which, for a symmetric jump distribution, reads \cite{Spitzer57,Wendel58}
\begin{align}
  \sum_{n=0}^\infty z^n\, \langle e^{-\alpha M_n+ik x_n}\rangle&= \exp\left[\sum_{n=1}^\infty \frac{z^n}{n}
   \int_{0}^\infty dy\, p_n(y) \left(e^{(ik- \alpha) y}+e^{-i k y}\right)\right]\,,\label{eq:spitzer}
\end{align}
where $p_n(y)$ is the propagator of the random walk given in (\ref{eq:PI}).
In order to extract the maximum of bridge random walks, which return to the origin after $n$ steps as in (\ref{eq:bridgec}), we will have to set $x_n=0$. Before doing this, let us first extract the expected maximum from the joint distribution in (\ref{eq:spitzer}). To do so, we take a derivative with respect to $\alpha$ and set $\alpha=0$. After a few steps, we obtain \cite{BSG21a}:
\begin{align}
  \sum_{n=0}^\infty z^n\, \langle  M_n e^{ik x_n}\rangle&=\sum_{l=0}^\infty\sum_{m=1}^\infty  z^{m+l} \frac{
  \hat f(k)^l}{m}\int_{0}^\infty dy\, y\, p_m(y) e^{ik y} \,.\label{eq:powz}
\end{align}
Upon identifying the powers in $z$, we get
\begin{align}
 \langle M_n\,e^{ik x_n}\rangle= \sum_{m=1}^n \frac{1}{m}
   \int_{0}^\infty dy\, y\, p_m(y)\, e^{ik y}\, \hat f(k)^{n-m}\,.\label{eq:powzid}
\end{align}
Note that by setting $k=0$ and using that $\hat f(0)=1$ by normalization, we recover the expression for the expected maximum in (\ref{eq:maxfree}). Performing an inverse Fourier transform on (\ref{eq:powzid}), setting $x_n=0$, and normalizing by $p_n(0)$ which sums over all the bridge trajectories, we obtain 
\begin{align}
   \langle M_n\rangle = \sum_{m=1}^n \frac{1}{m}
   \int_{0}^\infty\,dy\, y\,\frac{ p_m(y) \, p_{n-m}(y)}{p_n(0)}\,,\label{eq:M}
\end{align}
which is the result presented in the introduction in (\ref{eq:maxbridgeI}). The asymptotic behavior of $\langle M_n\rangle $ for large $n$ discussed in Section \ref{sec:expi} was obtained by carefully analyzing (\ref{eq:maxbridgeI}) up to the second leading order $n$. The details of the derivation can be found in the paper whose abstract is given on p.~\pageref{chap:A5}.

Let us end this section by illustrating our results on one application. We consider a bridge version of the ``lamb-lion'' problem \cite{Franke12,Majumdar19Smol,Krapivsky96Lamb,Redner99Lamb,redner2001guide} where the lions are constrained to return to their initial position after their hunt. The setting is illustrated in figure \ref{fig:lamblion}. An immobile lamb is located at the origin and $N$ lions, performing random walks, are initially uniformly distributed on the segment $[0,L]$. 
\begin{figure}[t]
  \begin{center}
    \includegraphics[width=0.5\textwidth]{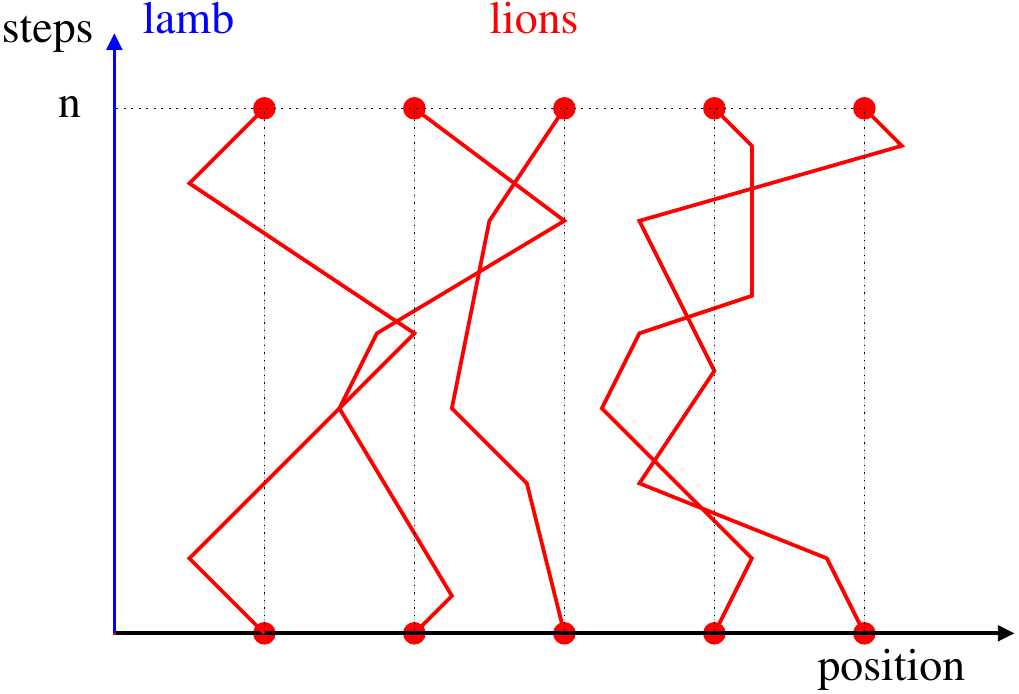}
    \caption{``Lamb-lion'' problem. The lamb is an immobile target located at the origin (blue). The lions, performing random walks, are initially uniformly distributed on the positive line with density $\rho_0$. The lions are further constrained to return to the origin at the $n^\text{th}$ step. The survival probability $S_n$ is the probability that none of the lions have encountered the lamb during $n$ steps.}
    \label{fig:lamblion}
  \end{center}
\end{figure}
Quite remarkably, in the limit $N\to \infty$ and $L\to \infty$ with the density $\rho_0=N/L$ fixed, the survival probability $S_n$ that none of the lions have encountered the lamb is related to the expected maximum of a random walk by the following relation \cite{Majumdar19Smol}
\begin{align}
  S_n = \exp(-\rho_0\, \langle M_n\rangle )\,.\label{eq:Snll}
\end{align}
For large $n$, using our first order results on the expected maximum of a bridge random walk, we find that the survival probability of the lamb decays like
\begin{align}
S_n \sim \exp(-\rho_0 \,  h_1(\mu) \, n^{\frac{1}{\mu}}) \quad, \quad n \to \infty \,, \label{eq:Snapprox}
\end{align}
where $h_1(\mu)$ is the amplitude given in (\ref{eq:h1intro}) and $\mu$ is the Lévy index of the jump distribution of the lions (\ref{eq:f}). Our results show that the second leading order correction to the expected maximum ${\mathbb E}[M_n]$ does not necessarily decay when $n$ is large. This means that the leading finite-size correction, therefore, plays an important role as it will contribute to the amplitude of the decay of the survival probability. For instance, if the jump distribution of the lions is a Cauchy distribution with scale $\gamma$ (\ref{eq:cauchyDist}), one needs to include the leading finite-size correction, given in (\ref{eq:MasIII}), to find that the survival probability decays as
\begin{align}
  S_n \sim n^{\frac{1}{2\pi}\,\rho_0\,\gamma}\,e^{-\frac{1}{8}\,\rho_0\,\gamma\,\pi\,n}\,,\label{eq:survll}
\end{align}
up to a constant prefactor which would require an asymptotic analysis of the expected maximum up to the third order to be determined. Further applications on the convex hull of tethered Rouse polymer chains and the run-and-tumble particle can be found in the paper whose abstract is given on p.~\pageref{chap:A5}.

\begin{figure}
\begin{center}
 \fboxsep=10pt\relax\fboxrule=1pt\relax
 \fbox{
   \begin{minipage}{\textwidth}
\hspace{2em}
\begin{center}
\LARGE \bf Expected maximum of bridge random walks \& Lévy
flights
\end{center}

\hspace{2em}
\begin{center}
B. De Bruyne, S. N. Majumdar and G. Schehr,
J.~Stat.~Mech., 083215 (2021).
\end{center}

\hspace{2em}
\begin{center}
  {\bf Abstract:} 
\end{center}

We consider one-dimensional discrete-time random walks (RWs) with arbitrary symmetric and continuous jump distributions $f(\eta)$, including the case of L\'evy flights. We study the expected maximum $\langle M_n\rangle$ of bridge RWs, i.e., RWs starting and ending at the origin 
after $n$ steps. We obtain an exact analytical expression for $\langle M_n\rangle$ valid for any $n$ and jump distribution $f(\eta)$, which we then analyze in the large $n$ limit up to second leading order term. For jump distributions whose Fourier 
transform behaves, for small $k$, as $\hat f(k) \sim 1 - |a\, k|^\mu$ with a L\'evy index $0<\mu \leq 2$ and an arbitrary length scale $a>0$,  
we find that, at leading order for large $n$, $\langle M_n\rangle\sim a\, h_1(\mu)\, n^{1/\mu}$. We obtain an explicit expression for the amplitude $h_1(\mu)$ and find that it carries the signature of the bridge condition, being different from its counterpart for the free random walk. For $\mu=2$, we find that the second leading order term is a constant, which, quite remarkably, is the same as its counterpart for the free RW. For generic $0< \mu < 2$, this second leading order term is a growing function of $n$, which depends non-trivially on further details  of $\hat f (k)$, beyond the L\'evy index $\mu$. 
Finally, we apply our results to compute the mean perimeter of the convex hull of the $2d$ Rouse polymer chain and of the $2d$ run-and-tumble particle, as well as to the computation of the survival probability in  
a bridge version of the well-known ``lamb-lion'' capture problem.
\end{minipage}
   }
\end{center}
\captionsetup{labelformat=empty}
\caption{\textbf{Abstract of article \themycounter} : Expected maximum of bridge random walks \& Lévy
flights.}
\label{chap:A5}
\addtocounter{mycounter}{1}
\end{figure}

\newpage

\section{Convex hull of Brownian motion in confined geometries}
\label{sec:con}
In the previous sections, we discussed some EVS of discrete-time random walks in one spatial dimension. In this section, we will study some EVS of continuous time stochastic processes in higher dimensions. We will focus on the $d$-dimensional Brownian motion $\bm{x}(t)$ with diffusion coefficient $D$, which, in its simplest form,  evolves according to the Langevin equation
\begin{align}
  \bm{\dot x}(t) &=\sqrt{2D}\, \bm{\eta}(t)\,,
  \label{eq:eomb}
\end{align}
where $\bm{x}(0)=\bm{0}$ is the initial position and where $\bm{\eta}(t)$ is a $d$-dimensional white noise with zero mean $\langle \bm{\eta}(t)\rangle=\bm{0}$ and delta correlated independent components $\langle \eta_i(t)\eta_j(t')\rangle=\delta_{i,j}\delta(t-t')$. As for discrete-time random walks, Brownian motion (BM) serves as a natural model to study the EVS of strongly correlated systems. For instance, the analog of the global maximum $M_n$ of discrete-time random walks for one-dimensional Brownian motion is the maximum $M(t)$ of the Brownian path $x(t)$ over the time interval $[0,t]$, defined as
\begin{align}
  M(t) = \max_{0\leq\tau\leq t} \left[x(\tau)\right]\,,\label{eq:Min} 
\end{align}
for which the cumulative probability distribution is given by a half-Gaussian \cite{bray2013persistence}
\begin{align}
  \text{Prob.}\left(M(t)<M\right) = \Theta(M)\,\text{erf}\left(\frac{M}{2 \sqrt{D t}}\right)\,.\label{eq:Pmz}
\end{align}
In particular, from the distribution (\ref{eq:Pmz}), one can easily obtain that the average maximum $M(t)$ of a free Brownian motion which is given by
\begin{align}
  \langle M(t) \rangle = \frac{2\sqrt{Dt}}{\sqrt{\pi}}\,.\label{eq:Mavgi}
\end{align}
EVS have found a particularly nice application in the description of the convex hull of a Brownian motion and other stochastic processes \cite{Kac54,Spitzer56,Snyder,Kabluchko17a,Kabluchko17b,Schawe17,Claussen15,Letac80,Dumonteil13,Reymbaut11,Chupeau15a,Chupeau15b,Dumonteil13,Majumdar21convex,Randon09Convex,Majumdar10Random,GrebenkovConvex17,Schawe18Large,Cauchy32}. It serves, for instance, as a simple model to describe the extension of the territory of foraging animals in behavioral ecology \cite{Berg93,Bartumeus05,Murphy92,Worton95,Giuggioli11}. This connection has been made possible due to a method developed in \cite{Letac80,Randon09Convex,Majumdar10Random} which relies on Cauchy's formula \cite{Cauchy32} for convex curves. To explain this connection, let us first briefly recall Cauchy's formula. This formula concerns the convex hull of a set of $n$ points $\{\bm{x}_1,\ldots,\bm{x}_n\}$ on the two-dimensional plane $\mathbb{R}^2$. The convex hull is defined as the smallest polygon enclosing all of the $n$ points (see figure \ref{fig:convex}).
\begin{figure}[t]
  \begin{center}
    \includegraphics[width=0.38\textwidth]{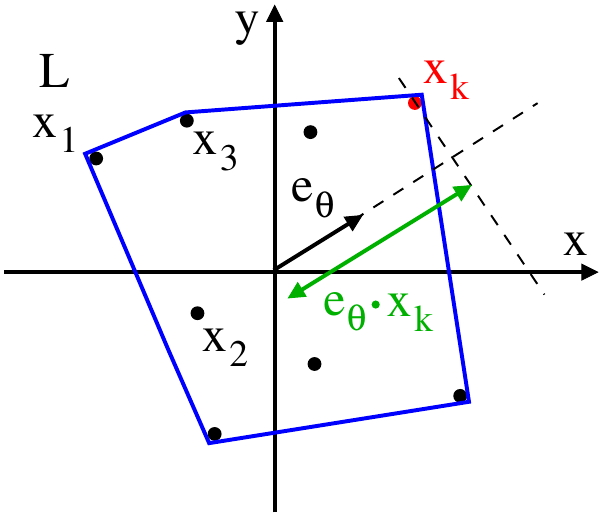}
    \caption{The convex hull of a set of $n=8$ points is shown. The length $L$ of the convex hull can be computed with Cauchy's formula in (\ref{eq:cau}). The formula involves the maximum projection $\max_{1\leq i\leq n} [\bm{x}_i\cdot \bm{e}_\theta]$ which is shown here for a particular direction $\bm{e}_\theta$ and is maximized at the point $\bm{x}_k$. }
    \label{fig:convex}
  \end{center}
\end{figure}
 The formula states that the length $L$ of the perimeter of the convex hull of the points $\{\bm{x}_1,\ldots,\bm{x}_n\}$ is given by
\begin{align}
  L = \int_0^{2\pi} d\theta \max_{1\leq i\leq n} [\bm{x}_i\cdot \bm{e}_\theta]\,,\label{eq:cau}
\end{align}
where $\mathbf{e}_\theta$ is the unit vector whose angle with the $x$-axis is $\theta$. Cauchy's formula in (\ref{eq:cau}) has been extended to stochastic trajectories and, in particular, states that the average length $\langle L(t)\rangle$ of the convex hull  of a stochastic process $\mathbf{x}(t)$ in $d=2$ can be expressed in terms of the expected maximum $\langle \max_{0\leq \tau\leq t}[\mathbf{x}(\tau) \cdot \mathbf{e}_\theta]\rangle$ in the direction $\mathbf{e}_\theta$ as \cite{Randon09Convex,Majumdar10Random}:
\begin{align}
\langle L(t) \rangle &=  \int_0^{2\pi} d\theta\left\langle \max_{0\leq \tau\leq t}[\mathbf{x}(\tau) \cdot \mathbf{e}_\theta]\right\rangle\,.\label{eq:LC}
\end{align}
Using the isotropy of BM in $d=2$, one finds directly from (\ref{eq:Mavgi}) and (\ref{eq:LC}) that the mean perimeter of a free BM grows as $\langle L(t) \rangle = 2\pi\langle \max_{0\leq \tau\leq t}[\mathbf{x}(\tau) \cdot \mathbf{e}_x]\rangle= \sqrt{8\pi D t}$ \cite{Letac80}. Most of the current results on the convex hull of BM concern isotropic processes in an unconfined two-dimensional geometry. Nevertheless, in many practical situations, the process takes place in the presence of boundaries that may limit the growth of the convex hull, such as a river in the context of foraging animals. Recently, the effect of such a boundary was studied for a planar Brownian motion in the presence of an infinite reflecting wall. It was shown that the presence of the wall breaks the isotropy of the process and induces a non-trivial effect on the convex hull of the Brownian motion \cite{Chupeau15a,Chupeau15b}. However, the question of the growth of the convex hull of a stochastic process, and more generally of its EVS, in a closed confining geometry remains largely open. Our main contribution to this line of research is to address this question for Brownian motion in one of the simplest, yet non-trivial, geometry, namely the $d$-dimensional ball.

We study a Brownian motion (\ref{eq:eomb}) confined in a $d$-dimensional ball of radius $R$ with reflecting boundaries. We investigate the growth of the maximum of the process $M_x(t)=\max_t[\mathbf{x}(t) \cdot \mathbf{e}_x]$ in an arbitrary direction (see figure \ref{fig:traj}), which we set to be the $x$-direction without loss of generality due to the rotational symmetry.
\begin{figure}[t]
  \begin{center}
    \includegraphics[width=0.4\textwidth]{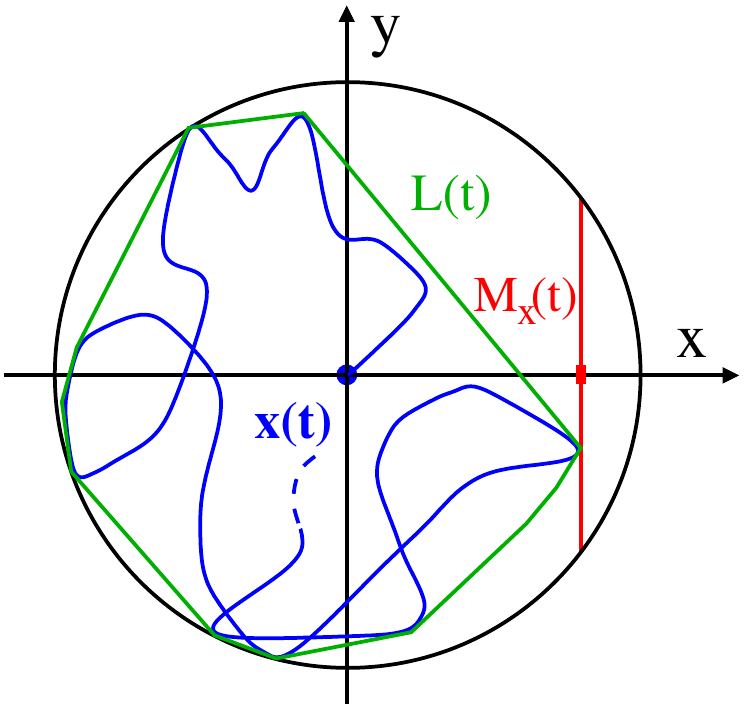}
    \caption{Schematic representation of a planar Brownian motion $\mathbf{x}(t)$ (blue line) in $d=2$ evolving in a disk (black line) of radius $R$ with reflecting boundary conditions. The length of the convex hull $L(t)$ is the length of the convex envelope of the motion up to time $t$ (green line). The maximum in the $x$-direction of the process $M_x(t)=\max_t[\mathbf{x}(t) \cdot \mathbf{e}_x]$ is depicted by the red dash. For $t\to \infty$, the maximum tends to the radius of the circle $M_x(t)\to R$. The distribution of the fluctuations $\Delta(t)=[R-M_x(t)]/R$ follows a non-trivial behavior (see expressions (\ref{eq:dist1i}) for $d=2$ and also (\ref{eq:Mx3di}) for $d\geq 3$). }
    \label{fig:traj}
  \end{center}
\end{figure}
It is clear that in the limit $t\to \infty$, the maximum will tend toward the radius of the ball, namely
 \begin{align}
   M_x(t) \to R\,,\quad t\to \infty\,.\label{eq:MlimR}
 \end{align}
However, for large but finite $t$, the maximum will fluctuate below this limiting value and the fluctuations $\Delta(t)$ can be described by the relative difference between the radius of the ball and the maximum: 
 \begin{align}
   \Delta(t) = \frac{R-M_x(t)}{R}\,.\label{eq:Delta}
 \end{align}
This observable is \textit{a priori} difficult to study for an arbitrary dimension $d$ as it cannot be reduced to a one-dimensional problem, due to the reflecting boundaries of the ball. Nevertheless, by establishing a connection with a similar albeit different problem, which concerns the narrow escape time \cite{Meyer11,Benichou10,Benichou08,Rupprecht15,Singer0,Singer1,Singer2,Singer3}, we obtain exact analytical expressions for the distribution of $\Delta(t)$ in the large $t$ limit. After deriving these results for an arbitrary dimension $d$, we will focus on the special case of $d=2$, where we will use Cauchy's formula to study the growth of the convex hull of a Brownian motion confined in a disk (see figure \ref{fig:traj}). 

 Let us first present our results on the growth of the maximum $M_x(t)$ in the direction $x$ for a Brownian motion confined in a $d$-dimensional ball. We find that the decay of the fluctuations (\ref{eq:Delta}) displays a rich behavior depending on the dimension $d$ of the ball. Our main results can be summarised as follows: 
 
 \begin{itemize}
   \item \textbf{Exponential decay in $d\!=\!2$.}
   In the case of $d=2$, the maximum $M_x(t)$ along the $x$-direction of a Brownian motion in a disk of radius $R$, starting from the origin, will reach $R$ in an infinite amount of time. We find that the typical fluctuations decay exponentially with time as
   \begin{align}
   \Aboxed{
 \Delta(t) \sim  A_2\,e^{-2\frac{Dt}{R^2} \chi_2}\,,\quad t\to \infty\,,}\label{eq:dist1i}
\end{align}
where $\chi_2$ is a random variable of order $O(1)$ whose probability distribution $f_2(\chi_2)$ is given by
\begin{align}
  \Aboxed{f_2(\chi_2) = \frac{1}{\chi_2^2}\,e^{-\frac{1}{\chi_2}}\,, \quad \chi_2 \geq 0\,.}\label{eq:chi2i}
\end{align}
In equation (\ref{eq:dist1i}), the amplitude $A_2$ is difficult to compute exactly. Below, we give a heuristic argument, leading to $A_2=2\,e^{\frac{1}{4}}$, which is in good agreement with our simulations (see figure \ref{fig:fchi2d}). Note that the asymptotic relation (\ref{eq:dist1i}) between the random variables $\Delta(t)$ and $\chi_2$ is valid ``in distribution''.  Despite the exponential decay of the fluctuations in (\ref{eq:dist1i}), we find that the \emph{average} fluctuations decay anomalously as a stretched exponential with time, i.e.,
  \begin{align}
   \Aboxed{\langle  \Delta(t)\rangle &\sim 2^{1/4}\sqrt{\pi } \,A_2\, \left(\frac{D t}{R^2}\right)^{1/4} \,e^{-2^{3/2} \sqrt{\frac{D t}{R^2}}}\,,\quad t\rightarrow \infty\,.}\label{eq:saddleMxi}
\end{align}
The average fluctuations (\ref{eq:saddleMxi}) therefore behave differently from the typical fluctuations (\ref{eq:dist1i}). This originates from the fact that the distribution $f_2(\chi_2)$ has a heavy tail $f_2(\chi_2) \sim \chi_2^{-2}$ for $\chi_2 \to \infty$ such that its first moment does not exist.
 \item \textbf{Power law decay in $d\!\geq\!3$.}
 In the case of $d\geq 3$, the maximum $M_x(t)$ along the $x$-direction of a Brownian motion in a $d$-dimensional ball of radius $R$, starting from the origin, will reach $R$ in an infinite amount of time. We find that the typical fluctuations decay algebraically as
\begin{align}
     \Aboxed{\Delta(t) \sim A_d  \left(\frac{R^2}{Dt}\right)^{\frac{2}{d-2}}\, \chi_d\,,\quad t\to \infty\,,}\label{eq:Mx3di}
\end{align}
where $\chi_d$ is a random variable of order $O(1)$ whose distribution $f_d(\chi_d)$ is given by
\begin{align}
  \Aboxed{f_d(\chi_d) = \frac{d-2}{2}\,\frac{e^{-\chi_d^{\frac{d-2}{2}}}}{\chi_d^{\frac{4-d}{2}}}\,,\quad \chi_d \geq 0\,. }\label{eq:chidi} 
\end{align}
Here also, the asymptotic relation (\ref{eq:Mx3di}) between the random variables $\Delta(t)$ and $\chi_d$ is valid ``in distribution''. For $d=3$, we have found that the amplitude is given by $A_3= \frac{\pi^2}{18}$ but we did not find an expression for $A_d$ for $d>3$. From the distribution (\ref{eq:Mx3di}), we find that the average fluctuations decay as a power law with time:
\begin{align}
  \Aboxed{\langle\Delta(t)\rangle \sim  A_d \,\Gamma\left(\frac{d}{d-2}\right)\left(\frac{R^2}{Dt}\right)^{\frac{2}{d-2}}\,,\quad t\to \infty\,,}\label{eq:Mx3ai}
\end{align}
where $\Gamma(z)$ is the gamma function. The case of $d\geq 3$ is therefore quite different from the case of $d\!=\!2$ as the average fluctuations (\ref{eq:Mx3ai}) and the typical ones (\ref{eq:Mx3di}) scale similarly since the first moment of $\chi_d$ is finite for $d\geq 3$. Furthermore, note that all moments of $\Delta(t)$  for $d\geq 3$ in (\ref{eq:Mx3di}) decay algebraically in time.

\end{itemize}
In the special case of $d=2$, by making use of Cauchy formula (\ref{eq:LC}), we studied the growth of the convex hull of a Brownian motion in a disk (see figure \ref{fig:traj}). We find that the average length of the convex hull $ \langle L(t)\rangle $ approaches slowly the perimeter of the disk $2\pi R$ as a stretched exponential:
\begin{align}
 \Aboxed{2\pi R- \langle L(t)\rangle \sim   2^{5/4}\pi^{3/2} \,A_2\, R \left(\frac{D t}{R^2}\right)^{1/4} \,e^{-2^{3/2} \sqrt{\frac{D t}{R^2}}}\,,\quad t\rightarrow \infty\,.}\label{eq:saddleLxi}
\end{align}

\begin{figure}[t]
  \centering
\includegraphics[width=0.4\textwidth]{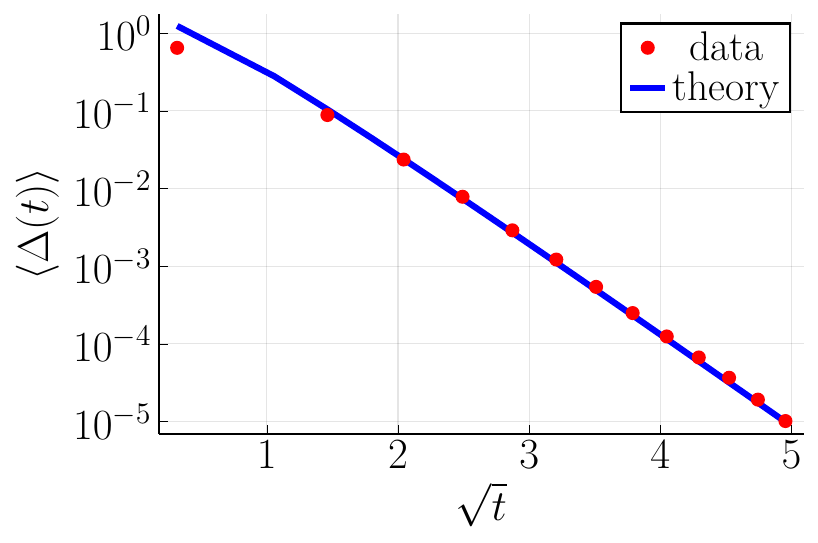}~
\includegraphics[width=0.4\textwidth]{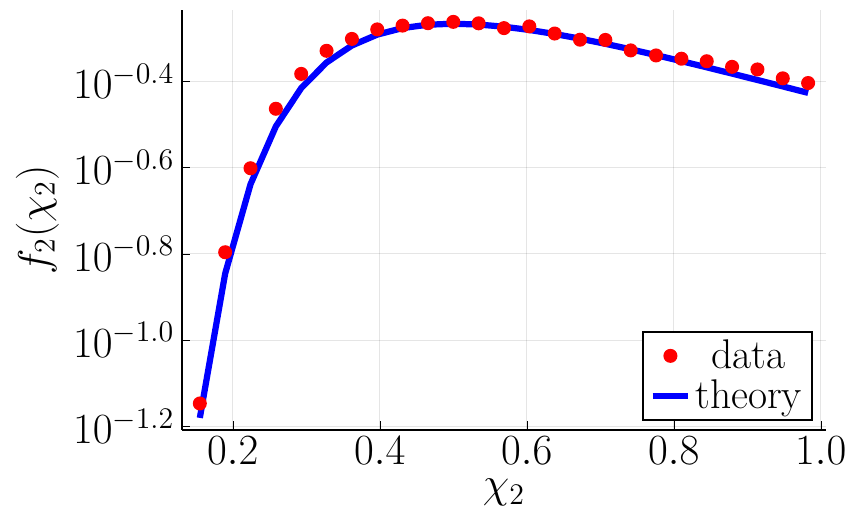}%
    \caption{\textbf{Left panel:} Evolution of the mean fluctuations $\langle \Delta(t )\rangle=\left\langle \frac{R-M_x(t)}{R}\right\rangle$ of a $d\!=\!2$ Brownian motion in a disk of radius $R$ with reflecting boundaries. The numerical data (red dots) have been obtained by averaging over $10^5$ trajectories with $dt=10^{-6}$ and are compared to the theoretical prediction (\ref{eq:saddleMxi}) with $D=1$, $R=1$ and $A_2=2\,e^{\frac{1}{4}}$.  \textbf{Right panel:} Distribution of the rescaled fluctuations $\chi_2=\frac{R^2}{2Dt}\left[\frac{1}{4}-\ln\left(\frac{\Delta(t)}{2R}\right)\right]$ computed numerically (red dots), averaged over $ 10^6$ realizations with $dt=10^{-9}$ and compared to the theoretical prediction (blue line), given in (\ref{eq:chi2i}) with $A_2=2\,e^{\frac{1}{4}}$ for $D=1$, $R=1$ and $t=8$. The plot is limited to $\chi_2<1$ where the discretization step is much smaller than the difference $R-M_x$. }
    \label{fig:fchi2d}
  \end{figure}
  
 Having presented our main results, let us now briefly explain how we derived them. Our results rely on recent works on the ``narrow escape problem'' \cite{Meyer11,Benichou10,Benichou08,Rupprecht15,Singer0,Singer1,Singer2,Singer3}, which is as follows. Consider a $d$-dimensional Brownian motion in a closed domain $\Omega$. Let the boundary of the domain $\partial \Omega$ be reflecting everywhere except for a small opening $\partial \Omega_a$, which is absorbing (see figure \ref{fig:circle}).
  \begin{figure}[t]
  \begin{center}
    \includegraphics[width=0.4\textwidth]{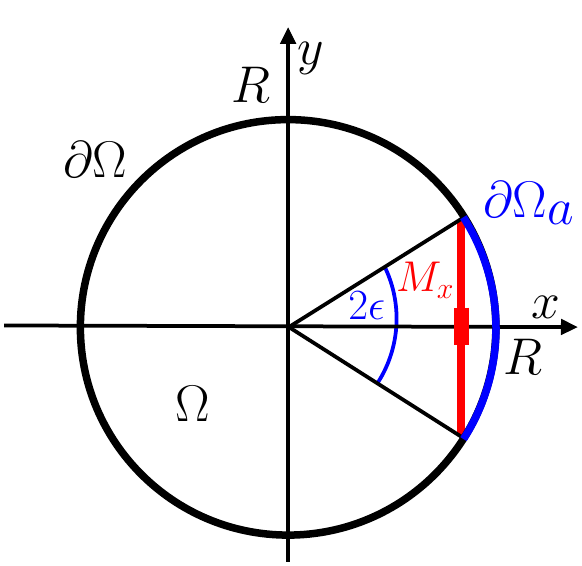}
    \caption{The narrow escape time $\langle T(\epsilon\,|\,\mathbf{x_0})\rangle$ for Brownian motion in a disk $\Omega$ of radius $R$ is the mean first-passage time for the process, starting from $\mathbf{x_0}\in \Omega$, to reach the arc $\partial\Omega_a$ spanned by the angle $2\epsilon$ (blue arc) in the limit of $\epsilon \to 0$, while the complementary boundary $\partial\Omega\backslash \partial\Omega_a$ is reflecting. To study the fluctuations of the maximum in the $x$-direction $M_x$, we assume that when $M_x\to R$, the first-passage time for the process to reach $M_x$ (red line) is asymptotically equal to the first-passage time to reach the arc spanned by the angle $2\epsilon$ with $\epsilon=\arccos\left(\frac{R^2-M_x^2}{R^2}\right)\sim\frac{\sqrt{2(R-M_x)}}{\sqrt{R}}$. }
    \label{fig:circle}
  \end{center}
\end{figure}
The narrow escape problem is then: ``What is the time required for a Brownian motion to escape through the small opening $\partial \Omega_a$?''. This time is known as the narrow escape time (NET) and has received much attention recently due to its importance in various applications such as in biochemical reactions \cite{Singer0,Kolesov07}. To describe the NET, it is convenient to introduce the ratio $\epsilon$ of the size of the opening window over the total size of the boundary:
\begin{align}
  \epsilon =  \frac{|\partial \Omega_a|}{|\partial \Omega|}\ll 1\,.\label{eq:epsilon}
\end{align} 
Clearly, as $\epsilon \to 0$, the mean time to absorption $\langle T(\epsilon\,|\,\mathbf{x_0})\rangle$ starting from $\mathbf{x_0}$ diverges:
\begin{align}
  \langle T(\epsilon\,|\,\mathbf{x_0})\rangle \to \infty\,,\quad \epsilon \to 0\,,\label{eq:Tinf}
\end{align}
when the initial position $\mathbf{x_0}$ of the Brownian walker is located sufficiently far away from the opening window.
As it was shown in a series of papers \cite{Meyer11,Benichou10,Benichou08,Singer0,Singer1,Singer2,Singer3,Rupprecht15}, one can obtain the asymptotic behavior of $\langle T(\epsilon\,|\,\mathbf{x_0})\rangle$ as $\epsilon \to 0$ for a wide range of geometries and in various dimensions. We summarize below the different cases that are relevant to the present work.

\begin{itemize}
  \item In $d=2$, it was found that for regular domains $\Omega$ that can be conformally mapped to a disk, the NET diverges logarithmically as \cite{Singer0}
\begin{align}
    \langle T(\epsilon\,|\,\mathbf{x_0})\rangle &= \frac{|\Omega|}{D \pi}\left[\ln\left(\frac{1}{\epsilon}\right)+O(1)\right]\,,\label{eq:genT}
\end{align}
where $|\Omega|$ is the size of the domain. In the particular case when $\Omega$ is a disk of radius $R$, it is possible to obtain the next-to-leading order correction in the asymptotic expansion (\ref{eq:genT}). This correction depends on the initial position of the process. When the process starts at the origin of the disk, the NET $ \langle T(\epsilon)\,|\,\mathbf{x_0}=\mathbf{0}\rangle$ is given by \cite{Singer0}
\begin{align}
  \langle T(\epsilon)\,|\,\mathbf{x_0}=\mathbf{0}\rangle = \frac{R^2}{D}\left[\ln\left(\frac{1}{\epsilon}\right)+\ln(2)+\frac{1}{4} + O(\epsilon)\right]\,.\label{eq:Td0}
\end{align} 
On the other hand, the NET averaged over an initial uniform distribution for $\mathbf{x_0}$ in the disk $\overline{\langle T(\epsilon\,|\,\mathbf{x_0})\rangle}$ is given by \cite{Singer0}
\begin{align}
  \overline{\langle T(\epsilon\,|\,\mathbf{x_0})\rangle} =  \frac{R^2}{D}\left[\ln\left(\frac{1}{\epsilon}\right)+\ln(2)+\frac{1}{8} + O(\epsilon)\right]\,.\label{eq:Tda}
\end{align}

\item In $d=3$, it was found that for regular bounded domains $\Omega$ with a smooth boundary, the NET through a small disk of radius $\epsilon R$ located on the boundary diverges algebraically as \cite{Singer1}
\begin{align}
    \langle T(\epsilon\,|\,\mathbf{x_0})\rangle = \frac{|\Omega|}{4DR\epsilon} + O[\ln(\epsilon)]\,,\quad \epsilon \to 0\,.\label{eq:Teps3gen}
\end{align}
This result was extended to higher dimensions $d>3$ in \cite{Benichou08} where it was shown that
\begin{align}
    \langle T(\epsilon\,|\,\mathbf{x_0})\rangle \sim  \frac{C_d\,|\Omega|}{DR^{d-2}\,\epsilon^{d-2} }\,,\quad \epsilon\to 0\,,\label{eq:Tepsd}
\end{align}
but the amplitude $C_d$ was not computed. 
\end{itemize}
Recently, it was also argued that the cumulative distribution $\text{Prob.}\left[ \tilde T(\epsilon\,|\,\mathbf{x_0})>t\right]$ of the mean first-passage time $\tilde T(\epsilon\,|\,\mathbf{x_0})$ to a small target of size $\epsilon$ located \emph{inside} a domain $\Omega$ behaves in the limit of $t\to \infty$ as \cite{Meyer11,Benichou10}
\begin{align}
  \text{Prob.}\left[ \tilde T(\epsilon\,|\,\mathbf{x_0})>t\right] \sim  \frac{\langle\tilde T(\epsilon\,|\,\mathbf{x_0})\rangle}{ \overline{\langle\tilde T(\epsilon\,|\,\mathbf{x_0})\rangle}}\,\exp\left(-\frac{t}{\overline{\langle\tilde T(\epsilon\,|\,\mathbf{x_0})\rangle}}\right)\,, \quad \epsilon \to 0\,,\quad t \rightarrow\infty\,,\label{eq:PepsT}
\end{align}
where $\langle\tilde T(\epsilon\,|\,\mathbf{x_0})\rangle$ is the mean first-passage time to the target from the initial position $\mathbf{x_0}$ and $\overline{\langle \tilde T(\epsilon\,|\,\mathbf{x_0})\rangle}$ is the same quantity but averaged over an initial uniform position in the domain $\Omega$. It is natural to extend this result to the CDF of the NET which can be considered as a first-passage time to a target located on the boundary of the domain, as it was done in \cite{Rupprecht15} for the case of a spherical domain. To make the connection with our setting, we assumed that the asymptotic behavior (\ref{eq:PepsT}) is also valid for the CDF of the NET $\text{Prob.}\left[ T(\epsilon\,|\,\mathbf{x_0})>t\right]$.

We now show how our problem is related to the NET. We will restrict ourselves to the case of $d=2$ as the case of $d\geq 3$ can be done in a similar way \cite{BSG21b}. To obtain the distribution of $M_x(t)$ for the case of a two-dimensional Brownian motion in a disk of radius $R$ with reflecting boundaries starting from the origin, we make the following observation: 
 the cumulative distribution $\text{Prob.}\left(M_x(t)<M_x\right)$ is equal to the probability that the diffusive particle did not reach $M_x$ up to time $t$:
\begin{align}
  \text{Prob.}\left(M_x(t)<M_x\right) = \text{Prob.}\left(T_{M_x}>t\right)\,,\label{eq:id1d}
\end{align}
where $T_{M_x}$ is the first-passage time to $M_x$ starting from the origin. We now use the NET results presented above to compute $\text{Prob.}\left(T_{M_x}>t\right)$ in the limit of $t\to \infty$ by relying on the following assumption (see figure \ref{fig:circle})
\begin{align}
  \text{Prob.}\left(T_{M_x}>t\right) & \sim \text{Prob.}\left[T\left(\epsilon=\sqrt{\frac{2(R-M_x)}{R}}\,\bigg|\,\mathbf{0}\right)>t\right]\,,\quad t\to \infty\,,\label{eq:id}
\end{align}
where $T_{M_x}$ is the first-passage time to $M_x$ starting from the origin and $T(\epsilon|\mathbf{x_0})$ is the NET to reach an arc of angle $2\epsilon$ in a circular domain, given that the Brownian motion started at $\mathbf{x_0}$. In other words, the assumption (\ref{eq:id}) can be stated as follows: the probability to hit the arc of angle $2\epsilon$ for the first time is asymptotically equal to the probability to hit the chord subtended by this angle for the first time in the limit of $\epsilon\to 0$ (see figure \ref{fig:circle}). The assumption (\ref{eq:id}) is an approximation for any finite $t$ but we expect it to be asymptotically exact in the limit $t\to \infty$. Under this assumption, we can further use the identity (\ref{eq:id1d}) and the CDF of the NET (\ref{eq:PepsT}) to obtain that
\begin{align}
  \text{Prob.}\left(M_x(t)<M_x\right) \sim \frac{\left\langle T\left(\epsilon\,|\,\mathbf{0}\right)\right\rangle}{\overline{\left\langle T\left(\epsilon\,|\,\mathbf{x_0}\right)\right\rangle}}\exp\left(-\frac{t}{\overline{\left\langle T\left(\epsilon\,\big|\,\mathbf{x_0}\right)\right\rangle}}\right)\Bigg\rvert_{\epsilon=\sqrt{\frac{2(R-M_x)}{R}}} \,,\quad t \rightarrow\infty\,, \label{eq:Fmt}
\end{align}
where $\overline{\langle T(\epsilon|\mathbf{x_0})\rangle}$ is the NET averaged over an initial uniform position in the disk. We now recall the expressions of the NET for a disk given in (\ref{eq:Td0})-(\ref{eq:Tda}) evaluated at $\epsilon=\frac{\sqrt{2(R-M_x)}}{\sqrt{R}}$:
\begin{subequations}
\begin{align}
  \left\langle T\left(\epsilon\,|\,\mathbf{0}\right)\right\rangle\bigg\rvert_{\epsilon=\sqrt{\frac{2(R-M_x)}{R}}}&\sim \frac{R^2}{2D} \ln\left(\frac{1}{R-M_x}\right)+O(1)\,,\quad M_x\rightarrow R\,,\label{eq:Tapprox1}\\[1em]
  \overline{\left\langle T\left(\epsilon\,|\,\mathbf{x_0}\right)\right\rangle}\bigg\rvert_{\epsilon=\sqrt{\frac{2(R-M_x)}{R}}} &\sim \frac{R^2}{2D} \ln\left(\frac{1}{R-M_x}\right)+O(1)\,,\quad M_x\rightarrow R\,.\label{eq:Tapprox2}
\end{align}
\label{eq:Tapprox}
\end{subequations}
Inserting the NETs (\ref{eq:Tapprox}) in the cumulative distribution (\ref{eq:Fmt}), we find
\begin{align}
  \text{Prob.}\left(M_x(t)<M_x\right) \sim \exp\left(-\frac{2Dt}{R^2\,\ln\left(\frac{1}{R-M_x}\right)+O(1)}\right)\,,\quad t\to \infty\,.\label{eq:Probf}
\end{align}
By taking a derivative with respect to $M_x$ of the cumulative distribution (\ref{eq:Probf}) and by denoting $\chi_2=[R^2\,\ln\left(\frac{1}{R-M_x}\right)+O(1)]/(2Dt)$, one obtains the distribution (\ref{eq:dist1i}) displayed in the introduction with an unknown amplitude $A_2$. In principle, this amplitude can be obtained from the order O(1) term in (\ref{eq:Tapprox2}), which is, unfortunately, not known. If we assume that this next-to-leading order term is the same as the one in the expansion of the NET given in (\ref{eq:Td0})-(\ref{eq:Tda}), we obtain that $A_2=2\,e^{\frac{1}{4}}$. However, it is not clear that we are allowed to do so as the first-passage time to reach the arc of angle $2\epsilon$ and the first-passage time to reach the chord subtended by this angle might differ by finite-size corrections in the limit of $\epsilon\to 0$ (see figure \ref{fig:circle}). Nevertheless, this result is in good agreement with numerical data (see the right panel in figure \ref{fig:fchi2d}). From the distribution (\ref{eq:dist1i}), one can compute the average value of the fluctuations which gives
\begin{align}
   \langle \Delta(t )\rangle &=  A_2\,\int_0^\infty \frac{d\chi_2}{\chi_2^2}\,\exp\left[-\left(\frac{2Dt}{R^2} \,\chi_2+\frac{1}{\chi_2}\right)\right]\,.\label{eq:avgDelta2}
\end{align}
   This integral can be computed in the limit $t\to \infty$ by the saddle point method (see \cite{BSG21b}) and we recover the stretched exponential decay in (\ref{eq:saddleMxi}) displayed in the introduction.
This result is in good agreement with numerical simulations (see the left panel in figure \ref{fig:fchi2d}). The derivation for arbitrary dimensions and other geometries, such as the ellipse, can be found in the paper whose abstract is given on p.~\pageref{chap:A10}.

\begin{figure}
\begin{center}
 \fboxsep=10pt\relax\fboxrule=1pt\relax
 \fbox{
   \begin{minipage}{\textwidth} 
\hspace{2em}
\begin{center}
\LARGE \bf Statistics of the maximum and the convex hull of a Brownian motion in confined geometries
\end{center}

\hspace{2em}
\begin{center}
B. De Bruyne, O. Bénichou, S. N. Majumdar and G. Schehr,
J.~Phys.~A:~Math.~Theor. {\bf 55}, 144002 (2021).
\end{center}

\hspace{2em}
\begin{center}
  {\bf Abstract:} 
\end{center}

 We consider a Brownian particle with diffusion coefficient $D$ in a $d$-dimensional ball of radius $R$ with reflecting boundaries. We study the maximum $M_x(t)$ of the trajectory of the particle along the $x$-direction at time $t$. In the long time limit, the maximum converges to the radius of the ball $M_x(t) \to R$ for $t\to \infty$. We investigate how this limit is approached
  and obtain an exact analytical expression for the distribution of the fluctuations $\Delta(t) = [R-M_x(t)]/R$ in the limit of large $t$ in all dimensions. We find that the distribution of $\Delta(t)$ exhibits a rich variety of behaviors depending on the dimension $d$. These results are obtained by establishing a connection between this problem and the narrow escape time problem. We apply our results in $d=2$ to study the convex hull of the trajectory of the particle in a disk of radius $R$ with reflecting boundaries. We find that the mean perimeter $\langle L(t)\rangle$ of the convex hull exhibits a slow convergence towards the perimeter of the circle $2\pi R$ with a stretched exponential decay $2\pi R-\langle L(t)\rangle \propto \sqrt{R}(Dt)^{1/4} \,e^{-2\sqrt{2Dt}/R}$. Finally, we generalize our results to other confining geometries, such as the ellipse with reflecting boundaries. Our results are corroborated by thorough numerical simulations.
\end{minipage}
   }
\end{center}
\captionsetup{labelformat=empty}
\caption{\textbf{Abstract of article \themycounter} : Statistics of the maximum and the convex hull of a Brownian motion in confined geometries.}
\label{chap:A10}
\addtocounter{mycounter}{1}
\end{figure}

\newpage

\chapter{Constrained stochastic processes}
\label{chap:con}
In the previous chapter, we discussed some analytical results on EVS in stochastic processes. The models and the observables that we considered were sufficiently simple to study them analytically. However, in many practical situations, it is not possible to obtain exact analytical results and one needs to resort to numerical simulations. When dealing with EVS, this can be particularly challenging as they are rare, by definition, and are therefore difficult to observe, even numerically. This highlights the need for specific numerical methods to study EVS. The goal of this chapter is to study some of these methods to generate  stochastic processes which are constrained to rare events. In Section \ref{sec:conB}, we will first review some existing methods to sample rare trajectories for Brownian motion. In Section \ref{sec:conD}, we will extend one of these methods to discrete-time random walks. Finally, in Section \ref{sec:gen}, we will generalize it to other stochastic processes and various rare trajectories. 

\section{Constrained Brownian motion}
\label{sec:conB}
Let us consider a Brownian motion $x(t)$ in one dimension which evolves according to the Langevin equation (\ref{eq:eomb}). Simulating a free Brownian motion is easy: one just discretizes the time with increments $\Delta t$ in the Langevin equation (\ref{eq:eomb}), which gives 
\bea \label{eq:discrete_L}
x(t+\Delta t) = x(t) + \sqrt{2D} \, \eta(t) \, \Delta t \;.
\eea
One then draws independently, at each step, a jump length $ \sqrt{2D} \eta(t) \Delta t$ distributed as a Gaussian with zero mean and variance $2 D \Delta t$. In the limit $\Delta t\to 0$, this will converge to a Brownian trajectory.

This simple procedure however does not work when the Brownian motion is constrained on some (rare) event. Examples of constrained Brownian motions are abundant. For instance, there have been several studies on Brownian bridges, Brownian excursions, Brownian meanders, reflected Brownian motion, etc \cite{Yor2000,majumdar2007brownian,MP2010,Dev2010,PY2018}. These constrained Brownian motions appear naturally in many applications, ranging from ecology to finance and statistics \cite{Giu,Randon09Convex,Majumdar10Random,Shepp79,bouchaud_satya,CB2012,Kol1933,Smirnov48}.  A prominent example is the Brownian bridge $X(t)$, which is the diffusive limit of the discrete-time bridge random walks in (\ref{eq:bridgec}). The Brownian bridge is a Brownian motion evolving according to (\ref{eq:eomb}) with the constraint that it must return to the origin at a fixed time $t_f$ (see figure \ref{fig:bridgeB}):
\begin{align}
  X(0) = X(t_f) = 0\,.
\end{align}
 These bridge trajectories are in principle difficult to obtain numerically as the probability that they occur is small. A natural question then arises: `How do we generate such Brownian bridges efficiently ?''  Fortunately, there is a simple way to generate a trajectory which consists in a path transformation from a free trajectory $x(t)$ to a bridge trajectory $X(t)$ given by
\begin{align}
 X(t) = x(t) - \frac{t}{t_f}\,x(t_f)\,,\quad t\in [0,t_f]\,,\label{eq:bcbm}
\end{align}
where $x(t)$ is a \textit{free} Brownian motion starting at the origin $x(0)=0$. Note that the bridge condition $X(t_f) = X(0)=0$ is manifestly satisfied by the construction (\ref{eq:bcbm}). In addition, it is easy to show from (\ref{eq:eomb}) that the covariance of the process $X(t)$ is $\langle X(t)X(t')\rangle=2D\min(t,t')-2Dtt'/t_f$, which corresponds to the Brownian bridge one. Because (\ref{eq:bcbm}) is a linear mapping between two Gaussian processes, this is sufficient to show that $X(t)$ is indeed a Brownian bridge. However, this construction is very specific to the continuous-time Brownian bridge and cannot be easily generalized to generate other constrained Brownian motions. It would then be nice to have a general method to generate constrained Brownian motions that are not specific to a particular type of constraint. Indeed, in probability theory, there exists the well-known Doob transforms \cite{Doob,Pitman} that provide a prescription to construct constrained Markov trajectories \cite{CT2013,Rose21}.  

Recently, for continuous-time Markov processes, this transform was explicitly studied and led to an effective Langevin equation with a constraint-force term \cite{MajumdarEff15,Orland,CT2013}. For instance, a Brownian bridge with $X(0)=X(t_f) =0$ is generated by the effective Langevin equation~\cite{MajumdarEff15,CT2013}  
\bea \label{bridge_eff}
\dot X(t) = -\partial_X U_\text{eff}(X,t) + \sqrt{2D}\,\eta(t) \;,
\eea
where the effective potential is given by
\begin{align}
  U_\text{eff}(X,t) = \frac{X^2}{2(t_f-t)}\,,\label{eq:Ueffb}
\end{align}
and where $\eta(t)$, as before, is a Gaussian white noise with zero mean and which is delta-correlated. On the right-hand side in (\ref{bridge_eff}), the first term is an effective constraint-force that drives
the particle to the final position $X(t_f)=0$ at time $t_f$. The potential in (\ref{eq:Ueffb}) is a confining harmonic potential whose intensity grows and ultimately pins the particle back at the origin. Trajectories can then be easily generated by time-discretizing the effective Langevin equation (\ref{bridge_eff}) as in the case of the free Brownian motion in (\ref{eq:discrete_L}).
 This construction of an effective Langevin equation is rather versatile and can be extended to other constrained continuous-time Markov processes \cite{CT2013,MajumdarEff15,Baldassarri2021}, such as Brownian excursions, Brownian meanders, Ornstein-Uhlenbeck bridges and more recently to non-intersecting Brownian motions \cite{Grela2021}. For instance, for Brownian excursions, which must return to the origin at $t=t_f$ and remain on the positive axis during the whole duration, the effective potential is given by \cite{MajumdarEff15}
 \begin{align}
   U_\text{eff}(X,t) = \frac{X^2}{2(t_f-t)} -2D\ln\left(\frac{X}{\sqrt{4D(t_f-t)}}\right)\,.\label{eq:Ueffe}
 \end{align}
 For Brownian meanders, which just remain positive during the whole duration, the effective potential is given by \cite{MajumdarEff15}
 \begin{align}
   U_\text{eff}(X,t) = -\ln\left[\text{erf}\left(\frac{X}{\sqrt{4D(t_f-t)}}\right)\right]\,.\label{eq:Ueffm}
 \end{align}
Numerical realizations of Brownian bridges, excursions, and meanders are given in figure \ref{fig:bridgeBn}.
\begin{figure}[t]
  \begin{center}
    \includegraphics[width=0.3\textwidth]{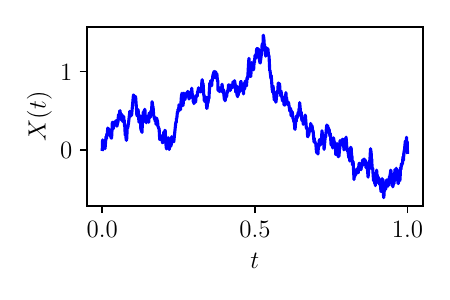}\includegraphics[width=0.3\textwidth]{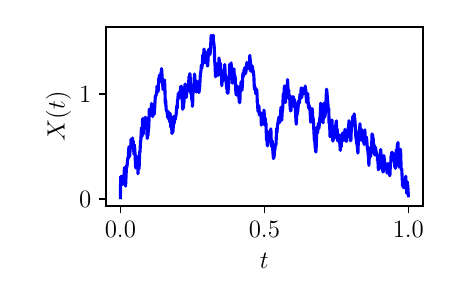}\includegraphics[width=0.3\textwidth]{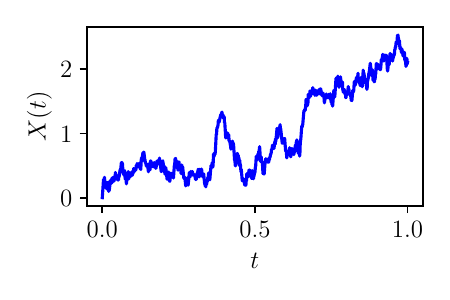}
    \caption{Realizations of a Brownian bridge (left panel), Brownian excursion (center panel), and Brownian meander (right panel), generated with the effective Langevin equation (\ref{bridge_eff}) along with the respective effective potentials in (\ref{eq:Ueffb}), (\ref{eq:Ueffe}), and (\ref{eq:Ueffm}). In the three panels, we have set $D=1$ and $t_f=1$.}
    \label{fig:bridgeBn}
  \end{center}
\end{figure}

Let us now briefly sketch the derivation of the effective Langevin equation (\ref{bridge_eff}) for the case of the Brownian bridge (the case of the meander and excursion can be done similarly). The main idea is as follows. We consider a bridge trajectory of duration $t_f$ and ask what is the probability that the process is located at $X$ at an intermediate time $t$? Given that the process is Markovian, we observe that this probability is the product of the probability of two independent events: (i) the event that the process reaches $X$ at time $t$ given that it started from the origin initially, (ii) the event that the process reaches the origin at time $t_f$ given that it was located at $X$ at time $t$. Due to the time invariance of Brownian motion, it can be written as (see figure \ref{fig:bridgeB})
\begin{align}
  P_b(X,t|t_f) =  \frac{P(X,t)Q(X,t_f-t)}{P(0,t_f)}\,,\label{eq:b}
\end{align}
where $P(x,t)$ and $Q(x,t)$ are respectively the forward and backward propagators of free Brownian motion. The index $b$ in $P_b(X,t|t_f)$ refers to the ``bridge'' propagator. The forward propagator $P(x,t)$ is the probability that a free Brownian motion is located at $x$ at time $t$ given that it started from the origin. It satisfies the well-known forward Fokker-Plank equation
\begin{align}
  \partial_t P(x,t) = D \partial_{x}^2 P(x,t)\,,\label{eq:forward}
\end{align}
with the initial condition $P(x,0)=\delta(x)$. The backward propagator $Q(x,t)$ is the probability that a free Brownian motion reaches the origin in a time $t$ given that it is located at $x$. By time and space reflection invariance, it is simply given by $Q(x,t)=P(x,t)$. Note that the name of forward and backward propagators refers to the fact that $x$ is the final position in the former one and is the initial position in the latter one. Upon taking a time derivative of the bridge propagator in (\ref{eq:b}) and using the equation (\ref{eq:forward}), we find that the bridge propagator also satisfies the Fokker-Plank equation given by \cite{MajumdarEff15}
\begin{align}
  \partial_t P_b(X,t|t_f) = D\partial_{X}\left[\partial_X P_b(X,t|t_f) - 2 P_b(X,t|t_f)\partial_X \ln(Q(X,t))\right]\,.
\end{align}
From this equation, we see that the Fokker-Plank equation originates from the following Langevin equation 
\begin{align}
  \dot X(t) =   2D\partial_X  \ln(Q(X,t)) + \sqrt{2D}\,\eta(t)\,.\label{eq:efflnQ}
\end{align}
This equation was initially obtained in the probability literature by Doob \cite{Doob}, and upon replacing $Q(X,t)=e^{-X^2/(4Dt)}/\sqrt{4\pi Dt}$, which is the solution of (\ref{eq:forward}), we recover the effective Langevin equation in (\ref{bridge_eff}) with the effective potential in (\ref{eq:Ueffb}). The effective potentials in (\ref{eq:Ueffe}) and (\ref{eq:Ueffm}) for the excursions and meanders can be obtained by inserting the appropriate expression for the backward propagator in (\ref{eq:efflnQ}).

\begin{figure}[t]
  \begin{center}
    \includegraphics[width=0.4\textwidth]{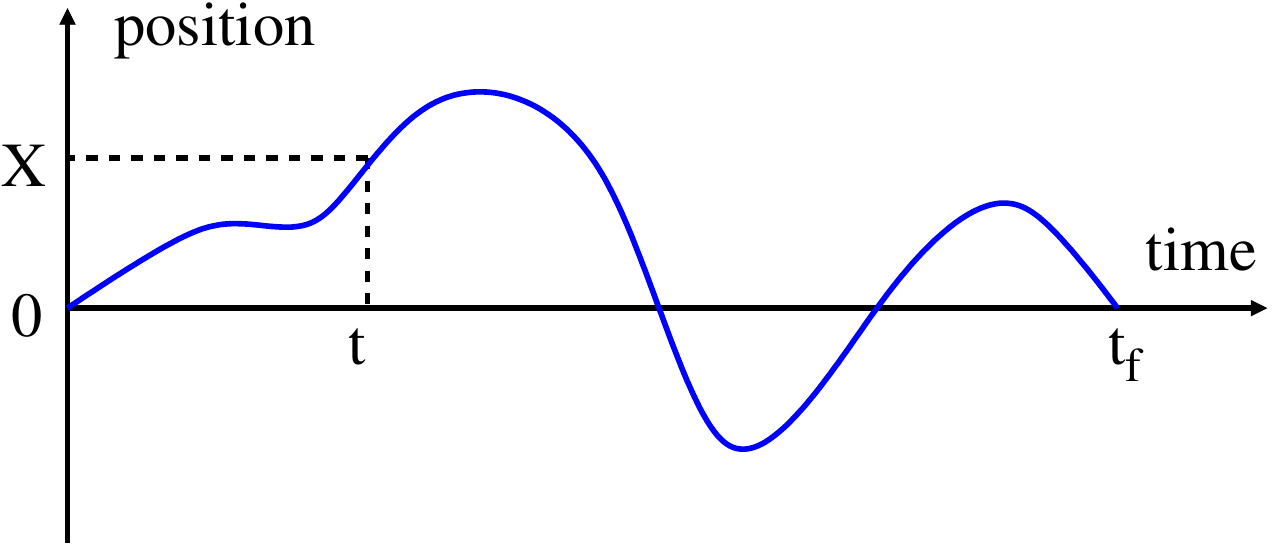}
    \caption{Schematic representation of a Brownian bridge of duration $t_f$. Due to the Markov property, a Brownian bridge can be decomposed into two independent parts: a left part over the interval $[0,t]$, where it propagates from $0$ to $X$, and a right part over the interval $[t,t_f]$, where it propagates from $X$ to $0$.}
    \label{fig:bridgeB}
  \end{center}
\end{figure}

This construction of an effective Langevin equation works very nicely for continuous-time Markov processes. However, it is not immediately clear how to extend it to discrete-time random walks. This will be the topic of the next section.

\section{Constrained discrete-time random walks}
\label{sec:conD}
In this section, we present the main idea that was used to obtain the results on bridge random walks discussed in Section \ref{sec:geni}. We consider a bridge discrete-time random walk $X_m$ locally evolving according to (\ref{eq:xm}) with the bridge constraint (\ref{eq:bridgec}). As in the continuous-time Brownian bridge, we show that discrete-time bridges can
be generated by an effective Markov jump process as in (\ref{eq:xm}), but the jumps $\eta_m$ have to be drawn from an effective
distribution that depends on the bare jump distribution $f(\eta)$ and that effectively accounts for the bridge constraint. The derivation of the effective jump process for discrete-time random walk bridges follows closely the approach used
for the continuous time Brownian bridge in section \ref{sec:conB}.

Consider a bridge random walk trajectory in figure \ref{fig:bridge} where the walk starts at the origin, returns to the origin after $n$ steps, and arrives at $X$ at an intermediate time $m$. Using the Markov property, this trajectory can be decomposed into a left part over the time interval $[0,m]$ and a right part over the interval $[m,n]$. Clearly, as in the continuous-time case in (\ref{eq:b}), the probability $P_b(X,m\,|\,n)$ that the bridge random walk is located at $X$ at step $m$  can then be written as
\begin{align}
    P_b(X,m\,|n) = \frac{P(X,m)\,Q(X,n-m)}{P(0,n)}\,, \label{eq:bridgePQ1}
 \end{align}
 where $P(x,m)$ and $Q(x,m)$ are respectively the forward and backward propagators of free discrete-time random walks. The forward propagator $P(x,m)$ is the probability that the random walk is located at $x$ at step $m$ given that it started from the origin. It satisfies the integral equation
\begin{align}
  P(x,m) = \int_{-\infty}^\infty dy P(y,m-1)f(x-y)\,,\label{eq:P0}
\end{align}
with the initial condition $P(x,0)=\delta(x)$. The equation (\ref{eq:P0}) can be obtained by stating that for the particle to be located at $x$ at step $m$, it must have been at some $y$ at step $m-1$ and jumped from $y$ to $x$. This event is then summed over all $y$ weighted by the jump distribution $f(x-y)$. The forward propagator in this section is denoted $P(x,m)$, instead of $p_m(x)$ as in (\ref{eq:prop}), for convenience of notation. The backward propagator $Q(x,m)$ is the probability that the random walk reaches the origin in $m$ steps given that it is located at $x$. It satisfies the backward Fokker-Plank equation
\begin{align}
Q(x,m) = \int_{-\infty}^\infty dy\, Q(y,m-1)f(y-x)\,,\label{eq:Q0}
\end{align}
which has to be solved with condition $Q(x,0)=\delta(x)$. The equation (\ref{eq:Q0}) can be obtained by stating that for the particle to
reach the origin from $x$ in $m$ steps, it must first jump to some $y$ and reach the origin in $m-1$ steps. This event is then summed over all $y$ weighted by the jump distribution $f(y-x)$. Note that, unlike in the forward case, where one varies the ``final position'', in the backward case, one varies the initial position. Note that both equations, (\ref{eq:P0}) and (\ref{eq:Q0}), are valid even when the jump distribution $f(\eta)$ is non-symmetric.

Our goal now is to write a forward Kolmogorov-type equation for the bridge propagator $P_b(X,m\,|n)$ in (\ref{eq:bridgePQ1}). To do so, we replace $Q(X,n-m)$ in (\ref{eq:bridgePQ1}) by using the backward equation in (\ref{eq:Q0}) and find that $P_b(X,m\,|n)$ satisfies the following integral equation \cite{BSG21ca}
\begin{align}
   P_b(X,m\,|\,n)  &=  \int_{-\infty}^\infty dY \,P_b(Y,m-1\,|\,n)\, \tilde f(X-Y\,|\,Y,m-1,n)\,,\label{eq:effInt}
\end{align}
where the effective jump distribution $\tilde f(X-Y\,|\,Y,m,n)$ at time $m$ of the bridge of length $n$ is given in (\ref{eq:eff}). One can then generate bridge random walks by using the effective Markov rule in (\ref{eq:bridgeeff}) with jumps drawn from the effective jump distribution $\tilde f(X-Y\,|\,Y,m-1,n)$. 

To illustrate this result, we now discuss a simple application of our method to the case of a lattice bridge random walk of $n$ steps. The free jump distribution is
\begin{align}
  f(\eta) = \frac{1}{2}\delta(\eta-1)+\frac{1}{2}\delta(\eta+1)\,,\label{eq:fSim}
\end{align}
as it jumps to $+1$ or $-1$ with probability $1/2$.
The backward propagator $Q(Y,m)$, in this case, is well-known and can be easily computed as follows. Let $n_+$ and $n_-$ denote the number of positive and negative jumps respectively that bring the walker from the initial position $Y$ to $0$ in $m$ steps. Clearly $n_+ + n _- = m$ and $n_+ - n_- = -Y$. Consequently $n_+ = (m-Y)/2$  and $n_- = (m+Y)/2$. Note that $Y$ has to be such that both $n_+$ and $n_-$ are integers. The probability that $n_-$ out of $m$ steps are negative is simply given by the binomial distribution $P(n_-|m) = \binom{m}{n_-} 2^{-m}$. Hence, replacing $n_-$ by $(m+Y)/2$ gives
the backward propagator
\begin{align}
  Q(Y,m) = \binom{m}{\frac{m+Y}{2}}\,2^{-m}\,, \label{eq:propsimple}
\end{align}
where $(m+Y)$ is even (otherwise $Q(Y,m)$ vanishes). We now substitute (\ref{eq:propsimple}) in (\ref{eq:eff}), giving the effective jump distribution
\begin{align}
   \tilde f(\eta\,|\,Y,m,n) = \frac{1}{2}\left(1-\frac{Y}{n-m}\right)\delta(\eta-1) + \frac{1}{2}\left(1+\frac{Y}{n-m}\right)\delta(\eta+1) \label{eq:effSim} \;.
  \end{align}
One can directly sample the jumps from the effective jump distribution in (\ref{eq:effSim}) to generate bridge trajectories (see the left panel in figure \ref{fig:bridgeSim}).
\begin{figure}[t]
 \includegraphics[width=0.5\textwidth]{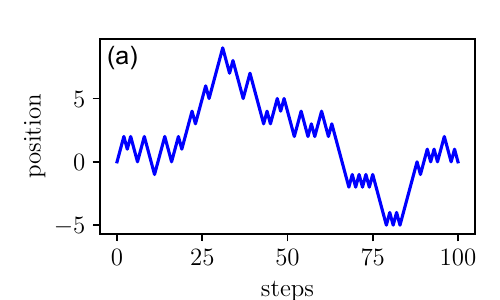}%
\hfill
  \includegraphics[width=0.5\textwidth]{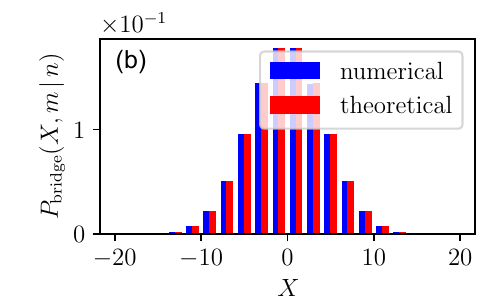}%
\hfill
\caption{\textbf{Left panel:} A typical trajectory of a lattice bridge random walk of $n=100$ steps generated by the effective jump distribution in (\ref{eq:effSim}). \textbf{Right panel:} Position distribution at $m=75$ for a lattice bridge of $n=100$ steps. The position distribution $P_{b}(X,m|n)$ obtained numerically by sampling $10^5$ trajectories using the effective  distribution (\ref{eq:effSim}) is compared with the theoretical prediction in (\ref{Pbridgelat}). }\label{fig:bridgeSim}
\end{figure}
 In the right panel in figure \ref{fig:bridgeSim}, we computed numerically the probability distribution of the position at some intermediate time by generating bridge trajectories from (\ref{eq:effSim}). This is compared to the theoretical position distribution for the bridge which can be easily computed by substituting the free propagators $P(X,m) = \binom{m}{\frac{X+m}{2} }2^{-m}$ and $Q(X,m)$ from (\ref{eq:propsimple}) in (\ref{eq:bridgePQ1}), which gives
\bea \label{Pbridgelat}
P_{b}(X,m|n) = \frac{\binom{m}{\frac{m+X}{2}} \binom{n-m}{\frac{n-m+X}{2}}}{\binom{n}{\frac{n}{2}}} \;,
\eea
which is nonzero only if $(m+X)$ as well as $n$ are both even numbers.  Note that, for the particular case of the lattice walk, the effective jump distribution (\ref{eq:effSim}) was already known in the mathematics literature, see for instance \cite{ChafaiRW12}. We discuss another example of a Cauchy bridge random walk in Section \ref{sec:geni}. 

The method outlined above extends nicely to other types of rare trajectories. Another natural constraint is the meander in which a random walk must stay above the origin (see figure \ref{fig:meander}). It is described by the equation of motion (\ref{eq:xm}) along with the constraints
\begin{subequations}
\begin{align}
  X_0 &= 0\,,\\
  X_m &\geq  0 \,,\quad m = 1,\ldots, n\,.
\end{align}
\label{eq:meaM}
\end{subequations}
\begin{figure}[ht]
  \begin{center}
    \includegraphics[width=0.48\textwidth]{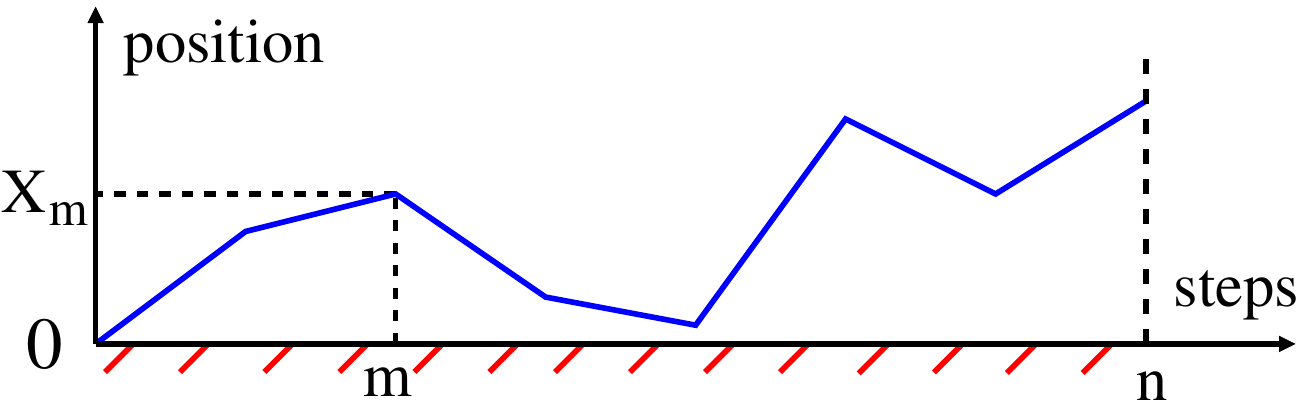}
    \caption{A meander random walk of $n$ steps is a random walk that is constrained to remain above the origin. Due to the Markov property, a meander random walk can be decomposed into two independent parts: a left part over the interval $[0,m]$, where it propagates from the $0$ to $X_m$ while staying positive, and a right part over the interval $[m,n]$, where it propagates to any position while staying positive. The horizontal line with red dashes just indicates that the origin at $X=0$ is absorbing.}
    \label{fig:meander}
  \end{center}
\end{figure}
One can show that meander trajectories can be generated by relying on an effective jump distribution $\tilde f_M(\eta\,|\,X,m,n)$ which is given by \cite{BSG21ca}
\begin{align}
\Aboxed{
   \tilde f_M(\eta\,|\,X,m,n) = f(\eta) \,\frac{S_{n-m-1}(X+\eta)}{S_{n-m}(X)}\,, }\label{eq:effGm}
\end{align}
where $f(\eta)$ is the free jump distribution and $S_m(X)$ is the survival probability of the free random walk, which is the solution of the Wiener-Hopf equation (\ref{eq:survi}). An interesting feature of (\ref{eq:effGm}) is that it has a well-defined limit when $n\to \infty$. Indeed by using the asymptotic results (\ref{eq:Snxa}) obtained in \cite{MMS17}, we find that one can generate trajectories that survive infinitely long by using the time-independent effective distribution
\begin{align}
\Aboxed{
  \tilde f_M(\eta\,|\,X) = \lim_{n\to \infty} \tilde f_M(\eta\,|\,X,m,n) = f(\eta) \,\frac{U(X+\eta)}{U(X)}\,,}\label{eq:inff}
\end{align}
where the function $U(X)$ is given in terms of its Laplace transform in (\ref{eq:UMou}). This function can be made explicit in a few cases. For instance for the double-sided exponential jump distribution $f(\eta)=e^{-|\eta|/a}/(2a)$, this function is given by $U(X)=(1+X/a)/\sqrt{\pi}$ \cite{MMS17}. Inserting this expression in (\ref{eq:inff}), we find that, in the case of the double-sided jump distribution, the effective jump distribution in (\ref{eq:inff}) is given by
\begin{align}
  \tilde f_M(\eta\,|\,X) = \frac{1}{2a}\,\frac{a+X+\eta}{a+X}e^{-\frac{|\eta|}{a}}\,,\quad \eta>-X\,.
\end{align}
Another interesting fact about the effective jump distribution (\ref{eq:inff}) is that for power-law distributions $f(\eta)\propto \eta^{-1-\mu}$ where $\mu$ is the Lévy index, the effective distribution has a heavier tail $\tilde f_M(\eta\,|X)\propto \eta^{-1-\mu/2}$ for $\eta\to \infty$. This can be seen from the asymptotic behavior $U(X)\propto X^{\mu/2}$ for $X\to \infty$ given in \cite{MMS17}.

Further examples of excursions and meanders bridge random walks can be found in the paper whose abstract is given on p.~\pageref{chap:A4}. In addition, an acceptance-rejection sampling method is also provided in order to numerically sample the effective jump distribution from the free jump distribution.

\begin{figure}
\begin{center}
 \fboxsep=10pt\relax\fboxrule=1pt\relax
 \fbox{
   \begin{minipage}{\textwidth}
\hspace{2em}
\begin{center}
\LARGE \bf Generating discrete-time constrained random walks and Lévy flights
\end{center}

\hspace{2em}
\begin{center}
B. De Bruyne, S. N. Majumdar and G. Schehr,
Phys.~Rev.~E. {\bf 104},  024117 (2021).
\end{center}

\hspace{2em}
\begin{center}
  {\bf Abstract:} 
\end{center}

We introduce a method to exactly generate \textit{bridge} trajectories for discrete-time random walks,
with arbitrary jump distributions, that are constrained to initially start at the origin and return to
the origin after a fixed time. The method is based on an effective jump distribution that implicitly
accounts for the bridge constraint. It is illustrated on various jump distributions and is shown to
be very efficient in practice. In addition, we show how to generalize the method to other types of
constrained random walks such as generalized bridges, excursions, and meanders.

\end{minipage}
   }
\end{center}
\captionsetup{labelformat=empty}
\caption{\textbf{Abstract of article \themycounter} : Generating discrete-time constrained random walks and Lévy flights.}
\label{chap:A4}
\addtocounter{mycounter}{1}
\end{figure}

\newpage

\section{Generalizations to other constrained stochastic processes}
\label{sec:gen}
In the previous section, we saw that the concept of an effective evolution equation for constrained processes is quite robust and can be extended to discrete-time random walks. Up to now, we have discussed rather local constraints, such as fixing the endpoint for the bridge processes. One might ask if it is possible to deal with more global constraints, such as fixing the total area below the trajectory? In addition, up to now, we have only dealt with Markovian processes and one might wonder if something can be said for non-Markovian ones? In this section, we work along these lines and provide further extensions of this method. We will first show that it allows us to generate Brownian trajectories with a fixed area below its curve, as well as trajectories with other ``global'' constraints. Then, we will show that the method can also be extended to a non-Markovian process, namely, the run-and-tumble particle, which we will introduce then.
\subsection{Brownian bridge and random walks with a fixed area}
Let us consider a one-dimensional Brownian motion $x(t)$ evolving according to (\ref{eq:eomb}). While the effective Langevin equation has proven to be a successful technique to generate constrained paths with local constraints, such as the initial and final points in the Brownian bridge (see Section \ref{sec:conB}), it is natural to ask if one can write an effective Langevin equation for paths with a global constraint, such as constraints on time-integrated quantities. One prominent example of a time-integrated quantity for Brownian motion is the area $A(t)$ under its trajectory
\begin{align}
  A(t) = \int_0^t dt'\, x(t')\,.\label{eq:Ac}
\end{align}
An interesting question to ask is: ``How to generate Brownian paths that return to the origin after a fixed amount of time $t_f$ with a fixed area $A_f$ under their trajectory?''. 
The area under a Brownian motion has received sustained interest and attention in various fields such as mathematics \cite{Takacs91,Janson07} and computer science due to its relation to algorithmic problems
\cite{KnuthThe98,FlajoletOn98,MajumdarExact02,majumdar2007brownian}. 
In physics, the area under a Brownian motion plays a central role in many problems, including 
$(1+1)$-dimensional fluctuating interfaces. Before explaining how to generate bridge Brownian trajectories with a fixed area, we will discuss further the motivations for such trajectories.

 An extensively studied model of $(1+1)$-dimensional fluctuating interfaces is governed by the celebrated Kardar-Parisi-Zhang (KPZ) equation which, in its simplest form, describes the spatio-temporal evolution of a height function $H(x,t)$ of an interface on a linear substrate of length $L$ \cite{KPZ}:
\begin{align}
  \partial_t H(x,t) = \partial^2_{x} H(x,t) + \lambda \left(\partial_x H(x,t)\right)^2 + \xi(x,t)\,,\label{eq:KPZ}
\end{align}
where $\xi(x,t)$ is a Gaussian white noise of zero mean with a correlator $\langle \xi(x,t)\xi(x',t') \rangle=\delta(x-x')\delta(t-t')$. When the non-linear term is absent ($\lambda=0$), the KPZ equation reduces to the well-known Edwards-Wilkinson (EW) interface model \cite{EW}. On a substrate of finite size $L$, the KPZ equation displays two regimes: (i) a growing regime for time $t \ll L^z$ (where the dynamical exponent is $z=3/2$) and (ii) a stationary regime when $t \gg L^z$ \cite{HHZ95,Krug97b}. While there have been extensive recent studies on the growing regime, which is connected to random matrix theory \cite{KK10,Cor12,HHT15}, here our focus is on the stationary regime. For $t \gg L^z$, 
the joint distribution of the heights $\{ H(x,t)\}$, for $0 \leq x \leq L$, does not reach a time-independent stationary state, since the mean height $\overline{H(x,t)}=\frac{1}{L}\int_0^L dx\, H(x,t)$ keeps growing with time. However, if one defines the relative heights as 
\begin{align}
  h(x,t) = H(x,t) - \overline{H(x,t)}\,, \label{eq:relh}
\end{align}
then the joint distribution of the relative heights $h(x,t)$ for $t \gg L^z$ does reach a stationary state. For periodic boundary conditions $h(0)=h(L)$, this stationary distribution is given by \cite{Majumdar04Flu1,Satya_Airy2}
\bea \label{stat_pdf}
P_{\rm stat}[\{h(x)\}] = \frac{1}{Z_L} e^{- \frac{1}{2} \int_0^L \left[\partial_x h(x)\right]^2 \, dx}\; \delta(h(0)-h(L)) \; \delta\left(\int_0^L h(x)\, dx\right) \;.
\eea
While the first delta-function represents the periodic boundary conditions, the second one reflects the constraint satisfied by the relative heights in (\ref{eq:relh}). 
Indeed, the definition in (\ref{eq:relh}) imposes the \textit{global} constraint that the total area under the relative heights is exactly zero. In (\ref{stat_pdf}), $Z_L$
is the partition function that normalizes the probability measure. Note that this stationary measure (\ref{stat_pdf}) holds both for the KPZ as well as the EW interface ($\lambda=0$). With the identification $h(x) \to x(t)$ and $x \in [0,L]$ transposing to $t \in [0,t_f]$ with $t_f=L$, the 
stationary measure in (\ref{stat_pdf}) corresponds to a Brownian bridge ($x(0) = x(t_f)$) with the global constraint that the area under the bridge is exactly zero. 
To sample the distribution $P_{\rm stat}[\{h(x)\}]$ in (\ref{stat_pdf}), one then needs to generate Brownian bridges constrained by the zero area condition. 
This is a concrete physical example of a Brownian bridge with a global constraint. This global constraint played a crucial role in the behavior of many stationary observables, such
as on the distribution of the maximal relative height \cite{Majumdar04Flu1,Satya_Airy2} and on the spatial persistence \cite{MD2006}. The effect of this global zero area constraint on the relative heights was also studied in various generalizations of interfaces with a non-Brownian stationary measure \cite{Majumdar01,gyorgyi}.  

Another generalization of the stationary measure of the Brownian interface with a zero area constraint (\ref{stat_pdf}) corresponds to studying $(1+1)$-dimensional solid-on-solid models 
on a discrete lattice of size $L$ with periodic boundary conditions of the form \cite{Schehr10Area}
\bea \label{stat_pdf_disc}
P_{\rm stat}[\{h_i\}] = \frac{1}{Z_L} e^{- K \sum_{i=1}^L |h_{i+1} - h_i|^\alpha}\;\delta(h_{L+1}-h_1) \;\delta\left( \sum_{i=1}^L h_i\right) \;,
\eea
where $h_i$ represents the stationary height of the interface at site $i$ and $\alpha > 0$. Here, instead of a Brownian motion in space, the interface height in the
stationary state performs a random walk in space
\bea \label{rw_hi}
h_{i+1} = h_i + \eta_i \;,
\eea 
where $\eta_i$'s are independent and identically distributed (IID) random noises, each drawn from a PDF $f(\eta) \propto e^{-K\,|\eta|^\alpha}$. More generally, the stationary
measure reads \cite{Schehr10Area}
\bea \label{stat_pdf_disc_gen}
P_{\rm stat}[\{h_i\}] = \frac{1}{Z_L} \left[\prod_{i=1}^L f(h_{i+1} - h_i)\right]\;\delta(h_{L+1}-h_1) \;\delta\left( \sum_{i=1}^L h_i\right) \;,
\eea
where $f(\eta)$ may have a heavy tail corresponding to a L\'evy interface in space. In these discrete cases also, to sample the stationary measure (\ref{stat_pdf_disc_gen}), one needs
to generate discrete time random walk bridges with the global zero area constraint, upon the identification $h_i \to x_i$ and the space index $i$ in the interface model identified with the time step of the random walk bridge. This is then a discrete-time random walk analog of its continuous-time counterpart, namely the Brownian bridge in the presence of the zero area constraint.    

Our main contribution to this line of work is to provide a method to generate Brownian and discrete-time random walk bridges with a fixed area below their trajectory. An effective Langevin equation for a continuous-time Brownian bridge path with a fixed total area $A$ under it up to time $t$ was recently derived by Mazzolo \cite{Mazzolo17a} by applying a technique developed for general continuous-time Gaussian processes by Sottinen and Yazigi \cite{SY2014}. However, for a discrete-time random walk bridge with arbitrary symmetric jump distribution, as required to study the solid-on-solid models discussed above, it is not clear how to generate an effective discrete Markov process for the paths with a fixed area $A$ under it. We proposed an alternative approach that allows us (i) to re-derive exactly the effective Langevin equation for the continuous-time Brownian bridge with a fixed area under it and (ii) to extend it to a discrete-time random walk bridge with arbitrary jump distributions. In addition, we showed how the effective Langevin equation can be generalized to other constraints, such as a fixed occupation time on the positive axis or a fixed quadratic area under the trajectory. 

We now consider a Brownian motion up to some fixed time $t_f$ and with a total fixed area under the curve $A_f = \int_0^{t_f} x_c(t) \, dt$ where 
$x_c(t)$ denotes the position of the Brownian motion at some intermediate time $t$ and the subscript $c$ refers to the fact that the motion is ``constrained''. To proceed, it is convenient to define a dynamical area variable $A_c(t) = \int_0^t x_c( t') \, d t'$. Now we consider the process $(x_c(t), A_c(t))$ jointly. The constraint on the trajectory is that it must start and return to the origin after a fixed amount of time $t_f$ with a given area $A_f$ under its trajectory, namely
\begin{align}
  x_c(0)=x_c(t_f)=0\,,\quad A_c(t_f)=A_f\,.\label{eq:constc}
\end{align}
Thus we can think of this joint process as a bridge in the plane, going from the initial value $(x_c(0)=0, A_c(0)=0)$ to the final value $(x_c(t_f)=0, A_c(t_f)=A_f)$. 

The derivation of the effective Langevin equation for this joint process then closely follows the derivation for a one-dimensional bridge in  
Section \ref{sec:conB} and can be found in \cite{BSG21cc}. We find that the effective Langevin equations read \cite{BSG21cc}
\begin{subequations}
\begin{align}
\Aboxed{
  \dot x_c(t) &= \sqrt{2\,D}\,\eta(t) - \frac{6\,(A_c(t)-A_f)}{(t_f-t)^2} -\frac{4\,x_c(t)}{t_f-t}\,,}\label{eq:effLBMa} \\
   \dot A_c(t) &= x_c(t)\,. \label{eq:effLBMb}
\end{align}
  \label{eq:effLBM}
\end{subequations}
This effective Langevin equation is the generalization of equation (\ref{eq:eff}) presented in the introduction, with the additional area constraint. This result coincides with the result of Mazzolo in \cite{Mazzolo17a} where this equation was derived using a different method. Note that this result was also obtained for the Ornstein-Uhlenbeck process in \cite{MazzoloOU}. By discretizing the effective Langevin equations (\ref{eq:effLBM}) over small time increments, it can be used to generate constrained trajectories (see the left panel in figure \ref{fig:constrained}). In the right panel in figure \ref{fig:constrained}, we computed numerically the marginal probability distributions of the position and the area at some intermediate time $t=t_f/2$, by generating trajectories from (\ref{eq:effLBM}). This is compared to the theoretical marginal distributions of the position and area for the bridge, which are computed in \cite{BSG21cc}.
\begin{figure}[t]
\centering
 \includegraphics[width=0.4\textwidth]{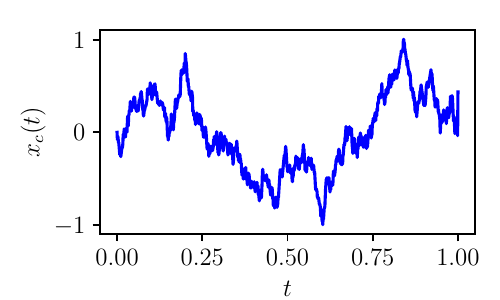}%
  \includegraphics[width=0.4\textwidth]{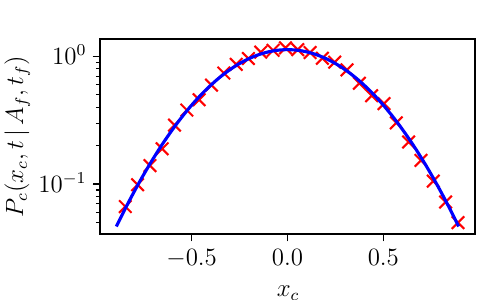}%
\\
 \includegraphics[width=0.4\textwidth]{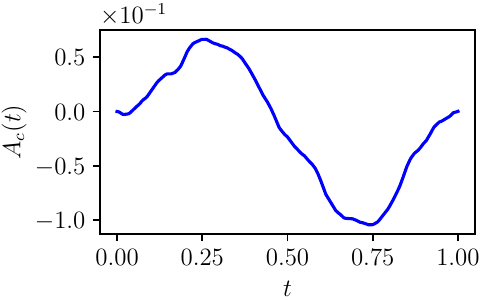}%
  \includegraphics[width=0.4\textwidth]{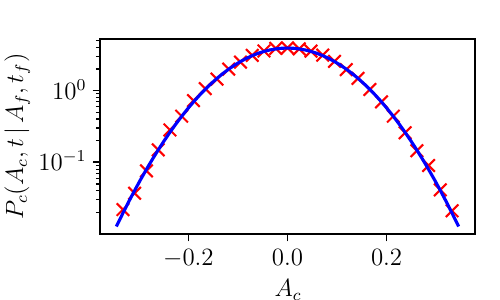}%
\caption{\textbf{Left panel:} A typical trajectory $x_c(t)$ vs $t$ (top) and $A_c(t)$ vs $t$ (bottom) for a bridge Brownian motion of duration $t_f=1$ with a zero area constraint $A_f=0$ generated by the effective Langevin equation (\ref{eq:effLBM}) for $D=1$. \textbf{Right panel:} Marginal position (top) and area (bottom) distributions at $t=t_f/2$ for a bridge Brownian motion of duration $t_f=1$ with a zero area constraint $A_f=0$. These marginal distributions, obtained numerically by sampling the trajectories from the effective Langevin equation (\ref{eq:effLBM}), are compared with the theoretical predictions given in \cite{BSG21cc}. }\label{fig:constrained}
\end{figure} 

Going beyond the area, one could ask the more general question: How to generate Brownian paths $x(t)$ of duration $t_f$ with the value of a general observable $\mathcal{O}(t_f)=\int_0^{t_f} dt\,V[x(t)]$ fixed where $V(x)$ can be any arbitrary function? Such observables are usually referred to as functionals of Brownian motion (see for instance \cite{majumdar2007brownian,Perret12}). While it seems difficult to provide an exact answer for an arbitrary $V(x)$, there exist two specific examples, beyond the area where $V(x)=x$, for which we can make analytical progress: the occupation time of Brownian motion on the positive axis, which corresponds to $V(x)=\Theta(x)$, where $\Theta(x)$ is the Heaviside step function, and a ``generalized area'' which corresponds to $V(x)=x^n$, where $n$ is an integer. Below, we show how to generate Brownian bridges with a fixed occupation time on the positive axis. The latter case, which is of interest in the context of characterizing the roughness of fluctuating $(1+1)$-dimensional interfaces \cite{Foltin94,Racz94}, is presented in \cite{BSG21cc}.

The occupation time on the positive axis $T(t)=\int_0^{t} dt'\,\Theta[x(t')]$ of a Brownian path $x(t')$ corresponds to the total amount of time it has spent on the positive axis. For a free Brownian motion, the distribution of this time follows the well-known ``L\'evy's arcsine law'' \cite{Levy} -- for generalizations to other stochastic processes see \cite{Lamperti58}. In physics, this observable is important in the context of stationary processes \cite{MB2002}, coarsening dynamics \cite{Dornic98,Newman98}, anomalous diffusive processes \cite{BBDG99,DeSmedt01,Dhar99}, blinking quantum dots \cite{Dahan03,MB05,barkai09} and spin glasses or disordered systems \cite{Majumdar02,MC02,SMC06,BB2007}. Below, we show how to generate exactly Brownian bridges $x_c(t)$ of duration $t_f$ with a fixed occupation time $T_c(t_f)=T_f$. The constraints on the trajectories read
\begin{align}
  x(0)=x(t_f)=0\,,\quad T_c(t_f) = T_f\,.\label{eq:constt}
\end{align}
Following similar steps as in Section \ref{sec:conB}, we find that the effective Langevin equations read \cite{BSG21cc}
\begin{subequations}
\begin{align}
  \Aboxed{\dot x_c(t) &= \sqrt{2\,D}\,\eta(t)+\,\frac{2\,\text{sign}(x_c(t))\sqrt{D}}{\sqrt{t_f-t}}  \times\mathcal{G}\left(y,\,\nu\right) \,,}\\
  \dot T_c(t) &= \Theta[x_c(t)]\,,
\end{align}
\label{eq:effLBMt}
\end{subequations}
where the function $\mathcal{G}(y,\nu)$ is given by
\begin{align}
 \mathcal{G}(y,\nu) = \frac{\sqrt{\pi \nu } y \left(y^2-6\right) e^{\frac{y^2}{4 \nu }} \text{erfc}\left(\frac{y}{2
   \sqrt{\nu }}\right)-2 \nu  \left(y^2-2\right)-4}{4 \nu  y-2 \sqrt{\pi\nu } \left(y^2-2\right)
   e^{\frac{y^2}{4 \nu }} \text{erfc}\left(\frac{y}{2 \sqrt{\nu }}\right)}\,,\label{eq:Fdimdm}
\end{align}
and where $y=\frac{|x_c(t)|}{\sqrt{D(t_f-t)}}$ and $\nu= \left(\frac{T_f-T_c(t)}{(t_f-t)-(T_f-T_c(t))}\right)^{\text{sign}(x_c(t))}$.
This effective Langevin equation can be used to generate constrained trajectories (see figure \ref{fig:constrainedo}).
\begin{figure}[t]
\centering
 \includegraphics[width=0.4\textwidth]{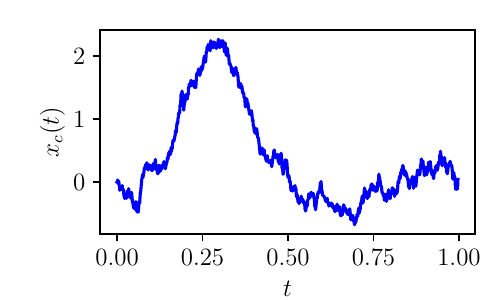}%
  \includegraphics[width=0.4\textwidth]{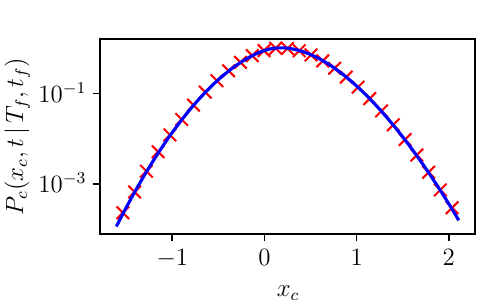}\\
 \includegraphics[width=0.4\textwidth]{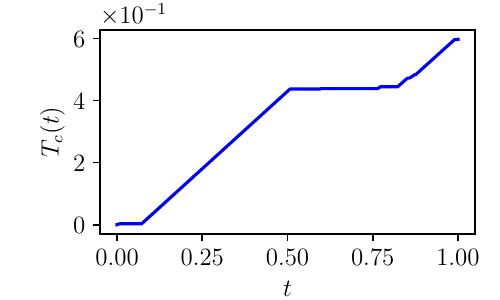}%
  \includegraphics[width=0.4\textwidth]{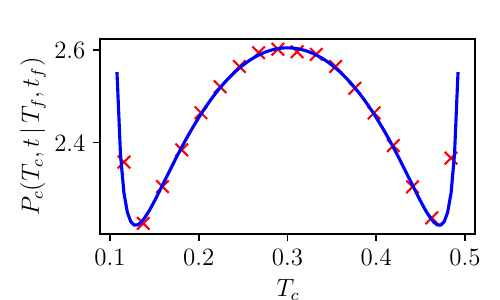}%
\caption{\textbf{Left panel:} A typical trajectory $x_c(t)$ vs $t$ (top) and $T_c(t)$ vs $t$ (bottom) of a bridge Brownian motion of duration $t_f=1$ with a fixed occupation time of $T_f=0.6$ generated by the effective Langevin equation (\ref{eq:effLBMt}) for $D=1$. \textbf{Right panel:} Marginal position (top) and occupation time (bottom) distributions at $t=t_f/2$ for a bridge Brownian motion of duration $t_f=1$ with a fixed occupation time of $T_f=0.6$. The distributions, obtained numerically by sampling the trajectories from the effective Langevin equation (\ref{eq:effLBMt}), are compared with the theoretical predictions given in \cite{BSG21cc}. }\label{fig:constrainedo}
\end{figure} In the right panel in figure \ref{fig:constrainedo}, we computed numerically the marginal probability distributions of the position and the occupation time at some intermediate time $t=t_f/2$, by generating trajectories from (\ref{eq:effLBMt}). This is compared to the theoretical marginal distributions of the position and the occupation time for the constrained process which are computed in \cite{BSG21cc}.

Further extensions to discrete-time random walks with global constraints and detailed computations can be found in the paper whose abstract is given on p.~\pageref{chap:A9}.
\newpage
\begin{figure}
\begin{center}
 \fboxsep=10pt\relax\fboxrule=1pt\relax
 \fbox{
   \begin{minipage}{\textwidth}

\hspace{2em}
\begin{center}
\LARGE \bf Generating stochastic trajectories with global dynamical constraints
\end{center}

\hspace{2em}
\begin{center}
B. De Bruyne, S. N. Majumdar, H. Orland and G. Schehr,
J.~Stat.~Mech.,  123204 (2021).
\end{center}

\hspace{2em}
\begin{center}
  {\bf Abstract:} 
\end{center}

 We propose a method to exactly generate Brownian paths $x_c(t)$ that are constrained to return to the origin at some future time $t_f$, with a given fixed area $A_f = \int_0^{t_f}dt\, x_c(t)$ under their trajectory. We derive an exact effective Langevin equation with an effective force that accounts for the constraint. In addition, we develop the corresponding approach for discrete-time random walks, with arbitrary jump distributions including L\'evy flights, for which we obtain an effective jump distribution that encodes the constraint. Finally, we generalize our method to other types of dynamical constraints such as a fixed occupation time on the positive axis $T_f=\int_0^{t_f}dt\, \Theta\left[x_c(t)\right]$ or a fixed generalized quadratic area $\mathcal{A}_f=\int_0^{t_f}dt \,x_c^2(t)$. 

 \end{minipage}
   }
\end{center}
\captionsetup{labelformat=empty}
\caption{\textbf{Abstract of article \themycounter} : Generating stochastic trajectories with global dynamical constraints.}
\label{chap:A9}
\addtocounter{mycounter}{1}
\end{figure}

 \clearpage
\subsection{Constrained run-and-tumble particles}
\label{sec:conRTP}
As we have seen above for Markov processes, such as the Brownian motion, the effects of constraints (e.g., bridges, excursions, meanders, etc) can be included in an effective Langevin equation (alternatively in effective transition probabilities for discrete-time processes). For non-Markovian processes, which are however abundant in nature \cite{Hanggi1995}, a similar effective Langevin approach is still lacking. For such processes, there are two levels of complexity: (i) the non-Markovian nature of the dynamics indicating temporal correlations in the history of the process and (ii) the effects of the additional geometrical constraints such as the bridge constraint. This two-fold complexity renders the derivation of an effective Langevin equation rather challenging for such processes. Our contribution to this line of work is to study an example of a non-Markovian process for which we show that the effective Langevin equation, ensuring the geometric constraints, can be derived exactly.  

Our example of a non-Markovian stochastic process is the celebrated run-and-tumble dynamics of a particle in one dimension, also known as the persistent random walk \cite{kac1974,weiss2002, masoliver2017}, which is of much current interest in the context of active matter \cite{berg08,marchetti13,cates15}. The run-and-tumble particle (RTP) is a simple model that describes self-propelled particles such as the \textit{E. coli} bacteria \cite{berg08}, that can move autonomously rendering them inherently different from the standard passive Brownian motion. Active noninteracting particles, including the run-and-tumble model, have been studied extensively in the recent past, both experimentally and theoretically \cite{berg08,marchetti13,cates15,bechinger16,tailleur08}. Even for such  noninteracting systems, a plethora of interesting phenomena have been observed, arising purely from the ``active nature'' of the driving noise, which typically induces a ballistic motion on short time scales, and diffusive motion on long time scales. These phenomena include, e.g., non-trivial density profiles \cite{Bijnens20,Martens12,Basu19,Basu20,Dhar19,Singh20,Santra20,Dean21}, dynamical phase transitions \cite{Doussal20,Gradenigo,MoriPRE}, anomalous transport properties \cite{Doussal20,Dor19,Demaerel19,Banerjee20}, or interesting first-passage and extremal statistics \cite{Orsingher90,Orsingher95,Lopez14,CinqueF20,CinqueS20,Foong92,Masoliver92,Angelani14,Angelani15,Artuso14,Evans18,Weiss87,Malakar18,Ledoussal19,MoriL20,MoriE20,DebruyneSur21,HartmannConvex20,Singh2019}. 

In its simplest form, a \textit{free} one-dimensional RTP moves (runs) with a fixed velocity $v_0$ in the positive direction during a random time $\Delta t$ drawn from an exponential distribution $p(\Delta t)=\gamma\, e^{-\gamma \Delta t}$ after which it changes direction (tumbles) and goes in the negative direction during another random time. The process continues and the particle performs this run-and-tumble motion indefinitely. The position of the particle $x(t)$ evolves according to the Langevin equation
\begin{align}
  \dot x(t)=v_0\,\sigma(t)\,,\label{eq:eomr}
\end{align}
where $\sigma(t)$ is a telegraphic noise that switches between the values $1$ and $-1$ with a \emph{constant} rate $\gamma$ (see figure \ref{fig:telegraphic}). During an infinitesimal time interval $dt$, the particle changes direction with probability $\gamma\, dt$ or remains in the same direction with the complementary probability $1- \gamma\, dt$:
\begin{align}
  \sigma(t+dt) = \left\{\begin{array}{rl}\sigma(t)\,  \quad & \text{with \, prob.~ }=1-\gamma\, dt\, , \\
  -\sigma(t)\,  \quad &\text{with \, prob.~ } =\gamma\, dt\, . \end{array}\right. \label{eq:telegraphic}
\end{align}
Consequently, the time between two consecutive tumbles $\Delta t$ is drawn independently from an exponential distribution $p(\Delta t)=\gamma \, e^{- \gamma \Delta t}$ and the sequence of tumbling times follows a Poisson process with a constant rate $\gamma$ (see figure \ref{fig:telegraphic}).
\begin{figure}[t]
        \centering
        \includegraphics[width=0.4\textwidth]{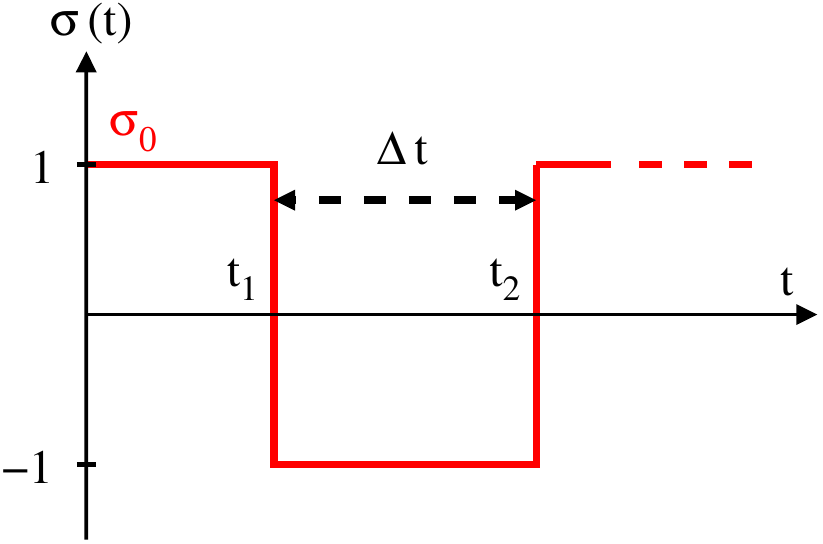}
    \caption{Telegraphic noise $\sigma(t)$ driving the sign of the velocity of the RTP. The signal switches with a constant rate $\gamma$. The time between two consecutive switches $\Delta t$ is drawn independently from an exponential distribution $p(\Delta t)=\gamma \, e^{- \gamma \Delta t}$. The sequence of tumbling times $t_1,\ldots,t_n$ follows a Poisson process of constant rate $\gamma$.}
            \label{fig:telegraphic}
\end{figure}
The autocorrelation function of the telegraphic noise is given by $\langle \sigma(t_1)\sigma(t_2) \rangle = e^{-2 \gamma (t_2-t_1)}$ and has a finite persistence time $ \gamma^{-1}$. This persistence, also called \emph{activity}, renders the process non-Markovian and hence does not fall into the universality class of the Brownian motion, which makes this process challenging to study. Nevertheless, it is possible to recover the Brownian diffusive regime by taking the scaling limit
\begin{align}
 \gamma\rightarrow \infty\,, \quad v_0\rightarrow \infty\,,\quad \frac{v_0^2}{2\, \gamma}\equiv D\, ,\label{eq:brownianlimit}
\end{align}such that the effective diffusion coefficient $D$ is finite. In this limit, the persistence time $ \gamma^{-1}$ tends to zero and the run-and-tumble particle behaves like a Brownian motion. Indeed, in this limit, the driving noise in the equation of motion (\ref{eq:eom}) becomes
\begin{align}
   \langle  v_0\,\sigma(t_1)\, v_0\,\sigma(t_2)\rangle = v_0^2\, e^{-2 \gamma (t_2-t_1)} \rightarrow 2D\, \delta(t_2-t_1)\, ,
\end{align}
which is the well-known uncorrelated white noise.

To generate a trajectory $x(t)$ of a free RTP starting from the origin with a given initial velocity
\begin{align}
  x(0)=0\,,\quad \dot x(0) = \sigma_0\,v_0\,,\label{eq:init}
\end{align}
where $\sigma_0=\pm 1$, one simply generates a sequence of tumbling times $t_1,\ldots,t_n$ that follow a \emph{homogeneous} Poisson process of constant rate $\gamma$:
\begin{align}
  t_{m+1} = t_{m} + \Delta t_m\,,\label{eq:tm}
\end{align}
where $\Delta t_m$ are independently drawn from an exponential distribution $p(\Delta t)=\gamma\, e^{-\gamma\,\Delta t}$. Then, the trajectory $x(t)$ of the particle is simply obtained by integrating the equation of motion (\ref{eq:eomr}) which yields the piecewise linear function:
\begin{align}
  x(t) = \sigma_0\,v_0\,(-1)^n\, (t-t_n)+\sum_{m=0}^{n-1} \sigma_0\, v_0\, (-1)^m \, (t_{m+1}-t_{m}) \,,\label{eq:xt}
\end{align}
where $n$ is such that $t_n$ is the latest tumbling time before $t$, i.e. such that $t_n<t<t_{n+1}$. The sum in (\ref{eq:xt}) accounts for all complete runs that happened before $t$ and the first term corresponds to the last run that is not yet completed at time $t$. This sampling method works well to generate \emph{free} run-and-tumble trajectories. However, as in the case of Brownian motion, some applications require sampling only specific trajectories, such as bridge trajectories where, in addition to satisfying the initial condition (\ref{eq:init}), the particle must also return to the origin after a fixed time $t_f$ with a given velocity $\sigma_f\,v_0$:
\begin{align}
  x(t_f)=0\,,\quad \dot x(t_f) = \sigma_f\, v_0\,,\label{eq:final}
\end{align}
where $\sigma_f =\pm 1$. Note that the final position need not necessarily be the origin but any fixed point in space -- here for simplicity we only consider the case where the final position coincides with the origin. One possible application of run-and-tumble bridge trajectories is in the context of animal foraging, where animals typically return to their nest after a fixed time, and one could study the persistence and memory effects in their trajectories
 \cite{Giu,Randon09Convex,Majumdar10Random,Murphy92,BLDS2009}. 
As in the case of Brownian motion, obtaining realizations of bridge trajectories using the free sampling method would be computationally wasteful. One, therefore, needs an efficient algorithm to generate run-and-tumble bridge trajectories, in a similar spirit as the effective Langevin equation (\ref{bridge_eff}) for Brownian motion. 

Using a similar path decomposition of a bridge trajectory (see figure \ref{fig:bridges}) as in the previous sections, as well as the fact that the RTP becomes a Markovian process in phase space, i.e.~in the $(x,\dot x)$ space, we derived an exact effective Langevin equation for RTPs to generate bridge trajectories efficiently. Let us simply present the final results, the full derivation is similar to the one in the previous sections (it can be found in \cite{BSG21cb}). We showed that the effective process, that automatically takes care of the bridge constraints (\ref{eq:init}) and (\ref{eq:final}) can be written as
\begin{align}
  \Aboxed{\dot x(t)=v_0\,\sigma^*(x,\dot x,t\,|\,\sigma_0,t_f,\sigma_f)\,,}\label{eq:effeom}
\end{align}
where $\sigma^*(x,\dot x,t\,|\,\sigma_0,t_f,\sigma_f)$ is now an effective telegraphic noise that switches between the values $1$ and $-1$ with a space-time dependent rate $ \gamma_B^*(x,\dot x,t\,|\,\sigma_0,t_f,\sigma_f)$, which we compute exactly. We find that the transition rates $\gamma_B^*(x,\dot x=\sigma v_0,t\,|\,\sigma_0,t_f,\sigma_f)$ for a particle that is located at $x$ with a velocity $\dot x=\sigma v_0$ at time $t$, given that it started at the origin with velocity $\dot x=\sigma_0 v_0$ and must return to the origin with velocity $\dot x=\sigma_f v_0$ at time $t_f$, are given by
\begin{subequations}
\begin{align}
\Aboxed{  \gamma_B^*(x,\dot x=+v_0,t\,|\,\sigma_0,t_f,\sigma_f) &= \gamma\, \frac{Q(x,\tau,-\,|\,\sigma_f)}{Q(x,\tau,+\,|\,\sigma_f)}\,,}\\[1em]
\Aboxed{  \gamma_B^*(x,\dot x=-v_0,t\,|\,\sigma_0,t_f,\sigma_f) &= \gamma\, \frac{Q(x,\tau,+\,|\,\sigma_f)}{Q(x,\tau,-\,|\,\sigma_f)}\,,}
\end{align}
\label{eq:effBQ}
\end{subequations}
where $\tau=t_f-t$ and $Q$ is the free backward propagator satisfying the backward Fokker-Planck equations:
\begin{subequations}
\begin{align}
  \partial_t Q(x,t,+)= +v_0\,\partial_x Q(x,t,+)-\gamma \,Q(x,t,+)+\gamma\, Q(x,t,-)\,,\\
  \partial_t Q(x,t,-)=- v_0\,\partial_x Q(x,t,-)-\gamma\, Q(x,t,-)+\gamma\, Q(x,t,+)\,.
\end{align}
\label{eq:Q}
\end{subequations}
The backward propagator $Q(x,t,\sigma|\sigma_f)$  is the probability density of the free particle to reach the origin at time $t$ with velocity $\dot x=\sigma_f\,v_0$ given that it started at $x$ with velocity $\dot x=\sigma\, v_0$. It can be obtained analytically by solving the differential equations (\ref{eq:Q}) on the real line along with the initial condition $Q(x,t=0,\sigma|\sigma_f)=\delta_{\sigma,\sigma_f}\,\delta(x)$. Physically, the effective tumbling rate in (\ref{eq:effBQ}) is the free tumbling rate that is modified in such a way that tumbling events that bring the particle closer to the origin are more likely to happen. Using the expression of the free backward propagator (given in \cite{BSG21cb}), we find the exact expressions of the transition rates. For example, when $\sigma_0=+1$ and $\sigma_f=-1$, we get
\begin{subequations}
\begin{align}
 \Aboxed{ \gamma_B^*(x,\dot x=+v_0,t\,|\,+,t_f,-) &= 2\,\gamma\,\delta[f(\tau,x)]+\,\gamma\,\sqrt{\frac{g(\tau,x)}{f(\tau,x)}} \frac{I_1[ h(\tau,x)]}{I_0[ h(\tau,x)]}\,,}\\[1em]
\Aboxed{  \gamma_B^*(x,\dot x=-v_0,t\,|\,+,t_f,-) &= \gamma\, \frac{1}{2\,\delta[f(\tau,x)]+\sqrt{\frac{g(\tau,x)}{f(\tau,x)}} \frac{I_1[ h(\tau,x)]}{I_0[ h(\tau,x)]}}\,,}
\end{align}
\label{eq:effB}
\end{subequations}
where $\tau=t_f-t$. In the expressions (\ref{eq:effB}), $I_0(z)$ and $I_1(z)$ denote the modified Bessel functions while the functions $f$, $g$, and $h$ are defined as   
\begin{align}
    f(t,x) = \gamma\,t-\frac{\gamma\,x}{v_0}\,,\quad g(t,x)= \gamma\,t+\frac{\gamma\,x}{v_0}\,, \quad h(t,x)=\sqrt{f(t,x)\,g(t,x)}\,. \label{eq:fgh}
  \end{align}  
  The Dirac delta terms in the effective rates (\ref{eq:effB}) enforce the particle to remain in the double-sided light cone defined as (see figure \ref{fig:bridges})
  \begin{figure}[t]
        \centering
        \includegraphics[width=0.4\textwidth]{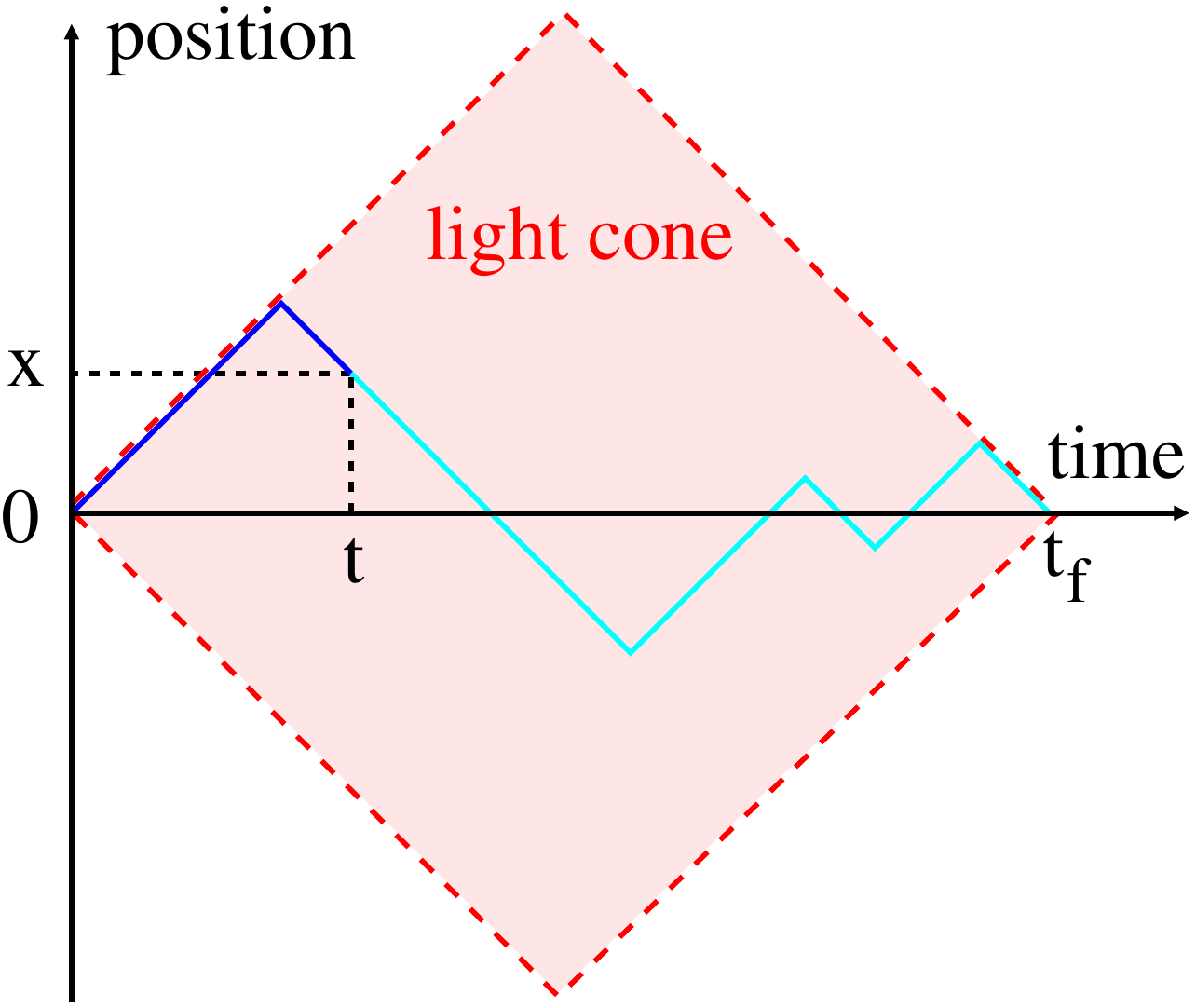}
    \caption{A sketch of a run-and-tumble bridge trajectory that starts at the origin with a positive velocity $\dot x=+v_0$ and returns to the origin at a fixed time $t_f$ with a negative velocity $\dot x=-v_0$.  Due to the Markov property in phase space $(x,\dot x)$, the bridge trajectory can be decomposed into two independent parts:  a left part over the time interval $[0,t]$ (dark blue), where the particle freely moves from the point $(0,+v_0)$ to the point $(x,-v_0)$ at time $t$ and a right part over the time interval $[t,t_f]$ (light blue), where it moves from the point $(x,-v_0)$ at time $t$ to the point $(0,-v_0)$ at time $t_f$. The combination of the finite velocity of the particle and the bridge condition induces a double-sided light cone in which the particle must remain (shaded red region). In addition to the double-sided light cone, the trajectory of the particle is further locally constrained by the initial and final conditions on its velocity.}
            \label{fig:bridges}
\end{figure}
 \begin{align}
   \left\{\begin{array}{ll} |x|\leq v_0\, t\,, &\text{when }\, 0\leq t \leq \frac{t_f}{2}\, ,\\ 
   |x|\leq v_0 \,(t_f-t) \,, &\text{when }\, \frac{t_f}{2}\leq t \leq t_f\, ,\label{eq:lc}
\end{array} \right.
 \end{align}
which is a natural boundary induced by the combination of the finite velocity of the particle along with the bridge constraint. In practice, when performing numerical simulations, these Dirac delta terms can be safely removed from the effective tumbling rates and can be replaced by hard constraints such that the particle must remain in the double-sided light cone (\ref{eq:lc}). 

RTPs with space and time dependent tumbling rates are relatively easy to simulate and there have been quite a few recent studies on them~\cite{Doussal20,Dor19,Singh20,Angelani14}. 
Unlike these models where the space and time dependency of the tumbling rates are ``put in by hand'', here we see from first principle how geometric constraints, such as the bridge condition, naturally generate space-time dependent tumbling rates. To generate trajectories of RTPs with space-time dependent tumbling rates, one proceeds as follows.  
Instead of generating a sequence of tumbling times that follow a \emph{homogeneous} Poisson process with a constant rate $\gamma$, as presented in the introduction, one needs to generate a sequence of times that follow a \emph{non homogeneous} Poisson process with a variable rate. There exist several methods to generate non homogeneous Poisson processes (see \cite{Lewis79} for a review). A quick and simple method is to discretize the effective equation (\ref{eq:effeom}) over small time increments $\Delta t$. One then obtains the bridge trajectories by repeating the following two steps
\begin{enumerate}
  \item evolve the discretized equation of motion according to
\begin{align}
  x_B(t+\Delta t) = x_B(t) + v_0\,\Delta t\,\sigma_B^*(x_B,\dot x_B,v_0,t)\,,\label{eq:eomd}
\end{align}
\item evolve the telegraphic signal according to
\begin{align}
 \!\!\!\!\!\!\!\sigma_B^*(x_B,\dot x_B,v_0,t+\Delta t) \!=\! \left\{\begin{array}{rl}\!\!\sigma_B^*(x_B,\dot x_B,v_0,t)\,   \!&\! \text{with \, prob.~ }\!\!\!\!\!=\!1-\gamma_B^*(x_B,\dot x_B,t)\, \Delta t\, , \\
  \!\!-\sigma_B^*(x_B,\dot x_B,v_0,t)\,  \!&\!\text{with \, prob.~ }\!\!\!\!\! =\!\gamma_B^*(x_B,\dot x_B,t)\, \Delta t\, , \end{array}\right. \label{eq:telegraphicN}
\end{align}
\end{enumerate}
where we have  omitted the conditional dependence in the telegraphic noise and switching rates.
This method is very simple to implement but nevertheless requires choosing the time increments $\Delta t$ sufficiently small such that the switching probabilities in (\ref{eq:telegraphicN}) do not exceed unity. It can be an issue if one is interested in regimes close to the light cone structure where the effective rates become large, as can be seen in (\ref{eq:effB}), and might require more advanced sampling techniques, such as events-based methods \cite{Lewis79}. Nevertheless, this method effectively generates run-and-tumble bridge trajectories and works well in practice (see the left panel in figure \ref{fig:bridgertp}) and is more efficient than the naive method.  In the right panel in figure \ref{fig:bridgertp}, we computed numerically the probability distribution of the position at some intermediate time $t=t_f/2$, by generating bridge trajectories from the effective tumbling rates (\ref{eq:effB}) and compared it to the theoretical position distribution for the bridge propagator given in \cite{BSG21cb}. 
\begin{figure}[t]
\centering
 \includegraphics[width=0.4\textwidth]{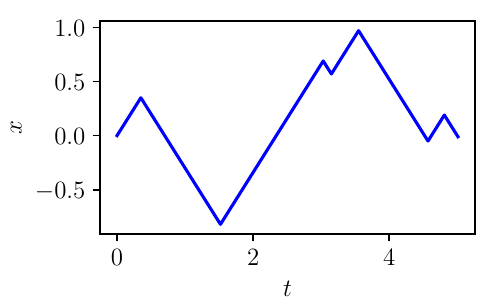}%
  \includegraphics[width=0.4\textwidth]{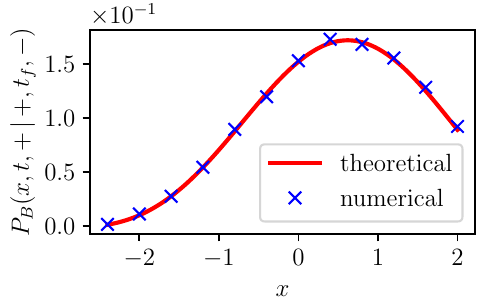}%
\caption{\textbf{Left panel:} A typical bridge trajectory of a RTP starting at the origin with a positive velocity $\dot x=+v_0$ and returning to the origin after a time $t_f=5$ with a negative velocity $\dot x=-v_0$. The trajectory was generated using the effective tumbling rates (\ref{eq:effB}) with $v_0=1$ and $\gamma=1$. \textbf{Right panel:} Position distribution at $t=t_f/2$ for a RTP starting at the origin with a positive velocity $\dot x=+v_0$ and returning to the origin after a time $t_f=5$ with a negative velocity $\dot x=-v_0$. The position distribution $P_B(x,t,+\,|\,+,t_f,-)$ obtained numerically by sampling from the effective tumbling rates (\ref{eq:effB}) is compared with the theoretical prediction given in \cite{BSG21cb}.  }\label{fig:bridgertp}
\end{figure}

Note that in the diffusive limit (\ref{eq:brownianlimit}), the effective tumbling rates (\ref{eq:effB}) both become the same constant $\gamma$ which is independent of $x$ and $t$. The signature of the bridge constraint can be found in the second-order term of this limit which gives 
 \begin{subequations}
\begin{align}
 \gamma_B^*(x,\dot x=+v_0,t\,|\,+,t_f,-)  &\sim \gamma+\frac{x}{\tau\,\sqrt{2\,D}}\,\gamma^{\frac{1}{2}}+O(\gamma^{-1}),\\
 \gamma_B^*(x,\dot x=-v_0,t\,|\,+,t_f,-)  &\sim \gamma-\frac{x}{\tau\,\sqrt{2\,D}}\,\gamma^{\frac{1}{2}}+O(\gamma^{-1})\,,
\end{align} 
\label{eq:effgbm}
 \end{subequations}
where we used the asymptotic expansion of the Bessel function $I_{0,1}(x)\sim e^{x}\,/\sqrt{2\pi x}$ for $x\rightarrow \infty$ and the expressions of the functions $f$, $g$, and $h$ defined in (\ref{eq:fgh}). Note that one needs to retain the subleading terms up to order $O(\sqrt{\gamma})$ to capture the nontrivial $x$-dependence, which indeed ensures the bridge condition. Upon using the asymptotic expansion of these rates, one can recover the effective Langevin equation for the Brownian bridges in (\ref{bridge_eff}) \cite{BSG21cb}. 

The detailed calculations as well as extensions to meander and excursion constraints can be found in the paper whose abstract is given on p.~\pageref{chap:A6}.

\newpage
\begin{figure}
\begin{center}
 \fboxsep=10pt\relax\fboxrule=1pt\relax
 \fbox{
   \begin{minipage}{\textwidth}

\hspace{2em}
\begin{center}
\LARGE \bf Generating constrained run-and-tumble trajectories
\end{center}

\hspace{2em}
\begin{center}
B. De Bruyne, S. N. Majumdar and G. Schehr,
J.~Phys.~A:~Math.~Theor. {\bf 54}, 385004 (2021).
\end{center}

\hspace{2em}
\begin{center}
  {\bf Abstract:} 
\end{center}

We propose a method to exactly generate bridge run-and-tumble
trajectories that are constrained to start at the origin with a given velocity and to
return to the origin after a fixed time with another given velocity. The method
extends the concept of effective Langevin equations, valid for Markovian stochastic
processes such as Brownian motion, to a non-Markovian stochastic process driven by a
telegraphic noise, with exponentially decaying correlations. We obtain effective space-
time dependent tumbling rates that implicitly account for the bridge constraint. We
extend the method to other types of constrained run-and-tumble particles such as
excursions and meanders. The method is implemented numerically and is shown to be
very efficient.

\end{minipage}
   }
\end{center}
\captionsetup{labelformat=empty}
\caption{\textbf{Abstract of article \themycounter} : Generating constrained run-and-tumble trajectories.}
\label{chap:A6}
\addtocounter{mycounter}{1}
\end{figure}

\clearpage
\section{Application to a trapping problem}
\label{sec:app}
Let us end this chapter with an application of rare trajectory sampling to a trapping problem. Trapping problems have a long-standing interest in the physics community and beyond as they govern the behavior of a variety of applications ranging from target searching strategies \cite{Oshanin02} to chemical kinetics and diffusion-limited reactions \cite{Smol16,Chand43,Rice85,Montroll(1965),Benson60,Krapivsky10}. Such problems have been studied with various dynamics for the particle and in a wide variety of static, dynamic and random environments \cite{Bramson88,Bray02a,Bray02b,Bray02c,Majumdar03,Bray03,Yuste08,Krapivsky2014,Ledoussal2009,Texier2009,Grabsch2014,Condamin05}. 
From a theoretical point of view, they have shown to exhibit quite a rich behavior with non-trivial features such as a slower-than-exponential decay of the survival probability in the case of randomly distributed traps \cite{Lifshitz63,Lifshitz65,Lifshitz88,Balagurov74,Donsker75,Donsker79}.

The trapping problem we study is as follows. We consider a diffusive particle, undergoing Brownian motion, in the presence of periodically distributed partially absorbing point traps with intensity $\beta$ and separated by a distance $L$. While the limit $\beta\to \infty$ corresponds to fully absorbing traps where the particle is absorbed upon its first encounter with a trap, the case of a finite $\beta$ corresponds to partially absorbing traps where the particle can cross a trap several times before being absorbed (see figure \ref{fig:modelG}). 
In \cite{GPBD1}, we initially studied the survival probability of the particle, namely the probability that it did not get absorbed up to a certain time. Then, we went beyond the survival probability and analyzed the transport properties of the surviving particle, i.e.~conditioned on the fact that it did not get absorbed, which turns out to exhibit rich features. In particular, we were interested in the effective diffusion coefficient $D_\text{eff}$ of the surviving particles, i.e.~
\begin{align}
  D_\text{eff} = \lim_{t\to \infty}\left[\frac{\langle x^2(t)\rangle -\langle x(t) \rangle^2}{2 t}\right]\,,\label{eq:DeffGen}
\end{align}
where $\langle x(t)\rangle$ and $\langle x^2(t)\rangle$ are respectively the conditional first moment and second moment given by
\begin{align}
  \langle x(t)\rangle &= \frac{\int_{-\infty}^\infty dx\, x\, p(x,t)}{S(t)}\,,\qquad \langle x^2(t)\rangle = \frac{\int_{-\infty}^\infty dx\, x^2\, p(x,t)}{S(t)}\,,\label{eq:sm}
\end{align}
where $S(t)=\int_{-\infty}^\infty dx \,p(x,t)$ is the survival probability of the particle at time $t$, which was studied in \cite{GPBD1}, and $p(x,t)$ is the unconditional probability distribution function of the position $x$ of the Brownian motion in the presence of the traps at time $t$. 
\begin{figure}[t]
  \begin{center}
    \includegraphics[width=0.5\textwidth]{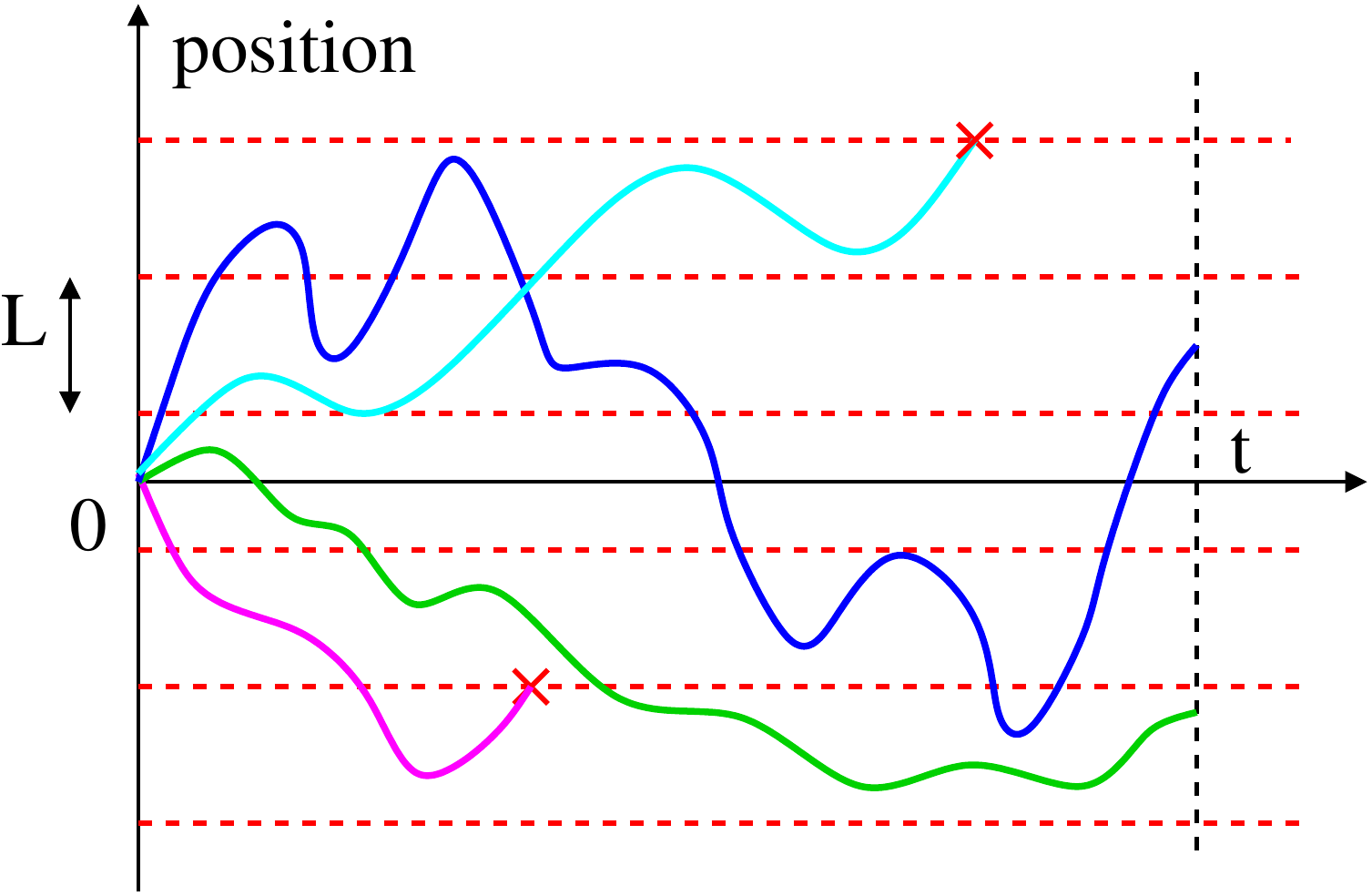}
    \caption{Schematic representation of trajectories of particles diffusing in the presence of periodically distributed partially absorbing point traps (red dashed lines) with intensity $\beta$ separated by a distance $L$ and evenly spaced around the initial position of the particles. As the traps are partially absorbing, the particles can cross a trap several times before being absorbed. Within the four depicted trajectories, only two particles (green and blue) survived up to time $t$. The effective diffusion coefficient in (\ref{eq:DeffGen}) is computed by averaging only over the surviving trajectories at time $t$. The surviving particles typically stay in-between the traps and the effective diffusion coefficient follows a non-trivial behavior $D_\text{eff}=\mathcal{H}(\beta L/D)$, where $\mathcal{H}$ is a scaling function given in (\ref{eq:Deffinf}).}
    \label{fig:modelG}
  \end{center}
\end{figure}
The propagator $p(x,t)$ satisfies the forward Fokker-Planck equation
\begin{align}
  \partial_t p(x,t) = D \partial_{x}^2 p(x,t)- \beta \sum_{m=-\infty}^{\infty} \delta\left(x-\frac{L}{2}-m\,L\right)p(x,t)\,,\label{eq:fp}
\end{align}
where the last term on the right-hand side accounts for the partially absorbing traps of intensity $\beta$ separated by a distance $L$ and evenly spaced around the initial position $x_0=0$. The differential equation (\ref{eq:fp}) must be solved with the initial condition $p(x,t=0)=\delta(x)$ as the particle starts from the origin $x_0=0$. This is a rather difficult task since the initial condition breaks the periodic symmetry $x\to x+L$. Nevertheless, by establishing a connection with a similar albeit different problem, which concerns the winding statistics of a Brownian motion on a ring \cite{Kundu(2014)}, we obtained an exact closed-form expression for the effective diffusion coefficient in (\ref{eq:DeffGen}). Furthermore, we provided a rejection-free algorithm, based on an effective Langevin equation, to generate surviving particles, which is particularly useful for numerical purposes. We showed that the point absorbers induce an effective repulsive potential on the surviving particles. We will first present our results and then sketch the derivation of the effective diffusion coefficient. The full derivation can be found in \cite{GPBD2}.

Let us first present the effective Langevin equation used to generate  trajectories that survive up to $t\to\infty$. Upon using similar ideas as in the previous sections, we found that the effective Langevin equation that governs the evolution of the surviving particles is given by \cite{GPBD2}
\begin{align}
   \dot x_s(t) = \sqrt{2D}\,\eta(t) - \partial_{x_s} U_\text{eff}(x_s)\,,\label{eq:effLf}
\end{align}
where the subscript $s$ in $x_s(t)$ refers to ``surviving'' trajectories and the effective potential $U_\text{eff}(x)$ induced by the infinite number of traps is defined as 
\begin{align}
 \Aboxed{ U_\text{eff}(x) = - 2D \ln\left[\cos\left(\frac{x}{L}\sqrt{\mathcal{G}\left(\mathcal{W}=\frac{\beta L}{D}\right)}\right)\right]\,,\qquad -\frac{L}{2}\leq x\leq \frac{L}{2}\,,}\label{eq:Ueff}
\end{align}
and is $L$-periodic $U_\text{eff}(x)=U_\text{eff}(x+L)$. The scaling variable $\mathcal{W}=\beta L/D$ in (\ref{eq:Ueff}) is the Sherwood number and the scaling function $\mathcal{G}(\mathcal{W})$ is given implicitly as the first zero of the transcendental equation
 \begin{align}
  \cot\left(\frac{\sqrt{\mathcal{G}(\mathcal{W})}}{2}\right) = \frac{2\sqrt{\mathcal{G}(\mathcal{W})}}{\mathcal{W}}\,.\label{eq:gu}
\end{align}
The Sherwood number $\mathcal{W}$ is a dimensionless number in fluid mechanics that represents the ratio of convective mass transfer over diffusive mass transport \cite{Sherwood}. Interestingly, the scaling function defined in (\ref{eq:gu}) is the same as the one found in \cite{GPBD1} which governs the decay rate of the survival probability. A plot of the function $\mathcal{G}(\mathcal{W})$, along with its asymptotic behavior, is given in figure \ref{fig:BMW}.
\begin{figure}[t]
  \begin{center}
    \includegraphics[width=0.4\textwidth]{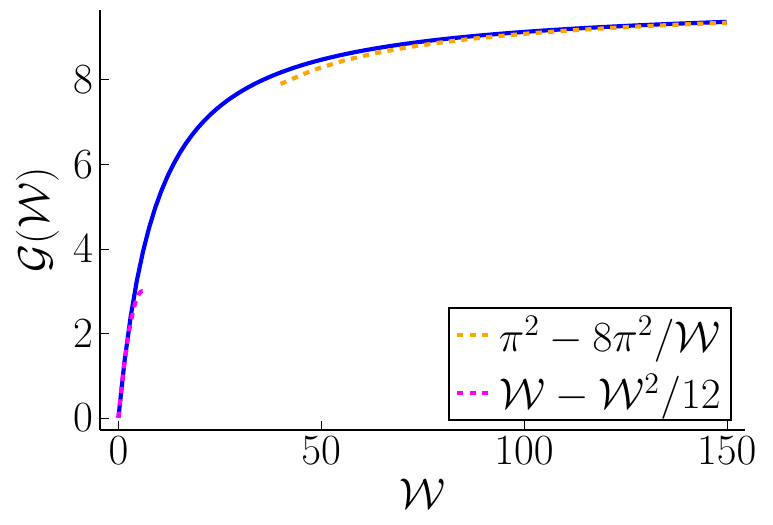}
    \caption{Scaling function $\mathcal{G}(\mathcal{W})$ in (\ref{eq:gu}) as a function of the Sherwood number $\mathcal{W}=\beta L/D$. Its asymptotic behaviors are shown in dashed lines. }
    \label{fig:BMW}
  \end{center}
\end{figure}
 The effective potential in (\ref{eq:Ueff}) takes a remarkably simple form with cusps located at the positions of the point absorbers (see figure \ref{fig:EffPot}). Indeed, close to $x^*=(2m+1)L/2$, with $m\in \mathbb{Z}$, the effective potential in (\ref{eq:Ueff}) behaves as 
 \begin{align}
   U_{\text{eff}}(x) \sim -2D\ln\left[\cos\left(\frac{1}{2}\sqrt{\mathcal{G}\left(\mathcal{W}\right)}\right)\right] -\beta\,\left|x-x^*\right|\,,\quad x\to x^*\,,
 \end{align}
 where $\mathcal{W}=\beta L/D$.
\begin{figure}[htbp]
  \begin{center}
    \includegraphics[width=0.6\textwidth]{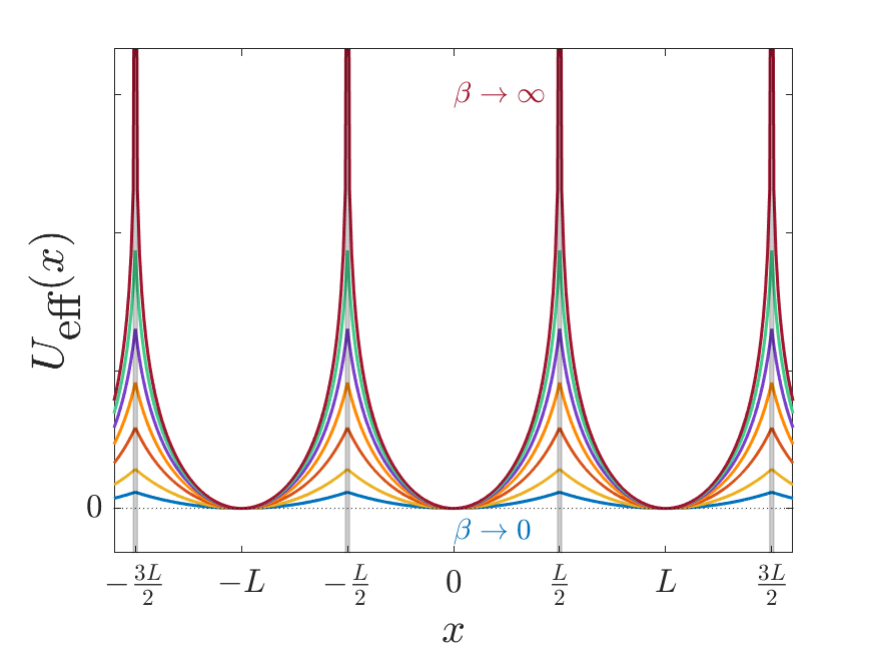}
    \caption{Effective potential $U_\text{eff}(x)$ induced by the periodically distributed partially absorbing traps on the surviving particles as a function of $x$ for several trap intensities $\beta$ (\ref{eq:Ueff}). As the trap intensity tends to infinity $\beta\to \infty$, the potential diverges at the location of the traps. In this plot, we have set $D=1$.}
    \label{fig:EffPot}
  \end{center}
\end{figure}

Let us now present our results on the effective diffusion coefficient (\ref{eq:DeffGen}) of the surviving particles. We find that it is given by
\begin{align}
 \Aboxed{ D_\text{eff} = D \mathcal{H}\left(\mathcal{W}=\frac{\beta L}{D}\right)\,,}\label{eq:Deffinf}
\end{align}
where the scaling function $\mathcal{H}\left(\mathcal{W}\right)$ is 
\begin{align}
\Aboxed{ \mathcal{H}(\mathcal{W}) =\left[\sin ^2\left(\frac{\sqrt{\mathcal{G}(\mathcal{W})}}{2}\right)\left(\frac{2}{\mathcal{G}(\mathcal{W})}+\frac{2}{\mathcal{W}}+\frac{\mathcal{W}}{2\,\mathcal{G}(\mathcal{W})}\right)\right]^{-1}\,,}\label{eq:Deff}
\end{align}
  and the scaling function $\mathcal{G}(\mathcal{W})$ is given in (\ref{eq:gu}). The asymptotic behaviors of $\mathcal{H}(\mathcal{W})$ are given by
\begin{align}
  \mathcal{H}(\mathcal{W}) \sim \left\{\begin{array}{ll}
    1\,,\qquad&\mathcal{W} \to 0\,,\\[1em]
   \dfrac{2\pi^2}{\mathcal{W}} \,,\qquad&\mathcal{W} \to \infty\,.
  \end{array}\right.
\end{align}
A plot of the scaling function (\ref{eq:Deff}) is shown in figure \ref{fig:Deff}. As one can see, the effective diffusion coefficient is always less than $D$, as the particle tends to stay in-between traps, hence having a smaller diffusion coefficient than the original one.
\begin{figure}[t]
  \begin{center}
    \includegraphics[width=0.5\textwidth]{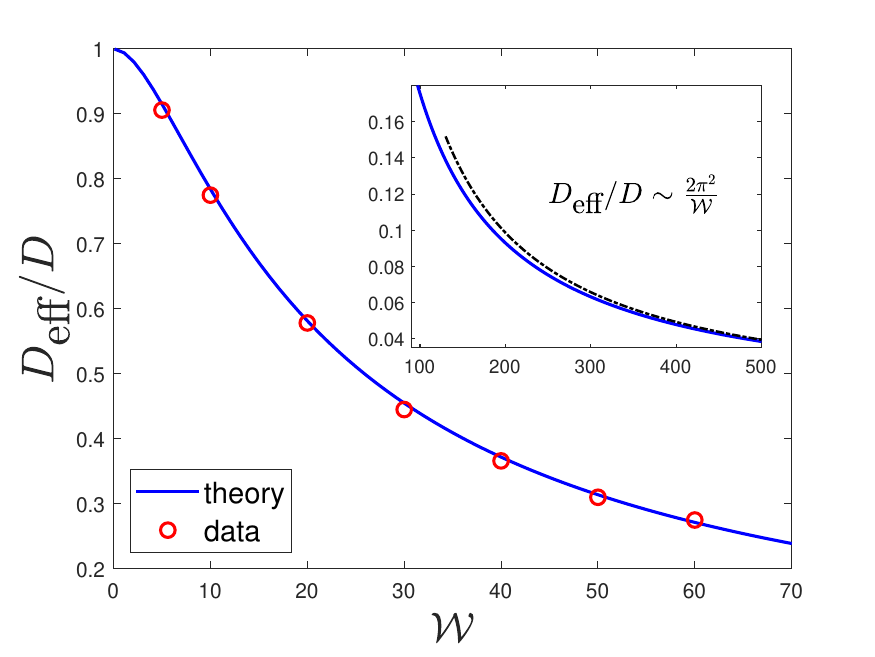}
    \caption{Normalized effective diffusion coefficient $D_\text{eff}/D$ as a function of the Sherwood number $\mathcal{W}=\beta L/D$ for a diffusive particle, starting from the origin, conditioned to survive in the presence of infinitely many partially absorbing traps with intensity $\beta$, periodically distributed with a period $L$ and evenly spaced around the origin. The blue line is the theoretical prediction in (\ref{eq:Deff}), whereas data points are obtained by simulating $3\cdot 10^8$ trajectories. The time increment is set to $\tau=0.001$, with $D=1$ and the point absorbers become trapping intervals of length $\beta\tau$ separated by a distance $L=10$. In the inset, the normalized effective diffusion coefficient is shown for large values of $\mathcal{W}$ and is compared to the theoretical prediction of its asymptotic behavior in the limit $\mathcal{W}\to \infty$. }
    \label{fig:Deff}
  \end{center}
\end{figure}

\begin{figure}[t]
  \centering
  \includegraphics[width=0.3\textwidth]{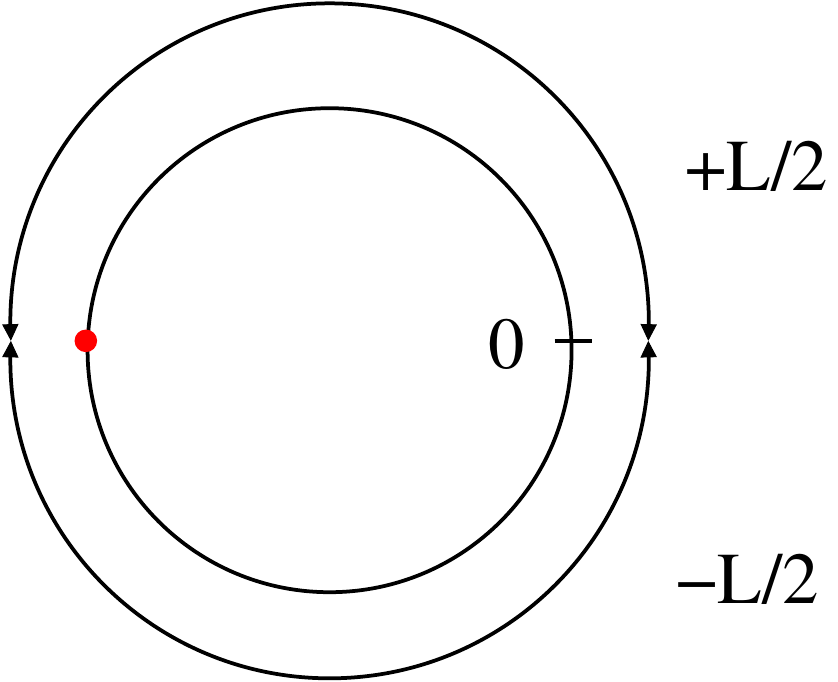}\hspace{2em}
    \includegraphics[width=0.4\textwidth]{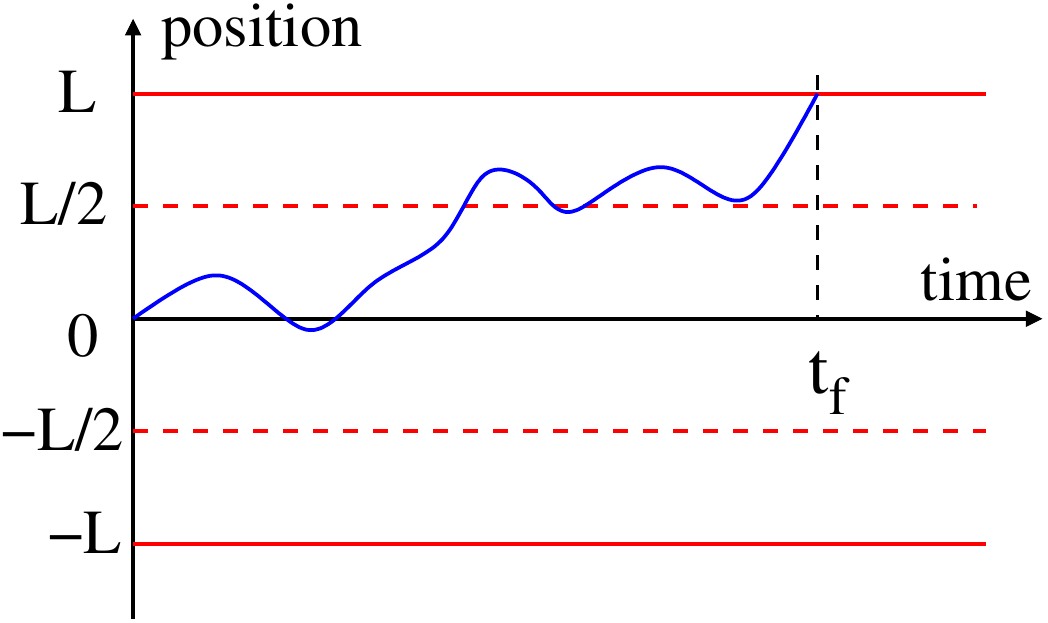}
    \caption{\textbf{Left panel:} Mapping of the periodic trapping environment in figure \ref{fig:modelG} to a circle of perimeter $L$ with a single partially absorbing trap (red dot) located at arc length $\pm L/2$. As time grows, the surviving particle crosses several traps in figure \ref{fig:modelG}, which corresponds to several turns around the circle. More precisely, when the particle is exactly in the middle of two point absorbers and propagates to the middle of one of the two neighboring regions in figure \ref{fig:modelG}, it makes a full turn around the circle and the process starts anew.  Depending on which direction it goes, the particle turns either in a clockwise or anti-clockwise fashion. We track the position on the $x$-axis in figure \ref{fig:modelG} by counting the net number of full revolutions around the circle. \textbf{Right panel:} Each complete turn of the circle in a time $t_f$ in the left panel is associated with a first-exit event of a box $[-L,L]$ at time $t_f$ with two partially absorbing traps located at $\pm L/2$. A full clockwise turn corresponds to exiting the box at $+L$, whereas a full anti-clockwise turn corresponds to exiting the box at $-L$. The distribution of the exit time $t_f$, either at $+L$ or $-L$, is denoted by $f(t_f)$. The survival probability at time $\tau$, where the particle does not exit the box nor gets trapped until time $\tau$, is denoted by $S_\text{box}(\tau)$.}
    \label{fig:mapping}
\end{figure}
We now briefly outline how we obtained the results in (\ref{eq:Deffinf}) and (\ref{eq:Deff}). We relied on an idea developed in \cite{Kundu(2014)} on a seemingly different problem concerning the winding statistics of a Brownian motion on a ring. The main idea is as follows. We map the periodic structure of the environment to a circle of perimeter $L$ with a single partially absorbing trap located at arc length $\pm L/2$ (see figure \ref{fig:mapping}). As time increases, the particle turns around the circle either in a clockwise or anti-clockwise fashion until it is absorbed by one of the traps. The presence of the point absorbers makes this problem inherently different from \cite{Kundu(2014)}. Using a similar notation as in the original paper \cite{Kundu(2014)}, we denote by $m_+(t)$ the total number of complete counter-clockwise turns and $m_-(t)$ the total number of complete clockwise turns after time $t$. Furthermore, we denote by $m(t)=m_+(t)+m_-(t)$ the absolute number of complete turns and by $k(t)=m_+(t)-m_-(t)$ the net number of full revolutions after time $t$. Then we introduce the probability distribution $R(k,t)$ of the net number of complete turns $k$ after time $t$:
\begin{align}
  R(k,t) = \text{Prob.}\left(k(t) = k\right)\,.\label{eq:Rkt}
\end{align}
 Due to the presence of the traps, the distribution $R(k,t)$ is not normalized to unity and is in fact normalized to the survival probability $S(t)$ with the initial position set equal to $x_0=0$. The normalization reads
\begin{align}
  S(t) = \sum_{k=-\infty}^\infty R(k,t)\,.\label{eq:SrelRk}
\end{align}
The net number of complete laps is directly related to the position of the particle on the real line by
\begin{align}
  k(t) = \left\lfloor \frac{x(t)}{L} \right\rfloor \,,\label{eq:kxrel}
\end{align}
where $\lfloor a\rfloor$ denotes the floor operation returning the highest integer lower or equal to $a$. In the long time limit, we expect that the unconditional second moment on the real line behaves as
\begin{align}
  \int_{-\infty}^\infty dx\,x^2\, p(x,t) \approx L^2\,\sum_{k=-\infty}^\infty k^2\,R(k,t)\,,\qquad t\to \infty\,.\label{eq:x2k2}
\end{align}
In fact, in this limit, one can neglect the dynamics inside a box of length $2L$ and record the trajectory on a large scale as jumps between the centers of these intervals, enumerated by the net winding number $k$.
By symmetry of the trapping environment, it is clear that $\langle x(t)\rangle =0$, and inserting the expressions (\ref{eq:SrelRk})-(\ref{eq:x2k2}) into (\ref{eq:sm}), we find that the effective diffusion coefficient (\ref{eq:DeffGen}) can be written as
\begin{align}
  D_{\text{eff}} = \lim_{t\to \infty}\frac{L^2\,\sum_{k=-\infty}^\infty k^2\,R(k,t)}{2\,t\,\sum_{k=-\infty}^\infty R(k,t)}\,.\label{eq:Deffint}
\end{align}
We will now show how to compute $R(k,t)$ using a similar approach as in \cite{Kundu(2014)}. It is first convenient to analyze a more general quantity, $P(m,k,t)$, which is the joint distribution of the absolute number $m$ and the net number $k$ of complete turns after time $t$:
\begin{align}
  P(m,k,t) = \text{Prob.}\left(m(t)=m,k(t)=k\right)\,.\label{eq:defpmkt}
\end{align}
 From this joint distribution, we obtain $R(k,t)$ by summing over all values of $m$, i.e.
\begin{align}
  R(k,t) = \sum_{j=0}^\infty P(2j+|k|,k,t)\,,\label{eq:RP}
\end{align}  
where we used the fact that the absolute number $m$ of turns can be written as a sum of an even number $2j$ of turns plus the absolute value of the net number of turns $|k|$. To obtain $P(m,k,t)$, we first count the total number of turns $m$ completed before the particle gets absorbed by a trap, say after a time $t$, and the corresponding durations $\{\tau_1,\ldots,\tau_m\}$ between consecutive windings. We denote by $\tau_{\text{last}}=t-\sum_{i=1}^m \tau_i$ the duration of the last incomplete turn. The number of absolute turns $m$ and the durations $\{\tau_1,\ldots,\tau_m,\tau_{\text{last}}\}$ are random variables whose joint probability distribution is given by
\begin{equation}
\mathcal{P}(m, \{\tau_1,\tau_2,\dots,\tau_m,\tau_{last}\}, t)=f(\tau_1)f(\tau_2)\dots f(\tau_m)S_\text{box}(\tau_{\text{last}})\delta \left(t-\tau_{last}-\sum_{i=1}^m\tau_i\right)\,,\label{eq:Pturn}
\end{equation}
 where $f(\tau)$ is the first-exit probability from a box $[-L,L]$, with a centered initial position and partially absorbing traps of strength $\beta$ located at $\pm L/2$, whereas $S_\text{box}(\tau)$ is the survival probability in the box, i.e. the probability that the particle did not exit the box and did not get trapped (see figure \ref{fig:mapping}). The first $m$ terms on the right-hand side of \eqref{eq:Pturn} refer to the first $m$ laps, the factor $S_\text{box}(\tau_{\text{last}})$ refers to the time spent after the last complete turn, and the final Dirac delta term imposes the constraint that $t=
\tau_{last}+\sum_{i=1}^m\tau_i$. The marginal distribution of the total number of complete turns $\mathcal{P}(m,t)$ can be obtained by integrating over all possible durations $\{\tau_1,\ldots,\tau_m,\tau_{\text{last}}\}$:
\begin{align}
  P(m,t)=\int_0^\infty d\tau_1\ldots d\tau_m d\tau_{\text{last}}f(\tau_1)f(\tau_2)\dots f(\tau_m)S_\text{box}(\tau_{\text{last}})\delta \left(t-\tau_{last}-\sum_{i=1}^m\tau_i\right)\,.\label{eq:Pturn2}
\end{align}
From the distribution of the absolute number of complete turns in (\ref{eq:Pturn2}), we obtain the joint distribution $P(m,k,t)$ with the net number of complete laps $k$ using the following argument. Given that the particle made $m$ absolute number of complete turns, the particle will have made $m_+$ counter-clockwise turns with probability $(1/2)^{m_+}$ and $m_-$ clockwise turns with probability $(1/2)^{m_-}$, as when the particle exits the box, it is located in the middle of the neighboring interval (see figure \ref{fig:mapping}). By using the relations $m_+=(m+k)/2$ and $m_-=(m-k)/2$, and summing over all possible combinations of $m$ and $k$, we find that
\begin{align}
  P(m,k,t) = P(k|m,t) P(m,t) = \binom{m}{\frac{m+k}{2}}\left(\frac{1}{2}\right)^m P(m,t)\,,\label{eq:relPmkt}
\end{align}
where the binomial coefficient counts the number of ways to make $m_+=(m+k)/2$ counter-clockwise turns among $m$ turns and the factor $\left(\frac{1}{2}\right)^m$ is the probability of each configuration. 
Inserting the expression (\ref{eq:Pturn2}) into (\ref{eq:relPmkt}) and noting that the convolution structure over time of this integral is well suited to a Laplace transform, we find that in Laplace domain $P(m,k,t)$ reads
\begin{align}
 \tilde P(m,k,s)= \int_0^\infty dt e^{-st} P(m,k,t) = \binom{m}{\frac{m+k}{2}}\left(\frac{1}{2}\right)^m\tilde f^m (s)\tilde S_\text{box}(s)\,,\label{eq:tildPms}
\end{align}
where $\tilde f (s)$ and $\tilde S_\text{box}(s)$ are the Laplace transforms of $f(t)$ and $S_\text{box}(t)$ respectively. Thus, taking a Laplace transform of the expression (\ref{eq:RP}), we find that in Laplace domain the probability $R(k,t)$ for a net winding number $k$ in a time $t$ is given by
\begin{equation}
\tilde{R}(k,s)=\sum_{m=0}^{\infty}\binom{2m+|k|}{m}\left(\frac{\tilde{f}(s)}{2}\right)^{2m+|k|}\tilde{S}_\text{box}(s)\,.
\end{equation}
More explicitly, using the identity \cite{wiki}
$$\sum_{m=0}^{\infty}\binom{2m+|k|}{m} z^m= \frac1{\sqrt{1-4z}}\left(\frac{1-\sqrt{1-4z}}{2z}\right)^{|k|}\,,\qquad |z|<\frac 1 4\,,$$
we get
\begin{align}
  \tilde R(k,s) = \frac{\tilde S_\text{box}(s)}{\sqrt{1-\tilde f^2(s)}}\left(\frac{1-\sqrt{1-\tilde f^2(s)}}{\tilde f(s)}\right)^{|k|}\,.\label{eq:tR}
\end{align}
 Using (\ref{eq:tR}), we find that the Laplace transform of the numerator and denominator in (\ref{eq:Deffint}) are given by
\begin{align}
  \int_0^\infty dt \,e^{-st}\sum_{k=-\infty}^\infty k^2 R(k,t) &= \sum_{k=-\infty}^\infty k^2 \tilde R(k,s) = \frac{\tilde f(s) \tilde S_\text{box}(s)}{[1-\tilde f(s)]^2}\,,\\
  \int_0^\infty dt e^{-st}\sum_{k=-\infty}^\infty R(k,t) &= \sum_{k=-\infty}^\infty \tilde R(k,s) =  \frac{\tilde S_\text{box}(s)}{1-\tilde f(s)}\,.
\end{align}
The expressions of the first-passage distribution $\tilde f(s)$ and the survival probability $\tilde S_\text{box}(s)$ can be computed straightforwardly from a Fokker-Planck approach \cite{GPBD2}.
By analyzing the asymptotic behavior of these expressions in the limit $s\to 0$, which corresponds to $t\to \infty$, we recover the effective diffusion coefficient in (\ref{eq:Deffinf}) \cite{GPBD2}.

The detailed derivation of these results and extensions to other trapping environments can be found in the paper whose abstract is given on p.~\pageref{chap:A15}.

\begin{figure}
\begin{center}
 \fboxsep=10pt\relax\fboxrule=1pt\relax
 \fbox{
   \begin{minipage}{\textwidth}

\hspace{2em}
\begin{center}
\LARGE \bf Transport properties of diffusive particles conditioned to survive in trapping environments
\end{center}

\hspace{2em}
\begin{center}
G. Pozzoli and B. De Bruyne,
J. Stat. Mech., 113205 (2022).
\end{center}

\hspace{2em}
\begin{center}
  {\bf Abstract:} 
\end{center}

We consider a one-dimensional Brownian motion with diffusion coefficient $D$ in the presence of $n$ partially absorbing traps with intensity $\beta$, separated by a distance $L$ and evenly spaced around the initial position of the particle. We study the transport properties of the process conditioned to survive up to time $t$. We find that the surviving particle first diffuses normally, before it encounters the traps, then undergoes a period of transient anomalous diffusion, after which it reaches a final diffusive regime. The asymptotic regime is governed by an effective diffusion coefficient $D_\text{eff}$, which is induced by the trapping environment and is typically different from the original one. We show that when the number of traps is \emph{finite}, the environment enhances diffusion and induces an effective diffusion coefficient that is systematically equal to $D_\text{eff}=2D$, independently of the number of the traps, the trapping intensity $\beta$ and the distance $L$. On the contrary, when the number of traps is \emph{infinite}, we find that the environment inhibits diffusion with an effective diffusion coefficient that depends on the traps intensity $\beta$ and the distance $L$ through a non-trivial scaling function $D_\text{eff}=D \mathcal{H}(\beta L/D)$, for which we obtain a closed-form. Moreover, we provide a rejection-free algorithm to generate surviving trajectories by deriving an effective Langevin equation with an effective repulsive potential induced by the traps. Finally, we extend our results to other trapping environments.
\end{minipage}
   }
\end{center}
\captionsetup{labelformat=empty}
\caption{\textbf{Abstract of article \themycounter} : Transport properties of diffusive particles conditioned to survive in trapping environments.}
\label{chap:A15}
\addtocounter{mycounter}{1}
\end{figure}

\clearpage

\chapter{Stochastic optimization problems}
\label{chap:sto}
In this section, we present some optimization problems involving stochastic processes and extreme value statistics. These problems are largely independent of each other but have some connections with the previous chapters of this thesis. In Section \ref{sec:opt}, we introduce resetting Brownian bridges as a search process and reveal an interesting mechanism, induced by resetting, which enhances the search performances of the process. In Section \ref{sec:res}, we present a general framework to determine the optimal policy to operate dynamical systems undergoing restart. In Section \ref{sec:fpr}, we combine the notion of first-passage and resetting to define an optimization problem in which one wants to operate a system close to maximum capacity without experiencing too many breakdowns.

\section{Optimal resetting bridges}
\label{sec:opt}

The first optimization problem we consider is in the context of search processes. These processes appear in a wide range of situations ranging from foraging animals \cite{Bartumeus09,Viswanathan11}, biochemical reactions \cite{Berg81,Coppey04,Ghosh18,Chowdhury19} and all the way to behavioral psychology \cite{Wolfe04,Bell91,Adam68}. Search problems exhibit rich features \cite{Montanari02,Gelenbe10,Snider12,Abdelrahman13,Chupeau17,Oshanin07} and finding an optimal search strategy in a given context is an interesting problem with multiple applications
across disciplines~\cite{Benichou11,Lomholt08}. In recent years, there has been a surge of interest in the effect of resetting in search processes (for a recent review see \cite{ESG20}). Stopping and starting from scratch has shown to be an efficient search strategy in several contexts such as in optimization algorithms \cite{Villen91,Luby93,Tong08,Avrachenkov13,Lorenz18}, chemical reactions \cite{Reuveni14,Reuveni15}, animal foraging \cite{BS14,MSS15,Boyer18,Maso19c,Pal20} and catastrophes in population dynamics \cite{Levikson77,Pakes78,Pakes97,Brockwell82,Brockwell85,Kyriakidis94,Manrubia99,Dharmaraja15}.
Perhaps, the effect of resetting is best seen in the simple model of diffusion introduced by Evans and Majumdar~\cite{ES11,ES11b}. 
In this resetting Brownian motion (RBM) model, the position $x(t)$ of a Brownian motion, e.g. in one dimension, is reset to the origin randomly in time according to a Poisson process with a constant rate $r$. In a time interval $dt$, the position $x(t)$ follows the stochastic rule
\begin{align}
  \hspace{-0.5em} x(t\!+\!dt)\!=\!\left\{\begin{array}{ll}
    \hspace{-0.4em}x(t)\!+\!\sqrt{2D}\,\eta(t)dt, &\text{with \, prob. } 1\!-\!r\,dt,\\
    \hspace{-0.4em}0,&\text{with \, prob. } \!r\,dt,
  \end{array} \quad \text{with}\quad x(0)=0\,,\right.\label{eq:eom}
\end{align}
where $D$ is the diffusion coefficient and $\eta(t)$ is an uncorrelated white noise with zero mean $\langle \eta(t)\rangle=0$ and delta correlator $\langle\eta(t)\eta(t')\rangle=\delta(t-t')$. The dynamics, therefore, consists of a combination of pure diffusion with intermittent resets to the origin (see the left panel in figure \ref{fig:RBM}). When $r=0$, we recover the purely diffusive dynamics in (\ref{eq:eomb}).
\begin{figure}[t]
  \begin{center}
   \raisebox{0.22\height}{\includegraphics[width=0.4\textwidth]{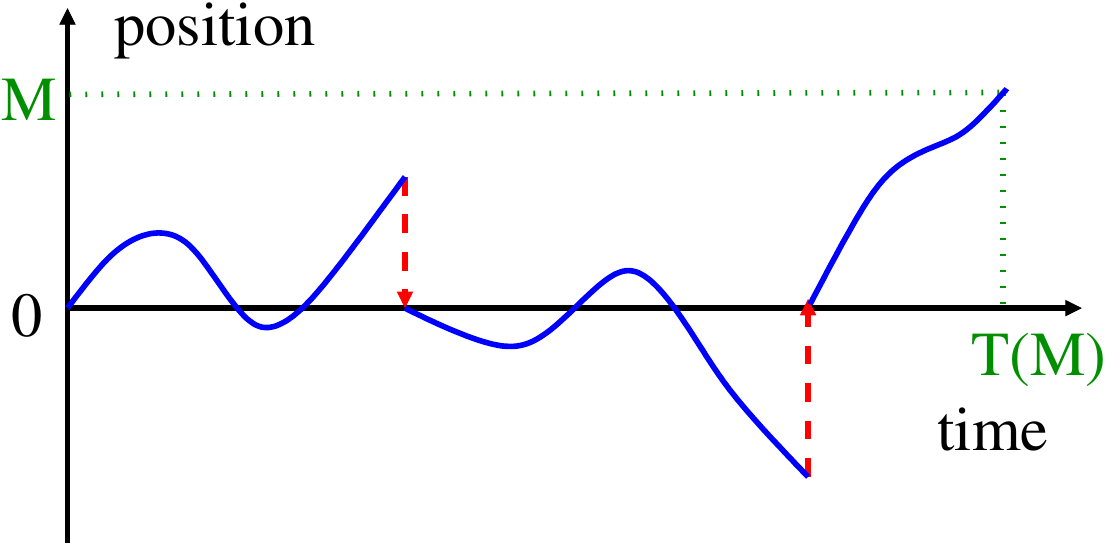}}\hspace{1em}\includegraphics[width=0.35\textwidth]{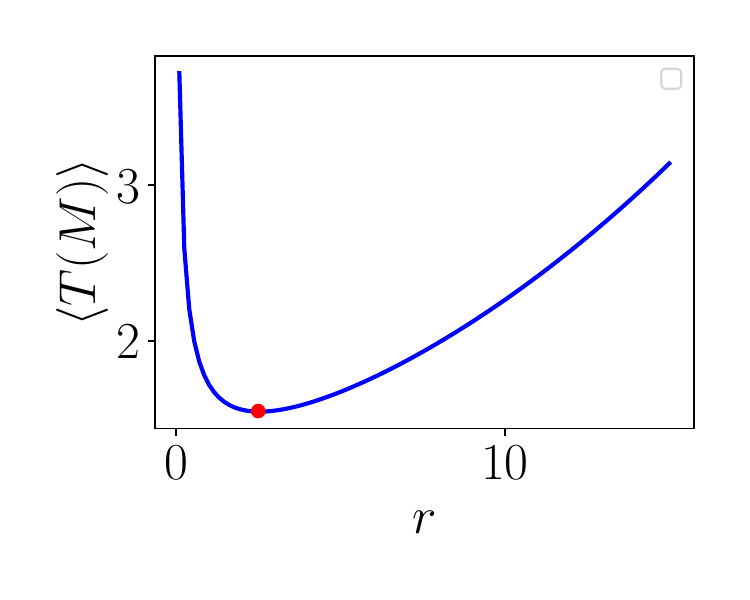}
    \caption{\textbf{Left panel:} Schematic representation of a resetting Brownian motion evolving according to (\ref{eq:eom}). At a constant rate $r$, the process is reset to the origin (dashed red arrows). The first-passage time $T(M)$ is the first time that the process reaches the level $M$. \textbf{Right panel:} Mean first-passage time $\langle T(M)\rangle$ to the level $M$ as a function of the resetting rate $r$ (\ref{eq:TMRBM}). The mean first passage time is minimized at a critical resetting rate (red dot). In this plot, we have set $D=M=1$. }
    \label{fig:RBM}
  \end{center}
\end{figure}
The effect of resetting on the search process can be simply measured by the mean first-passage time $\langle T(M)\rangle$ to a level $M$, which is the mean time the searcher takes to find a target located at a position $M$.  For pure diffusion without resetting, it is well-known that this quantity is infinite \cite{redner2001guide,bray2013persistence}. In contrast, resetting leads to the striking result that the mean first-passage time $\langle T(M)\rangle$ becomes finite and is given by \cite{ES11} 
\begin{align}
 \langle T(M) \rangle = \frac{1}{r}\left(e^{\sqrt{\frac{r}{D}}M}-1\right)\,.\label{eq:TMRBM}
\end{align}
Not only does it become finite, but it also becomes minimal at
an optimal resetting rate $r^*$ (see the right panel in figure \ref{fig:RBM}). The mechanism behind this result is that resetting suppresses the trajectories that diffuse far away from the target and makes them restart from the origin, hence increasing their chances to find the target. This model is straightforward to generalize to higher dimensions and an optimal resetting rate has been shown to exist in all dimensions \cite{Evans14}. Since the original model, the existence of an optimal resetting rate has been studied extensively for various stochastic processes, leading to a tremendous amount of activities \cite{ES11b,Montero16,Evans14,CS15,Evans18,Nagar16,Pal16,EM16,Kusmierz14,CM15,Bressloff20a,Bressloff20b,R16,PR17,Sokolov18,Prasad19a,Prasad19b,DeBruyne20a,DeBruyne20b,DeBruyneMori21} -- see \cite{ESG20} for a review. The existence of this optimal resetting rate has also been confirmed in experiments with optical traps in both one and two dimensions~\cite{Tal20,Besga20,Faisant21}. 

In this section, we study a Brownian bridge model with a fixed duration $t_f$, but
in the presence of {\em resetting} at a constant rate $r$ to its initial position.
We call this a {\em resetting Brownian bridge} (RBB), with ${x}_B(t)$ denoting its coordinate at time $0\leq t\leq t_f$ with
the bridge conditions ${x}_B(0)=0$ and ${x}_B(t_f)={x}_f$ (see figure \ref{fig:B}).
The general question we address is: is resetting still a good search strategy in the presence
of a bridge constraint? In other words, does the paradigm of an optimal resetting rate
$r^*$ still hold for RBB? We find, rather surprisingly, that there
is an interesting trade-off between resetting and the bridge constraint such that
a small resetting rate, in the presence of a bridge constraint, actually \textit{enhances}
bridge fluctuations, rather than \textit{reducing} it as naive expectations would suggest.
\begin{figure}[t]
\centering
 \includegraphics[width=0.4\textwidth]{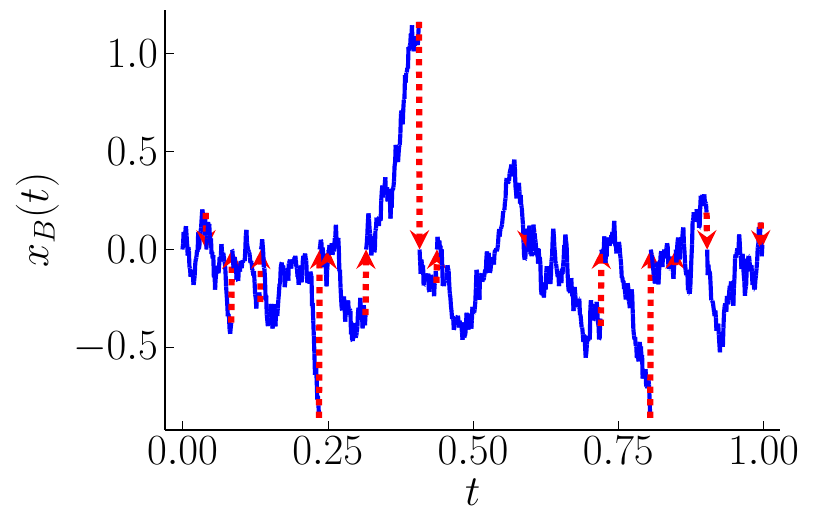}%
    \caption{A typical RBB trajectory $x_B(t)$ with resetting rate $r=10$, diffusion constant $D=1$ and duration $t_f=1$. The resetting events are denoted by red dashed lines with arrows. The trajectories have been generated with a time step of $dt=10^{-3}$ and by using an effective Langevin equation given in \cite{BMSOpt}.}
    \label{fig:B}
\end{figure}

We compute three observables illustrating the ``enhanced fluctuations mechanism'' (EFM) exhibited by the RBB as a search process.
Let us start with the mean-square displacement. Alike the case of Brownian bridges in (\ref{eq:b}), the probability density function (PDF) of the position $x_B(t)$ of an RBB at some intermediate time $0\le t \le t_f$
is given by
{
\begin{align}
  P_B(x,t\,|\,t_f) =\frac{P_r(x,t)Q_r(x,t_f-t)}{P_r(0,t_f)} \;,\label{eq:Pb}
\end{align}
}
where $P_r(x,t)$ is the forward propagator of RBM, namely the probability that it reaches $x$ at time $t$ given that it started from the origin, and $Q_r(x,t)$ is the backward propagator of RBM, namely the probability that it reaches the origin at a time $t$ given that it started from $x$. The denominator is just a normalization constant that ``counts'' all the trajectories of the RBM of duration $t_f$, starting and ending at $0$. {The two quantities $P_r(x,t)$ and $Q_r(x,t)$ satisfy respectively} the forward and backward Fokker-Plank equations of RBM respectively given by (see \cite{BMSOpt} for details)
\begin{subequations}
\begin{align}
  \partial_t P_r(x,t) &\!=\! D \partial_{x}^2 P_r(x,t)\!-\!r P_r(x,t)\!+\!r \delta(x),\label{eq:fP}\\
   \partial_t Q_r(x,t) &\!=\! D \partial_{x}^2 Q_r(x,t)\!-\!r Q_r(x,t)\!+\!r Q_r(0,t),\label{eq:bP}
\end{align}
\label{eq:fbP}
\end{subequations}
\hspace*{-0.2cm}with the initial conditions $P_r(x,0)=\delta(x)$, $Q_r(x,0)=\delta(x)$. The mean position $\langle x_B\rangle (t\,|\,t_f)$ vanishes by symmetry. Hence the minimal quantity that characterizes the spatial fluctuations is the second moment of the PDF, i.e., the mean-square displacement $\langle x_B^2\rangle (t|t_f)$. We compute $\langle x_B^2\rangle (t|t_f)$ from~(\ref{eq:Pb}) analytically, leading to
\begin{align}
\Aboxed{\langle x_B^2\rangle (t\,|\,t_f)= \int_{-\infty}^\infty dx\,x^2 P_B(x,t|t_f) =2D\, t_f \, f\left(a=\frac{t}{t_f}\,\bigg|\,R= r\, t_f\right) ,}
\label{msd_rbb.2}
\end{align}
where the scaling function $f(a|R)$ can be obtained explicitly (see \cite{BMSOpt}). 
A plot of the function $f(a|R)$ vs. $a\in [0,1]$,
for different values of $R$, is given in the left panel in figure \ref{fig.far_1}. 
As the rescaled resetting rate $R= r \,t_f$ varies from $0$ to $\infty$, the function $f(a|R)$, crosses over from a 
parabolic to a flat shape, i.e., 
\begin{align}
f(a|R\to 0)  = a(1-a) \quad  \text{and} \quad f(a|R\to \infty) \approx 1/R\,.
\end{align} 
For a general $R$, the function $f(a|R)$ is not symmetric around $a=1/2$,
since resetting breaks the time-reversal symmetry. 
For a given $R$, the function $f(a|R)$ has a unique maximum at $a=a^*(R)$ and
this maximal mean square displacement
$f(a^*(R)|R)$ varies nonmonotonically with $R$:
it first increases with increasing $R$, achieves a maximum at
$R=R^*\approx 0.895$ and then decreases again with increasing $R$ (see figure \ref{fig.far_1}). 
\begin{figure}[t]
\centering
 \includegraphics[width=0.4\textwidth]{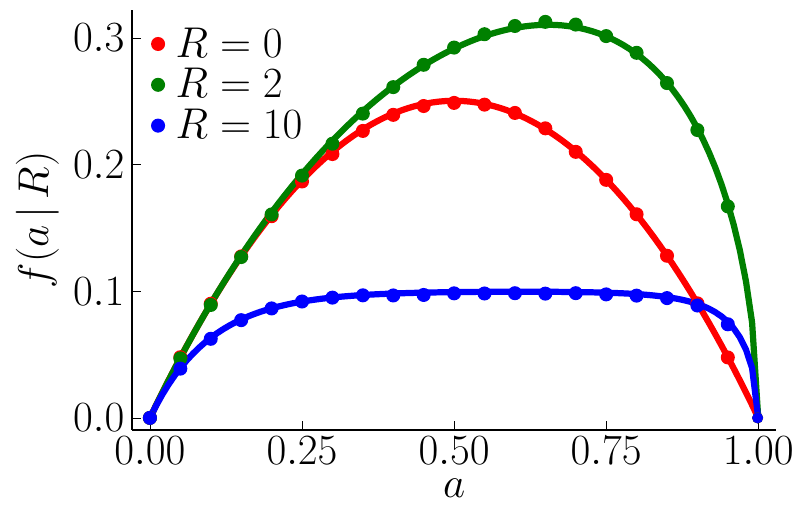}%
  \includegraphics[width=0.4\textwidth]{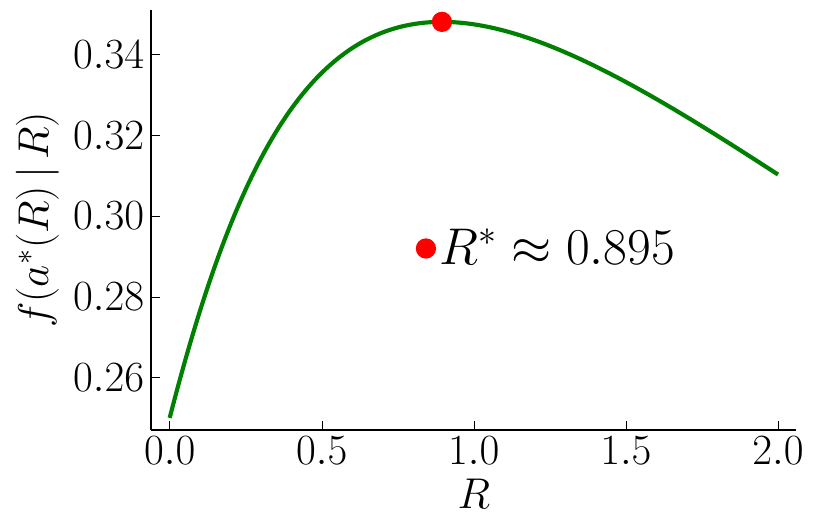}%
\caption{\textbf{Left panel:} The function $f(a|R)$ plotted vs.~$a$ is evaluated numerically by using the effective Langevin equation in \cite{BMSOpt} (symbols) and is compared to the theoretical prediction (plain lines), given in (\ref{msd_rbb.2}), for different values of $R$.
This function is clearly asymmetric around $a=1/2$. Only when $R\to 0$,
it approaches to the symmetric form $f(a|R\to 0)=a(1-a)$. For any $R$, the function $f(a|R)$ has a unique maximum at $a=a^*(R)$. \textbf{Right panel:}  The maximal value $f(a^*(R)|R)$ plotted vs. $R$.
It has a unique maximum at $R^*\approx 0.895$ (red dot).}
\label{fig.far_1}
\end{figure}
Thus, interestingly, a nonzero resetting rate, when it is not too large, actually enhances the bridge fluctuations and maximizes them at an optimal value $R^*$, thus enabling
the particle to explore more space. This is a rather surprising result, as one would naively think that resetting would suppress fluctuations, but this is only true for large resetting rates. Before explaining the reasons for such enhancement, let us first check that other search observables can also be enhanced by resetting. 

To illustrate further the EFM in the context of a search of a target located at $M$, we next compute the hitting probability, i.e., the probability that the RBB (searcher) finds the target at $M$ before time $t_f$. The hitting probability can be computed from the relation
\begin{align}
  p_{\text{hit}}(t_f,M) = \int_0^{t_f} dt F_B(t\,|\,M,t_f)\,,\label{eq:phit}
\end{align}
where $F_B(t\,|\,M,t_f)$ is the first-passage probability density of the RBB at level $M$ with $t \leq t_f$. This can be computed by decomposing the RBB trajectories into two parts: one in the time interval $[0,t]$ where it first hits the level $M$ at a time $t<t_f$, another one in the time interval $[t,t_f]$ where it propagates from $M$ to the origin. One gets
\begin{align}
  F_B(t\,|\,M,t_f) = \frac{F_r(t\,|\,M) Q_r(M,t_f-t)}{P_r(0,t_f)}\,,\label{eq:FB}
\end{align}
where $F_r(t\,|\,M)$ is the first-passage time distribution of a RBM \cite{ES11}, $Q_r(x,t)$ is the backward propagator satisfying (\ref{eq:bP}) and the denominator is a normalization factor that ``counts'' all the bridge trajectories. Using the known results for $F_r(t\,|\,M)$  and the propagator \cite{ES11} we get
\begin{align}
  \Aboxed{ p_{\text{hit}}(t_f,M) &= h\left(R=rt_f,m=\frac{M}{\sqrt{2D t_f}}\right)\,,}\label{eq:phits}
\end{align}
where the scaling function $h(R,m)$ can also be computed analytically \cite{BMSOpt}. 
When $R=0$, we recover the hitting probability of a Brownian bridge $h(R=0,m)=e^{-2m^2}$. For a given target position $m$, the function $h(R,m)$ {varies nonmonotonically with $R$} and achieves a maximum at $R=R^*(m)$. A plot of $h(R,m)$ vs $R$ for $m=1$ is shown in the left panel in figure \ref{fig:probh}. Thus the paradigm of an optimal resetting rate $R^*(m)$ is also manifest in the behavior of the hitting probability. Another observable that also confirms this optimal paradigm is the expected maximum of the RBB as a function of the rescaled resetting rate $R$ that we have computed exactly in \cite{BMSOpt} (as shown in the right panel in figure \ref{fig:probh}).

\begin{figure}[t]
\centering
 \includegraphics[width=0.4\textwidth]{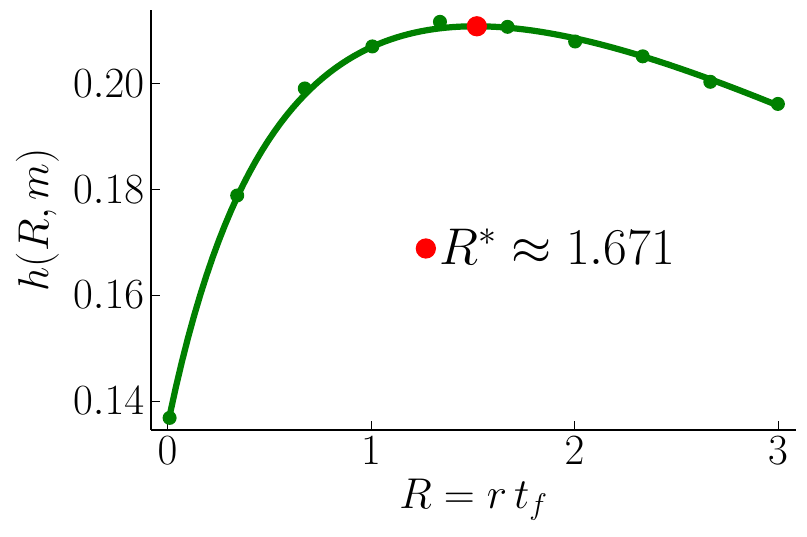}%
  \includegraphics[width=0.4\textwidth]{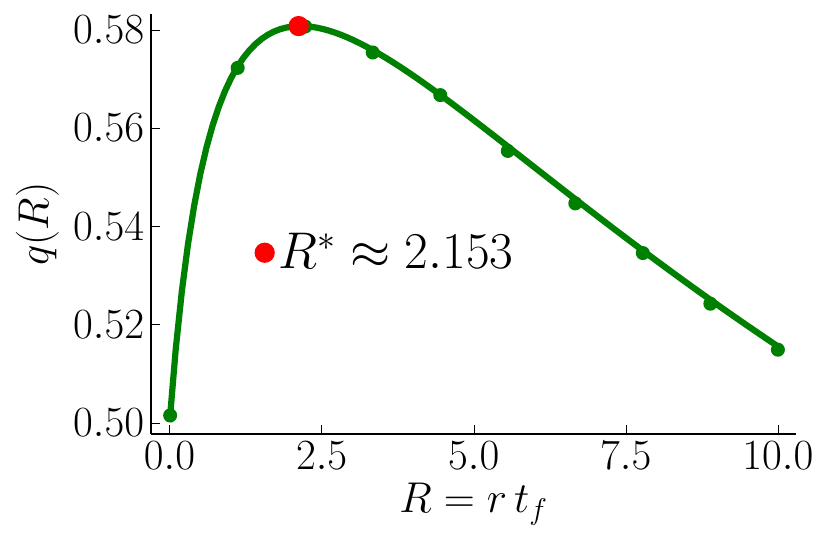}%
\caption{\textbf{Left panel:} The theoretical prediction (solid line) of the hitting probability in (\ref{eq:phits}) with $m=1$ (more explicit form given in \cite{BMSOpt}), compared with the one evaluated numerically from the effective Langevin equation (symbols) with $D=1$ and $t_f=1$. For a given $m$, it exhibits a unique maximum at $R=R^* \approx 1.671$ (red dot).  \textbf{Right panel:} The theoretical prediction (solid line) of the rescaled expected maximum $\langle M(t_f)\rangle=\sqrt{\pi D\,t_f}\, q(R=r t_f)$ given in \cite{BMSOpt} compared with the one evaluated numerically (symbols) with $D=1$ and $t_f=1$. The function $q(R)$ has a maximum at $R^*\approx 2.153$. In both panels, we sampled $10^6$ trajectories. }
\label{fig:probh}
\end{figure}

The origin of EFM can be qualitatively understood as follows. In the absence of resetting, 
the particle cannot go too far away from the origin, since it has to come back to the final position close to the origin at time $t_f$, by a slow diffusing process. However, when a small amount of
resetting rate $r$ is switched on, the particle can go further away from the origin since it can come back close to the origin at time $t=t_f$ by a ``last minute'' instantaneous resetting. 
Hence there is a subtle trade-off between the resetting and the bridge constraint. Clearly, this argument, which is different from the one in the free RBM, is rather general and is expected to hold in any dimension, as illustrated in \cite{BMSOpt}. Further details on this mechanism, as well as on effective Langevin equations to generate RBB trajectories in arbitrary dimensions, can be found in the paper whose abstract is given on p.~\pageref{chap:A12}.

 \begin{figure}
\begin{center}
 \fboxsep=10pt\relax\fboxrule=1pt\relax
 \fbox{
   \begin{minipage}{\textwidth}

\hspace{2em}
\begin{center}
\LARGE \bf Optimal Resetting Brownian Bridges
\end{center}

\hspace{2em}
\begin{center}
B. De Bruyne, S. N. Majumdar and G. Schehr,
Phys.~Rev.~Lett. {\bf 128}, 200603 (2022).
\end{center}

\hspace{2em}
\begin{center}
  {\bf Abstract:} 
\end{center}

We introduce a resetting Brownian bridge as a simple model to study search processes where the total search time $t_f$ is finite and the searcher returns to its starting point at $t_f$. This is simply a Brownian motion with a Poissonian resetting rate $r$ to the origin which is constrained to start and end at the origin at time $t_f$. We unveil a surprising general mechanism that enhances fluctuations of a Brownian bridge, by introducing a small amount of resetting. This is verified for different observables, such as the mean-square displacement, the hitting probability of a fixed target and the expected maximum. This mechanism, valid for a Brownian bridge in arbitrary dimensions, leads to a finite optimal resetting rate that minimizes the time to search a fixed target. The physical reason behind an optimal resetting rate in this case is entirely different from that of resetting Brownian motions without the bridge constraint. We also derive an exact effective Langevin equation that generates numerically the trajectories of a resetting Brownian bridge in all dimensions via a completely rejection-free algorithm.
\end{minipage}
   }
\end{center}
\captionsetup{labelformat=empty}
\caption{\textbf{Abstract of article \themycounter} : Optimal Resetting Brownian Bridges.}
\label{chap:A12}
\addtocounter{mycounter}{1}
\end{figure}

\clearpage
\section{Resetting in stochastic optimal control}
\label{sec:res}
In the previous section, we saw that the resetting rate can be tuned to optimize a given search observable. In this section, we discuss a slightly more general problem in which the resetting rate $r(x,t)$ is allowed to depend on the position $x$ of the system at time $t$. The scalar optimization problem now becomes a functional one. A convenient framework to address such a problem is optimal control theory, which we will introduce now. 

Since the seminal works of Pontryagin \cite{P87} and Bellman \cite{BK65}, optimal control theory has received renewed interest due to its applications in a wide range of contexts, such as artificial intelligence \cite{RN} and finance \cite{P09}. In a typical setting, optimal control considers a system whose state at time $t$ can be represented by a $d$-dimensional vector $\bm{x}(t)$. For instance, the state $\bm{x}(t)$ could correspond to the degrees of freedom of an autonomous robot or the asset values in a financial portfolio. The system typically evolves in time following a deterministic law, e.g., the laws of motion for mechanical systems or the law of supply and demand for financial markets. Oftentimes, the mathematical modeling of these laws is prohibitively expensive and one introduces a stochastic contribution to account for the missing information on the environment. Given the laws of motion, optimal control aims at operating the system in the best possible way by using an external control, e.g., actuators for robots or market orders in finance.

One of the simplest ways to describe analytically the evolution in time of the system $\bm{x}(t)$ is a first-order differential equation of the form $\dot{\bm{x}}(t)=\bm{f}(\bm{x},t)$. This law is often a simplified description of the system and a source of Gaussian white noise $\bm{\eta}(t)$ is introduced to capture the fluctuations around the deterministic model. In addition, the external control on the system is usually modeled as a drift $\bm{u}(\bm{x},t)$. Summing up these contributions, the full mathematical description of the system is given by
\begin{equation}
\dot{\bm{x}}(t)=\bm{f}(\bm{x},t)+\sqrt{2 D}~\bm{\eta}(t)+\bm{u}(\bm{x},t)\,,
\label{eq:langevin}
\end{equation}
where $\sqrt{2D}$ is the strength of the noise. The external control $\bm{u}(\bm{x},t)$ can be tuned to achieve a given goal, e.g., performing a task for a robot or generating profits in finance. Of course, controlling the system will generate operating costs, such as electrical consumption or transaction fees. Optimally controlling the system requires balancing a trade-off between high rewards, measured over time by a function $R(\bm{x},t)$, and low operating costs, often taken to be proportional to $\bm{u}^2(\bm{x},t)$. To be precise, for a system located at position $\bm{x}$ at time $t$, the reward in a small time interval $dt$ is $R(\bm{x},t)dt$ and the cost is $\bm{u}^2(\bm{x},t)dt/2$.

In principle, solving this optimization problem is intrinsically difficult due to the high dimensionality of the space of solutions. Remarkably, Bellman introduced a general way to solve this problem, known as dynamical programming, which consists in breaking down the optimization into simpler subproblems in a recursive manner such that the present action is taken to maximize the future outcome. In doing so, the key quantity to keep track of is the optimal payoff $J(\bm{x}, t)$, defined as the expected payoff for an optimally controlled system located at ${\bm x}$ at time $t$. Using this framework, one can show that the optimal control is simply given by $\bm{u^*}(\bm{x},t)=\nabla_{\bm{x}} J(\bm{x},t)$, driving the system towards the direction in which the payoff increases the most. The optimal payoff $J(\bm{x},t)$ satisfies the celebrated Hamilton-Jacobi-Bellman (HJB) equation \cite{S93}
\begin{equation}
-\partial_t J(\bm{x},t) =D \Delta_{\bm{x}} J(\bm{x},t)+\bm{f}(\bm{x},t)\cdot\nabla_{\bm{x}} J(\bm{x},t)+\frac12\left(\nabla_{\bm{x}} J(\bm{x},t)\right)^2+R(\bm{x},t)\,,
\label{HJB}
\end{equation}
where $\Delta_{\bm{x}}$ and $\nabla_{\bm{x}}$ are  respectively the Laplacian and the gradient operators. The quadratic term $(\nabla_{\bm{x}} J)^2$ renders this equation nonlinear and difficult to solve for arbitrary reward functions. Nevertheless, there exist few analytically solvable cases. For instance, in the case of $d\!=\!1$, where $f(x,t)\!=\!0$ and $R(x,t)\!=\!-\alpha(x-x_f)^2\delta(t-t_f)/2$, the optimal control has the simple form $u^*(x,t)\!=\!-\alpha(x-x_f)/[1+\alpha(t_f-t)]$, which, in the limit $\alpha\to\infty$, is reminiscent of the effective force to generate bridge Brownian motion \cite{MajumdarEff15}. This optimal control continuously drives the system to maximize the final reward by arriving close to the target $x_f$ at time $t_f$. In more realistic systems, one has to rely on numerical methods to solve \eqref{HJB} \cite{Kappen05}. 
 
 Ideas from optimal control have also been proven successful in different areas of physics \cite{SS07,BSG15,BSS21}. Moreover, stochastic optimal control has been applied to a variety of systems, such as supply-chain planning \cite{DIS19}, swarms of drones \cite{GTS16}, and fluctuating interfaces \cite{KNO10}. These systems all have in common that the optimal control can be orchestrated as a coordinated sequence of infinitesimal local changes. However, numerous systems do not fall in this class and require global changes to be optimally managed. Examples of such instances arise in the contexts of search processes, both in the cases of computer algorithms \cite{MZ02} and time-sensitive rescue missions \cite{S06}. In the latter situations, a common and widely observed strategy is to stop and resume operations from a predefined location.  Unfortunately, the HJB framework is not well suited to study such resetting protocols. Indeed, as we have seen in the previous chapter, resetting is known to be quite different from a local force and exhibits interesting features, including out-of-equilibrium dynamics \cite{MC16,EM16,MMS21}, dynamical phase transitions \cite{Kusmierz14,MSS15,CM15}, and nontrivial first-passage properties \cite{ES11,ES11b}. This observation naturally called into question the existence of an analytical framework to devise the optimal resetting control policy.

In our work, we combined the notions of stochastic resetting and optimal control into \emph{resetting optimal control}, which provides a natural framework to operate systems through stochastic restarts. Our goal  is not to provide an
accurate description of a specific system, but rather
to consider a minimal model to explore resetting in optimal control. To model resetting policies, we exchange the control force $\bm{u}(\bm{x},t)$ for a resetting rate $r(\bm{x},t)$. In a small time interval $dt$, the state of the system is reset to a predefined location $\bm{x}_{\text{res}}$ with probability $r({\bm x},t)dt$ and evolves freely otherwise. In sum, the dynamical system evolves according to
\begin{equation}
  \! {\bm x}(t\!+\!dt) \!= \! \left\{\begin{array}{ll}
  \!\!    \bm{x}_{\text{res}}\,,  & \text{prob.}\!=\!r({\bm x},t)dt\,,\\
   \!\!{\bm x}(t) \!+\!\bm{f}(\bm{x},t)dt\!+\!\sqrt{2D}{\bm\eta}(t)dt, \!\!\!\!  &\text{prob.}\!=\!1\!\!-\!r({\bm x},t)dt\,,
  \end{array}\right.\label{eq:eomrs}
\end{equation}
where the subscript ``res'' in $\bm{x}_{\text{res}}$ stands for ``resetting location''. Note that (\ref{eq:eomrs}) is a generalization of the resetting dynamics in (\ref{eq:eom}). As in the HJB framework, we aim at finding the optimal resetting rate $r(\bm{x},t)$, as a function of $\bm{x}$ and $t$, that balances the trade-off between high rewards, measured over time by the function $R(\bm{x},t)$, and low operating costs,  which depend on $r(\bm{x},t)$. To pose the optimization problem, we naturally extend the HJB paradigm and define the following payoff functional
\begin{equation}
\label{eq:F_resetting}
\!\!\!\mathcal{F}_{\bm{x_0},t}\left[r\right]\!=\!\left\langle\int_{t}^{t_f}\!\!d\tau\left[ R\left(\bm{x}(\tau),\tau\right)\!-\!c\!\left(\bm{x}(\tau),\tau\right)r\!\left(\bm{x}(\tau),\tau\right)\right]\right\rangle_{\!\bm{x_0}}\!\!,
\end{equation}
where $c(\bm{x},\tau)$ is the cost associated with resetting, $t_f$ is the time horizon up to which the system is controlled, and the symbol $\langle\ldots \rangle_{\bm{x_0}}$ indicates the average over all stochastic trajectories starting from $\bm{x_0}$ at time $t$ and evolving according to \eqref{eq:eom}. Note that the payoff $\mathcal{F}_{\bm{x_0},t}\left[r\right]$ is a functional and depends on the full function $r(\bm{x},t)$.

We find that the optimal resetting policy $r^*(\bm{x},t)$ that maximizes $\mathcal{F}_{\bm{x_0},t}\left[r\right]$ is \emph{bang-bang} resulting in an impulsive control strategy \cite{S93}: the system is reset with probability one if its state is outside of a time-dependent domain $\Omega(t)$ and evolves freely otherwise
\begin{equation}
  r^*\left({\bm x},t\right) dt =  \begin{cases}
  0 \,\,\,\,\,\,\text{ if } &\bm{x}\in \Omega(t)\,,\\
  1\,\,\,\,\,\,\text{ if } &\bm{x}\notin \Omega(t)\,.\\  
    \end{cases}\label{eq:rs}
\end{equation}
The domain $\Omega(t)$ evolves according to 
\begin{align}
\Aboxed{  \Omega(t) = \{{\bm x} : J({\bm x},t)\geq J(\bm{x}_{\text{res}},t) -c({\bm x},t) \}\,,}\label{eq:Omegad}
\end{align}
where the optimal payoff function $J(\bm{x},t)=\max_{r}\mathcal{F}_{\bm{x},t}\left[r\right]$ is the solution of the differential equation 
\begin{align}
\Aboxed{-\partial_t J(\bm{x},t)=D \Delta_{\bm{x}} J(\bm{x},t)+\bm{f}(\bm{x},t)\cdot\nabla_{\bm{x}}J(\bm{x},t)+R(\bm{x},t)\,,\,\,\,\,\,\,\bm{x}\in \Omega(t)\,.}
\label{HJB_resetting}
\end{align}
Note that \eqref{HJB_resetting} must be solved jointly with \eqref{eq:Omegad} starting from the final condition $J(\bm{x},t_f)=0$ and evolving backward in time with the Neumann boundary condition $\nabla_{\bm x} J({\bm x},t) \cdot {\bm n}({\bm x}) = 0$, where ${\bm n}({\bm x}) $ is the normal unit vector to the boundary. Outside of the domain $\Omega(t)$, the solution is given by $J({\bm x},t)= J(\bm{x}_{\text{res}},t) -c({\bm x},t)$. The definition of the domain $\Omega(t)$ in \eqref{eq:Omegad} has a clear interpretation: at any given time the optimal policy is to restart the system if its expected payoff is less than the one at the resetting location minus the cost incurred for a restart. The emerging framework outlined in \eqref{eq:rs}, \eqref{eq:Omegad}, and \eqref{HJB_resetting} is the main result of our work and provides a general method to optimally control stochastic systems through restarts. The derivation of this result is presented in \cite{DeBruyneMori21}.

Going from \eqref{eq:F_resetting} to \eqref{HJB_resetting}, we have reduced a functional optimization problem to a partial differential equation, which is often easier to solve. Note however that the mathematical problem in \eqref{eq:Omegad} and \eqref{HJB_resetting} is of a special kind as the evolution of the domain of definition $\Omega(t)$ is coupled to the solution $J(\bm{x},t)$ of the differential equation. This kind of equations belongs to a class known as Stefan problems \cite{Crankbook}, which often arise in the field of heat transfer, where one studies the evolution of an interface between two phases, e.g., ice and water on a freezing lake. In this context, one must solve the heat equation for the temperature profile with an interface between the water and ice phases, which moves according to the temperature gradient. The interface is therefore to be considered as an additional unknown function, which must be jointly solved with the differential equation. To draw an analogy with our case, the optimal payoff function $J(\bm{x},t)$ plays the role of the temperature and the boundary of the domain $\Omega(t)$ corresponds to the water-ice interface. Note however that the two Stefan problems have different boundary conditions. The domain $\Omega(t)$ can be obtained by solving the Stefan problem in \eqref{eq:Omegad} and \eqref{HJB_resetting} numerically. This can be achieved, for instance, by using an Euler explicit scheme with a space-time discretization and updating the domain $\Omega(t)$ at each time step according to \eqref{eq:Omegad}. 

The Stefan problem in \eqref{HJB_resetting} is the analog of the HJB equation \eqref{HJB} for a resetting control. Despite the moving boundary, \eqref{eq:Omegad} and \eqref{HJB_resetting} have a linear dependence in $J$. One might therefore wonder if exactly solvable models exist within this framework. Interestingly, we have found a time-independent formulation in the infinite time horizon limit $t_f\to\infty$ which allows for exact analytical solutions to be found. This is achieved by considering discounted rewards and costs of the form $R\left({\bm x},t\right) = e^{-\beta t}\mathcal{R}\left({\bm x}\right)$ and $c\left({\bm x},t\right) = e^{-\beta t}\mathcal{C}\left({\bm x}\right)$, where $\beta>0$ is the discount rate. Accordingly, we also consider the drift to be time-independent $\bm{f}(\bm{x},t)=\bm{f}(\bm{x})$. Discounted payoffs are common in the control theory literature \cite{FS06} and capture situations in which the effect of the payoff decays over a typical timescale $1/\beta$. Such an effect is for instance observed in financial contexts, where $\beta$ is related to interest rates and is used to compare future and present benefits. Using the ansatz $J\left({\bm x},t\right) = e^{-\beta t}\mathcal{J}\left({\bm x}\right)$, we find that \eqref{HJB_resetting} becomes a time-independent ordinary differential equation of the form
\begin{align}
\Aboxed{
 \beta \mathcal{J}\left({\bm x}\right)= D \Delta_{\bm{x}}\mathcal{J}\left({\bm x}\right) +\bm{f}\left({\bm x}\right)\cdot\nabla_{\bm{x}}\mathcal{J}\left({\bm x}\right) +\mathcal{R}\left({\bm x}\right)\,,\quad {\bm x}\in \Omega\,,}\label{eq:Jdiffd}
\end{align}
where the domain $\Omega = \{{\bm x} : \mathcal{J}({\bm x})\geq \mathcal{J}(\bm{x}_{\text{res}}) -\mathcal{C}({\bm x}) \}$ is also independent of time. This equation can be explicitly solved in the absence of external forces by choosing a quadratic reward $\mathcal{R}(x)=-\alpha x^2$ and a constant resetting cost $\mathcal{C}(x)=c$ to the origin $x_{\text{res}}=0$. Note that $\mathcal{R}(x)\leq0$ and is maximized at $x = 0$, rewarding the system for being close to the origin. Solving \eqref{eq:Jdiffd}, we obtain, for $\beta\!=\!D\!=\!1$, the exact expression for the optimal payoff
\begin{align}
\Aboxed{
\mathcal{J}(x) =\alpha\left[-2-x^2 + \frac{2u(v)\cosh(x)}{\sinh(u(v))}\right]
\,,\quad x\in \Omega\,,}
\label{eq:gyv}
\end{align}
where $u(v)$ is the boundary of the symmetric domain $\Omega$, i.e.,
\begin{equation}
\Omega=\left\{x: \,|x|<u\!\left(v\right)\right\}\,.
\end{equation}
The boundary $u(v)$ is the unique positive solution of the transcendental equation $v-u^2(v) +2u(v)\tanh \left(u(v)/2\right)=0$, where $v=c/\alpha$ is the cost-reward ratio. The optimal strategy thus corresponds to resetting the system if $|x|>u(v)$. When $v\ll 1$, the cost of resetting is much smaller than the reward, therefore the boundary is close to the origin $u(v) \sim \sqrt{2}\,(3v)^{1/4}$, allowing the state of the system to remain close to the optimal location $x=0$. On the other hand, when $v\gg 1$, the cost of resetting is much larger than the running cost and the boundary is set far away from the origin $  u(v) \sim \sqrt{v}+1$. Beyond the optimal resetting policy, our approach predicts the function $\mathcal{J}(x)$, measuring the expected payoff upon starting from $x$ and following the optimal strategy. In figure \ref{fig:gyv}, $\mathcal{J}(x)$ is shown for $|x|<u(v)$ and for various values of the cost-reward ratio $v$. As a function of $x$, $\mathcal{J}(x)$ has a symmetric bell shape centered around the origin, where the reward is maximal. As $|x|$ increases, $\mathcal{J}(x)$ decreases since the reward decreases and the resetting boundary comes closer.

\begin{figure}[t]
\centering
\includegraphics[scale=0.5]{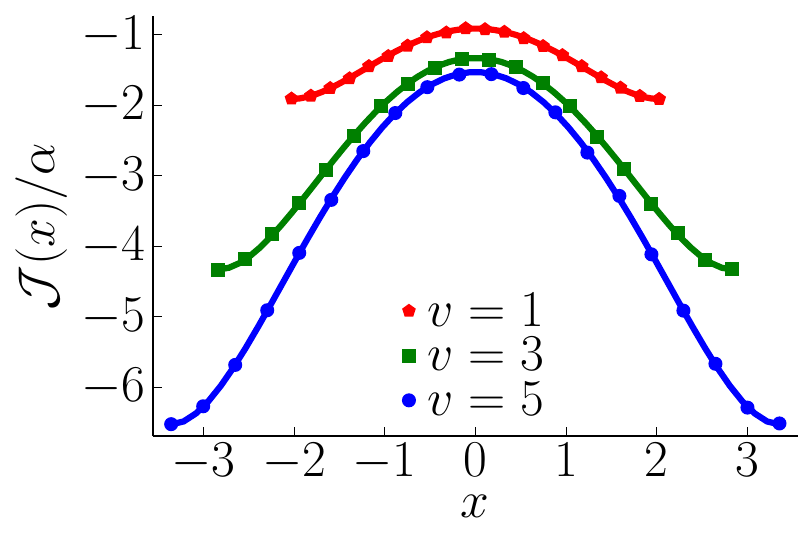} 

\caption{The optimal expected payoff $\mathcal{J}(x)$ for a one-dimensional random walk starting at $x$ for the discounted reward $\mathcal{R}(x)=-\alpha x^2 $ and cost $\mathcal{C}(x)=c$ and for different values of the cost-reward ratio $v\!=\!c/\alpha$. The continuous lines correspond to the exact result in \eqref{eq:gyv} while the symbols correspond to numerical simulations performed with $\beta=D=1$. The optimal strategy is to reset for $|x|>u(v)$, where $u(v)$ is defined in the text. For $|x|<u(v)$, the system is let free to evolve and its payoff depends continuously on $x$.
\label{fig:gyv}}
\end{figure}

\begin{figure}[t]
\centering
\includegraphics[scale=0.5]{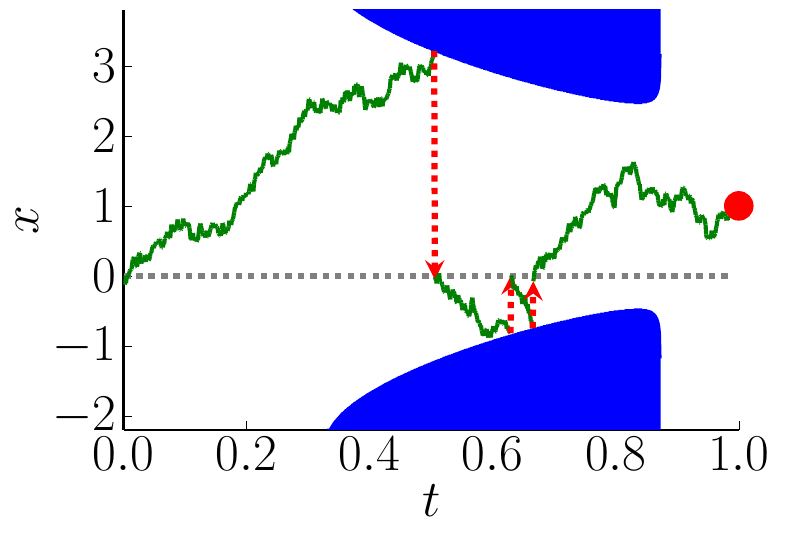}
\caption{Space-time illustration of the optimal resetting policy for a one-dimensional Brownian motion (green line) to reach the target location $x_f=1$ (filled circle) exactly at time $t_f=1$. A reward $\alpha$ is received upon reaching the location $x_f$ at the final time while a unit cost is incurred upon resetting (red dashed arrows) to the origin $x_{\text{res}}=0$ (grey dashed line). The optimal strategy that maximizes the expected payoff is to reset upon touching the blue region and to evolve freely in the white region, which we denote $\Omega(t)$ in the text. The domain $\Omega(t)$ is obtained by numerical integration of the Stefan problem in \eqref{eq:Omegad} and \eqref{HJB_resetting}, with $R(x,t)=\alpha\delta(t-1)\delta(x-1)$, $f(x,t)\!=\!0$, $\alpha=10$ and $D\!=\!1$.  The boundary of the domain $\Omega(t)$ guides the particle to the location $x_f$ at time $t_f$ while avoiding resetting as much as possible. The shape of $\Omega(t)$ depends nontrivially on the reward value $\alpha$. Further explanations for this shape are given in the text.
\label{fig:delta}}
\end{figure}

Our previous example focused on the case of an infinite time horizon, corresponding to $t_f\to \infty$. We now investigate the effect of a finite time horizon. One of the simplest settings where such an effect can be studied is the case of a one-dimensional Brownian motion with a Dirac delta final reward $R(x,t)=\alpha\,\delta(x-x_f)\delta(t-t_f)$, with $\alpha>0$, and a constant resetting cost $c(x,t)=c$. Such parameters correspond to rewarding the system by a finite amount $\alpha$ for arriving at the target position $x_f$ at time $t_f$, while penalizing it with a constant cost $c$ for each resetting event. Note that the system needs to arrive at the location $x_{f}$ exactly at time $t_f$ to get the reward while earlier visits do not provide any additional benefits. Before presenting the optimal policy for this problem, it is instructive to consider two limiting cases. For $\alpha\to 0$, one should never reset as the reward is not worth the resetting cost. On the other hand, for $\alpha\to\infty$, the cost of resetting becomes negligible and the optimal strategy is to reset if restarting would bring the system closer to $x_f$, independently of time. Interestingly, we observe that the crossover between these two regimes occurs as a sharp transition at the critical value $\alpha=\alpha_c$, where $\alpha_c=x_f c \sqrt{2\pi e}\approx 4.13273 ~x_f c$, which we predict analytically (see \cite{DeBruyneMori21}). For $\alpha<\alpha_c$, the optimal policy is to never reset, corresponding to $\Omega(t)=\mathbb{R}$ for all $0\leq t\leq t_f$. The situation is more subtle for $\alpha>\alpha_c$, where resetting is only favorable in a specific time window before $t_f$. To describe this window, it is convenient to introduce the backward time $\tau=t_f-t$. We find that no boundary is present for $\tau<\tau^*$, where $\tau^*$ is the smallest positive solution of the transcendental equation $\alpha e^{-x_f^2/(4D\tau^*)}=c\sqrt{4\pi D\tau^*}$.
At $\tau=\tau^*$, a boundary appears and one must resort to numerical integration techniques to find the solution for $\tau>\tau^*$ (see figure \ref{fig:delta}). We observe numerically that the boundary evolves with $\tau$, i.e., backward in time, in a non-monotonic way and eventually disappears. This optimal policy can be understood as follows. Close to $t_f$, where $\tau<\tau^*$, it is unlikely for the system to reach the target location $x_f$ from the origin in the remaining time. Thus, it is not convenient to reset. On the other hand, for very early times it is not yet necessary to reset since, after resetting, the system would typically evolve away from the target.

It is possible to extend our framework to other resetting dynamics and cost functions. The details of these extensions can be found in the paper whose abstract is given on p.~\pageref{chap:A11}.

\begin{figure}
\begin{center}
 \fboxsep=10pt\relax\fboxrule=1pt\relax
 \fbox{
   \begin{minipage}{\textwidth}
\hspace{2em}
\begin{center}
\LARGE \bf Resetting in Stochastic Optimal Control
\end{center}

\hspace{2em}
\begin{center}
B. De Bruyne and F. Mori,
Phys.~Rev.~Res. {\bf 5}, 013122  (2023).
\end{center}

\hspace{2em}
\begin{center}
  {\bf Abstract:} 
\end{center}

 ``When in a difficult situation, it is sometimes better to give up and start all over again''. While this empirical truth has been regularly observed in a wide range of circumstances, quantifying the effectiveness of such a heuristic strategy remains an open challenge. In this Letter, we combine the notions of optimal control and stochastic resetting to address this problem. The emerging analytical framework allows not only to measure the performance of a given restarting policy but also to obtain the optimal strategy for a wide class of dynamical systems. We apply our technique to a system with a final reward and show that the reward value must be larger than a critical threshold for resetting to be effective. Our approach, analogous to the celebrated Hamilton-Jacobi-Bellman paradigm, provides the basis for the investigation of realistic restarting strategies across disciplines. As an application, we show that the framework can be applied to an epidemic model to predict the optimal lockdown policy.
 
\end{minipage}
   }
\end{center}
\captionsetup{labelformat=empty}
\caption{\textbf{Abstract of article \themycounter} : Resetting in Stochastic Optimal Control.}
\label{chap:A11}
\addtocounter{mycounter}{1}
\end{figure}

 \clearpage
 
\section{Optimization in first-passage resetting}
\label{sec:fpr}
In this section, we study a simple optimization problem in which one would like to operate a system close to maximum performance without having too many breakdowns. To do so, we combine the notions of first passage and
resetting into \emph{first-passage resetting}, in which the particle is reset
whenever it reaches a specified threshold.  Contrary to standard resetting,
the time at which first-passage resetting occurs is defined by the motion of
the diffusing particle itself rather than being imposed
externally~\cite{ES11,ES11b,ESG20}. Feller
showed that such a process is well-defined mathematically and provided
existence and uniqueness theorems~\cite{feller1954diffusion}, while similar
ideas were pursued in~\cite{sherman1958limiting}.  First-passage resetting
was initially treated in the physics literature for the situation in which
two Brownian particles are biased toward each other and the particles are
reset to their initial positions when they encounter each
other~\cite{falcao2017interacting}. In addition, first-passage resetting shares some similarities with stress rearrangement mechanisms in elastoplastic models (such as the Hébraud–Lequeux model) \cite{HL98,ABM15,BGT16,EB17,DEJ17}.

We treat first-passage resetting on a finite interval, which has a
natural application to reliability theory.  Here the particle is restricted
to the interval $[0,L]$ where $x=L$ is again the boundary where resetting
occurs and the particle is immediately reinjected at $x=0$ when it reaches
$x=L$ (see figure \ref{fig:modelS}).  We may view this mechanism as characterizing the performance of a
driven mechanical
system~\cite{ross2014introduction,gnedenko2014mathematical,317656,867175},
with the coordinates $x=0$ and $x=L$ indicating poor and maximal performance,
respectively.  While one ideally wants to operate the system close to its
maximum performance level ($x=L$), there is a risk of overuse, leading to
breakdowns whenever $x=L$ is reached.  Subsequently, the system has to be
repaired and then restarted from $x=0$.  This dynamics corresponds to
resetting that is induced by a first passage to the boundary $x=L$.  While the dynamics of the
operating coordinate is typically complicated and dependent on multiple
parameters, we view the coordinate $x$ as undergoing a drift-diffusion
process for the sake of parsimonious modeling, so that it evolves inside the interval as 
\begin{align}
  \dot x(t) = \sqrt{2D}\eta(t) +v\,,\label{eq:eoms}
\end{align} 
where $D$ is the diffusion coefficient and $v$ is the drift.
 For the system to be close to
$x=L$, the drift should be positive.  On the other hand, breakdowns of the
system are to be avoided because a cost is incurred with each breakdown.
This suggests that the drift velocity should be negative.  The goal is to
determine the optimal operation that maximizes the performance of the system
as a function of the cost for each breakdown and the drift velocity.
Although the analogy between first-passage resetting and a mechanical system
is naive, this formulation allows us to determine the optimal operation in a
concrete way.  A preliminary account of some of these results was given
in~\cite{DeBruyne20a}.

\begin{figure}[ht]
  \center{ \includegraphics[width=0.6\textwidth]{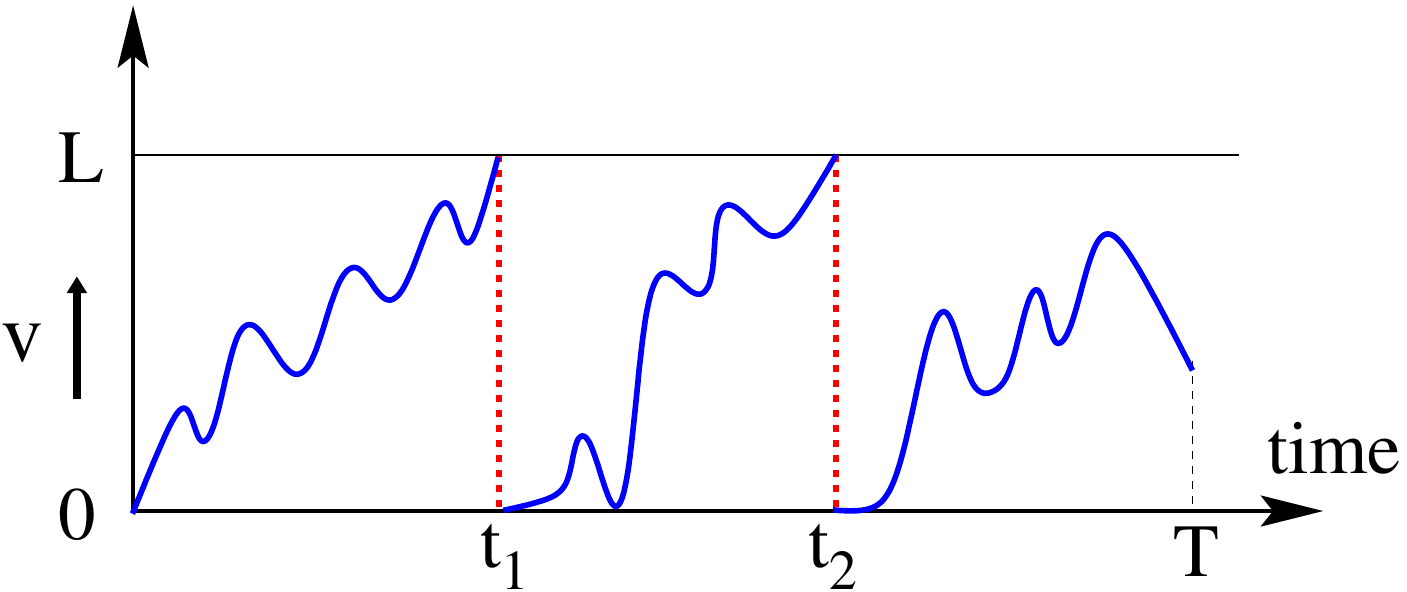}}
  \caption{Schematic illustration of first-passage resetting in the interval $[0,L]$ with a reflecting boundary at $0$ and a resetting boundary at $L$. Inside the interval, the particle undergoes drift-diffusion with velocity $v$ as in (\ref{eq:eoms}). Each time the particle reaches the
    threshold $L$, it is reset to the origin, which represents a breakdown (dashed red lines). In this illustration, the particle has been reset $\mathcal{N}(T)=2$ times. The times of the resetting
    events are denoted by $t_1, t_2,\ldots$.}
\label{fig:modelS}
\end{figure}

The basic control parameter is the magnitude of the drift velocity $v$.  If the
velocity is large and negative, the system is under-exploited because it
operates far from its maximum capacity.  Conversely, if the velocity is large
and positive, the system breaks down often.  We seek the optimal operation by
maximizing an objective function $\mathcal{F}(v)$ that rewards high performance
and penalizes breakdowns.  A natural choice for $\mathcal{F}(v)$ is
\begin{align}
\label{F}
\mathcal{F}(v) =\lim_{T\to\infty} \frac{1}{T}\left\langle  \left(\frac{1}{L}\int_0^T x(t)\,dt\right) - C\, \mathcal{N}(T)\right\rangle\,,
\end{align}
where $T$ is the total operation time, $\mathcal{N}(T)$ is the number
of breakdowns within a time $T$, and $C$ is the cost of each breakdown. The dependence on $v$ on the right-hand side lies in the average over all trajectories evolving according to (\ref{eq:eoms}) inside the interval with a reflecting boundary at $0$ and a resetting boundary at $L$ (see figure \ref{fig:modelS}).  As defined, this objective
function rewards operation close to the maximum point $L$ and penalizes
breakdowns.

We now compute explicitly the average in (\ref{F}). To do so, we use the linearity of the expectation value and find
\begin{align}
  \mathcal{F}(v) = \lim_{t\to\infty}\frac{\langle x(t)\rangle}{L} - C\,\lim_{T\to \infty}\frac{\langle \mathcal{N}(T)\rangle}{T}\,.\label{eq:F2}
\end{align}
The first term on the right-hand side in (\ref{eq:F2}) is the average position of the particle in its steady-state, which can be computed by standard Fokker-Plank techniques and is given by \cite{DeBruyne20a,DeBruyne20b}
\begin{align}
\label{eq:xavg}
\lim_{t\to\infty}\frac{\langle x(t)\rangle}{L} =
 \frac{\left(2\text{Pe}^2-2\text{Pe}+1\right)\,e^{2\text{Pe}}-1}{ 2\text{Pe}\left[(2\text{Pe}-1)\,e^{2\text{Pe}}+1\right]} \,,
\end{align}
where $\text{Pe}=vL/(2D)$ is the Péclet number (the
dimensionless bias velocity). The average number of resets $\langle \mathcal{N}(T)\rangle$ can be computed using a renewal approach, which is common in stochastic resetting computations \cite{ES11,ES11b}. We note that $\langle \mathcal{N}(T)\rangle$ satisfies the equation
\begin{align}
  \langle \mathcal{N}(T)\rangle = \int_{0}^T dt' F(L,t') [1+\langle \mathcal{N}(T-t')\rangle]\,,\label{eq:renw}
\end{align}
where $F(L,t')$ is the distribution of the first-passage time to $L$. The renewal equation (\ref{eq:renw}) states that to have an average of $\langle \mathcal{N}(T)\rangle$ resets at time $T$, there must have been a first reset at a time $t'$, and then the process starts anew and will have $\langle \mathcal{N}(T-t')\rangle$ average number of resets. The renewal equation (\ref{eq:renw}) can be solved by standard Laplace transform techniques, and gives 
\begin{align}
\langle\mathcal{N}(T)\rangle &\sim \frac{4\text{Pe}^2}{2\text{Pe}-1+ e^{-2\text{Pe}}}\,\,\frac{T}{L^2/D}\,,\quad T\to\infty\,.\label{eq:Ninf}
\end{align}
Inserting (\ref{eq:xavg}) and (\ref{eq:Ninf}) into the objective function gives
\begin{align}
\Aboxed{
  \mathcal{F}(v) = \frac{\left(2\text{Pe}^2-2\text{Pe}+1\right)\,e^{2\text{Pe}}-1}{ 2\text{Pe}\left[(2\text{Pe}-1)\,e^{2\text{Pe}}+1\right]} -  C' \,\frac{4\text{Pe}^2}{2\text{Pe}-1+ e^{-2\text{Pe}}}\,,}\label{eq:Fpe}
\end{align}
where $\text{Pe}=vL/(2D)$ is the dimensionless bias velocity, and $C'=CD/L^2$ is the dimensionless cost per breakdown. Representative plots of the objective function versus $\text{Pe}$
are shown in figure ~\ref{fig:F}.  For a given cost of a breakdown, there is
an optimal drift velocity or optimal P\'eclet number.  The higher this cost,
the smaller the optimal bias and the value of $\mathcal{F}(v)$.  Moreover, the
optimal bias is not necessarily negative. Indeed, if the cost of a breakdown
is relatively small, then it is advantageous to operate the system close to
its limit $L$ and absorb the (small) cost of many breakdowns.  On the
contrary, if the cost of a breakdown is high, it is better to run the system
at low level and with a negative bias to avoid breakdowns.

\begin{figure}[ht]
     \centering
   \includegraphics[width=0.475\textwidth]{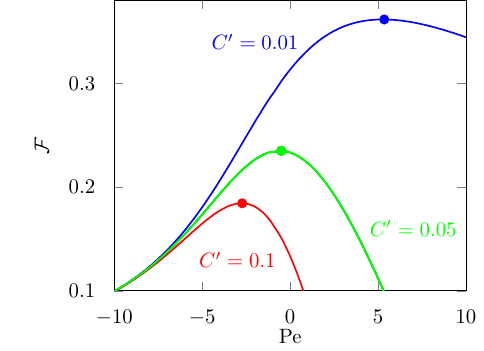}
\caption{The objective function (\ref{eq:Fpe}) versus the P\'eclet number $\text{Pe}=vL/(2D)$ for different
  normalized cost values $C'=C/(L^2/D)$.  Indicated on each curve is the optimal operating point. }
     \label{fig:F}
\end{figure}

Generalizations to other more realistic models, including delay as well as higher dimensions, can be found in the paper whose abstract is given on p.~\pageref{chap:A11}.

\begin{figure}
\begin{center}
 \fboxsep=10pt\relax\fboxrule=1pt\relax
 \fbox{
   \begin{minipage}{\textwidth}

\hspace{2em}
\begin{center}
\LARGE \bf Optimization and growth in first-passage resetting
\end{center}
\hspace{2em}
\begin{center}
B. De Bruyne, J. Randon-Furling and S. Redner, 
J.~Stat.~Mech., 013203 (2020).
\end{center}
\hspace{2em}
\begin{center}
  {\bf Abstract:} 
\end{center}
We combine the processes of resetting and first-passage to define \textit{first-
passage resetting}, where the resetting of a random walk to a fixed position is triggered
by a first-passage event of the walk itself. In an infinite domain, first-passage
resetting of isotropic diffusion is non-stationary, with the number of resetting events
growing with time as $\sqrt{t}$. We calculate the resulting spatial probability distribution
of the particle analytically, and also obtain this distribution by a geometric path
decomposition. In a finite interval, we define an optimization problem that is controlled
by first-passage resetting; this scenario is motivated by reliability theory. The goal is
to operate a system close to its maximum capacity without experiencing too many
breakdowns. However, when a breakdown occurs the system is reset to its minimal
operating point. We define and optimize an objective function that maximizes the
reward (being close to maximum operation) minus a penalty for each breakdown. We
also investigate extensions of this basic model to include delay after each reset and
to two dimensions. Finally, we study the growth dynamics of a domain in which the
domain boundary recedes by a specified amount whenever the diffusing particle reaches
the boundary after which a resetting event occurs. We determine the growth rate of
the domain for the semi-infinite line and the finite interval and find a wide range of
behaviors that depend on how much the recession occurs when the particle hits the
boundary.

\end{minipage}
   }
\end{center}
\captionsetup{labelformat=empty}
\caption{\textbf{Abstract of article \themycounter} : Optimization and growth in first-passage resetting.}
\label{chap:A1}
\addtocounter{mycounter}{1}
\end{figure}

\clearpage

\chapter{Conclusion}

In this thesis, we investigated the extreme value statistics of some class of stochastic processes. In Chapter \ref{chap:ext}, we obtained several exact analytical results for the statistics of various extreme observables. Let us summarize them and discuss perspectives for further research. In Section \ref{sec:ord}, we studied a one-dimensional discrete-time random walk model where the walker starts at the origin and its position is incremented at each step independently by adding a noise drawn from a symmetric and continuous distribution $f(\eta)$. This includes L\'evy flights, where $f(\eta) \sim |\eta|^{-1-\mu}$ has a power law tail, with the L\'evy index $0<\mu \leq 2$. 
We observe the walk up to $n$ steps and order the positions at different times (discarding the initial position $x_0=0$) and denote the ordered positions by $M_{1,n} > M_{2,n} > \ldots > M_{n,n}$. Our first interesting result was to show that in the limit $n \to \infty$, the distribution of the gap $\Delta_{k,n} = M_{k,n} - M_{k+1,n}$ approaches an $n$ independent limit, which still depends on $k$. In the stationary limit, we could also compute the distribution of the typical gap, with size $\Delta_k = O(k^{1/\mu-1})$, in the scaling limit for large $k$, for all $1\leq \mu\leq 2$. We unveiled the existence of an anomalous ``condensate part'' in the distribution that describes large atypical gaps of order $O(k^{1/\mu})$, which are much larger than the typical gaps of size $O(k^{1/\mu-1})$. An interesting question is to investigate the gap statistics when the gaps are deep inside the ``bulk'', i.e., when $k = \alpha n$, with $\alpha \in [0,1]$ fixed. Here, we have focused on the ``edge'' limit, where we took the limit where $n \to \infty$ but keeping $k$ fixed of order $O(1)$. The bulk limit would correspond to taking both $k$ and $n$ large simultaneously, with their ratio $\alpha = k/n$ fixed,  as it was done for the case of finite variance jump distributions in \cite{Lacroix}. 

In Section \ref{sec:exp}, we first obtained an explicit formula for the expected maximum of bridge discrete-time random walks of length $n$ with arbitrary jump distributions. This formula nicely extends an existing formula for free random walks. We then derived the asymptotic limit of the expected maximum for large $n$ up to second leading order and found a rich phase diagram depending on the jump distribution. In particular, we showed that, contrary to free random walks, bridge random walks with infinite first moment jump distributions with a Lévy index $\mu<1$ have a well-defined expected maximum. We have also demonstrated that the leading finite size correction displays a rich behavior depending on the power law tail of the jump distribution. Going beyond the expected value of the global maximum studied here, it would be interesting to investigate the full distribution of the global maximum as well as its order statistics in the limit of large but finite $n$. Furthermore, it would be interesting to generalize our results to bridge random walks in higher dimensions, where one would for instance measure the maximum of the radial extent of the walk, and study how the leading finite size corrections are affected.

In Section \ref{sec:con}, we studied the evolution of the maximum of a diffusive particle in confined environments in arbitrary dimensions. We first focused on the case of a particle confined in a $d$-dimensional ball of radius $R$. By relying on results on the NET, we showed that the behavior of the fluctuations of the maximum for $t\to \infty$ and close to $R$ exhibits a rich variety of behaviors depending on the dimension $d$. We then focused on the particular case of $d=2$ and applied our results to study the growth of the convex hull of Brownian motion in a disk with reflecting boundaries. Interestingly, we showed that it converges slowly to $2\pi R$ with a stretched exponential behavior.  It would be interesting to investigate further the extreme value statistics of Brownian motion in confined geometries. For instance, one could study the effect of confinement on the growth of the area of the convex hull. Another possible extension of this work would be to extend our results to the case of Lévy flights and study how the fluctuations of the maximum are affected.

In Chapter \ref{chap:con}, we introduced an efficient method to sample rare trajectories numerically. In Section \ref{sec:conD}, we studied discrete-time random walk bridges, where the random walk starts at the origin and is constrained to return to the origin after a fixed number $n$ of steps. We have derived an exact formula for an effective jump distribution, which accounts for the bridge constraint, and computed it explicitly for a few examples of bare jump distributions. Going beyond the simple bridges, we have further extended our method to other constrained discrete-time random walks, such as excursions and meanders. 
One interesting application of our method is in the context of extreme value statistics for constrained discrete-time random walks. For such walks, there have been a lot of interesting analytical results that have been derived recently for arbitrary jump distribution. Another example is the exact distribution of the maximal relative height of a one-dimensional discrete solid-on-solid model in the stationary state 
with periodic boundary condition \cite{SOS_Airy}.  In order to verify such analytical predictions numerically, one needs to generate efficiently the discrete-time bridge trajectories with the correct statistical weight. The method presented in this section will be useful for this purpose. Finally, it would be interesting to extend the sampling method to generate constrained trajectories in higher dimensions and particular geometries, such as two-dimensional cones \cite{KR10,KR12,BRS14,KR16,KR19}.

In Section \ref{sec:gen}, we extended our sampling method to continuous-time and discrete-time bridge random walks in the presence of a time-integrated constraint. In the case of continuous-time, we developed a new approach to derive the effective Langevin equation to generate Brownian bridges with fixed time-integrated constraints, such as the area, or the occupation time on the positive axis. We note that our approach has recently been generalized by Monthus \cite{Monthus} to provide a general framework for the so-called ``micro-canonical conditioning'' of Markov processes on time-additive observables, which led to further applications where the processes are conditioned with absorbing boundary conditions, first encounter times, killing rates, and local times \cite{Monthus2,Monthus3,Monthus4,Monthus5,Monthus6}. Finally, we studied run-and-tumble bridge trajectories, which is a prominent example of a non-Markovian constrained process. We provided an efficient way to generate them numerically by deriving an effective Langevin equation for the constrained dynamics. We showed that the tumbling rate of the RTP acquires a space-time dependency that naturally encodes the bridge constraint. We derived the exact expression of the effective tumbling rate and showed how it yields to an efficient sampling of run-and-tumble bridge trajectories.

In Section \ref{sec:app}, we applied our method to sample a one-dimensional Brownian particle conditioned to survive in the presence of periodically distributed point absorbers with a fixed trap intensity $\beta\,$. We showed that, in the long time limit, the effective diffusion coefficient of the surviving particles is a non-trivial function of the trapping strength $\beta$. Going beyond the asymptotic regime, it would be interesting to characterize the transient regime at intermediate times, and in particular the existence of extrema in the evolution of the mean-square displacement. 

In Chapter \ref{chap:sto}, we studied various optimization problems involving stochastic processes and extreme value statistics. In Section \ref{sec:opt}, we introduce resetting Brownian bridges as a search process. The main result of this work was the uncovering of an unexpected mechanism of enhanced fluctuations, caused by the combined effect of
the bridge condition and a small resetting rate. This mechanism is very general and holds in arbitrary dimensions. This enhanced fluctuation mechanism 
also leads to the existence of a resetting rate $r^*$ that optimizes the search process. This optimal paradigm holds in all dimensions but the mechanism for it
is different from that of a standard resetting Brownian motion. It would be interesting to investigate other types of constrained resetting Brownian motion and see if similar enhancing mechanisms are present.

In Section \ref{sec:res}, we combined optimal control and stochastic resetting to address the effectiveness of restarting policies. The emerging framework provided a unifying paradigm to obtain the optimal resetting strategy for a wide class of dynamical systems. It would be interesting to investigate extensions to optimal stopping problems and to study cost functions that are first-passage functionals \cite{majumdar2007brownian}, for instance where the time horizon is a first-passage time. This would be particularly relevant in the context of search processes.

In Section \ref{sec:fpr}, we defined an optimization problem that
describes, in a schematic way, aspects of the repeated breakdown of a driven
mechanical system.  The operation domain of the system is a finite interval.  The
resetting boundary corresponds to the system reaching its operating limit, after which a breakdown occurs and the system has to be
restarted from scratch.  The control parameter is the bias velocity, which may either drive the system
toward breakdown or toward minimal-level operation.  We showed that there
exists an optimal bias velocity that optimizes the performance of the system.
This optimum balances the gain by operating close to maximum performances while minimizing
the number of breakdowns. 

Some of the results presented in this thesis lead to new questions on which it would be interesting to work in the future. Our results on the order statistics in Chapter \ref{chap:ext} seem to suggest that there exists a stationary process describing the random walk close to its global maximum in the limit of a large number of steps. It would be interesting to unveil the existence of such process and to describe its statistical properties. Moreover, one could numerically study this process using the tools from Chapter \ref{chap:con} and in particular rely on the effective jump distribution (\ref{eq:inff}) of meander random walks in the limit of a large number of steps. 

Another research direction concerns the extension of the results presented in Section \ref{sec:app} to the transport properties of Brownian motion conditioned to survive in a random environment. It would be interesting to generalize our results to the case where the trap intensities or distances are themselves random variables.

 Finally, the analytical approach of effective random walks developed in Section \ref{sec:ord} has shown to be successful in describing the EVS of random walks and may be extended further to other stochastic processes and extreme observables.

\appendix
\chapter{Appendix}
\label{chap:app}
\section{Universal survival probability of a run-and-tumble
particle in an open linear half-space}
\label{app:convRTP}
In this appendix, we show how the generalization of Sparre Andersen theorem in (\ref{eq:kab}) can be used to extend some recent results on the universal survival probability of the run-and-tumble particle (RTP) in $d$ dimensions. In \cite{MoriL20,MoriE20}, it was shown that the survival probability that the $x$-component of the RTP does not change sign up to time $t$ is independent on the dimension $d$ for any finite time $t$, as a consequence of the Sparre Andersen theorem in (\ref{eq:SAres}). Moreover, it was shown that the universality holds for a large class of RTP models in which the speed $v$ of the particle after each tumbling is drawn from an arbitrary probability distribution $W(v)$. 

Here, we will compute the survival probability $S_{d}(t)$ that the particle remains in an open linear half-space up to time $t$ and show that it is independent of the speed distribution $W(v)$ but depends on the dimension $d$. In contrast to \cite{MoriL20,MoriE20}, we do not focus on the $x$-component of the particle, but instead, focus on the full trajectory of the particle and ask that it remains in an open linear half-space. Note that the half-space is not fixed, the trajectory just has to remain in any half-space for the particle to survive.

The RTP was introduced in its simplest version in one dimension $d=1$ in Section~\ref{sec:conRTP}. Here we consider a more general class of RTP in arbitrary $d$ dimensions in which the particle tumbles at a constant rate $\gamma$, after which it chooses a new direction uniformly at random and moves at a constant velocity $v$ drawn from a probability distribution $W(v)$ (see figure \ref{fig:convRTP}). We are interested in the survival probability $S_{d}(t)$ that the particle remains in an open linear half-space up to time $t$. 
\begin{figure}[t]
  \begin{center}
    \includegraphics[width=0.4\textwidth]{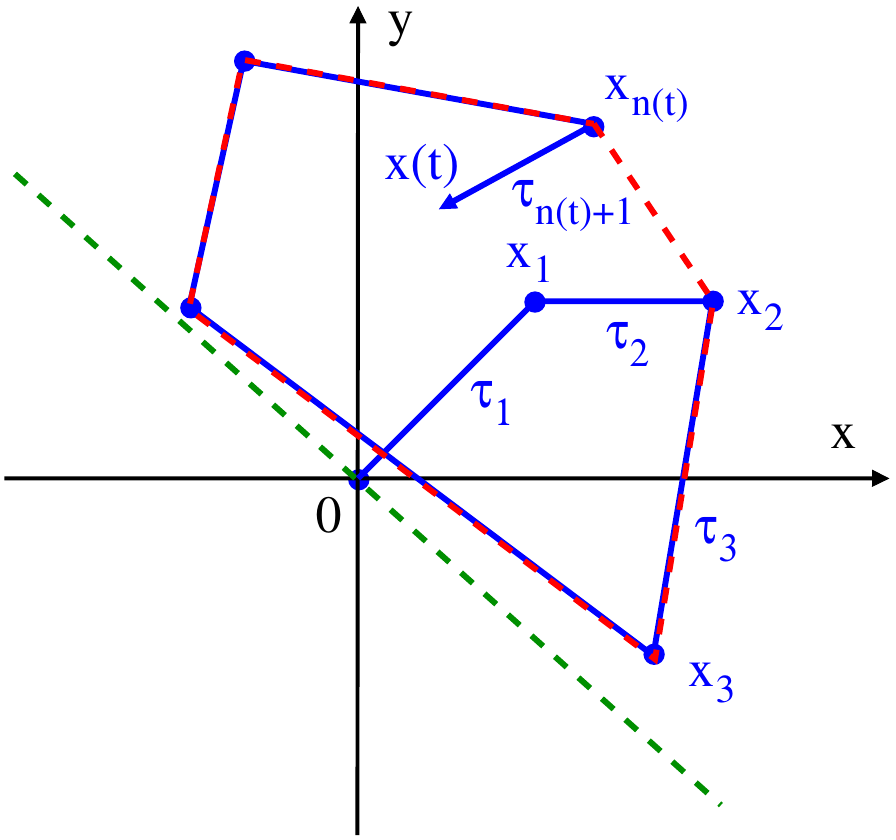}
    \caption{Illustration of a RTP trajectory in $d=2$ dimensions (blue curve). The RTP tumbles at a constant Poisson rate $\gamma$ and chooses a new direction uniformly at random and travels at a random constant velocity $v$. Up to time $t$, the particle will have tumbled $n(t)$ times at the tumbling locations $\bm{x}_1,\ldots,\bm{x}_{n(t)}$, and will be currently located at $\bm{x}(t)$. The running times $\tau_1,\ldots,\tau_{n(t)+1}$ are i.i.d.~random variables following a Poisson process with rate $\gamma$. The survival probability $S_d(t)$ is the probability that the trajectory remains in an open linear half-space. The illustrated particle is still alive as its trajectory is inside a linear half-space, for instance, the one defined by the green dashed line. This event is equivalent to the event that the origin is not included in the convex hull of the tumbling locations and the current position, $\text{Conv}\{\bm{x}_1,\ldots,\bm{x}_{n(t)},\bm{x}(t)\}$ (red dashed line).  }
    \label{fig:convRTP}
  \end{center}
\end{figure}
As shown in figure \ref{fig:convRTP}, this is equivalent to the probability that the origin is not included in the convex hull of the tumbling locations $\bm{x}_1,\ldots,\bm{x}_{n(t)}$ and the current position $\bm{x}(t)$:
\begin{align}
  S_d(t) = \text{Prob.}\left(\bm{0}\notin \{\bm{x}_1,\ldots,\bm{x}_{n(t)},\bm{x}(t)\}\right)\,,\label{eq:Sdta}
\end{align}
where $n(t)$ is the number of tumblings that occurred up to time $t$.
By relying on the generalization of Sparre Andersen theorem in (\ref{eq:kab}), we will show that $S_d(t)$ is independent of $W(v)$ and is given by
\begin{align}
\Aboxed{
  S_d(t)  =\sum_{n=1}^\infty \frac{\gamma ^{n-1} t^{n-1} e^{-\gamma t}}{2^{n-1} n!(n-1)!} \sum_{k=1}^{\lceil d/2 \rceil}B(n,d-2k+1)\,,}\label{eq:Sdt}
\end{align}
where $B(n,k)$ is defined in (\ref{eq:Bnk}). Moreover, we will show that the survival probability (\ref{eq:Sdt}) behaves, for $t\to \infty$, as
\begin{align}
  S_d(t) \sim \frac{1}{2^{d-2}(d-1)!\sqrt{\pi }}\,\frac{\ln(\gamma t)^{d-1}}{\sqrt{\gamma t}}\,,\quad t\to \infty\,.\label{eq:Sdtas}
\end{align}

We will now sketch the derivation of the result (\ref{eq:Sdt}). The derivation is similar to the one in \cite{MoriL20}. As the running times $\tau_1,\ldots,\tau_{n(t)+1}$ are i.i.d.~random variables following a Poisson process with rate $\gamma$ (see figure \ref{fig:convRTP}), their joint distribution is given by
\begin{align}
  P(\{\tau_i\},n\,|\,t) = \left[\prod_{i=1}^{n}\gamma e^{-\gamma \tau_i}\right] e^{-\gamma \tau_{n+1}} \delta\left(t-\sum_{i=1}^{n+1} \tau_i\right)\,,\label{eq:Ptaui}
\end{align}
where the terms in bracket correspond to the $n$ runs after which the particle has tumbled, and the term $e^{-\gamma \tau_{n+1}}=\int_{\tau_{n+1}}^\infty d\tau\gamma e^{-\gamma \tau}$ is the probability of no tumbling events during the last run. The joint distribution of the running times and the positions $\bm{x}_1,\ldots,\bm{x}_{n(t)},\bm{x}(t)$ can be obtained by using (\ref{eq:Ptaui}) and integrating over the speed distribution $W(v)$:
\begin{align}
  P(\{\bm{x}_i\},\{\tau_i\},n\,|\,t) = \gamma^n\int_0^\infty \left[\prod_{i=1}^{n+1}dv_i\, W(v_i) e^{-\gamma \tau_i}\,\frac{\delta(||\bm{x}_{i}-\bm{x}_{i-1}||-\tau_i \,v_i)}{\Omega_d (\tau_i\,v_i)^{d-1}}\right] \delta\left(t-\sum_{i=1}^{n+1} \tau_i\right)\,,\label{eq:Ptauxi}
\end{align}
where $\Omega_d=2\pi^{d/2}/\Gamma(d/2)$ is the surface area of the unit sphere in $d$ dimensions, and where $\bm{x}_0=\bm{0}$ and $\bm{x}_{n+1}=\bm{x}(t)$. The fraction in (\ref{eq:Ptauxi}) states that if the particle travels at a speed $v_i$ during a time $\tau_i$ in a direction chosen uniformly at random, the probability distribution of its new position $\bm{x}_{i+1}$ is a sphere of radius $v_i \tau_i$ centered on its previous position $\bm{x}_i$. The denominator is the normalization coefficient of this probability distribution. By integrating over the $\tau_i$'s in (\ref{eq:Ptauxi}), we obtain the joint distribution of the positions $\bm{x}_i$'s and the number $n$ of tumbling events
\begin{align}
 P(\{\bm{x}_i\},n\,|\,t) =   \int_0^\infty \left[\prod_{i=1}^{n+1}d\tau_i\right] P(\{\bm{x}_i\},\{\tau_i\},n\,|\,t) \,.\label{eq:inttaui}
\end{align}
Recognizing the convolution structure in the integration over the $\tau_i$'s in (\ref{eq:inttaui}), we take a Laplace transform of $ P(\{\bm{x}_i\},n\,|\,t)$ with respect to $t$ which gives
\begin{align}
   \int_0^\infty dt e^{-st}\,P(\{\bm{x}_i\},n\,|\,t) =\frac{1}{\gamma}\left(\frac{\gamma}{\gamma+s}\right)^{n+1} \prod_{i=1}^{n+1} \tilde p_{s}(\bm{x_i}-\bm{x_{i-1}})\,,\label{eq:pprop}
\end{align}
where we defined
\begin{align}
  p_s(\bm{\eta}) =  (\gamma+s)\int_0^\infty d\tau\, e^{-(\gamma+s) \tau}\,\int_0^\infty dv\, W(v)\,\frac{\delta(||\bm{\eta}||-\tau \,v)}{\Omega_d (\tau\,v)^{d-1}}\,.\label{eq:psr}
\end{align}
Note that we added a $(\gamma+s)$ prefactor in (\ref{eq:psr}) so that $p_s(\bm{\eta})$ can be interpreted as a probability distribution as it is normalized:
\begin{align}
\int_{\mathbb{R}^d} d\bm{\eta}\, p_s(\bm{\eta})=1\,.
\end{align}
The survival probability (\ref{eq:Sdta}) can now be obtained by integrating over all possible positions and summing over all number of tumbling events:
\begin{align}
S_d(t) = \sum_{n=1}^\infty \int_{\mathbb{R}^d} \left[\prod_{i=1}^{n+1} \bm{dx}_i\right]\mathrm{1}_{\bm{0}\notin \{\bm{x}_1,\ldots,\bm{x}_{n+1}\}}\,P(\{\bm{x}_i\},n\,|\,t)\,,\label{eq:Sdtp}
\end{align}
where $\mathrm{1}_{\bm{0}\notin \{\bm{x}_1,\ldots,\bm{x}_{n+1}\}}$ such that it is equal to unity if its argument is true, and zero otherwise.
Taking a Laplace transform of (\ref{eq:Sdtp}) and inserting (\ref{eq:pprop}) gives
\begin{align}
 \int_0^\infty dt e^{-st}\, S_d(t) = \sum_{n=0}^\infty \frac{1}{\gamma}\left(\frac{\gamma}{\gamma+s}\right)^{n} \int_{\mathbb{R}^d} \left[\prod_{i=1}^{n} \bm{dx}_i\right]\mathrm{1}_{\bm{0}\notin \{\bm{x}_1,\ldots,\bm{x}_{n}\}}\prod_{i=1}^{n} \tilde p_{s}(\bm{x_i}-\bm{x_{i-1}})\,,\label{eq:multint}
\end{align}
where we have shifted the summation index $n$ by one. We now interpret the multiple integrals in (\ref{eq:multint}) as the probability that the origin is not inside the convex hull of a discrete-time random walk of $n$ steps with a jump distribution $\tilde p_{s}(\bm{\eta})$ indexed by $s$, which also depends on $W(v)$ [see equation \ref{eq:psr}]. By the generalization of the Sparre Andersen theorem in (\ref{eq:kab}), this probability is universal, i.e.~it does not depend on $s$ or on $W(v)$, and is given by 
\begin{align}
  \int_{\mathbb{R}^d} \left[\prod_{i=1}^{n} \bm{dx}_i\right]\mathrm{1}_{\bm{0}\notin \{\bm{x}_1,\ldots,\bm{x}_{n}\}}\prod_{i=1}^{n} \tilde p_{s}(\bm{x_i}-\bm{x_{i-1}}) =  \frac{2}{2^n n!}\sum_{k=1}^{\lceil d/2 \rceil}B(n,d-2k+1)\,,\label{eq:kabrtp}
\end{align}
where $B(n,k)$ is given in (\ref{eq:Bnk}). Inserting (\ref{eq:kabrtp}) in (\ref{eq:multint}) gives
\begin{align}
  \int_0^\infty dt e^{-st}\, S_d(t) = \sum_{n=0}^\infty \frac{1}{\gamma}\left(\frac{\gamma}{\gamma+s}\right)^{n}\frac{2}{2^n n!}\sum_{k=1}^{\lceil d/2 \rceil}B(n,d-2k+1)\,.
\end{align}
By inverting this Laplace transform, we recover the result announced in (\ref{eq:Sdt}). By using the asymptotic expansion (\ref{eq:askab}), we obtain the asymptotic behavior (\ref{eq:Sdtas}).

\section{Derivation of the Pollaczek-Spitzer formula}
\label{app:PSf}
In this appendix, we sketch the derivation of the Pollaczek-Spitzer formula (\ref{eq:SAres2}) for a discrete-time random walk (\ref{eq:xm}) with a symmetric jump distribution $f(\eta)$. To do so, it is convenient to introduce the concave majorant of a random walk, which is the top part of its convex hull \cite{Steele02}. We denote by $y_1,\ldots, y_l$ and $k_1,\ldots,k_l$ the increments and the durations respectively of the $l$ faces of the concave majorant (see figure \ref{fig:concave_majoratn}).
\begin{figure}[h]
    \centering
    \includegraphics[width=0.7\textwidth]{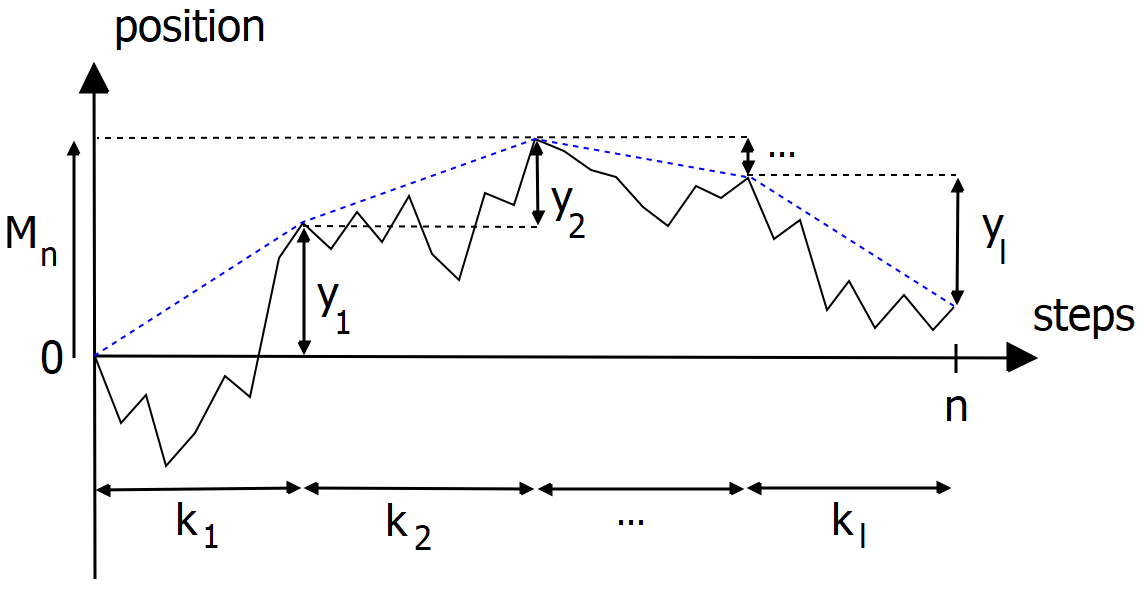}
    \caption{Concave majorant of a random walk of $n$ steps (blue dashed line). The number of faces of the concave majorant is denoted by $l$. The increments and the durations of the faces are respectively denoted $y_1,\ldots,y_l$ and $k_1,\ldots,k_l$, with $\sum_{i=1}^l k_i=n$.  The maximum of the random walk up to step $n$ is denoted by $M_n$.}
    \label{fig:concave_majoratn}
\end{figure}
 By definition of the concave majorant, one has that the sequence of slopes of the faces are decreasing : 
 \begin{align}
 \frac{y_1}{k_1} > \frac{y_2}{k_2}>\ldots>\frac{y_l}{k_l}\,.
 \end{align}
  Due to the Markovian property of the random walk, the joint distribution $p_{k_1,\ldots,k_l}(y_1,\ldots, y_l)$ of the increments, the durations, and the number of the faces is given by 
  \begin{align}
      p_{k_1,\ldots,k_l}(y_1,\ldots, y_l) = \prod_{i=1}^l q_{k_i}(y_i)\,, \quad \text{with}\quad \frac{y_1}{k_1} > \frac{y_2}{k_2}>\ldots>\frac{y_l}{k_l}\,,\label{eq:pklA}
  \end{align}
  where $q_k(y)$ is the probability that the random walk propagates from $0$ to $y$ in $k$ steps while remaining below the straight line between $(0,0)$ and $(k, y)$. By cyclic permutation symmetry, one can show that 
\begin{align}
    q_k(y) = \frac{1}{k}\, p_{k}(y)\,,\label{eq:qky}
\end{align}
where $p_k(y)$ is the free propagator of the random walk (\ref{eq:PI}), i.e.~the probability that it propagates from $0$ to $y$ in $k$ steps without any constraints. Inserting (\ref{eq:qky}) in (\ref{eq:pklA}) gives
\begin{align}
     p_{k_1,\ldots,k_l}(y_1,\ldots, y_l) = \prod_{i=1}^l\frac{1}{k_i}\, p_{k_i}(y_i)\,, \quad \text{with}\quad \frac{y_1}{k_1} > \frac{y_2}{k_2}>\ldots>\frac{y_l}{k_l}\,.\label{eq:pklB}
\end{align}
One can check that, upon summing over all possible number of faces $l$ and their durations and increments, this distribution is normalized, i.e.~
\begin{align}
\sum_{l=1}^n \sum_{k_1+\ldots k_l=n}\int_{\frac{y_1}{k_1}> \ldots >\frac{y_l}{k_l}} dy_1\ldots dy_l\,p_{k_1,\ldots,k_l}(y_1,\ldots, y_l) = 1\,.\label{eq:normpkl}
\end{align}
Using these variables, the maximum $M_n$ of the random walk is simply given by (see figure \ref{fig:concave_majoratn})
 \begin{align}
     M_n=\sum_{i=0}^l y_i\, \Theta(y_i)\,,
 \end{align}
  where $\Theta(x)$ is the Heaviside function such that $\Theta(x)=1$ if $x>0$ and $\Theta(x)=0$ otherwise.
The distribution of the maximum therefore reads
\begin{align}
\langle \delta(M_n - m)\rangle = \sum_{l=1}^n \sum_{k_1+\ldots k_l=n}\int_{\frac{y_1}{k_1}\geq \ldots \geq \frac{y_l}{k_l}} dy_1\ldots dy_l \,p_{k_1,\ldots,k_l}(y_1,\ldots, y_l) \,\delta\left(m - \sum_{i=1}^l y_i \,\Theta(y_i)\right)\,.\label{eq:distM}
\end{align}
Inserting (\ref{eq:pklA}) in (\ref{eq:distM}), changing coordinates $w_i=y_i/k_i$, and taking a Laplace transform gives
\begin{align}
    \langle e^{-sM_n}\rangle  = \sum_{l=1}^n  \sum_{k_1+\ldots k_l=n}  \int_{w_1\geq \ldots \geq w_l} dw_1\ldots dw_l \prod_{i=1}^l \left[ p_{k_i}(k_i w_i)e^{-s w_i k_i \Theta(w_i)}\right]\,.\label{eq:distMs}
\end{align}
Using the permutation symmetry of the integrand gives
\begin{align}
      \langle e^{-sM_n}\rangle  &= \sum_{l=1}^n  \frac{1}{l!}\sum_{k_1+\ldots k_l=n}  \prod_{i=1}^l\left[\int_{-\infty}^{\infty} dw\, p_{k_i}(k_i w)e^{-s w k_i \Theta(w)}\right]\,.\label{eq:distMs}
\end{align}
Taking a generating function with respect to $n$ gives
\begin{align}
   \sum_{n=0}^\infty z^n \langle e^{-sM_n}\rangle
      &= 1 +\sum_{l=1}^\infty   \frac{1}{l!} \left[\sum_{k=1}^\infty z^k\int_{-\infty}^{\infty} dw p_{k}(k w)e^{-s w k \Theta(w)}\right]^l\,,\\
&= \exp\left(\sum_{k=1}^\infty z^k\int_{-\infty}^{\infty} dw p_{k}(k w)e^{-s w k \Theta(w)}\right)\,.\,
\end{align}
where the additional term $1$ in the first line originates from the term $n=0$ in the left-hand side, and where we recognized the exponential function in the series to go to the second line.
Changing coordinates $y=wk$ and separating the integral over the negative and positive axis gives
\begin{align}
    \sum_{n=0}^\infty z^n \langle e^{-sM_n}\rangle
      &=\frac{1}{\sqrt{1-z}}\exp\left(\sum_{k=1}^\infty  \frac{z^k}{k}\int_{0}^{\infty} dy p_{k}(y)e^{-s y }\right)\,,\label{eq:genPS}
\end{align}
where we used that $\int_{-\infty}^0 dy\, p_k(y)=1/2$ and that  $\sum_{k=1}^\infty z^k/k = -\ln(1-z)$. Upon using the Fourier transform of the propagator (\ref{eq:PI}), the expression (\ref{eq:genPS}) becomes
\begin{align}
    \sum_{n=0}^\infty z^n \langle e^{-sM_n}\rangle
      &=\frac{1}{\sqrt{1-z}}\exp\left(\sum_{k=1}^\infty  \frac{z^k}{k} \int_{-\infty}^\infty \frac{dq}{2\pi} \frac{[\hat f(q)]^k}{s+iq} \right)\,,\nonumber\\
       &=\frac{1}{\sqrt{1-z}}\exp\left(- \int_{-\infty}^\infty \frac{dq}{2\pi} \frac{\ln[1-\hat f(q)]}{s+iq} \right)\,,\nonumber\\
       &=\frac{1}{\sqrt{1-z}}\exp\left(-s \int_{0}^\infty \frac{dq}{\pi} \frac{\ln[1-\hat f(q)]}{s^2+q^2} \right)\,,\label{eq:genPS2}
\end{align}
where we used again that $\sum_{k=1}^\infty z^k/k = -\ln(1-z)$ and separated again the integral over the positive and negative axis. By relying on the identity (\ref{eq:idmaxS}) between the maximum and the survival probability, we recover the Pollaczek-Spitzer formula (\ref{eq:SAres2}). Note that the derivation above is original and has
not been seen elsewhere up to our knowledge.

\chapter*{Résumé en français}
\label{chap:resfr}
\addcontentsline{toc}{section}{Résumé en français}

Un des faits remarquables de la Physique est que l'on peut tenter de trouver des lois simples pour décrire un phénomène naturel à une échelle donnée sans connaître les lois microscopiques régissant ses constituants à des échelles plus petites. Cela nous permet de faire progresser notre compréhension des systèmes à l'échelle macroscopique, et de réaliser des percées technologiques dans la société, sans avoir une compréhension complète des lois fondamentales qui régissent le monde dans lequel nous vivons. Ce beau concept est l'une des principales raisons grâce auxquelles de nombreuses lois en physique fonctionnent remarquablement bien pour décrire ce que nous observons. La nature universelle de ces lois est plutôt fascinante et est l'une des raisons pour lesquelles la Physique vaut la peine d'être étudiée.

Les raisons qui sous-tendent le concept d'universalité sont assez profondes et fascinent la communauté de la Physique depuis longtemps. Les travaux fondateurs de Kadanoff en 1966, puis poursuivis par Wilson en 1971, ont permis d'éclairer ce phénomène et de l'expliquer comme la conséquence d'un très grand nombre d'éléments en interaction \cite{KLP66,KGW71}. Leur découverte a constitué une avancée majeure dans notre compréhension des phases de la matière et des transitions entre elles, comme lorsque l'eau bout et se transforme en vapeur. Cette universalité émergente a été largement observée dans une variété de systèmes complexes et constitue une pierre angulaire de la physique statistique moderne. Anderson le résume très bien dans son article intitulé ``More is different'' \cite{PWA72}, où il montre que le comportement des grands systèmes ne peut pas être simplement extrapolé par les propriétés de ses constituants, et que des propriétés entièrement nouvelles apparaissent à chaque niveau de complexité. Il souligne particulièrement l'importance de la symétrie dans les lois de la nature.

Des systèmes vastes et complexes apparaissent dans une grande variété de domaines des sciences naturelles ainsi que des sciences appliquées. Bien qu'intrinsèquement différents les uns des autres, ils partagent en commun le fait d'être composés de nombreux constituants en interaction avec un nombre élevé de degrés de liberté. Bien qu'il s'agisse de systèmes déterministes à l'échelle microscopique, c'est-à-dire au niveau des parties en interaction, ils se comportent effectivement de manière stochastique au niveau agrégé. Extraire des informations pertinentes sur le comportement macroscopique d'un système à partir de ses propriétés microscopiques est l'une des tâches plutôt difficiles auxquelles le domaine de la physique statistique est consacré. Alors que le domaine était à l'origine axé sur les systèmes issus de la physique, il est maintenant devenu un domaine interdisciplinaire avec des applications allant de la biologie à la finance. Un exemple paradigmatique de systèmes complexes sont les marchés financiers qui sont devenus un domaine de recherche actif dans la communauté \cite{BJP18}.

Alors que les systèmes complexes se comportent généralement de manière typique, ils présentent parfois des comportements atypiques qui peuvent donner lieu à des événements extrêmes tels que des tremblements de terre, des inondations extrêmes et de grands incendies de forêt. Ces événements, qui sont omniprésents dans la nature, peuvent avoir des conséquences dévastatrices. Des questions naturelles que l'on peut se poser sont: (a) quelle est la grandeur du plus grand des éléments  ?~(b) quand se produit-il ?~(c) est-il isolé, ou y a-t-il beaucoup d'autres éléments de même grandeur ? Le domaine des statistiques des valeurs extrêmes (SVE) est consacré à l'étude de ces questions et a trouvé une grande variété d'applications allant des sciences de l'environnement \cite{gumbel,katz} à la finance \cite{embrecht,bouchaud_satya}. Les SVE jouent également un rôle clé en physique, notamment dans la description des systèmes désordonnés \cite{bm97,PLDCecile,Dahmen,sg}, des interfaces fluctuantes \cite{Shapir,GHPZ,Majumdar04Flu1,Satya_Airy2,SOS_Airy} et des matrices aléatoires \cite{ TW,SMS14} (pour une revue récente, voir \cite{SMS14r,Vivo15,reviewMPS}).
Alors que les SVE des variables aléatoires i.i.d.~a été étudiée en profondeur \cite{Arnold,Nagaraja}, on en sait beaucoup moins sur les SVE des variables aléatoires fortement corrélées, qui apparaissent souvent dans des contextes pratiques. Plusieurs modèles spécifiques de variables aléatoires corrélées ont été étudiés et ont montré des comportements très riches ainsi que des caractéristiques universelles \cite{Feller,dean_majumdar,PLDCecile,pld_carpentier,Satya_Airy2,gyorgyi,satya_yor,comtet_precise,schehr_rsrg,SM12,MMS13,MMS14,BM17,Lacroix,BSG21a,BSG21b,PT20,PT21,Mori1,Mori2,Mori3}.
En particulier, il a été montré que les marches aléatoires unidimensionnelles en temps discret constituent un terrain de jeu très utile pour étudier les SVE de variables aléatoires fortement corrélées \cite{SM12,MMS13,MMS17,MMS14,BM17,Lacroix,PT20,PT21,Feller,BSG23,Erdos46,Darling56,VVI94,Pollackzek,Wendel,Spitzer57}.

Lorsqu'il est trop difficile d'obtenir des résultats analytiques sur les SVE, une façon naturelle de procéder est de les étudier numériquement. Cependant, ce n'est pas une tâche facile car leurs occurrences sont rares par définition. Une question naturelle qui se pose alors est : ``Comment échantillonner des événements rares de manière efficace ?''. En général, les trajectoires rares sont importantes car elles capturent des informations sur le système qui ne peuvent pas être vues dans les trajectoires typiques où les observables se concentrent autour de leur moyenne. Par exemple, dans le contexte des verres, les trajectoires rares sont essentielles pour comprendre la dynamique lente de relaxation structurale proche de la transition vitreuse où les fluctuations sont importantes \cite{Gar2018}. Les méthodes numériques pour les échantillonner sont d'un intérêt primordial et plusieurs méthodes, comme celles de Monte-Carlo par chaines de Markov et l'échantillonnage d'importance, ont été développées pour les systèmes à l'équilibre et hors équilibre \cite{Metro,DLB20,AKH02,HAK11,APN18,BCDG2002,GKP2006,GKLT2011,KGGW2018,Causer21,Doob,Pitman,CT2013,Rose21,MajumdarEff15,BSG21ca,BSG21cb,BSG21cc,Das21,Oakes20}. 
.

Étroitement lié aux questions de valeurs extrêmes, le concept de \textit{premier passage} a été largement étudié à la fois en mathématiques \cite{aurzada2015persistence,SA} et en physique \cite{bray2013persistence,redner2001guide,benichou2011intermittent,majumdar2007brownian,majumdar2010universal,Hanggi90}. Ce concept fait référence au problème plutôt général de trouver le temps qu'il faut pour qu'un événement particulier se produise. Il joue un rôle crucial dans divers phénomènes tels que les réactions chimiques, les animaux à la recherche de nourriture, les actifs financiers atteignant un prix limite ou les rivières débordant de leur lit. Comme pour les SVE, cette observable est généralement assez difficile à calculer analytiquement. Cependant, lorsque cela est possible, il présente généralement des comportements riches. Par exemple, des comportements intéressants apparaissent déjà au niveau de l'un des processus stochastiques les plus simples, à savoir le mouvement Brownien unidimensionnel, pour lequel l'événement de premier passage est certain mais prendra en moyenne un temps infini pour se produire. Ce paradoxe apparent provient du fait que la distribution de probabilité du temps de premier passage est normalisée mais a une queue de loi de puissance telle que le premier moment est infini \cite{bray2013persistence,redner2001guide}. Une façon de modifier ce comportement est d'introduire une \textit{remise à zéro} dans la dynamique, dans laquelle le processus est réinitialisé à sa position de départ à un taux constant \cite{ES11,ES11b,ESG20}. Cela rend le temps de premier passage moyen fini et même minimisé à un taux de réinitialisation critique. Ce comportement a trouvé des applications naturelles dans l'optimisation de processus de recherche, où la recherche recommence si la cible est
introuvable dans un certain délai \cite{KG14,KG15,CM15,BBR16,EM16}. Plus généralement, la réinitialisation modifie le mouvement de manière fondamentale et a suscité de nombreuses travaux sur ses conséquences \cite{BS14,CS15,Reuveni15,MSS15,R16,PR17,B18,BCS19}.

Dans cette thèse, nous obtenons de nouveaux résultats analytiques sur les SVE des processus stochastiques, qui sont des modèles paradigmatiques de variables fortement corrélées. Dans certains cas, on obtient des résultats universels, qui ont le mérite de rester valables pour une large gamme de modèles. De plus, ils permettent une meilleure compréhension des caractéristiques pertinentes qui régissent les SVE et révèlent parfois des transitions de phase inattendues. De plus, nous fournissons de nouvelles méthodes pour échantillonner numériquement des trajectoires rares pour une large classe de processus stochastiques. Ces méthodes sont illustrées sur de nombreux exemples et se révèlent très efficaces en pratique. Enfin, nous présentons quelques applications des SVE dans certains problèmes d'optimisation stochastique. Tous les nouveaux résultats qui ont été publiés dans cette thèse sont entourés d'un encadré. La plupart de ces résultats étant analytiques, ils nécessitent parfois des calculs assez longs. Nous avons décidé de ne pas fournir tous les détails des dérivations dans cette thèse mais plutôt de fournir un aperçu et quelques perspectives sur les résultats. Nous nous référerons régulièrement aux articles publiés où se trouvent les calculs détaillés.

Cette thèse est organisée comme suit. Dans la suite du chapitre \ref{chap:int}, nous présentons une sélection des principaux résultats de cette thèse. Nous présentons quelques résultats analytiques sur les statistiques d'ordre des marches aléatoires en temps discret. Nous révélons leur comportement universel asymptotique dans la limite d'un grand nombre de pas. Ensuite, nous présentons d'autres résultats analytiques sur le maximum moyen de marches aléatoires avec une contrainte dite de ``pont''. Nous discutons sa limite asymptotique, ainsi que sa correction de taille finie, qui présente des caractéristiques riches en fonction de la queue de la distribution des sauts. Enfin, nous introduisons une méthode efficace pour générer des marches aléatoires avec cette contrainte, et discutons des généralisations à d'autres types de trajectoires rares.

Dans le chapitre \ref{chap:ext}, nous nous concentrons sur les SVE de processus stochastiques. Dans la section \ref{sec:int}, nous fournissons une introduction aux résultats classiques à propos des SVE. Dans la section \ref{sec:ord}, nous étudions les statistiques d'ordre des marches aléatoires en temps discret et esquissons la dérivation de leur comportement universel dans la limite d'un grand nombre d'étapes. Dans la section \ref{sec:exp}, nous étudions le maximum moyen de marches aléatoires en temps discret de pont et discutons son comportement asymptotique riche. Dans la section \ref{sec:con}, nous nous tournons vers les processus en temps continu et dérivons la distribution de la longueur de l'enveloppe convexe du mouvement Brownien dans un domaine confiné. Les résultats de ce chapitre ont donné lieu à plusieurs publications dont les résumés se trouvent p.~\pageref{chap:A14}, p.~\pageref{chap:A5} et p.~\pageref{chap:A10}.

Dans le chapitre \ref{chap:con}, nous nous intéressons à l'échantillonnage numérique des trajectoires rares. Dans la section \ref{sec:conD}, nous introduisons une méthode efficace pour échantillonner des marches aléatoires en temps discret. Nous illustrons notre méthode et l'appliquons à différents exemples. Dans la section \ref{sec:gen}, nous généralisons notre méthode à d'autres types de trajectoires rares et l'étendons à d'autres types de processus stochastiques, à la fois markoviens et non markoviens. Dans la section \ref{sec:app}, nous appliquons notre méthode à l'échantillonnage de particules diffusives en présence d'un environnement de piégeage périodique. Nous discutons brièvement des propriétés de transport effectives des particules survivantes. Les résultats de ce chapitre ont donné lieu à plusieurs publications dont les résumés se trouvent p.~\pageref{chap:A4}, p.~\pageref{chap:A9}, p.~\pageref{chap:A6} et p.~\pageref{chap:A15}.

Dans le chapitre \ref{chap:sto}, nous discutons de plusieurs problèmes d'optimisation dans les processus stochastiques. Dans la section \ref{sec:opt}, nous introduisons un pont Brownien à réinitialisation comme modèle simple pour étudier les processus de recherche en présence d'une contrainte de pont. Nous mettons en évidence un mécanisme surprenant induit par la réinitialisation qui amplifie les fluctuations du processus. Dans la section \ref{sec:res}, nous combinons la notion de réinitialisation et de contrôle optimal dans un cadre analytique, analogue au paradigme Hamilton-Jacobi-Bellman, pour contrôler de manière optimale les systèmes dynamiques soumis à une politique de réinitialisation. Nous illustrons notre méthode sur différents exemples. Dans la section \ref{sec:fpr}, nous étudions la diffusion classique avec la fonctionnalité supplémentaire qu'une particule diffusante est réinitialisée à son
point de départ chaque fois que la particule atteint un seuil spécifié. Nous définissons et résolvons une optimisation non triviale dans laquelle un coût est encouru chaque fois que la particule est réinitialisée et une
récompense est obtenue lorsque la particule reste près du point de réinitialisation. Les résultats de ce chapitre ont donné lieu à plusieurs publications dont les résumés se trouvent p.~\pageref{chap:A12}, p.~\pageref{chap:A11} et p.~\pageref{chap:A1}.

En raison du grand nombre d'articles écrits au cours de cette thèse, il n'aurait pas été raisonnable de les présenter tous. Il fallait faire un choix et certains articles ont été laissés de côté en annexe. Nous les mentionnons brièvement ici et ne les aborderons pas davantage dans cette thèse.

\begin{itemize}
   \item L'article intitulé ``Survival probability of a run-and-tumble particle in the presence of a drift'' traite de la probabilité de survie d'une marche aléatoire persistante avec une distribution de vitesse arbitraire, pas nécessairement symétrique, en présence d'une frontière absorbante. Nous obtenons une formule générale, que nous appliquons au cas d'une particule à deux états de vitesse $\pm v_0$ en présence d'une force constante, et obtenons des comportements riches avec trois phases distinctes selon l'intensité de la force. Le résumé se trouve en p.~\pageref{chap:A2}.
  
   \item L'article intitulé ``Survival probability of random walks leaping over
traps'' étudie la probabilité de survie d'une marche aléatoire en présence de pièges de taille finie par-dessus lesquels elle peut sauter. Nous montrons que le taux de décroissance de la probabilité de survie dépend non trivialement de la taille du piège. Nous généralisons le modèle aux marches aléatoires avec des temps d'attente distribués en loi de puissance et dérivons certaines limites diffusives du modèle. Le résumé se trouve en p.~\pageref{chap:A8}.
  
   \item L'article intitulé ``A Tale of Two (and More) Altruists'' présente un modèle dynamique minimaliste d'évolution et de partage des richesses entre $N$ agents. Nous comparons les effets d'une politique altruiste par rapport à une politique individualiste. Nous montrons que la meilleure politique dépend du critère choisi. Alors que l'altruisme conduit à une richesse médiane plus globale, les individualistes qui vivent le plus longtemps accumulent la majeure partie de la richesse et vivent plus longtemps que les altruistes. Le résumé se trouve en p.~\pageref{chap:A7}.
  
    \item L'article intitulé ``First-Passage-Driven Boundary Recession'' étudie un problème de frontière mobile pour une particule Brownienne sur la
ligne semi-infinie dans laquelle la frontière se déplace d'une distance proportionnelle au temps entre les collisions successives de la particule et de la frontière. Nous constatons que la queue de la distribution du $n$-ième temps de frappe devient progressivement plus épaisse à mesure que $n$ augmente. De plus, nous constatons une double croissance logarithmique lente du nombre de rencontre avec la frontière. Le résumé se trouve en p.~\pageref{chap:A13}.
  
   \item L'article intitulé ``Wigner function for noninteracting fermions in hard wall potentials'' traite des fluctuations quantiques dans l'espace des phases de $N$ fermions sans interaction dans une boîte de dimension $d$. Nous obtenons des fonctions d'échelle de ces fluctuations proches de la ``Fermi surf'' dans la limite de $N\to \infty$ et montrons qu'elles sont universelles par rapport à la dimension $d$ de la boîte. Le résumé se trouve en p.~\pageref{chap:A3}.
\end{itemize}

{\let\clearpage\relax \chapter*{Funding acknowledgments}
\addcontentsline{toc}{section}{Funding acknowledgments}
This work was partially supported by the Luxembourg National Research Fund (FNR) (App. ID 14548297).}

{\chapter*{Abstracts of articles not discussed in this thesis}
\addcontentsline{toc}{section}{Abstracts of articles not discussed in this thesis}
\begin{figure}[!htb]
\begin{center}
 \fboxsep=10pt\relax\fboxrule=1pt\relax
 \fbox{
   \begin{minipage}{\textwidth}
\hspace{-1em}

\begin{center}
\LARGE \bf Survival probability of a run-and-tumble particle in the presence of a drift
\end{center}

\hspace{-1em}

\begin{center}
B. De Bruyne, S. N. Majumdar and G. Schehr, 
J.~Stat.~Mech., 043211 (2021).
\end{center}

\hspace{-1em}

\begin{center}
  {\bf Abstract:} 
\end{center}

We consider a one-dimensional run-and-tumble particle, or persistent random walk, in the presence of 
an absorbing boundary located at the origin. After each tumbling event, which occurs at a constant rate $\gamma$, the (new) velocity 
of the particle is drawn randomly from a distribution $W(v)$. We study the survival probability $S(x,t)$ of a particle starting from $x \geq 0$ up to time $t$
and obtain an explicit expression for its double Laplace transform (with respect to both $x$ and $t$) for an \textit{arbitrary} velocity distribution $W(v)$, not necessarily symmetric. This result is obtained as a consequence of Spitzer's formula, which is well-known in the theory of random walks and can be viewed as a generalization of  the Sparre Andersen theorem. We then apply this general result to the specific case of 
a two-state particle with velocity $\pm v_0$, the so-called persistent random walk (PRW), and in the presence of a constant drift $\mu$ and obtain an explicit expression for $S(x,t)$, for which we present more detailed results. Depending on the drift $\mu$, 
we find a rich variety of behaviours for $S(x,t)$, leading to three distinct cases: (i) \textit{subcritical} drift $-v_0\!<\!\mu\!<\! v_0$, (ii) \textit{supercritical} drift $\mu < -v_0$ and (iii) \textit{critical} drift $\mu=-v_0$. In these three cases, we obtain exact analytical expressions for the survival probability $S(x,t)$ and establish connections with existing formulae in the mathematics literature. Finally, we discuss some applications of these results to record statistics and to the statistics of last-passage~times. 

\end{minipage}
   }
\end{center}
\captionsetup{labelformat=empty}
\caption{\textbf{Abstract of article \themycounter} : Survival probability of a run-and-tumble particle in the presence of a drift.}
\label{chap:A2}
\addtocounter{mycounter}{1}
\end{figure}

\begin{figure}
\begin{center}
 \fboxsep=10pt\relax\fboxrule=1pt\relax
 \fbox{
   \begin{minipage}{\textwidth}

\hspace{2em}
\begin{center}
\LARGE \bf Survival probability of random walks leaping over
traps
\end{center}

\hspace{2em}
\begin{center}
G. Pozzoli and B. De Bruyne,
J.~Stat.~Mech.,  123203 (2021).
\end{center}

\hspace{2em}
\begin{center}
  {\bf Abstract:} 
\end{center}

We consider one-dimensional discrete-time random walks (RWs) in the presence of finite-size traps of length $\ell$ over which the RWs can jump. We study the survival probability of such RWs when the traps are periodically distributed and separated by a distance $L$. We obtain exact results for the mean first-passage time and the survival probability in the special case of a double-sided exponential jump distribution. While such RWs typically survive longer than if they could not leap over traps, their survival probability still decreases exponentially with the number of steps. The decay rate of the survival probability depends in a non-trivial way on the trap length $\ell$ and exhibits an interesting regime when $\ell\rightarrow 0$ as it tends to the ratio $\ell/L$, which is reminiscent of strongly chaotic deterministic systems. We generalize our model to continuous-time RWs, where we introduce a power-law distributed waiting time before each jump. In this case, we find that the survival probability decays algebraically with an exponent that is independent of the trap length. Finally, we derive the diffusive limit of our model and show that, depending on the chosen scaling, we obtain either diffusion with uniform absorption, or diffusion with periodically distributed point absorbers.

\end{minipage}
   }
\end{center}
\captionsetup{labelformat=empty}
\caption{\textbf{Abstract of article \themycounter} : Survival probability of random walks leaping over
traps.}
\label{chap:A8}
\addtocounter{mycounter}{1}
\end{figure}

\begin{figure}
\begin{center}
 \fboxsep=10pt\relax\fboxrule=1pt\relax
 \fbox{
   \begin{minipage}{\textwidth}

\hspace{2em}
\begin{center}
\LARGE \bf A Tale of Two (and More) Altruists
\end{center}

\hspace{2em}
\begin{center}
B. De Bruyne, J. Randon-Furling and S. Redner,
J.~Stat.~Mech.,  103405 (2021).
\end{center}

\hspace{2em}
\begin{center}
  {\bf Abstract:} 
\end{center}

We introduce a minimalist dynamical model of wealth evolution and wealth
sharing among $N$ agents as a platform to compare the relative merits of altruism
and individualism. In our model, the wealth of each agent independently evolves by
diffusion. For a population of altruists, whenever any agent reaches zero wealth (that
is, the agent goes bankrupt), the remaining wealth of the other $N-1$ agents is equally
shared among all. The population is collectively defined to be bankrupt when its total
wealth falls below a specified small threshold value. For individualists, each time an
agent goes bankrupt (s)he is considered to be “dead” and no wealth redistribution
occurs. We determine the evolution of wealth in these two societies. Altruism leads
to more global median wealth at early times; eventually, however, the longest-lived
individualists accumulate most of the wealth and are richer and more long lived than
the altruists.

\end{minipage}
   }
\end{center}
\captionsetup{labelformat=empty}
\caption{\textbf{Abstract of article \themycounter} : A Tale of Two (and More) Altruists.}
\label{chap:A7}
\addtocounter{mycounter}{1}
\end{figure}

\begin{figure}
\begin{center}
 \fboxsep=10pt\relax\fboxrule=1pt\relax
 \fbox{
   \begin{minipage}{\textwidth}

\hspace{2em}
\begin{center}
\LARGE \bf First-Passage-Driven Boundary Recession
\end{center}

\hspace{2em}
\begin{center}
B. De Bruyne, J. Randon-Furling and S. Redner,
J.~Phys.~A:~Math.~Theor. {\bf 55}, 354002 (2022).
\end{center}

\hspace{2em}
\begin{center}
  {\bf Abstract:} 
\end{center}

We investigate a moving boundary problem for a Brownian particle on the
semi-infinite line in which the boundary moves by a distance proportional to the time
between successive collisions of the particle and the boundary. Phenomenologically
rich dynamics arises. In particular, the probability for the particle to first reach the
moving boundary for the n$^{\text{th}}$ time asymptotically scales as $t^{-(1+2^{-n})}$. Because the tail
of this distribution becomes progressively fatter, the typical time between successive
first passages systematically gets longer. We also find that the number of collisions
between the particle and the boundary scales as $\ln \ln t$, while the time dependence of
the boundary position varies as $t/ \ln t$.

\end{minipage}
   }
\end{center}
\captionsetup{labelformat=empty}
\caption{\textbf{Abstract of article \themycounter} : First-Passage-Driven Boundary Recession.}
\label{chap:A13}
\addtocounter{mycounter}{1}
\end{figure}

\begin{figure}
\begin{center}
 \fboxsep=10pt\relax\fboxrule=1pt\relax
 \fbox{
   \begin{minipage}{\textwidth}

\hspace{2em}
\begin{center}
\LARGE \bf Wigner function for noninteracting fermions in hard wall potentials
\end{center}

\hspace{2em}
\begin{center}
B. De Bruyne, D. S. Dean, P. Le Doussal, S. N. Majumdar and G. Schehr,
Phys.~Rev.~A. {\bf 104},  013314 (2021).
\end{center}

\hspace{2em}
\begin{center}
  {\bf Abstract:} 
\end{center}

The Wigner function $W_N({\bf x}, {\bf p})$ is a useful quantity to characterize the quantum fluctuations of an $N$-body system in its phase space. 
Here we study $W_N({\bf x}, {\bf p})$ for $N$ noninteracting spinless fermions in a $d$-dimensional spherical hard box of radius $R$ at temperature $T=0$. 
In the large $N$ limit, the local density approximation (LDA) predicts that $W_N({\bf x}, {\bf p}) \approx 1/(2 \pi \hbar)^d$ inside a finite region of the $({\bf x}, {\bf p})$ plane, namely 
for $|{\bf x}| < R$ and $|{\bf p}| < k_F$ where $k_F$ is the Fermi momentum, while $W_N({\bf x}, {\bf p})$ vanishes outside this region, or ``droplet'', on a scale determined by quantum fluctuations. In this paper we investigate systematically, in this quantum region, the structure of the Wigner function along the edge of this droplet, called the Fermi surf. In one dimension, we find that there are three distinct edge regions along the Fermi surf and we compute exactly the associated nontrivial scaling functions in each regime. We also study the momentum distribution $\hat \rho_N(p)$ and find a striking algebraic tail for very large momenta $\hat \rho_N(p) \propto 1/p^4$, well beyond $k_F$, reminiscent of a similar tail found in interacting quantum systems (discussed in the context of Tan's relation). We then generalize these results to higher $d$ and find, remarkably, that the scaling function close to the edge of the box is universal, i.e., independent of the dimension~$d$.

\end{minipage}
   }
\end{center}
\captionsetup{labelformat=empty}
\caption{\textbf{Abstract of article \themycounter} : Wigner function for noninteracting fermions in hard wall potentials.}
\label{chap:A3}
\addtocounter{mycounter}{1}
\end{figure}
}

\end{document}